\documentclass[11pt]{article}
\usepackage{color,amsmath,amsfonts,amssymb}
\usepackage{gbmacros}
\usepackage{times,mathpazo,mathrsfs,charter,graphicx,enumitem,cite,soul}
\usepackage[colorlinks=true,linkcolor=blue,citecolor=red,bookmarksnumbered=true]{hyperref}

\def\maketitle{\par\noindent{\LARGE\bf\sffamily\thetitle}\\[1.4ex]
{\large\theauthor}\\[0.6ex]
\textit{\thetextinfo}\\[0.2ex]
{\small\today}\par\vglue1.4\bigskipamount}
\def\title#1{\def\thetitle{#1}}
\def\author#1{\def\theauthor{#1}}
\def\textinfo#1{\def\thetextinfo{#1}}
\let\eref=\eqref
\let\eqref=\undefined
\allowdisplaybreaks

\makeatletter
\let\tru@int=\int
\def\int{\mathop{\textstyle\tru@int}\limits}
\def\overl@ss#1#2{\vcenter{\offinterlineskip
        \ialign{$\m@th#1\hfil##\hfil$\crcr#2\crcr<\crcr } }}
\def\overgr@at#1#2{\vcenter{\offinterlineskip
        \ialign{$\m@th#1\hfil##\hfil$\crcr#2\crcr>\crcr } }}
\def\gl{\mathrel{\mathpalette\overl@ss>}}
\def\lg{\mathrel{\mathpalette\overgr@at<}}
\def\d{\mathrm{d}}
\def\Natural{\mathbb{N}}
\def\Integer{\mathbb{Z}}
\def\Real{\mathbb{R}}
\def\Complex{\mathbb{C}}
\def\Re{\mathop{\rm Re}\nolimits}
\def\Im{\mathop{\rm Im}\nolimits}
\def\arg{\mathop{\rm arg}\nolimits}
\def\arcsech{\mathrm{arcsech}}
\def\arccsch{\mathrm{arccsch}}
\def\pvint{\mathop{\int\kern-0.94em-\kern0.2em}\limits}
\let\^=\hat
\let\==\bar
\let\@=\mathbf

\let\ooriginal=\o
\let\le=\leqslant
\let\ge=\geqslant
\def\d{\mathrm{d}}
\def\e{\mathrm{e}}
\def\o{\mathrm{o}}
\def\Wr{\mathrm{Wr}}
\def\H{\mathcal{H}}
\def\P{\mathscr{P}}
\def\eff{\mathrm{eff}}
\def\O#1{^{(#1)}}
\def\~#1{\tilde{#1}}
\makeatother

\advance\textwidth -2mm
\advance\hoffset 1mm
\advance\textheight 4mm
\advance\voffset -2mm

\def\be{\begin{equation}}
\def\ee{\end{equation}}
\def\bse{\begin{subequations}}
\def\ese{\end{subequations}}
\def\brho{{\boldsymbol\rho}}
\def\bsigma{{\boldsymbol\sigma}}

\def\re{\mathrm{re}}
\def\im{\mathrm{im}}

\definecolor{darkred}{rgb}{0.9,0,0}
\definecolor{darkblue}{rgb}{0,0,0.8}
\definecolor{darkorange}{rgb}{0.7,0.2,0}
\definecolor{darkgreen}{rgb}{0.0, 0.3, 0.0}

\usepackage{tocloft}
\begin{document}

\title{\color{darkblue}Inverse scattering transform for two-level systems with\break nonzero background}
\author{Gino Biondini$^1$, Ildar Gabitov$^2$, Gregor Kova\v{c}i\v{c}$^3$ and Sitai Li$^4$}
\textinfo{{\small1:} State University of New York at Buffalo, Department of Mathematics, Buffalo, NY 14260, USA\\
{\small2:} University of Arizona, Department of Mathematics, Tucson, AZ 85721, USA\\
{\small3:} Rensselaer Polytechnic Institute, Department of Mathematical Sciences, Troy, NY 12180, USA\\
{\small4:} University of Michigan, Department of Mathematics, Ann Arbor, MI 48109, USA}
\maketitle

\begingroup
\small
\noindent
\textbf{Abstract}
\\[1ex]
We formulate the inverse scattering transform for the scalar Maxwell-Bloch system of equations 
describing the resonant interaction of light and active optical media in the case when the light intensity does not vanish at infinity.  
We show that pure background states in general do not exist with a nonzero background field.
We then use the formalism to compute explicitly the soliton solutions of this system.
We discuss the initial population of atoms and show that
the pure soliton solutions do not correspond to a pure state initially.
We obtain a representation for the soliton solutions in determinant form, 
and explicitly write down the one-soliton solutions.
We next derive periodic solutions and rational solutions from the one-soliton solutions.
We then analyze the properties of these solutions, 
including discussion of the sharp-line and small-amplitude limits, 
and thereafter show that the two limits do not commute.  
Finally, we investigate the behavior of general solutions, 
showing that solutions are stable (i.e., the radiative parts of solutions decay) 
only when initially atoms in the ground state dominant,
i.e., initial population inversion is negative. 
\par
\endgroup


\bigskip
\section{Introduction}

Resonant interaction between pulses of monochromatic light and an active optical medium is one of the most useful phenomena studied in applied optics, 
having lead to such indispensable devices as lasers and optical amplifiers~\cite{ISI:A1960ZQ06900019,Mccall67,Allen87,Basharov90,Hau99}.
The optical medium is often assumed to have two working atomic levels,  
with the transition frequency between these two levels roughly equal to the frequency of the impinging light, 
which is known as the \emph{two-level medium} approximation~\cite{Allen87}.   
In addition to lasing and amplification, the interaction between light and two-level media has given rise to several classical nonlinear-optics effects including self-induced transparency~\cite{Mccall67,Mccall69,PhysRevA.5.1634,PhysRevLett.29.1211} 
(undisturbed propagation of pulses with sufficiently high amplitudes through the medium and absorption of weaker pulses),    
superfluorescence (generation of short optical pulses from material polarizability fluctuations in an excited medium~\cite{lamb71,lamb74,PhysRevA.20.2047,PhysRev.93.99,PhysRevLett.30.309,PhysRevA.11.1507,PhysRevA.12.587,PhysRevA.19.1192,gabitov84,PhysRevLett.36.1035,PhysRevLett.39.547}), 
optical nutation 
(a phenomenon during which the material properties exhibit oscillations akin to nutation in rigid bodies~\cite{Allen87}), 
and photon echo~\cite{kurnit64,lamb69,lamb71,PhysRevLett.20.1087,PhysRev.179.294} 
(generation of a third pulse from the injection of a pair of pulses into the medium).

The wealth of physical phenomena exhibited by light interacting with two-level active optical media has provided an abundant source of fundamental mathematical problems used in their description.  
This description is afforded by various versions of the two-level Maxwell-Bloch equations (MBEs)~\cite{PhysRev.93.99,feynman57,jaynes63,davis63,risken:4662,Allen87}.   
Frequently, such equations are derived by assuming unidirectional propagation and using the \emph{slowly-varying envelope approximation}, 
i.e., separating the fast-oscillating plane wave representing the color of the light from the slowly-varying envelope representing its amplitude and phase, 
and the \emph{rotating wave approximation}, 
i.e., averaging over the fast oscillations~\cite{risken:4662}.  
Mathematically, these approximations are equivalent to the more general method of multiple scales~\cite{kevcole}.  
The properties of the material are typically described in terms of the quantum density matrix that yields a Bloch vector analogous to that used in spin dynamics~\cite{feynman57}.   
Depending on the values of their parameters and inputs, 
the two-level MBEs exhibit a rich variety of dynamical regimes.
While some regimes of laser operation are chaotic~\cite{PhysRevA.41.3826,PhysRevA.33.1842,lneba85,ISI:A1984TY33700007,ISI:A1985AWU8900005}, 
and can be described by the Lorenz equations~\cite{Haken197577,PhysRevLett.57.2804}, 
or are even turbulent~\cite{PhysRevA.55.751,iom89,PhysRevLett.84.1894,Lvov1998317}, 
and have a finite-dimensional attractor~\cite{0951-7715-2-2-003,0951-7715-11-3-006,0951-7715-12-6-304,PhysRevLett.58.2205} or a slow manifold~\cite{menon:315}, 
other regimes of light propagating through rarefied gases may be approximated by completely-integrable, soliton-type equations%
~\cite{lamb71,lamb74,ablowitz74,kaup77,zakhharov80,zakharov82,manakov82,manakov86,gabitov83,gabitov84,gabitov85,Basharov90}.  

The integrable MBEs are derived under the above-mentioned approximations, 
and have addressed phenomena including self-induced transparency~\cite{lamb71,lamb74,ablowitz74}, area theorem~\cite{kaup77}, photon echo~\cite{zakharov82}, 
amplification~\cite{zakhharov80,manakov82,manakov86}, and superfluorescence~\cite{gabitov83,gabitov84,gabitov85}.   
Typically, these phenomena are addressed from the viewpoint of the initial-boundary-value or signaling problem for a finitely or infinitely long narrow tube containing the active medium, 
into which a narrow optical pulse is injected at one end, 
and whose initial state is given far in the past.  
The MBEs can be represented in terms of a Lax pair~\cite{ablowitz74,gabitov85}, 
thus, the initial-value problem can be solved using the inverse-scattering transform (IST)~\cite{ablowitz74,ablowitz81,gabitov85,novikov84}.  
The temporal piece of the Lax pair is in the AKNS form~\cite{zakharov72,ablowitz74b},
so just as for the nonlinear Schr\"odinger equation appearing in~\cite{hasegawa1995solitons,Agrawal2000NLFO},
which stretches from the infinite past to the infinite future, 
and pulses are evolved along the spatial variable, 
describing its propagation from the location, 
where the pulse is launched into the medium.   
The evolution operator in the Lax pair of the MBEs has an unusual form containing a Hilbert transform, 
  which complicates the derivation and form of the equations describing the evolution of the spectral data, 
as well as reflects the fact that the phenomena described by the MBEs are irreversible~\cite{ablowitz74}.    

All the above classic works on the IST for the integrable two-level MBEs addressed narrow optical pulses, i.e., pulses that decay both forward and backward in time.   
A few studies have discussed pulses riding on top of a continuous-wave (CW) background
\cite{msr1988,pb2001,xphc2016,hxpcd2016,RVB20051,RVB20052,hxp20121,wlqz2015,hxp20122,lhp2013}.
However, these phenomena have so far attracted much less attention.   
In this paper, 
we make a systematic attempt to fill this gap by developing the IST for studying pulses that asymptotically approach plane waves with equal amplitudes and frequencies in both backward and forward times, 
i.e., for MBEs with nonzero background (NZBG).  

The direct and inverse scattering components of our study closely parallel those in our work on the focusing nonlinear Schr\"odinger (NLS) equation with nonzero boundary conditions (NZBC)~\cite{JMP55p031506}.    
More generally, the study of nonlinear systems with NZBG has also received considerable interest thanks in part to its connections to the theory of modulational instability,
rogue waves and integrable turbulence~\cite{ZO2009,kffmdgad2010,srkj2007,PhysRevLett.96.014503,z2009,az2015,gbdm2016prl116,gbdm2017cpam,biondini2016PRELi,biondiniSIAMRev2018}.   
The continuous spectra for the scattering problems of both the NLS and MBEs comprise the real axis plus a symmetric interval along the imaginary axis with length twice that the amplitude of the CW background. 
This symmetric interval is the branch cut for the double-sheeted Riemann surface on which both the scattering and inverse-scattering take place.  

The main difficulty of the IST for MBEs lies in evaluating the evolution of the spectral data.   
Deriving this evolution is aided by the Liouville form of the equation for the material variables and the dual role of the spectral parameter as the frequency detuning~\cite{ablowitz74,gabitov85}.   
Especially intricate is finding the time-dynamics of the reflection coefficients, which requires careful evaluation of Fourier-type integrals using the Riemann-Lebesgue lemma and results in an inhomogeneous linear differential equation.   
While the result is somewhat similar to that derived in~\cite{gabitov85} for MBEs with vanishing background, 
the details of the derivation are significantly more involved.
We find that already the single-soliton solutions of the MBEs with NZBG  bring some surprises.   
The electric field envelope of their counterparts on zero background (ZBG) has a very simple, 
sech-like shape, possibly multiplied by an oscillating complex phase factor.   
On NZBG, this shape only remains for the solitons corresponding to the eigenvalue lying on the imaginary axis above the branch cut 
(and, of course, its complex conjugate, which we ignore from now on). 
An eigenvalue in the general position in the upper half-plane gives rise to a traveling breather-like structure, with an exponentially localized envelope containing sinusoidal ripples that move faster than the envelope
in the same direction.  
The entire structure is again multiplied by an oscillating complex phase factor.  
Away from its center, the structure rapidly approaches the background plane wave. 
Moreover, as the corresponding eigenvalue approaches the branch cut, the width of the solution envelope spreads to infinity, 
and what remains is a periodic traveling wave.
Moreover, the eigenvalue at the branch point produces  a traveling wave with a rational form.  
If we consider the spectral parameter to be a bifurcation parameter, 
these families of solitons represent the unfolding of the solutions around the spectral singularity at the branch point.  
The four types of one-soliton solutions have been previously reported in~\cite{lbkg2017}.
Finally, we investigate soliton solutions in the case in which the frequency of the impinging light is detuned from the peak frequency of the atomic transition in the medium, 
and describe further unfolding properties of the solutions also in terms of the detuning as a parameter.

The outline of this work is the following:
In Section~\ref{s:preliminaries} we present the MBEs and its Lax pair, 
and we discuss its background solutions and the treatment of the branch cut in the complex plane in the formulation of the IST.
In Section~\ref{s:directprob} we formulate the direct scattering.
We derive all scattering data including the transmission and reflection coefficients,
discrete spectrum and norming constants.
We then discuss the symmetries and asymptotics of eigenfunctions and scattering data,
and the asymptotic behavior of the density matrix.
In Section~\ref{s:propagation} we determine propagation equations for all the relevant quantities,
including the scattering data and the continuous-wave background.
In Section~\ref{s:inverseprob} we formulate the inverse problem in terms of a Riemann-Hilbert problem (RHP).
We then reconstruct general solutions of the MBEs from the solutions of the RHP.
Moreover, we derive a ``trace" formula which recovers the analytic scattering coefficient,
and derive a ``theta" condition which yields the asymptotic phase difference of the solution.
In particular, we write down an explicit formula for pure $N$-soliton solutions,
for which the reflection coefficient is identically zero.
In Section~\ref{s:solutions}, we focus on one-soliton solutions.
We show that there are four types of solitons depending on the location of the corresponding discrete eigenvalue.
They are traveling-wave solitons, oscillatory solitons, periodic solutions and rational solutions.
We also discuss their properties in detail and we analyze the stability of soliton solutions.
In Section~\ref{s:soliton_shift_lorentzian} we discuss soliton solutions resulting from a shifted Lorentzian as the spectral-line shape.
Such a shift originates from a detuning between the resonance frequency of the atoms (corresponding to the transition between the ground state and the excited state)
and the frequency of the incident light pulse.
Importantly, we show that the presence of such a shift produces markedly different properties for the resulting soliton solutions.
In particular, we will show that some of the solutions actually become rogue-wave in character, in the sense that they appear to 
arise from virtually nothing, attain a peak value which is several times larger than the average,
and eventually revert to the background.
We conclude this work with some final remarks in Section~\ref{s:conclusion}.

Auxiliary material is confined to the background.
In Appendix~\ref{s:notations} we describe some frequently used notation.
In Appendix~\ref{A:symmetries} we derive the symmetries of various quantities appearing in the direct problem of the IST.
In Appendix~\ref{A:asymptotics} we compute the asymptotics of the eigenfunctions.
In Appendix~\ref{A:Rpm} we derive the explicit formula for an auxiliary matrix that appears in the propagation part of the IST.
In Appendix~\ref{A:evolution} we derive the propagation equations for the norming constants and for the reflection coefficient.
In Appendix~\ref{A:RHPsoln} we show how to solve the RHP that appears in the inverse problem.
In Appendix~\ref{s:trace} we derive the trace formula and the ``theta" condition for the solutions of the MBEs with NZBG.
In Appendix~\ref{A:Rpm_explicit} we calculatie the auxiliary matrix explicitly with a Lorentzian as the spectral-line shape.
In Appendix~\ref{s:2ndIST} we briefly discuss an alternative formulation of the IST,
and prove the equivalence between the two versions of the IST.

\section{Preliminaries}
\label{s:preliminaries}

\subsection{The Maxwell-Bloch system of equations and its Lax pair}
\label{s:MBE}

Up to rescalings of the dependent and independent variables, 
the scalar MBEs are, in component form
\begin{gather}
\partial q/\partial z = -\int P(t,z,\lambda)g(\lambda)\d\lambda\,,
\label{e:MBEqscalar}
\\
\partial D /\partial t = 2\Re(q^* P)\,,\qquad
\partial P /\partial t - 2i\lambda\,P = -2 q D\,.
\label{e:MBErhoentries}
\vspace*{-1ex}
\end{gather}
where
$z= z_\mathrm{lab}$ is the propagation distance,
$t= t_\mathrm{lab} - z_\mathrm{lab}/c$ is a retarded time ($c$ is the speed of light in vacuum),
$q(t,z)$ is the optical field, i.e., the electric field envelope
corresponding to the transitions between the ground state and the excited state,
level inversion $D(t,z,\lambda)$ and medium polarization envelope $P(t,z,\lambda)$ are the entries of a $2\times2$ Hermitian matrix representing the density matrix,
and $g(\lambda)$ is the (known) shape of the spectral line.
All integrals in this work run from $-\infty$ to $\infty$ unless indicated otherwise.

The MBEs~\eref{e:MBEqscalar} and~\eref{e:MBErhoentries} can be written compactly as 
\begin{gather}
\rho_t= [i\lambda \sigma_3 + Q,\rho]\,,
\label{e:drhodt}
\\
Q_z= -\frac{1}{2} \int[\sigma_3,\rho]\,g(\lambda)\d\lambda\,,
\label{e:dqdz}
\end{gather}
where subscripts $t$ and $z$ denote partial differentiation,
$[A,B]= AB-BA$ is the matrix commutator,
the $2\times2$ optical matrix and density matrix are
\[
\nonumber
Q(t,z) = \begin{pmatrix}
 0 & q \\ -q^* & 0
\end{pmatrix}\,,\qquad
\rho(t,z,\lambda) = \begin{pmatrix}
D & P \\ P^* & -D
\end{pmatrix}\,,
\]
and $\sigma_1, \sigma_2, \sigma_3$ are the Pauli matrices
\[
\sigma_1= \begin{pmatrix}0 &1\\ 1& 0\end{pmatrix},\qquad
\sigma_2= \begin{pmatrix}0 &-i\\ i& 0\end{pmatrix},\qquad
\sigma_3= \begin{pmatrix}1 &0\\ 0& -1\end{pmatrix}.
\label{e:Paulidef}
\]
The Lax pair for the MBEs~\eref{e:MBEqscalar} and~\eref{e:MBErhoentries} is:
\begin{gather}
\phi_t= (i\lambda \sigma_3 + Q)\,\phi\,,
\label{e:scatteringproblem}
\\
\phi_z= V\,\phi\,,
\label{e:zevolution} 
\end{gather}
where 
Eq.~\eref{e:scatteringproblem} and $q(t,z)$ are referred to as the scattering problem and as the scattering potential,
respectively, and $\phi=\phi(t,z,\lambda)$ is the wave function, 
with
\vspace*{-1ex}
\[
\nonumber
V(t,z,\lambda)= \frac{i\pi}2\H_\lambda[\rho(t,z,\lambda')g(\lambda')]\,,
\]
where $\H_\lambda[f(\lambda')]$ is the Hilbert transform of the function $f(\lambda)$ defined by
\[
\H_\lambda[f(\lambda')]= \frac1\pi \pvint \frac{f(\lambda')}{\lambda'-\lambda}\d\lambda'\,.
\label{e:Hilbert}
\]
Throughout this work, primes in integration variables do \textit{not} denote differentiation.

The IST for the MBEs~\eref{e:drhodt} and~\eref{e:dqdz} with $q(t,z)\to0$ as $t\to\pm\infty$ ---
hereafter referred to as ZBG ---
was carried out in \cite{ablowitz74,gabitov85}
(while some earlier results appeared in \cite{lamb71,lamb74}).
Here we develop the IST for the scalar MBEs~\eref{e:drhodt} and~\eref{e:dqdz} with NZBG.
Namely, we solve system~\eref{e:drhodt} and~\eref{e:dqdz} with $q\to q_\pm$ as $t\to\pm\infty$.
We assume $|q_\pm|= q_o>0$.
Similarly to what has been done in~\cite{JMP55p031506} for the focusing nonlinear Schr\"odinger (NLS) equation with NZBC, 
we will formulate the IST such that it allows one to take the reduction $q_o\to0$ explicitly throughout.

Note that:
(i) System~\eref{e:drhodt} and~\eref{e:dqdz} implies $\partial_t(\tr\rho)= \partial_t(\det\rho) = 0$,
(ii) System~\eref{e:drhodt} and~\eref{e:dqdz} is invariant under the transformation
 $\rho(t,z,\lambda)\mapsto\~\rho(t,z,\lambda)= \rho(t,z,\lambda) + f(z,\lambda)\,I$, where $f$ is any scalar function independent of $t$.
Hence, without loss of generality we can take $\rho$ to be traceless and with determinant equal to $-1$.
Moreover, we assume that this normalization has been made for all $z\ge0$.

\subsection{Background solutions}
\label{s:background}

Before we formulate the IST, we investigate whether there exist exact 
``constant'' solutions of the MBEs~\eref{e:drhodt} and~\eref{e:dqdz} with NZBG. 
That is, we look for solutions $q(t,z) = q_o(z)$ $\forall t\in\Real$, with $q_o\ne0$.
The right-hand side (RHS) of Eq.~\eref{e:drhodt} contains the matrix $X_o = i\lambda \sigma_3 + Q_o$,
where $Q_o$ is the same as $Q$, but with $q_o$ instead of $q$.
Since $Q_o$ is independent of $t$, the solution of Eq.~\eref{e:drhodt} is simply
\[
\rho(t,z,\lambda) = \e^{X_ot}\,C\,\e^{-X_ot}\,,
\label{e:rhobackground}
\]
for any $2\times2$ matrix $C$ independent of $t$.
The eigenvalues of $X_o$ are $\pm i\gamma$, where $\gamma^2 = \lambda^2 + q_o^2$.
For now we limit our discussion to 
real values of $\lambda$.
(The extension of $\gamma$ to complex values of $\lambda$ will be discussed in Section~\ref{s:uniformization}.)
In order to be able to take the limit $q_o\to0$ continuously 
we need to choose
\vspace*{-1ex}
\[
\label{e:defgamma}
\gamma(\lambda) = \sign(\lambda)\,\sqrt{\lambda^2 + q_o^2}\,,\qquad
\lambda \in \Real\,. 
\]
Of course, with this choice $\gamma$ has a sign discontinuity 
across $\lambda=0$.
We can write an eigenvector matrix of $X_o$ compactly as
\[
Y_o = I + (i/\zeta) \,\sigma_3Q_o\,, 
\label{e:Ybackground}
\]
where we introduced the shorthand notation 
\vspace*{-0.4ex}
\[
\zeta = \lambda + \gamma\,,
\label{e:uniformization}
\]
which will be used throughout this work.
Explicitly,
$X_oY_o = Y_o \,i\gamma \sigma_3$. 
Thus,
\[
\rho(t,z,\lambda) = Y_o\,\e^{i\gamma t\sigma_3}\rho_o\,\e^{-i\gamma t\sigma_3}Y_o^{-1}\,, 
\label{e:rhobackgroundexplicit}
\]
where $\rho_o = Y_o^{-1}C Y_o$.
The inverse transformation to Eq.~\eref{e:uniformization} is 
\[
\lambda = \half\,(\zeta-q_o^2/\zeta)\,,\qquad
\gamma = \half\,(\zeta+q_o^2/\zeta)\,. 
\label{e:unifinverse}
\]
For future reference, note that 
\[
\nonumber
Y^{-1}_o = [I-(i/\zeta)\sigma_3Q_o]\,/\det Y_o\,,\qquad
\det Y_o = 2\gamma/\zeta =1+q_o^2/\zeta^2\,.
\]
Also note that $\det\rho= \det\rho_o$ and $\tr\rho= \tr\rho_o$.
Hence, equations $\rho^\dag= \rho$, $\tr\rho=0$ and $\det\rho= -1$ imply that
$\rho_o^\dag = \rho_o$, $\tr\rho_o=0$ and $\det\rho_o=-1$.
Therefore, we can write $\rho_o = \@h\cdot\bsigma$, 
where $\bsigma = (\sigma_1,\sigma_2,\sigma_3)^T$ and
$\@h\in\Real^3$, with
$\@h\cdot\@h = - \det\rho = 1$.
The superscript $T$ denote matrix transpose.
So, we have
\[
\nonumber
\rho(t,z,\lambda) = \@h\cdot\brho\,, 
\]
where $\brho = (\rho_1,\rho_2,\rho_3)^T$ and
\[
\rho_j(t,z,\lambda) = Y_o\,\e^{i\gamma t \sigma_3}\sigma_j\,\e^{-i\gamma t\sigma_3} Y_o^{-1}\,,
\label{e:rhojbackground}
\]
with $\rho_j^\dag= \rho_j$, $\tr\rho_j=0$, and $\det\rho_j = -1$.
Explicitly, 
\begin{gather}
\label{e:rhojbackgroundexplicit1}
\rho_1 = \frac1{2\gamma \zeta} \big( \zeta^2\,\e^{i\gamma t\sigma_3}\sigma_1\e^{-i\gamma t\sigma_3}
  - 2\zeta \Im(\e^{-2i\gamma t}q_o)\,\sigma_3
  + \e^{-i\gamma t\sigma_3}Q_o\sigma_1Q_o\e^{i\gamma t\sigma_3}\big)\,,
\\
\label{e:rhojbackgroundexplicit2}
\rho_2 = \frac1{2\gamma \zeta} \big( \zeta^2\,\e^{i\gamma t\sigma_3}\sigma_2\e^{-i\gamma t\sigma_3}
  - 2\zeta\Re(\e^{-2i\gamma t}q_o)\,\sigma_3
  + \e^{-i\gamma t\sigma_3}Q_o\sigma_2Q_o\e^{i\gamma t\sigma_3}\big)\,,
\\
\label{e:rhojbackgroundexplicit3}
\rho_3 = \frac1\gamma (\lambda \sigma_3 - iQ_o)\,.
\end{gather}
All these results reduce to the correct behavior in the ZBG limit (i.e., $q_o\to0$).
In particular, $\gamma\to \lambda$, $\zeta\to2\lambda$ and $Y\to I$ when $q_o\to0$,
implying
$\brho \to \e^{i\lambda t\sigma_3}\bsigma\,\e^{i\lambda t\sigma_3}$
in the limit $q_o\to 0$.
Remarkably, even when $q_o\ne0$, $\rho_3$ is always independent of $t$ as in the case with ZBG, 
even though the ODEs~\eref{e:drhodt} 
are more complicated when $Q_o\ne0$. 

Now we insert the behavior described by Eqs.~\eref{e:rhojbackgroundexplicit1},~\eref{e:rhojbackgroundexplicit2} and~\eref{e:rhojbackgroundexplicit3} into Eq.~\eref{e:dqdz}.
It is trivial to see that $[\sigma_3,\rho]= \@h\cdot[\sigma_3,\brho]$.
Moreover, direct calculation yields
\begin{gather}
\label{e:[J,rho]1}
[\sigma_3,\rho_1] = (1/\gamma \zeta)\,\e^{-i\gamma t\sigma_3}
    \big( i\zeta^2 \sigma_2 + \sigma_3\,Q_o\sigma_1Q_o \big)\,\e^{i\gamma t\sigma_3}\,,
\\
\label{e:[J,rho]2}
[\sigma_3,\rho_2] = (1/\gamma \zeta)\,\e^{-i\gamma t\sigma_3}
    \big( - i\zeta^2 \sigma_1 + \sigma_3\,Q_o\sigma_2Q_o \big)\,\e^{i\gamma t\sigma_3}\,,
\\
\label{e:[J,rho]3}
[\sigma_3,\rho_3] = -(i/\gamma)[\sigma_3,Q_o]\,. 
\end{gather}
The third commutator above is $t$-independent, leaving the field invariant.
Explicitly, 
\[
\nonumber
\int[\sigma_3,\rho_3]g(\lambda)\d\lambda = -iw_o[\sigma_3,Q_o]\,,
\]
where 
\[
\label{e:wo}
w_o = \int g(\lambda)/\gamma\,\d\lambda\,.
\]
On the other hand, the two commutators in Eqs.~\eref{e:[J,rho]1} and~\eref{e:[J,rho]2} contain $t$-dependent oscillating exponentials,
which in general do not cancel even upon integration.
Hence, in general the material polarization does not preserve the $t$-independence of the optical field,
i.e., they are not consistent with a constant solution $q(t,z) = q_o(z)$.
Accordingly, 
the only self-consistent background solution is the one obtained with $h_1=h_2=0$ and $h_3 = \pm1$, resulting in
\[
\label{e:backgroundstates_general}
q(t,z) = q_o(z) = q_o(0)\,\e^{-ih_3\,w_oz}\,,\qquad
\rho(t,z,\lambda) = h_3 (\lambda \sigma_3 - iQ_o)/\gamma\,.
\]
With a generic spectral-line shape $g(\lambda)$, it is clear from Eq.~\eref{e:wo} that $w_o\ne0$.
However, the simplest situation occurs when the distribution of atoms is even with respect to the normalized resonant frequency $\lambda = 0$.
In such a case, the evenness of the function $g(\lambda)$ and oddness of the function $\gamma(\lambda)$ imply $w_o=0$.
The above background solution~\eref{e:backgroundstates_general} reduces to
\[
\label{e:backgroundstates}
q(t,z) = q_o\,,\qquad
\rho(t,z,\lambda) = h_3 (\lambda \sigma_3 - iQ_o)/\gamma\,.
\]

Importantly, note that $\rho(t,z,\zeta)$ is discontinuous at $\lambda=0$.
Also
recall that in the case of the MBEs with ZBG,
$h_3 =-1$ implies the atoms are in the ground state, 
whereas, $h_3 = 1$ implies all the atom are in the excited state.
However, in the case with NZBG, the situation is seemingly more complicated.
Because of the presence of background radiation,
a polarization-free state does not exist in the background solutions~\eref{e:backgroundstates},
i.e., the medium is always polarized.
This polarization is reflected in the off-diagonal entries of the density matrix $\rho$, which are never vanish,
i.e., $P = -ih_3 q_o/\gamma$, which arises from the background radiation $q_o(z)$.
In fact, later we will show that this statement holds for general solutions as well.
Nonetheless, we will also show that a natural decomposition still exists for the density matrix.
We will also show in Section~\ref{s:boundaryvalues} that,
when $q(t,z)$ depends on $t$, 
as long as $q\to q_\pm$ as $t\to\pm\infty$, 
the asymptotic behavior of the density matrix [i.e., the solutions of Eq.~\eref{e:drhodt}]
is consistent with the above expressions obtained with a background value of the potential.

\subsection{Riemann surface and uniformization variable}
\label{s:uniformization}

Before we can start to develop the IST for the MBEs~\eref{e:drhodt} and~\eref{e:dqdz},
it is useful to discuss the branching of the eigenvalues of the scattering problem.
The asymptotic scattering problem as $t\to\pm\infty$ is $\phi_t = X_\pm\,\phi$, where $X_\pm= i\lambda \sigma_3 + Q_\pm$.
The eigenvalues of $X_\pm$ are $\pm i\gamma$, with $\gamma^2 = q_o^2 + \lambda^2$, where $q_o= |q_\pm|$.
As with the NLS equation \cite{zakharov73}, to address the branching, 
we introduce the two-sheeted Riemann surface defined by the complex square root
\[
\gamma(\lambda) = (q_o^2 + \lambda^2)^{1/2}\,.
\label{e:gammadef}
\]
Specifically, the Riemann surface is obtained by 
introducing two copies of the complex plane, called $\Complex_I$ and $\Complex_\II$, 
in which $\gamma(\lambda)$ takes either of the two possible signs of the complex square root.
The branch points are the values of $\lambda$ for which $\gamma(\lambda)=0$, i.e., $\lambda= \pm iq_o$.
Let $\lambda + iq_o = r_1\,\e^{i\theta_1}$ and $\lambda-iq_o = r_2\,\e^{i\theta_2}$. 
In this way we can write 
\[
\gamma(\lambda) = \sqrt{r_1r_2}\,\e^{i\Theta}\,,
\qquad
\Theta= (\theta_1+\theta_2)/2 + m\pi\,,
\label{e:gammadef2}
\]
where 
$m=0,1$ respectively on sheet I and II.
Now take $-\pi/2\le\theta_j<3\pi/2$ for $j=1,2$.
With these conventions, 
the discontinuity of $\gamma$ (which defines the location of the branch cut) occurs on the segment
$i[-q_o,q_o]$.
The Riemann surface is then obtained by gluing the two copies of the complex plane along the~cut.
Formally, $\^\Complex = \Complex_\I\cup\Complex_\II$.
Along the real $\lambda$ axis we then have $\gamma(\lambda)= \pm\sign(\lambda)\sqrt{q_o^2 + \lambda^2}$, 
where the plus and minus signs apply respectively on sheet I and sheet II of the Riemann surface, 
and where the square root sign denotes the principal branch of the real-valued square root function.
For later use, we use the subscripts $\mathrm{I}$ and $\mathrm{II}$ to denote that a quantity is evaluated on the first or second sheet, respectively.

The reason why 
we take the branch cut along $[-iq_o,iq_o]$ is that, in this way, 
$\gamma(\lambda)$ reduces to \eref{e:defgamma} for $\lambda\in\Real$
which in turn reduces to $\lambda$ as $q_o\to0$ .
Not only does this allow us to take the limit $q_o\to0$ throughout,
but it also allows us to choose the initial state for the MBEs motivated by the physical intuition gained from the case of ZBG.
Of course, one could just as well formulate the IST with the branch cut on $i\Real\backslash [-iq_o,iq_o]$.
In Appendix~\ref{s:2ndIST}, we show that doing so is equivalent to a redefinition of the initial states of the MBEs. Note that Eqs.~\eref{e:unifinverse} allows us to can express all $\lambda$ dependence in terms of $\zeta$,
and we will do so throughout this work.

\begin{figure}[t!]
    \centering
    \includegraphics[width=0.8\textwidth]{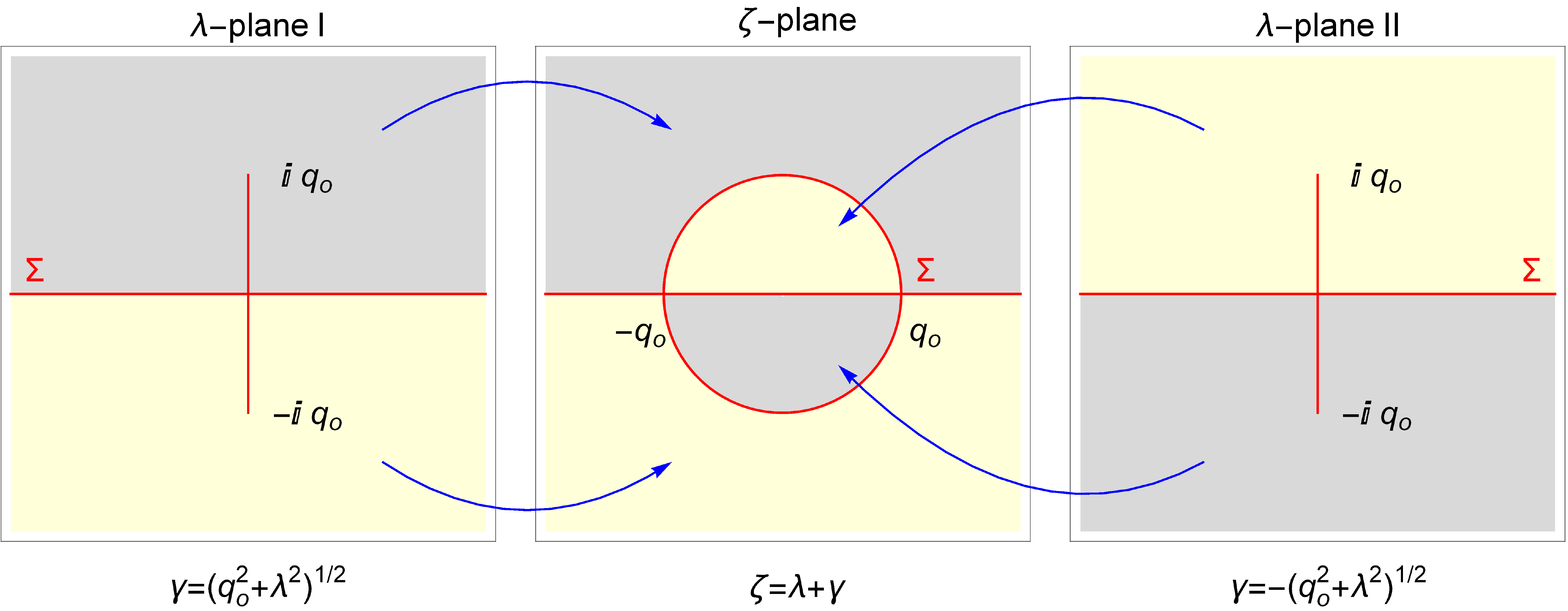}
    \caption{
        Left and right plots show the two complex $\lambda$-planes,
        in which the square root takes different signs.
        Center plot shows the complex $\zeta$-plane.
        The red contour $\Sigma$ in all three plots corresponds to the continuous spectrum $\gamma\in\Real$.
        The gray region and yellow region correspond to where $\Im \gamma>0$ and $\Im \gamma<0$, 
        i.e., the regions $\Gamma^\pm$~\eref{e:Cpmdef}, respectively.
        The blue curves denotes the uniformization mapping.}
    \label{f:domains}
\end{figure}

It will also be convenient to introduce a uniformization variable.
As in \cite{FT1987},
this is easily done by defining $\zeta = \lambda + \gamma$ as in Eq.~\eref{e:uniformization}, 
which is inverted by Eq.~\eref{e:unifinverse}.
Let $C_o$ be the circle of radius $q_o$ centered at the origin in the complex $\zeta$-plane.
With these definitions:
(i) the branch cut on either sheet is mapped onto $C_o$;
(ii) in particular: $\zeta(\pm iq_o) = \pm iq_o$ from either sheet, 
$\zeta(0_\I\pm\epsilon) \to \pm q_o$ and $\zeta(0_\II\pm\epsilon) \to \mp q_o$ as $\epsilon\to0^+$; 
(iii) $\Complex_\I$ is mapped onto the exterior of $C_o$;
(iv) $\Complex_\II$ is mapped onto the interior of $C_o$;
(v) in particular: $\zeta(\infty_I) = \infty$ and $\zeta(\infty_\II) = 0$;
(vi) the first/second quadrant of $\Complex_\I$ are respectively mapped into the first/second quadrant outside $C_o$;
(vii) the first/second quadrant of $\Complex_\II$ are respectively mapped into the second/first quadrant inside $C_o$;
(viii) note also $\zeta_\I \zeta_\II = -q_o^2$.
The uniformization map can be seen clearly from Fig.~\ref{f:domains}.

Note that, in general, 
the density matrix $\rho(t,z,\lambda)$ is only defined for $\lambda\in\Real$.
Writing it as a function of the uniformization parameter $\zeta$, 
we can evaluate it for all $\zeta\in\Real$, specifically
for $|\zeta|>q_o$ on sheet I, and for $|\zeta|<q_o$ on sheet II, where obviously $\rho(t,z,-q_o^2/\zeta) = \rho(t,z,\zeta)$
[since $\rho(t,z,\lambda)$ is a single-valued function of $\lambda$].
But since a priori we do not have any guarantee that $\rho(t,z,\lambda)$ can be analytically extended off the real $\lambda$ axis,
we will not be allowed to evaluate $\rho(t,z,\zeta)$ off the real $\zeta$ axis either.

We next discuss the difference between the two sheets.
Recall in the case with ZBG, 
the background density-matrix solution is $\rho = h_3 \,\sigma_3$ with $ h_3 =\pm1$.
The value $h_3 = \mp1$ indicates that all atoms are in the ground state or in the excited state, respectively.
Therefore the background solution~\eref{e:backgroundstates} with NZBG should reduce to $\rho = h_3\,\sigma$ as $q_o\to0$.
On sheet I, this is obviously true.
However, on sheet II, the solution~\eref{e:backgroundstates} reduces to $\rho = -h_3\,\sigma_3$ as $q_o\to0$.
Hence, \textit{to maintain consistency with the ZBG reduction, 
one must choose opposite values for $h_3$ on sheet II in the background solution~\eref{e:backgroundstates}}.
We will discuss this situation in detail in Section~\ref{s:boundaryvalues}.

Hereafter, with some abuse of notation we will rewrite all the $\lambda$ dependence as dependence on $\zeta$.
Note that Eq.~\eref{e:gammadef2} implies that $\Im\,\gamma\gl0$ respectively in $\Gamma^+$ and $\Gamma^-$,
where
\[
\Gamma^+=\{\zeta\in\Complex:(\Im \zeta)(|\zeta|^2-q_o^2)>0\}\,,\qquad
\Gamma^-=\{\zeta\in\Complex:(\Im \zeta)(|\zeta|^2-q_o^2)<0\}\,.
\label{e:Cpmdef}
\]
These two regions can be seen in Fig.~\ref{f:domains} with gray and yellow colors.
As we show next, this property will determine the analyticity regions of the Jost eigenfunctions.

\section{Direct problem}
\label{s:directprob}

We are now ready to start formulating the IST for the MBEs~\eref{e:drhodt} and~\eref{e:dqdz} with NZBG.

\subsection{Jost solutions, analyticity and scattering matrix}
\label{s:jost}

We first go back to the asymptotic scattering problem, 
i.e., the ODE~\eref{e:scatteringproblem} with $t\to\pm\infty$.
We find it convenient to consider the wave function $\phi(t,z,\lambda)$ as a $2\times2$ matrix (instead of a 2-component vector).
On either sheet, we can write the asymptotic eigenvalue matrix and eigenvector matrices as $i\gamma \sigma_3$ and 
\[
Y_\pm =  I + (i/\zeta)\sigma_3Q_\pm\,,
\label{e:Ydef}
\]
respectively, where $I$ denotes the $2\times2$ identity matrix and $Q\to Q_\pm$ as $t\to\pm\infty$, 
so that
\[
X_\pm Y_\pm = Y_\pm\,\,i\gamma \sigma_3\,,
\nonumber
\]
with
\[
\nonumber
X_\pm = i\lambda \sigma_3 + Q_\pm\,.
\] 
Similarly to what happened in Section~\ref{s:background},
one can easily calculate that
\begin{gather}
\det Y_\pm = 2\gamma/\zeta =1+q_o^2/\zeta^2\,,
\label{e:detYpm}
\\ 
Y^{-1}_\pm = \big(I-(i/\zeta)\sigma_3Q_\pm\big)\,/\det Y_\pm\,.
\label{e:YdetYinv}
\end{gather}
By definition~\cite{JMP55p031506}, the continuous spectrum $\Sigma_\lambda$ consists of all values of $\lambda$ (on either sheet) such that
$\gamma(\lambda)\in\Real$.
Based on the above choices~\eref{e:gammadef} [equivalent to Eq.~\eref{e:defgamma}], 
it is $\Sigma_\lambda= \Real\cup i[-q_o,q_o]$.
The corresponding set in the complex $\zeta$-plane
is $\Sigma_\zeta= \Real\cup C_o$ from the definition~\eref{e:uniformization}.
Hereafter we will omit the subscripts on $\Sigma$; the intended meaning should be clear from the context.

We expect that, for all $\zeta\in\Sigma$, 
the scattering problem~\eref{e:scatteringproblem} admits solutions $\phi_\pm$ which behave as
\begin{gather}
\phi_-(t,z,\zeta) = Y_-\e^{i\gamma t\sigma_3} + o(1) \qquad\mathrm{as}~t\to-\infty\,,
\\
\phi_+(t,z,\zeta) = Y_+\e^{i\gamma t\sigma_3} + o(1) \qquad\mathrm{as}~t\to\infty\,.~~
\end{gather}
As in Ref.~\cite{JMP55p031506}, we introduce modified eigenfunctions by removing the asymptotic exponential oscillations:
\[
\mu(t,z,\zeta)= \phi(t,z,\zeta)\,\e^{-i\gamma t\sigma_3}\,,
\label{e:mudef}
\]
so that $\lim_{t\to\pm\infty}\mu_\pm(t,z,\zeta)= Y_\pm$.
One can then formally integrate the resulting ODEs for $\mu_\pm$ found from the scattering problem~\eref{e:scatteringproblem}, 
and obtains integral equations for $\mu_\pm$ as follows
\begin{gather}
\label{e:muinteqns1}
\mu_-(t,z,\lambda)= Y_- + 
\int_{-\infty}^t Y_-\e^{i\gamma (t-\tau)\sigma_3}Y_-^{-1}\Delta Q_-(\tau,z)\mu_-(\tau,z,\zeta)\,\e^{-i\gamma (t-\tau)\sigma_3}\,\d \tau\,,
\\[-0.4ex]
\label{e:muinteqns2}
\mu_+(t,z,\lambda)= Y_+ - 
\int_t^\infty Y_+\e^{i\gamma (t-\tau)\sigma_3}Y_+^{-1}\Delta Q_+(\tau,z)\mu_+(\tau,z,\zeta)\,\e^{-i\gamma (t-\tau)\sigma_3}\,\d \tau\,,
\end{gather}
where $\Delta Q_\pm = Q-Q_\pm$.
In particular, after some calculations, 
the integrand in Eqs.~\eref{e:muinteqns1} and~\eref{e:muinteqns2} can be written as
\begin{gather*}
\frac1{\det Y_\pm}\,Y_\pm\,(G_\pm^{(1)}(t-\tau)\Delta Q_\pm\mu_{\pm,1}\,,\,G_\pm^{(2)}(t-\tau)\Delta Q_\pm\mu_{\pm,2})\,,
\\
\noalign{\noindent where}
G_\pm^{(1)}(s) = \begin{pmatrix} 1 & -iq_\pm/\zeta \\
   -i\e^{-2i\gamma s}q_\pm^*/\zeta &\e^{-2i\gamma s} \end{pmatrix},
\qquad
G_\pm^{(2)}(s) = \begin{pmatrix} \e^{2i\gamma s} & - i\e^{2i\gamma s}q_\pm/\zeta \\                                        
  -iq_\pm^*/\zeta & 1 \end{pmatrix},
\end{gather*}
and where the subscripts 1 and 2 on $\mu_\pm$ identify the matrix columns, 
i.e., $\mu_\pm= (\mu_{\pm,1},\mu_{\pm,2})$.
Note that the limits of integration imply that $s = t-\tau$ is either always positive 
(for $\mu^-$) or always negative (for $\mu^+$).
Recall that $\Im\,\gamma\gl0$ respectively in $\Gamma^+$ and $\Gamma^-$~\eref{e:Cpmdef} 
(and, in contrast, they are not sign-definite in $\Complex_\I$ and $\Complex_\II$).

Requiring boundedness as $\tau\to\pm\infty$, 
one can then show that the eigenfunctions can be analytically extended in the complex $\lambda$-plane
into the following regions:
\[
\label{e:analyticregion}
\mu_{+,1},~ \mu_{-,2}:\quad \Gamma^+\,,
\qquad
\mu_{-,1},~ \mu_{+,2}:\quad \Gamma^-\,.
\]
The above arguments are made rigorous using Neumann series, as in \cite{apt2004}.
The analyticity properties of the columns of the eigenfunction matrices $\phi_\pm$ follow trivially from those of $\mu_\pm$.
Hereafter, we will consistently use the \textit{subscripts} $\pm$ to denote limiting values as $t\to\pm\infty$, 
whereas the \textit{superscripts} $\pm$ will denote analyticity (or more in general meromorphicity) 
in the regions $\Gamma^\pm$.
The analyticity region of all eigenfunctions can be seen in Fig.~\ref{f:analyticity}(left).

\paragraph{Scattering matrix.}

If the wave matrix $\phi(t,z,\zeta)$ solves the scattering problem~\eref{e:scatteringproblem}, Abel's theorem implies
\[
\partial_t(\det\phi) = \tr(i\lambda \sigma_3 + Q) \,\det\phi = 0\,.
\nonumber
\]
Also for any spectral-parameter value $\zeta\in\Sigma$, 
we have
$\lim_{t\to\pm\infty}\phi_\pm(t,z,\zeta)\,\e^{-i\gamma t\sigma_3} = Y_\pm$.
Hence for any $\zeta\in\Sigma$ we have
\[
\det\phi_\pm(t,z,\zeta) = \det Y_\pm\,\qquad
\forall t\in\Real\,,
\label{e:detphipm}
\]
where $\det Y_\pm$ was given in Eq.~\eref{e:detYpm}.
So for any $\lambda\in\Sigma\backslash\{\pm iq_o\}$, 
$\phi_-$ and $\phi_+$ are two fundamental matrix solutions of the scattering problem.
As in the case of NLS equation with NZBC, 
the behavior at the branch points can be singular.
(See for example~\cite{FT1987} for a discussion of the possible situations that can arise.
A rigorous formulation of the IST for the NLS equation with NZBC considering the singularity at the brunch points is recently presented in~\cite{2017arXiv171006568B}. 
For now we ignore any such situations.)
Hence we can write, for any $\zeta\in\Sigma$,
\[
\phi_+(t,z,\zeta) = \phi_-(t,z,\zeta)\,S(\zeta,z)\,,
\label{e:scattering}
\]
where $S(\zeta,z) = (s_{i,j})$ is the scattering matrix.
The entries $s_{i,j}$ are called the scattering coefficients.
It is convenient to introduce the notation~ 
$\phi_\pm = (\phi_{\pm,1},\phi_{\pm,2})$, as before.
With this notation, the scattering relation~\eref{e:scattering} becomes the columns system
\[
\nonumber
\phi_{+,1} = s_{1,1}\phi_{-,1} + s_{2,1}\phi_{-,2}\,,\qquad
\phi_{+,2} = s_{1,2}\phi_{-,1} + s_{2,2}\phi_{-,2}\,,
\]
Moreover, Eqs.~\eref{e:detYpm} and \eref{e:detphipm} also imply
\[
\nonumber
\det S(\zeta,z) = 1\,. 
\]
The reflection coefficients that will be needed in the inverse problem are
\[
b(\zeta,z) = s_{2,1}/s_{1,1}\,,\qquad
\~b(\zeta,z) = s_{1,2}/s_{2,2}\,,\qquad
\zeta\in\Sigma\,.
\label{e:reflcoeffdef}
\]
Note that $\Wr\phi_- = \Wr Y_\pm= 2\gamma(\gamma-\lambda)/q_o^2 \ge 0$ from Eq.~\eref{e:detphipm},
where $\Wr$ denotes the Wronskian of a matrix.
Using the scattering relation~\eref{e:scattering}, we obtain
\begin{gather} 
\label{e:Swronskian1}
s_{1,1}(\zeta,z) = \Wr(\phi_{+,1},\phi_{-,2})/\det Y_\pm\,,\qquad
s_{1,2}(\zeta,z) = \Wr(\phi_{+,2},\phi_{-,2})/\det Y_\pm\,,
\\
\label{e:Swronskian2}
s_{2,1}(\zeta,z) = \Wr(\phi_{-,1},\phi_{+,1})/\det Y_\pm\,,\qquad
s_{2,2}(\zeta,z) = \Wr(\phi_{-,1},\phi_{+,2})/\det Y_\pm\,.
\end{gather}
So, 
from the analyticity properties~\eref{e:analyticregion} of the eigenfunctions, 
one can show that the diagonal entries of the scattering matrix are analytic in the following regions in the complex plane,
\[
\label{e:sanalytic}
s_{1,1}:\quad \Gamma^+\,,\qquad s_{2,2}: \quad \Gamma^-\,.
\]
As usual~\cite{JMP55p031506}, the off-diagonal scattering coefficients are nowhere analytic in general.

We can also write an integral representation of the scattering matrix $S(\zeta,z)$ from Eq.~\eref{e:scattering},
taking the limit of the relation~\eref{e:scattering} 
as $t\to-\infty$ and using the integral equations~\eref{e:muinteqns1} and~\eref{e:muinteqns2}, to get
\[
S(\zeta,z) = \lim_{t\to-\infty}
  \e^{-i\gamma t\sigma_3}Y_-^{-1} Y_+ \e^{i\gamma t\sigma_3}\bigg( I
     - \int_t^\infty \e^{-i\gamma\tau \sigma_3}Y_+^{-1}\Delta Q_+(\tau,t)\phi_+(\tau,t,\zeta)\,\d \tau \bigg)\,.
\label{e:Sintrepres}
\]
Note however that care is necessary to evaluate Eq.~\eref{e:Sintrepres}, 
since (unlike the case of ZBG) one cannot separate the two terms in parentheses 
(because the individual terms do not admit separate limits).

\subsection{Symmetries and discrete spectrum}
\label{s:symmetries}

\paragraph{Symmetries.}
The symmetries of the eigenfunctions and scattering coefficients 
are complicated by: 
(i) the presence of the Riemann surface, 
because, while with ZBG one simply deals with $\lambda\mapsto\lambda^*$, 
here one must also keep track of the sheets of the Riemann surface; 
(ii) after removing the asymptotic oscillations, 
the Jost solutions do not tend to the identity matrix as $t\to\pm\infty$.

Recall from Sections~\ref{s:background} and~\ref{s:uniformization} that~ $\gamma_\II(\lambda)= - \gamma_\I(\lambda)$,~
$\zeta=\lambda+\gamma$,~ $q_o^2/\zeta= \gamma-\lambda$,~ 
$\gamma= (\zeta+q_o^2/\zeta)/2$,~ $\lambda= (\zeta-q_o^2/\zeta)/2$.
One can consider several transformations compatible with Eq.~\eref{e:gammadef}.
As we show next, three of them correspond to actual symmetries of the scattering problem.
Of these, two are independent:
\\
\hbox{ }~ 1. $\zeta\mapsto \zeta^*$~ (upper/lower half plane), ~implying~ $(\lambda,\gamma)\mapsto(\lambda^*,\gamma^*)$~ (same sheet);
\\
\hbox{ }~ 2. $\zeta\mapsto -q_o^2/\zeta$~ (outside/inside $C_o$), ~implying~ $(\lambda,\gamma)\mapsto(\lambda,-\gamma)$~ (opposite sheets).

In Appendix~\ref{A:symmetries}, we show the following is true:

1. If $\phi(t,z,\zeta)$ is a solution of the scattering problem~\eref{e:scatteringproblem}, we have
\[
\nonumber
\phi_\pm^{-1}(t,z,\zeta) = \phi_\pm^\dag(t,z,\zeta^*)\,,\qquad
\zeta\in\Sigma\,,
\]
where $\dagger$ denotes Hermitian adjoint.
Explicitly, the columns satisfy, 
\[
\phi_{\pm,1}(t,z,\zeta)= \sigma_*\phi_{\pm,2}^*(t,z,\zeta^*)\,,\qquad 
\phi_{\pm,2}(t,z,\zeta)= -\sigma_*\phi_{\pm,1}^*(t,z,\zeta^*)\,,\qquad \forall\zeta\in\Sigma\,,
\label{e:phisymm1a}
\]
where $\sigma_*$ is defined by
\[
\sigma_*= \begin{pmatrix} 0 &1\\ -1 &0\end{pmatrix}\,. 
\label{e:sigma*}
\]
Similarly, for the scattering matrix, we have 
\[
S^\dag(\zeta^*,z) =  S^{-1}(\zeta,z)\,,\qquad \zeta\in\Sigma\,.
\label{e:S1symm1}
\]
We therefore have the following relations between the scattering coefficients:
\[
s_{2,2}(\zeta,z) = s_{1,1}^*(\zeta^*,z)\,,\qquad
s_{1,2}(\zeta,z) = -s_{2,1}^*(\zeta^*,z)\,,\qquad \forall\zeta\in\Sigma\,.
\label{e:S1symmcoeff}
\]
2. If $\phi(t,z,\zeta)$ is a solution of the scattering problem,
$\forall \zeta\in\Sigma$ it is
\[
\phi_\pm(t,z,\zeta)= (i/\zeta)\,\phi_\pm(t,z,-q_o^2/\zeta)\,\sigma_3Q_\pm\,.
\label{e:phisymm2}
\]
Recalling Eq.~\eref{e:sQJs}, 
we have, for the columns:
\[
\phi_{\pm,1}(t,z,\zeta)= (iq_\pm^*/\zeta)\,\phi_{\pm,2}(t,z,-q_o^2/\zeta)\,,\qquad
\phi_{\pm,2}(t,z,\zeta)= (iq_\pm/\zeta)\,\phi_{\pm,1}(t,z,-q_o^2/\zeta)\,.
\label{e:phisymm2a}
\]
Again for the scattering matrix we have, $\forall \zeta\in\Sigma$,
\[
S(-q_o^2/\zeta,z) = \sigma_3Q_-\, S(\zeta,z)\, (\sigma_3Q_+)^{-1}\,.
\label{e:S2symm}
\]
Recalling Eq.~\eref{e:sQJs}
we then have, elementwise,
\begin{gather}
\label{e:S2symmcoeff1}
s_{1,1}(\zeta,z) = (q_+^*/q_-^*)\,s_{2,2}(-q_o^2/\zeta,z)\,,\qquad
s_{1,2}(\zeta,z) = (q_+/q_-^*)\,s_{2,1}(-q_o^2/\zeta,z)\,,\\
\label{e:S2symmcoeff2}
s_{2,1}(\zeta,z) = (q_+^*/q_-)\,s_{1,2}(-q_o^2/\zeta,z)\,,\qquad
s_{2,2}(\zeta,z) = (q_+/q_-)\,s_{1,1}(-q_o^2/\zeta,z)\,,
\end{gather}
In Appendix~\ref{A:symmetries}, we also consider the combination of the two symmetries, i.e., 1+2.
The result for the scattering matrix is
$S^*(\zeta^*,z) = - \sigma_*(\sigma_3Q_-)^{-1}\,S(-q_o^2/\zeta,z)\,\sigma_3Q_+\sigma_*$. 
Or, elementwise,
\begin{gather}
\label{e:S3symmcoeff1}
s_{1,1}^*(\zeta^*,z) = (q_+/q_-)\,s_{1,1}(-q_o^2/\zeta,z)\,,\qquad
s_{1,2}^*(\zeta^*,z) = -(q_+^*/q_-)\,s_{1,2}(-q_o^2/\zeta,z)\,,\\
\label{e:S3symmcoeff2}
s_{2,1}^*(\zeta^*,z) = -(q_+/q_-^*)\,s_{2,1}(-q_o^2/\zeta,z)\,,\qquad
s_{2,2}^*(\zeta^*,z) = (q_+^*/q_-^*)\,s_{2,2}(-q_o^2/\zeta,z)\,.
\end{gather}
Even though Eqs.~\eref{e:S1symm1} and~\eref{e:S2symm} are only valid when $\zeta\in\Sigma$, 
the scattering coefficients $s_{1,1}$ and $s_{2,2}$ in Eqs.~\eref{e:S1symmcoeff},~\eref{e:S2symmcoeff1},~\eref{e:S2symmcoeff2},~\eref{e:S3symmcoeff1} and~\eref{e:S3symmcoeff2} are analytic in appropriate regions 
[cf. Eq.~\eref{e:sanalytic}].
Hence those three equalities with $s_{1,1}$ and $s_{2,2}$
can be extended uniquely to the appropriate regions shown in Eq.~\eref{e:sanalytic} of the $\zeta$-plane via the Schwartz reflection principle.

The above symmetries yield immediately the symmetries for the reflection coefficients:
\[
b(\zeta,z) = - \~b^*(\zeta^*,z) = (q_-/q_-^*)\,\~b(-q_o^2/\zeta,z)= -(q_-^*/q_-)\,b^*(-q_o^2/\zeta^*,z)\,
\qquad\forall \zeta\in\Sigma\,.
\label{e:bsymm}
\]
Recall that the two reflection coefficients $b$ and $\tilde b$ are defined in Eq.~\eref{e:reflcoeffdef}.

Note that:
\\[1ex]
(i) Unlike the case of ZBG, where there is only one symmetry, here there are two. 
\\
One of them, $\zeta\mapsto \zeta^*$, is the same as for ZBG; the other, $\zeta\mapsto-q_o^2/\zeta$, relates to the definition of $\gamma$.
\\[1ex]
(ii) Unlike the case of ZBG, and unlike the defocusing NLS with NZBC, 
here even the symmetries of the non-analytic scattering coefficients involve the map $\zeta\mapsto \zeta^*$.
This is because here the continuous spectrum is not just a subset of the real $\zeta$-axis.
\\[1ex]
(iii) The second involution, $\zeta\mapsto-q_o^2/\zeta$, simply expresses the switch from one sheet to the other.
(The corresponding transformation for the defocusing NLS has no minus sign because of the different location of the branch points.)
This transformation does not affect~$\lambda$.
That is, if $f(\lambda)$ is any single-valued function of $\lambda$, it is $f_\I(\lambda)= f_\II(\lambda)$.
Thus, $f$, when expressed as a function of $\zeta$, satisfies the symmetry $f(\zeta)= f(-q_o^2/\zeta)$.
That is because $f$ depends not on $\zeta$ directly, but only through the combination $\lambda= (\zeta-q_o^2/\zeta)/2$.
So, we have $\rho(t,z,\zeta)= \rho(t,z,-q_o^2/\zeta)$ and $V(t,z,\zeta)= V(t,z,-q_o^2/\zeta)$. 
More generally, Eqs. \eref{e:phisymm2},~\eref{e:S2symmcoeff1} and~\eref{e:S2symmcoeff2} relate the values of the 
Jost eigenfunctions and scattering coefficients on opposite sheets of the Riemann surface.

\paragraph{Discrete spectrum and residue conditions.}

The discrete spectrum of the scattering problem is the set of all values $\zeta\in\Complex\setminus\Sigma$ such that 
decaying eigenfunctions exist.
As usual, the discrete spectrum plays an important role in the inverse problem (see Section~\ref{s:rhp}).
We next show that these values are the zeros of $s_{1,1}(\zeta,z)$ in $\Gamma^+$ and those of $s_{2,2}(\zeta,z)$ in~$\Gamma^-$.
Note that, unlike defocusing NLS, we cannot exclude the possible presence of zeros along $\Sigma$.
Such zeros would give rise to the so-called ``embedded'' eigenvalues.
In this work we ignore this possibility.

Let $\zeta_1,\dots,\zeta_N$ be the zeros of $s_{1,1}(\zeta,z)$ on the portion of $\Gamma^+$ lying in the upper-half plane (UHP).
We assume these $N$ zeros are simple.
That is, we assume $s_{1,1}(\zeta_n,t)=0$ and $s'_{1,1}(\zeta_n,t)\ne0$, 
with $|\zeta_n|>1$ and $\Im \zeta_n>0$ for $n=1,\dots,N$,
and where the prime denotes differentiation with respect to $\zeta$.

Recalling the asymptotic behavior of the individual columns of $\phi_\pm$ as $t\to\pm\infty$ and the
fact that $\Im\gamma(\zeta)\gl0$ for $\zeta\in \Gamma^\pm$, 
we have that $\forall \zeta\in \Gamma^+$, $\phi_{+,1}(t,z,\zeta)\to0$ as $t\to\infty$, 
and $\phi_{-,2}(t,z,\zeta)\to0$ as $t\to-\infty$.
Recalling Eq.~\eref{e:Swronskian1}, however, if $s_{1,1}(\zeta,z)=0$ at $\zeta=\zeta_n$ the eigenfunctions 
$\phi_{+,1}$ and $\phi_{-,2}$ at $\zeta=\zeta_n$ must be proportional, i.e.:
\[
\phi_{+,1}(t,z,\zeta_n) = b_n\,\phi_{-,2}(t,z,\zeta_n)\,
\label{e:norming1a}
\]
with $b_n$ dependent on $z$ in general, but independent of $t$ and $\zeta$.
We therefore have an eigenfunction that decays as $ t\to\pm\infty$.

\begin{figure}[t!]
    \centering
    \includegraphics[width=0.35\textwidth]{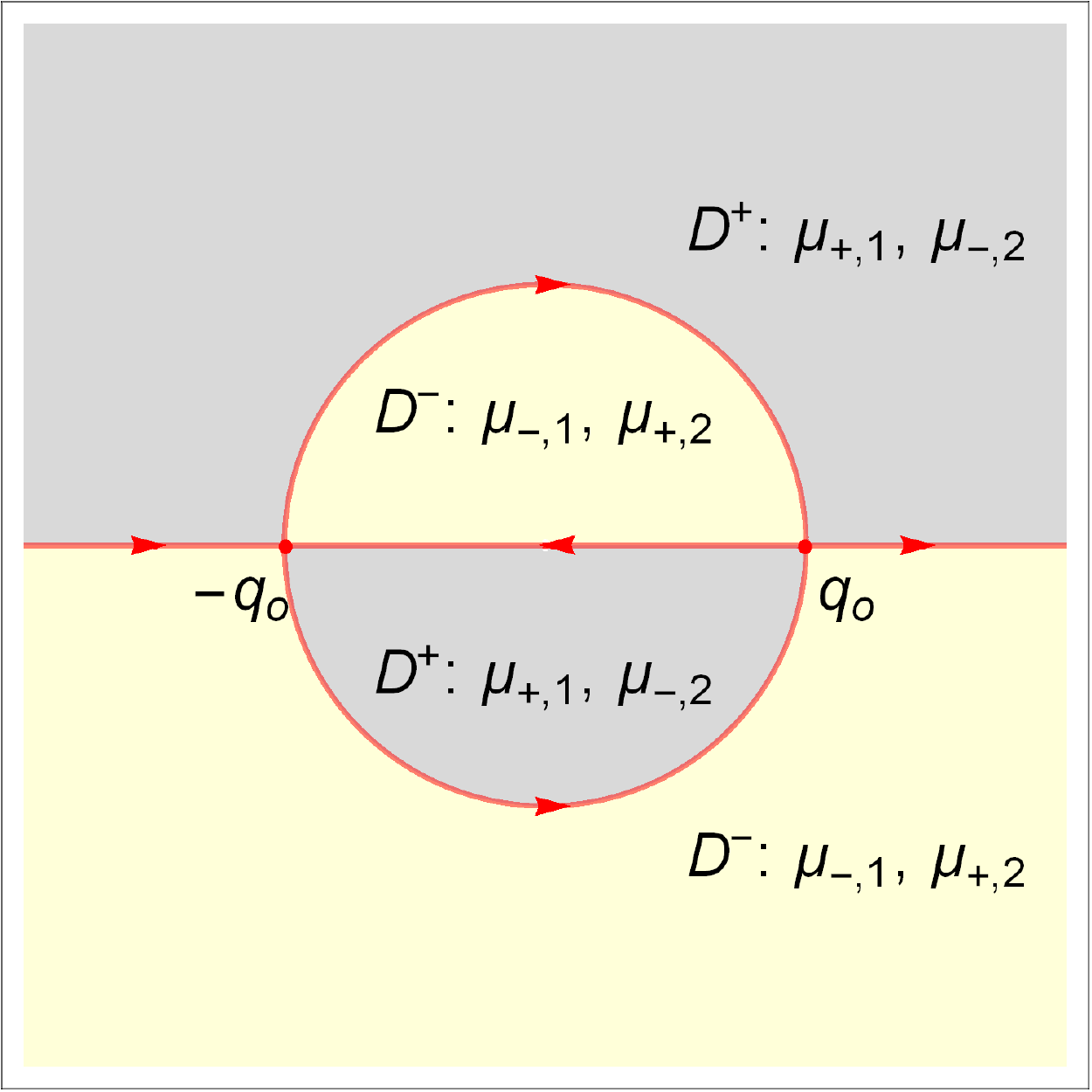}\hspace{8ex}
    \includegraphics[width=0.35\textwidth]{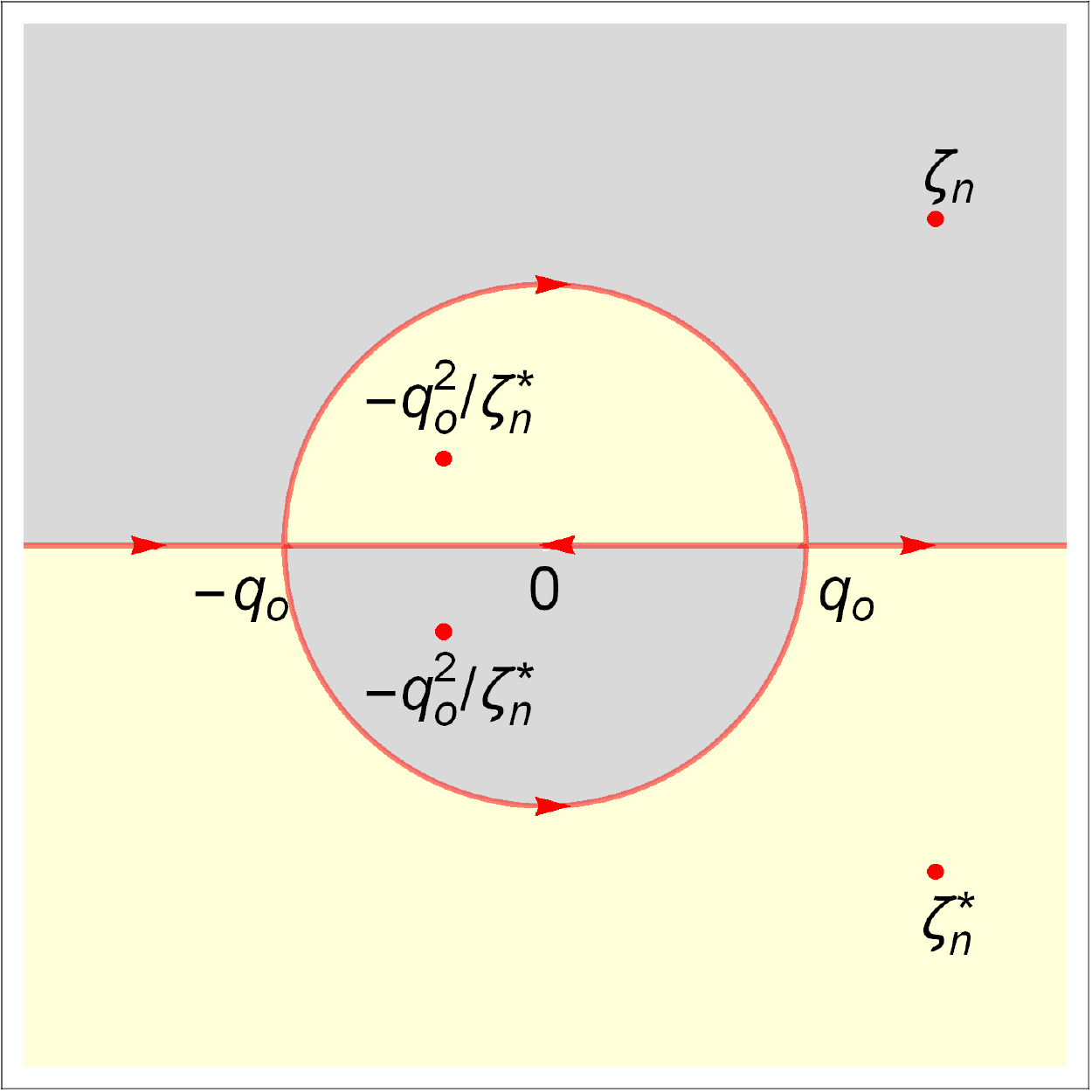}
    \caption{
        In both plots, gray and yellow regions correspond to $\Gamma^\pm$, respectively.
        Left: the analyticities of all eigenfunctions in the $\zeta$-plane.
        Right: a quartet of discrete eigenvalues $\zeta_n$.
    }
    \label{f:analyticity}
\end{figure}

Owing to the symmetries~\eref{e:S1symmcoeff},~\eref{e:S2symmcoeff1},~\eref{e:S2symmcoeff2},~\eref{e:S3symmcoeff1} and~\eref{e:S3symmcoeff2}, we have that 
\[
\nonumber
s_{1,1}(\zeta_n,t)=0 ~\Leftrightarrow~ 
s_{2,2}(\zeta_n^*,t)=0 ~\Leftrightarrow~ 
s_{2,2}(-q_o^2/\zeta_n,t)=0 ~\Leftrightarrow~ 
s_{1,1}(-q_o^2/\zeta_n^*,t)=0.
\]
For each $n=1,\dots,N$ we therefore have a quartet of discrete eigenvalues.
That is, the discrete spectrum is the set
\[
\nonumber
\{\zeta_n,\zeta_n^*,-q_o^2/\zeta_n,-q_o^2/\zeta_n^*\}_{n=1}^N\,.
\]
This is similar to what happens to the focusing and defocusing vector NLS with NZBC and scalar focusing NLS with NZBC.
[Instead, for the scalar NLS, in the focusing case (with ZBC) and in the defocusing case (with NZBC)
one has symmetric pairs, respectively in the $\lambda$ plane and in the $\zeta$ plane.]
The symmetries among a quartet can be seen in Fig.~\ref{f:analyticity}(right).

Next we derive the residue conditions.
We can write relation~\eref{e:norming1a} equivalently as
\[
\nonumber
\mu_{+,1}(t,z,\zeta_n) = b_n\e^{-2i\gamma(\zeta_n)t}\mu_{-,2}(t,z,\zeta_n)\,.
\]
Thus, 
\[
\Res_{\zeta=\zeta_n}\big[ \mu_{+,1}(t,z,\zeta)/s_{1,1}(\zeta,z) \big] =  
  C_n\,\e^{-2i\gamma(\zeta_n)t}\mu_{-,2}(t,z,\zeta_n)\,,
\label{e:Res1}
\]
where $C_n = b_n/s_{1,1}'(\zeta_n,z)$ is called the norming constant.
Again, prime denotes differentiation with respect to $\zeta$.
Equation~\eref{e:Res1} is the first of the residue conditions that will be used in the inverse problem.
Similarly, from Eq.~\eref{e:Swronskian2} we have that, 
if $s_{2,2}(\zeta_n^*,z)=0$ it is
\[
\phi_{+,2}(t,z,\zeta_n^*) = \~b_n\,\phi_{-,1}(t,z,\zeta_n^*)\,.
\label{e:norming1b}
\]
Or, equivalently, 
$\mu_{+,2}(t,z,\zeta_n^*) = \~b_n\e^{2i\gamma(\zeta_n^*)t}\mu_{-,1}(t,z,\zeta_n^*)$,
and as a result with $\~C_n = \~c_n/s_{2,2}'(\zeta_n^*,z)$,
\[
\nonumber
\Res_{\zeta=\zeta_n^*}\big[\mu_{+,2}(t,z,\zeta)/s_{2,2}(\zeta,z) \big]
  = \~C_n\,\e^{2i\gamma(\zeta_n^*)t}\mu_{-,1}(t,z,\zeta_n^*)\,.
\]
The above norming constants are related by the symmetries of the problem.
Applying relation~\eref{e:phisymm1a} to Eq.~\eref{e:norming1a} and comparing with Eq.~\eref{e:norming1b} 
one easily obtains $\~b_n = - b_n^*$.
It is also easy to see that symmetry~\eref{e:S1symmcoeff} implies $s_{1,1}'(\zeta_n,z)= (s_{2,2}'(\zeta_n^*,z))^*$.
Hence we have, $\forall n=1,\dots,N$,
\[
\~C_n = -C_n^*\,.
\label{e:Csymm}
\]
Finally, we need to discuss the remaining two points of the eigenvalue quartet.
Applying symmetry~\eref{e:phisymm2a} to Eqs.~\eref{e:norming1a} and~\eref{e:norming1b} we have the relations
\begin{gather*}
\phi_{+,2}(t,z,-q_o^2/\zeta_n) = (q_-/q_+^*)\,b_n\,\phi_{-,1}(t,z,-q_o^2/\zeta_n)\,,
\\
\phi_{+,1}(t,z,-q_o^2/\zeta_n^*) = (q_-^*/q_+)\,\~b_n\,\phi_{-,2}(t,z,-q_o^2/\zeta_n^*)\,.
\end{gather*}
Moreover, differentiating symmetries~\eref{e:S2symmcoeff1} and~\eref{e:S2symmcoeff2}, 
using  Eq.~\eref{e:S1symmcoeff}, and evaluating at $\zeta=\zeta_n$ or $\zeta=\zeta_n^*$, 
we have
\begin{gather*}
s_{1,1}'(-q_o^2/\zeta_n^*,z) = (\zeta_n/q_o)^2(q_-/q_+)\,(s_{1,1}'(\zeta_n,z))^*\,,\\
s_{2,2}'(-q_o^2/\zeta_n,z) = (\zeta_n/q_o)^2(q_-/q_+)^*\,(s_{2,2}'(\zeta_n^*,z))^*\,.
\end{gather*}
Combining these relations, we then have
\begin{gather}
\kern-0.2em
\Res_{\zeta=-q_o^2/\zeta_n^*}\big[\mu_{+,1}(t,z,\zeta)/s_{1,1}(\zeta,z) \big] = C_{N+n}\,\e^{-2i\gamma(-q_o^2/\zeta_n^*)t}\mu_{-,2}(t,z,-q_o^2/\zeta_n^*)\,,
\label{e:Res3a}
\\
\kern-0.2em
\Res_{\zeta=-q_o^2/\zeta_n}\big[ \mu_{+,2}(t,z,\zeta)/s_{2,2}(\zeta,z) \big] = \~C_{N+n}\,\e^{2i\gamma(-q_o^2/\zeta_n)t}\mu_{-,1}(t,z,-q_o^2/\zeta_n)\,,
\label{e:Res3b}
\end{gather}
where for brevity we defined 
\[
C_{N+n} = (q_o/\zeta_n^*)^2(q_-^*/q_-)\,\~C_n\,,\qquad 
\~C_{N+n} = (q_o/\zeta_n)^2(q_-/q_-^*)\,C_n\,,\qquad
n=1,\dots,N\,.
\label{e:symmnormconstdef}
\]
Note that $\~C_{N+n} = - C_{N+n}^*$, consistently with symmetry~\eref{e:Csymm}.

Hereafter it will be convenient to define $\zeta_{N+n} = - q_o^2/\zeta_n^*$ for $n=1,\dots,N$,
so that $\zeta_1,\dots,\zeta_{2N}$ are all the discrete eigenvalues in $\Complex^+$.

\subsection{Asymptotics as $\zeta\to0$ and $\zeta\to\infty$}
\label{s:asymp}

As usual, the asymptotic properties of the eigenfunctions and the scattering matrix
are instrumental in properly normalizing the RHP when we formulate the inverse problem later.
Moreover, the next-to-leading-order behavior of the eigenfunctions will allow us to 
reconstruct the potential from the solution of the RHP.
Again, the asymptotics with NZBG is more complicated than in the case of ZBG.
The calculations are somewhat simpler and cleaner, however, in the uniformization variable.
Note that the limit $\lambda\to\infty$ corresponds to $\zeta\to\infty$ in $\Complex_\I$ and to $\zeta\to0$ in $\Complex_\II$.
It is necessary to study both of these limits to normalize the RHP.

Recall from Eqs.~\eref{e:Ydef} and~\eref{e:YdetYinv} that the asymptotic eigenvector matrices are
$Y_\pm = I+(i/\zeta)\sigma_3 Q_\pm$ and 
$Y_\pm^{-1} = (I - (i/\zeta) \sigma_3Q_\pm)/(1+(q_o/\zeta)^2)$.
Also recall that $\gamma(\zeta)= \half(\zeta+q_o^2/\zeta)$.
Consider now the following formal expansion for $\mu_-(t,z,\zeta)$:
\begin{equation}
\label{e:muasymp}
\mu_-(t,z,\zeta) = \sum_{n=0}^\infty \mu\O{n}(t,z,\zeta)\,,
\end{equation}
with
\begin{gather}
\mu\O0(t,z,\zeta)= Y_-\,,\\
\mu\O{n+1}(t,z,\zeta) = \int_{-\infty}^t\! Y_-\e^{i\gamma(\zeta)(t-y)\sigma_3}Y_-^{-1}\Delta Q_-(y,t)\mu\O{n}(y,t,\zeta)\e^{i\gamma(\zeta)(y-t)\sigma_3}\,\d y\,.
\end{gather}
Let $A_d$ and $A_o$ denote respectively the diagonal and off-diagonal 
parts of a matrix~$A$.

In Appendix~\ref{A:asymptotics},
we prove that Eq.~\eref{e:muasymp} provides an asymptotic expansion for the columns of $\mu_-(t,z,\zeta)$
as $\zeta\to\infty$ in the appropriate region of the $\zeta$-plane described by Eq.~\eref{e:analyticregion}, 
with $\forall m\in\Natural$ and
\begin{gather}
\label{e:muasymp_zeta2infty1}
\mu_d\O{2m} = O(1/\zeta^m)\,,\qquad
\mu_o\O{2m} =  O(1/\zeta^{m+1})\,,\\
\label{e:muasymp_zeta2infty2}
\mu_d\O{2m+1} = O(1/\zeta^{m+1})\,,\qquad
\mu_o\O{2m+1} =  O(1/\zeta^{m+1})\,.
\end{gather}
Explicitly, the result holds with $\Im \zeta\le0$ for the first column and $\Im \zeta\ge0$ for the second column.
A similar asymptotic behavior to Eqs.~\eref{e:muasymp_zeta2infty1} and~\eref{e:muasymp_zeta2infty2} holds for $\mu^+$, 
and can be derived in a similar way.

Next we consider the asymptotics as $\zeta\to0$.
We use the same formal expansion~\eref{e:muasymp}.
In Appendix~\ref{A:asymptotics},
we prove that Eq.~\eref{e:muasymp} provides an asymptotic expansion for the columns of $\mu_-(t,z,\zeta)$ as $\zeta\to0$
in their analytic regions of the $\zeta$-plane,
with $\forall m\in\Natural$,
\begin{gather}
\label{e:muasymp_zeta2zero1}
\mu_o\O{2m} = O(\zeta^{m-1})\,,\qquad
\mu_d\O{2m} =  O(\zeta^m)\,,\\
\label{e:muasymp_zeta2zero2}
\mu_o\O{2m+1} = O(\zeta^m)\,,\qquad
\mu_d\O{2m+1} =  O(\zeta^m)\,.
\end{gather}
Again, similar asymptotic behavior to Eqs.~\eref{e:muasymp_zeta2zero1} and~\eref{e:muasymp_zeta2zero2} holds for $\mu_+$, 
and can be derived in a similar way.

In particular, by computing explicitly the first five terms in expansion~\eref{e:muasymp} we have that, 
as $\zeta\to\infty$,
\begin{multline}
 \mu_-(t,z,\zeta) = I + (i/\zeta)\sigma_3Q(t,z)\\
  + (i/\zeta)\!\int_{-\infty}^t \!\big( [\sigma_3Q_-,\Delta Q_-(y,t)]
  + \Delta Q_-(y,t)\sigma_3\Delta Q_-(y,t) \big)\,\d y + O(1/\zeta^2)\,.
\label{e:muasymp3}
\end{multline}
Equation~\eref{e:muasymp3} will provide the means to reconstruct the scattering potential $Q(t,z)$ from the solution of the RHP in the inverse problem.

Finally, 
inserting the above asymptotic expansions for the Jost eigenfunctions into the Wronskian representations~\eref{e:Swronskian1} and~\eref{e:Swronskian2} one shows that,
as $\zeta\to\infty$ in the appropriate regions of the complex $\zeta$-plane,
\[
\nonumber
S(\zeta,z)= I + O(\zeta)\,.
\]
Explicitly, the above estimate holds with $\Im \zeta\ge0$ and $\Im \zeta\le0$ for $s_{1,1}$ and $s_{2,2}$, respectively, and
with $\Im \zeta=0$ for $s_{1,2}$ and $s_{2,1}$.
Similarly, one shows that, as $\zeta\to0$,
\[
S(\zeta,z)= \diag(q_-/q_+,q_+/q_-) + O(1/\zeta)\,,
\label{e:Sk20asymp}
\]
again in regions of the $\zeta$-plane from Eq.~\eref{e:sanalytic}.

\subsection{Asymptotics of the density matrix}
\label{s:boundaryvalues}

One can see (by direct substitution) that, as in the case of ZBG, 
if $\phi(t,z,\zeta)$ is any fundamental matrix solution of the scattering problem, it is
\[
\nonumber
\partialderiv{ }t[\phi^{-1}(t,z,\zeta)\rho(t,z,\zeta)\,\phi(t,z,\zeta)] = 0\,.
\]
Hence, the quantity in square brackets must be independent of $t$.
So, $\forall \zeta\in\Sigma$, we can define
\[
\rho_\pm(\zeta,z) = \phi_\pm^{-1}(t,z,\zeta)\rho(t,z,\zeta)\,\phi_\pm(t,z,\zeta)\,.
\label{e:rhopmdef}
\]
Conversely, 
\[
\rho(t,z,\zeta) = \phi_\pm(t,z,\zeta)\rho_\pm(\zeta,z)\,\phi_\pm^{-1}(t,z,\zeta)\,.
\label{e:rhophipmrel}
\]
Then, taking the limit as $t\to\pm\infty$ and using Eqs.~\eref{e:muinteqns1} and~\eref{e:muinteqns2}, we obtain
\[
\rho_\pm(\zeta,z) =  \lim_{t\to\pm\infty}\e^{-i\gamma t\sigma_3}Y_\pm^{-1}(\zeta)\rho(t,z,\zeta)Y_\pm(\zeta)\,\e^{i\gamma t\sigma_3}\,.
\label{e:rhopmasymp}
\]
However, note that the density matrix $\rho(t,z,\zeta)$ by itself does not have a limit. 
In other words, one should remember that despite the $\pm$ signs on the left-hand side of Eq.~\eref{e:rhopmdef},
$\rho_\pm(\zeta,z)$ are not simply the limits of $\rho(t,z,\zeta)$ as $t\to\pm\infty$.
Note also that, even though $\rho$ is a single-valued function of $\lambda$, 
the Jost solutions $\phi_-$ and $\phi_+$ are not.
Thus, like $\phi_\pm$, in general $\rho_\pm$ is only single-valued on $\Real$, 
unlike in the case of ZBG.
(In other words, $\rho_{\pm,\II}\ne\rho_{\pm,\I}$ in general.) 

One can also see by direct calculation that~ $S\,\rho_+ = \rho_-S = \phi_-^{-1}\rho\,\phi_+$.
Hence,
\[
\rho_+(\zeta,z) = S^{-1}(\zeta,z)\,\rho_-(\zeta,z)\,S(\zeta,z)\,,\qquad \forall \zeta\in\Sigma\,.
\label{e:rhopmrel}
\]
Equation~\eref{e:rhopmrel} relates the asymptotic values of the density matrix as $t\to\pm\infty$,
and allows one to obtain $\rho_+$ from the knowledge of $\rho_-$ and $S$
[which in turn is completely determined by $q(t,z)$]. 
Thus, one can only pick one of $\rho_\pm$,
after which the other one is fixed.
Due to the causality, we choose $\rho_-$.

Note that properties of the density matrix $\tr\rho=0$ and $\det\rho= -1$ imply that 
$\tr\rho_\pm=0$ and $\det\rho_\pm= -1$.
Note also that since $\rho^\dag=\rho$ if $\zeta\in\Real$, 
it is also $\rho_\pm^\dagger = \rho_\pm$ if $\zeta\in\Real$.
In the rest of this work, 
we assume that $\rho^\dagger = \rho(\zeta^*)$ when $\zeta\in C_o$,
which implies $\rho_\pm^\dagger = \rho_\pm(\zeta^*)$ for all $\zeta\in\Sigma$.
[Notice that this assumption will not affect our final solution $\rho$ since it is defined only on the real axis.]
Thus, we denote the entries of $\rho_\pm$ as
\begin{gather*}
\rho_\pm(\zeta,z) = \begin{pmatrix}
D_\pm & P_\pm \\ P^*_\pm & -D_\pm
\end{pmatrix}\,,\qquad\zeta\in\Sigma\,,
\end{gather*}
with $P_\pm(\zeta) = P_\pm(\zeta^*)$.
Note however that $D_\pm$ and $P_\pm$ are \textit{not} the limits of $D$ and $P$ as $t\to\pm\infty$.
(Indeed, as we show below, such limits do not exist in general.)
Combining the symmetry~\eref{e:phisymm2} and definition~\eref{e:rhopmdef}, we have
\[
\label{e:rhopmsymmetry}
\rho_\pm(-q_o^2/\zeta,z) = \sigma_3 Q_\pm\rho_\pm(\zeta,z)Q_\pm^{-1}\sigma_3\,,\qquad
\zeta\in\Sigma\,.
\]
This implies that one does not have freedom to pick the asymptotic state $\rho_\pm$ for all values of $\lambda$ (or $\zeta$).
One can only pick $\rho_\pm$ for $\lambda\in\Real$ on the first sheet [or $\zeta\in(-\infty,-q_o]\cup[q_o,\infty)$].
Thus, let us denote $\rho_{\pm,\I}$ and $\rho_{\pm,\II}$ to the values of the asymptotic state on the first or second sheet, respectively.
Then, we define the asymptotic state on the two $\lambda$-sheets as
\begin{gather}
\label{e:rhopm=cdotsigma1}
D_{\pm,\I}\in\Real\,,\qquad
P_{\pm,\I}\in\Complex\,,\qquad
D_{\pm,\I}^2 + |P_{\pm,\I}|^2 = 1\,,\\
\label{e:rhopm=cdotsigma2}
D_{\pm,\II} = -D_{\pm,\I}\,,\qquad
P_{\pm,\II} = q_\pm^2 P_{\pm,\I}^*/q_o^2\,,
\end{gather}
or, in the $\zeta$-plane,
\begin{gather}
\label{e:rhopm=cdotsigma3}
D_\pm\in\Real\,,\qquad
P_\pm\in\Complex\,,\qquad
D_\pm^2 + |P_\pm|^2 = 1\,,\\
\label{e:rhopm=cdotsigma4}
D_\pm(-q_o^2/\zeta,z) = -D_\pm(\zeta,z)\,,\qquad
P_\pm(-q_o^2/\zeta,z) = q_\pm^2 P_\pm^*(\zeta,z)/q_o^2\,,
\end{gather}
Equation~\eref{e:rhopmrel} can then be used to obtain $D_+$ and $P_+$ from $D_-$ and $P_-$.
Note that, in principle, $D_\pm$ and $P_\pm$ depend on $\lambda$ on each sheet.  
Here however we make the assumption that the preparation of the medium on each $\lambda$-sheet is
independent of $\lambda$, 
which is the simplest possible situation from a physical point of view.

Because the asymptotic state~\eref{e:rhopm=cdotsigma1},~\eref{e:rhopm=cdotsigma2},~\eref{e:rhopm=cdotsigma3} and~\eref{e:rhopm=cdotsigma4}
 satisfies the symmetry~\eref{e:rhopmsymmetry},
we know that the calculations performed on both $\lambda$-sheets are equivalent.
Hereafter, we will focus on the first $\lambda$-sheet.

Equation~\eref{e:rhopmasymp} gives the explicit relation between $\rho_\pm$ and $\rho$.
In component form, it is
\begin{gather}
\label{e:DPpm1}
D_\pm = \frac{1}{\gamma}\lim_{t\to\pm\infty}[\lambda D - \Im(Pq^*)]\,,\\
\label{e:DPpm2}
P_\pm = \frac{1}{2\gamma}\lim_{t\to\pm\infty}\e^{-2 i \gamma  t}
\big(q^2 P^*/\zeta + 2i D q + \zeta P\big)\,.
\end{gather}
Conversely,
Eq~\eref{e:rhopmsymmetry} also implies that 
$\rho(t,z,\zeta) = Y_\pm\e^{i\gamma t \sigma_3}\rho_\pm(\zeta,z)\e^{-i\gamma t \sigma_3}Y_\pm^{-1} + o(1)$
as $t\to\pm\infty$,
which in component form, as $t\to\pm\infty$, is,
\begin{gather}
\label{e:DP=DPpm1}
D = \frac{\lambda}{\gamma}D_\pm - \frac{1}{\gamma}\Im(\e^{-2i\gamma t}P_\pm^* q_\pm) + o(1)\,,\\
\label{e:DP=DPpm2}
P = -\frac{i}{\gamma}q_\pm D_\pm + \frac{\zeta}{2\gamma}\e^{2i\gamma t}P_\pm + \frac{q_\pm}{2\gamma\zeta}\e^{-2i\gamma t}P_\pm^* + o(1)\,.
\end{gather}

Let us now discuss the implications of the relations \eref{e:DPpm1},~\eref{e:DPpm2},~\eref{e:DP=DPpm1} and~\eref{e:DP=DPpm2}.   
First, just as in the case of ZBG, 
Eq.~\eref{e:DP=DPpm2} implies that, in general, 
$P$ does not have a limit as $t\to\pm\infty$, 
but instead oscillates in time.  
Unlike in the case of ZBG, however, Eq.~\eref{e:DP=DPpm1} implies that, in general,  
due to the nonzero background radiation $q_\pm$,  
the quantity $D$ also does not have a limit in time and instead oscillates.  
Second, unlike in the case of ZBG, 
$P \ne0$ as $t\to\pm\infty$ even in the particular case in which $D$ and $P$ are time-independent and do tend to a limit as $t\to\infty$ or $t\to-\infty$, 
which is when $P_+=0$ or $P_-=0$, respectively.
The nonzero contribution arises from the polarization induced by the limiting electric field.  
(If one takes the limit $q_o\to0$, 
one of course recovers the relation $\lim_{t\to\pm\infty}D = D_\pm$ that holds in the case of ZBG.)   

To further elaborate on the above, let us focus on the asymptotic state at $t\to-\infty$. 
Recall that the normalization $\det\rho_- = -1$ implies the constraint 
\[\label{e:probone}
D_-^2 + |P_-|^2 = 1,
\]
which implies that one does not have the freedom to assign $D_-$ and $P_-$ independently.
In particular, in the special case when $\rho_-$ is diagonal, 
i.e., $D_- = \pm1$ and $P_- = 0$, we have
\[
\label{e:DPsoliton}
D = \frac{\lambda}{\gamma}D_- + o(1)\,,\qquad
P = -\frac{i}{\gamma}q_- D_- + o(1)\,,\qquad
t\to-\infty\,,
\]
which yields $P = -i q_- D/\gamma + o(1)$ as $t\to-\infty$.  
Therefore, 
the polarization in this case is entirely \emph{induced} by the limiting values of the electric field.   
On the other hand, the time-dependent terms in Eq.~\eref{e:DP=DPpm2} describe the \emph{intrinsic} part of the polarization, 
which is due to the excitation of the medium as $t\to-\infty$, 
and has parts proportional to $P_-$.   
Note that, generally, the induced polarization is still present in the initial state for the density matrix $\rho$ even in the presence of nonzero intrinsic polarization, 
except that it is reduced in magnitude due to the constraint~\eref{e:probone} together with $D_-$. 

Finally, we remark that, even in the absence of intrinsic initial polarization,
the medium is not in a pure state,
because of the initial induced polarization, which must exist (and so $\lim_{t\to-\infty}|P|\ne0$).   Instead,
\textit{initially the medium is necessarily in a superposition of two states}.
One can also obtain this result directly from 
$\lim_{t\to-\infty}|D| = |\lambda/\gamma|<1$ for any $q_o>0$.

\paragraph{The initial amount of population inversion.}

The above discussion shows that although the quantity $D_-$ is not the direct limit of $D$ even in an initial state with no intrinsic polarization,
it still provides an important indicator of the initial state of the medium.
In fact, \textit{$D_-$ is directly proportional to the amount of the initial population inversion in the medium},
i.e., it describes the amount of the population inversion allowed initially by the background electric field.    
More generally, \textit{$D_-$ is  proportional to the time-averaged amount of the initial population inversion in the medium},
i.e., it describes the amount of the time-averaged population inversion allowed initially by the background electric field and the intrinsic polarization.
Recall that the population inversion is defined as the difference between the populations of the atoms in the excited state and in the ground state.   
Thus, if $-1\le D_-<0$, initially there are more atoms in the ground state than in the excited state on average, 
while if $0 < D_- \le1$, initially there are more atoms in the excited state than in the ground state on average.   
As we show in Section~\ref{s:stability1}, the former condition leads to stable and the latter to unstable electric-field propagation through the medium.   
This is consistent with what happens to the case of ZBG ($D_\pm = \lim_{t\to\pm\infty}D$).

We further will see in Section~\ref{s:solitons} that pure soliton solutions can only exist when the initially intrinsic polarization of the medium vanishes.   
In the case when the ground state is more occupied initially than the excited state these solitons are subluminal and stable, and thus physical, 
while in the opposite case they are superluminal and unstable, and thus unphysical.

\section{Propagation}
\label{s:propagation}

Recall that in the MBEs, 
the traditional role of evolution variable is played by the 
the physical propagation distance $z$.  
As a result, 
we will refer to the $z$-dependence as the propagation.

\subsection{Evolution of the background}

Let us first discuss the $z$ dependence of the asymptotic values of the optical field.
The propagation of these values is given by the limit as $t\to\pm\infty$ of Eq.~\eref{e:dqdz}.
That is, 
\[
\nonumber
\partial Q_\pm/\partial z = -\half\lim_{t\to\pm\infty}\int[\sigma_3,\rho(t,z,\lambda)]\,g(\lambda)\,\d\lambda\,.
\]
Using Eq.~\eref{e:rhopmdef}, we can express the solution of Eq.~\eref{e:drhodt} as
\[
\rho(t,z,\zeta) = Y_\pm\,\e^{i\gamma t\sigma_3}\rho_\pm(\zeta,z)\,\e^{-i\gamma t\sigma_3}Y_\pm^{-1} + o(1)\,,\qquad
t\to\pm\infty\,.
\label{e:rho0def}
\]
Note that the off-diagonal elements of $\rho$ can never be zero as $t\to\pm\infty$,
even when the off-diagonal elements of $\rho_\pm$ are identically zero.
This is because of the background-induced polarization from Eqs.~\eref{e:DP=DPpm1} and~\eref{e:DP=DPpm2}.
Denote the diagonal and off-diagonal part of a matrix the subscripts $d$ and $o$, respectively.
In particular, let $\rho_\pm= \rho_{\pm,d}+\rho_{\pm,o}$.
Note that $[\sigma_3,\rho_{\pm,d}]= 0$ and $[\sigma_3,\rho_{\pm,o}]= 2\sigma_3\rho_{\pm,o}$.  
(Also, $Q_\pm \sigma_3+\sigma_3Q_\pm=0$.)
Using definitions \eref{e:rho0def} and~\eref{e:Ydef} we then have
\[
\nonumber
[\sigma_3,\rho]= [\sigma_3,Y_\pm\rho_{\pm,d}Y_\pm^{-1}] + 
[\sigma_3,Y_\pm\,e^{i\gamma t \sigma_3}\rho_{\pm,o}\e^{-i\gamma t \sigma_3}Y_\pm^{-1}] + o(1)\,,\qquad
t\to\pm\infty\,.
\]
All entries in the second term on the RHS, however, 
contain oscillating exponentials, which, when integrated, tend to zero in the limit
$t\to\pm\infty$ thanks to the Riemann-Lebesgue lemma.
Therefore, the effective part of $[\sigma_3,\rho]$ as $t\to\pm\infty$ is
\[
\nonumber
[\sigma_3,\rho]_\eff = [\sigma_3,Y_\pm\rho_{\pm,d}Y_\pm^{-1}] = 
  \frac1{\det Y_\pm} \big(\, [\sigma_3,\rho_{\pm,d}] + i\,[\sigma_3,[\sigma_3Q_\pm,\rho_{\pm,d}]] /\zeta
    + [\sigma_3,\sigma_3Q_\pm\rho_{\pm,d}\sigma_3Q_\pm]/\zeta^2\,\big) \,.
\]
Notice that the first and last commutators on the RHS vanish, because $\sigma_3$, 
$\rho_{\pm,d}$ and $\sigma_3Q_\pm\rho_{\pm,d}\sigma_3Q_\pm$ 
are all diagonal.
Finally, note that $[\sigma_3,[\sigma_3Q_\pm,\rho_{\pm,d}]]= 2[Q_\pm,\rho_{\pm,d}]$, 
and recall $\det Y_\pm= 2\gamma/\zeta$.

Evaluating Eq.~\eref{e:dqdz} in the limit $t\to\pm\infty$ 
(assuming the derivative with respect to $z$ and the limits as $t\to\pm\infty$ commute)
we then obtain the propagation of $Q_\pm$ with respect to~$z$:
\[
\partial Q_\pm/\partial z = iw_\pm[\sigma_3,Q_\pm]\,,
\label{e:dQpmdx}
\]
where
\[
w_\pm(z) = \half\, \int D_\pm(\lambda,z)\,g(\lambda)/\gamma\,\,\d\lambda\,.
\label{e:wpmdef}
\]
Recall that $\tr\,\rho_\pm= \tr\,\rho=0$ and $D_\pm$ has opposite signs on sheet I and sheet II from symmetries~\eref{e:rhopm=cdotsigma1},~\eref{e:rhopm=cdotsigma2},~\eref{e:rhopm=cdotsigma3} and~\eref{e:rhopm=cdotsigma4}.
Since $\gamma$ also takes opposite signs on both sheets,
it is easy to verify that $w_\pm$ is single-valued, as it should be.
Equation~\eref{e:dQpmdx} can be integrated to obtain 
\[
\label{e:q-timeevolution}
Q_\pm(z)= \e^{iW_\pm \sigma_3} Q_\pm(0)\,\e^{-iW_\pm \sigma_3}\,,\qquad
W_\pm(z) = \int_0^z\,w_\pm(z')\,\d z'\,,
\]
or simply,
\[
\nonumber
q_\pm(z) = \e^{2i W_\pm(z)} q_\pm(0)\,.
\]
Note that $|q_\pm(z)|^2 = |q_\pm(0)|^2= q_o^2$, $\forall z\in\Real^+$.
In particular, if $w_-$ is independent of $z$, 
taking $q_-(0) = q_o$ without loss of generality
we simply have
\[
\label{e:q-(z)}
q_-(z) = q_o \e^{2i w_- z}\,.
\]
Note also that Eq.~\eref{e:q-timeevolution} is similar to Eq.~\eref{e:backgroundstates_general}.
The difference is that Eq.~\eref{e:q-timeevolution} describes the background as $t\to-\infty$,
whereas Eq.~\eref{e:backgroundstates_general} describe the entire background solution for $\forall t\in\Real$.
The corresponding propagation of the asymptotic eigenvector matrices is given by
\[
\nonumber
Y_\pm(\zeta,z) = I+(i/\zeta)\sigma_3Q_\pm(z) = \e^{iW_\pm \sigma_3} Y_\pm(\zeta,0)\,\e^{-iW_\pm \sigma_3}\,,
\]
and the asymptotic behavior of the Jost solutions is 
\[
\phi_\pm(t,z,\zeta) = \e^{iW_\pm \sigma_3} Y_\pm(\zeta,0)\,\e^{i(\gamma t-iW_\pm)\sigma_3} + o(1)\,,\qquad
\mathrm{as}~t\to\pm\infty\,.
\label{e:phipmevol}
\]
Using relation~\eref{e:rhopmrel}, one can then express $w_+$ in terms of $\rho_{-,d}$.
In general, $w_+\ne w_-$, and therefore $q_+$ evolves at a different rate than $q_-$.
However, 
\[
\nonumber
(S^{-1}\rho_-S)_d= \rho_{-,d} + (s_{2,1}s_{2,2}P_-  - s_{1,1}s_{1,2}P_-^* )\,\sigma_3\,,
\]
which shows $w_+= w_-$ for reflectionless solutions (i.e., when $S_o=0$) or when $\rho_{-,o}=0$.
(We will see later that indeed $\rho_{-,o}=0$ is necessary for radiation to remain zero as a function of $z$.)

\subsection{Evolution of reflection coefficients and norming constants}
\label{s:time}

We now discuss the propagation of the reflection coefficients and norming constants.
As in the case with ZBG, this is the place where the
theory deviates the most from the IST for AKNS systems such as the focusing NLS equation with NZBC~\cite{JMP55p031506}.
Here the situation is further complicated by the nontrivial behavior of the 
Jost solutions as $t\to\pm\infty$.

\paragraph{Simultaneous solutions of the Lax pair and auxiliary matrix.}

Since the asymptotic behavior of $\phi_\pm(t,z,z)$ as $t\to\pm\infty$ is fixed,
in general they will not be solutions of Eq.~\eref{e:zevolution}.
Because both $\phi_+$ and $\phi_-$ are fundamental matrix solutions of the scattering problem,
any other solution $\Phi(t,z,\zeta)$ can be written as
\[
\Phi(t,z,\zeta) = \phi_+(t,z,\zeta)\,C_+(\zeta,z) = \phi_-(t,z,\zeta)\,C_-(\zeta,z)\,,\qquad \zeta\in\Sigma\,,
\label{e:Phidef}
\]
where $C_\pm(\zeta,z)$ are $2\times2$ matrices independent of $t$.
Suppose that $\Phi(t,z,\zeta)$ is a simultaneous solution of both parts of the Lax pair~\eref{e:scatteringproblem} and~\eref{e:zevolution},
so $\Phi_z = V\,\Phi$. 
Then some algebra shows that,
\[
\partialderiv{C_\pm}z = \txtfrac i2 R_\pm C_\pm\,, \qquad \forall \zeta\in\Sigma\,,
\label{e:dCdz0}
\]
where the auxiliary matrices $R_\pm$ are given by 
\[
\txtfrac i2 R_\pm(\zeta,z) = \phi_\pm^{-1}[V\,\phi_\pm - (\phi_\pm)_z]\,. 
\label{e:Rpmdef}
\]
It is not obvious a priori that the RHS of Eq.~\eref{e:Rpmdef} is independent of~$t$,
but Eq. \eref{e:dCdz0} shows it must be.
Also, even though $g(\lambda)$ and $\rho(t,z,\lambda)$ are only defined for $\lambda\in\Real$, 
definition~\eref{e:Rpmdef} can be evaluated $\forall \zeta\in\Sigma$.
But since $\phi_\pm$ are not single-valued functions of $\lambda$, neither are $C_\pm$ and $R_\pm$.
That is, in general we have $C_{\pm,\II}\ne C_{\pm,\I}$ and $R_{\pm,\II}\ne R_{\pm,\I}$.

Moreover, recalling definition \eref{e:rhopmdef}, we can write $V(t,z,\zeta)$ as
\[
V(t,z,\zeta) = \txtfrac{i\pi}2\H_\lambda[\phi_\pm(t,z,\zeta')\rho_\pm(\zeta',z)\phi_\pm^{-1}(t,z,\zeta')g(\lambda')]\,,
\label{e:Vrhopm}
\]
with shorthand notation $\zeta' = \zeta(\lambda')$.
As we show below,
Eqs.~\eref{e:Rpmdef} and~\eref{e:Vrhopm} allow one to compute the propagation of the reflection coefficients 
and norming constants, as in the case with ZBG.
Note that, even though the individual terms on the RHS of Eq.~\eref{e:Vrhopm} are only single-valued on the real line,
the whole RHS is a single-valued function for~$\lambda\in\Complex$.
In fact, one can use the symmetries of $\phi_\pm$ and $\rho_\pm$ to verify that
$V(t,z,\zeta)= V(t,z,-q_o^2/\zeta)$.

We next express $R_\pm$ in terms of known quantities for all $z\ge0$.
Recalling the definition~\eref{e:Rpmdef}, and under the assumption 
that $z$-derivatives and the limits as $t\to\pm\infty$ commute, 
we have
\begin{multline}
R_\pm(\zeta,z) = -2i\lim_{t\to\pm\infty} 
   \phi_\pm^{-1}[V\,\phi_\pm - (\phi_\pm)_z]
\\
  = \lim_{t\to\pm\infty}\bigg\{\e^{-i\gamma t\sigma_3}Y_\pm^{-1}(\lambda) 
     \H_\lambda
        [Y_\pm(\lambda')\e^{i\gamma't\sigma_3}\rho_\pm(\lambda',z)\e^{-i\gamma't\sigma_3}Y_\pm^{-1}(\lambda')g(\lambda')]\,Y_\pm(\lambda)\,\e^{i\gamma t\sigma_3}
\\
        - (\phi_\pm)^{-1}(\phi_\pm)_z\bigg\}\,,
\label{e:Rpmlim}
\end{multline}
with $\gamma' = \gamma(\lambda')$.
Above and throughout this subsection, 
the prime will denote the functional argument in the integrand of the Hilbert transform, \textit{not} derivative with respect to $\zeta$.

Note that: (i) unlike in the case of ZBG, the last term in the curly bracket does not vanish;
(ii) one cannot evaluate the terms in the curly brackets individually.
The detailed calculation is presented in Appendix~\ref{A:Rpm}, where we show that 
$R_\pm = R_{\pm,d} + R_{\pm,o}$, with 
\begin{gather}
R_{\pm,d}(\zeta,z) = 
 \pi\gamma\,\H_{\lambda}[ \rho_{\pm,d}'\,g(\lambda')/\gamma'] + 2w_\pm\,\sigma_3\,,\qquad \zeta\in\Complex\,,
\label{e:R_d_axis}
\\ 
R_{\pm,o}(\zeta,z) = 
\begin{cases}
\mp i\,\nu \pi\,g(\lambda)\rho_{\pm,o}(\lambda,z)\,\sigma_3\,, & \zeta\in\Real\,,\\
0\,, & \zeta\in C_o\,,
\end{cases}
\label{e:R_offd_axis}
\end{gather}
where $\nu = 1$ for $\zeta\in(-\infty,-q_o]\cup[q_o,\infty)$ and $\nu = -1$ for $\zeta \in (-q_o,q_o)$,
and where the subscripts $d$ and $o$ denote diagonal and off-diagonal parts, as before.
Equations~\eref{e:R_d_axis} and~\eref{e:R_offd_axis} in component form is
\begin{gather}
\label{e:Rij1}
R_{\pm,1,1} = \pvint \frac{\gamma + \lambda' - \lambda}{\gamma'(\lambda'-\lambda)}D_\pm'g(\lambda')\d \lambda'\,,\qquad
R_{\pm,2,2} = - R_{\pm,1,1}\,,\\
\label{e:Rij2}
R_{\pm,1,2} = \begin{cases}
\pm i \nu \pi g(\lambda)P_\pm\,,& \zeta\in\Real\,,\\
0\,, & \zeta\in C_o\,,
\end{cases}\qquad
R_{\pm,2,1} = \begin{cases}
\mp i \nu \pi g(\lambda)P_\pm^*\,,& \zeta\in\Real\,,\\
0\,, & \zeta\in C_o\,,
\end{cases}
\end{gather}
where $R_{\pm,i,j}$ the $(i,j)$-elements of $R_\pm$.
One should notice that in Eq.~\eref{e:R_d_axis}, 
both $\rho_{\pm,d}'$ and $\gamma'$ take opposite signs on sheet I and II,
so the Hilbert transform achieves a unique result on the two $\lambda'$-sheets, as it should.
Therefore, we conclude that the matrix $R_{\pm,d}$ is determined independently of 
the choice of the integration variable.

Note that, even though the calculation is considerably more involved than that in the case of ZBG, 
the final result appears relatively simple.
As in the case of ZBG, the matrices~\eref{e:R_d_axis} and~\eref{e:R_offd_axis} enable one to 
calculate the propagation equations for the reflection coefficients and norming constants.
Importantly, however, note that, since $R_\pm(\zeta,z)$ are defined in terms of a principal value integral, 
even when it admits an extension to the complex $\lambda$-plane,
their values are discontinuous across the real $\lambda$-axis,
both from above and from below.

In general, it is not possible to extend all the entries of $R_\pm$ off the continuous spectrum.
Some of the entries, however, \textit{can} be extended, as we show next.
Recalling Eq.~\eref{e:Rpmlim} and following the calculations in Appendix~\ref{A:Rpm},
the off-diagonal entries of $R_{\pm}$ can be written as
\[
\label{e:Ppm_o}
R_{\pm,o}(\zeta,z) = i\pi\nu\lim_{t\to\pm\infty}\H_\lambda \big[g(\lambda')\,
  \e^{-i(\gamma-\gamma')t\sigma_3}\rho_{\pm,o}(t,z,\lambda')\e^{i(\gamma-\gamma')t\sigma_3} \big]\,.
\]
We point out that Eq.~\eref{e:Ppm_o} is equivalent to Eq.~\eref{e:R_offd_axis} via Lemma~\ref{l:integral}.
For each matrix element, 
the Hilbert transform on the RHS of Eq.~\eref{e:Ppm_o} is analytic and bounded in the complex $\zeta$-plane wherever
the exponential inside tends to zero as $t\to\pm\infty$.
Hence, looking at the regions where $\Im\gamma\gl0$,
we have:
\begin{gather}
\label{e:RpmUHP1}
R_{+,1,2}(\zeta,z) = R_{-,2,1}(\zeta,z) = 0\,,\quad \forall \zeta\in\Complex_-\backslash C_o\,,\\
\label{e:RpmUHP2}
R_{-,1,2}(\zeta,z) = R_{+,2,1}(\zeta,z) = 0\,,\quad \forall \zeta\in\Complex_+\backslash C_o\,.
\end{gather}
Below we will show that, just as in the case of ZBG, these conditions are all that is necessary
to determine the propagation equations for the reflection coefficient and norming constants.

\paragraph{Evolution equations for the reflection coefficient.}

Since $\phi_+= \phi_-\,S$, 
Eq.~\eref{e:Phidef} implies,
\[
\nonumber
S(\zeta,z) = C_-(\zeta,z)C_+^{-1}(\zeta,z)\,,\qquad \forall \zeta\in\Sigma\,.
\]
Some algebra then shows that, 
\[
\partialderiv Sz = \frac i2 (R_-S - SR_+)\,,\qquad \forall \zeta\in\Sigma\,.
\label{e:dSdz}
\]
Recall that the reflection coefficients are $b= s_{2,1}/s_{1,1}$ and $\~b= s_{1,2}/s_{2,2}$,
with $\~b(\zeta,z) = - b^*(\zeta,z)$.
Let us introduce the matrix
\[
\nonumber
B(\zeta,z) = S_o S_d^{-1} = \begin{pmatrix} 0 &\~b \\ b &0 \end{pmatrix}\,, 
\]
where once more 
the subscripts $d$ and $o$ denote the diagonal and off-diagonal parts, respectively.
In Appendix~\ref{A:evolution} we show that
\[
 -2i\partialderiv Bz = \begin{cases}
  2R_{-,o} + [R_{-,d},B] + i\nu\,\pi\,g\,[B,\rho_{-,d}]\,\sigma_3\,, & \zeta\in\Real\,,\\
  [R_{-,d},B]\,, & \zeta\in C_o
\end{cases}
\label{e:dBdz2}
\]
where $\nu=1$ for $\zeta\in(-\infty,-q_o]\cup[q_o,\infty)$ and $\nu = -1$ for $\zeta\in(-q_o,q_o)$,
similarly to Eqs.~\eref{e:R_d_axis} and~\eref{e:R_offd_axis}.
(Note that this result is formally identical to that of the case with ZBG on sheet I.)
In particular,
\[
\partialderiv{b}z = -i A\,b - \nu\pi\,g(\lambda)\,P_-^*\,,
\label{e:dbdz}
\]
where $\nu = 1$ if $\zeta\in(-\infty,-q_o]\cup[q_o,\infty)$, 
$\nu = 0$ if $\zeta\in C_o$, and $\nu = -1$ if $\zeta\in(-q_o,q_o)$, and
\[
\label{e:Ypmdef}
A(\zeta,z) = \pvint \frac{\gamma +\lambda' - \lambda}{\gamma'(\lambda'-\lambda)}
    D_-'\, g(\lambda')\d\lambda' + i\,\nu\,\pi\, g(\lambda) D_-\,.
\]
Note that the propagation equation for the reflection coefficient is completely determined 
by the asymptotic values as $t\to-\infty$, as it should be.
Notice also that because of the symmetry of the reflection coefficient~\eref{e:bsymm},
in general, $b(\zeta,z)$ with $\zeta\in(-q_o,q_o)$ evolves differently from $b(\zeta,z)$ with $\zeta\in(-\infty,-q_o]\cup[q_o,\infty)$.
This can also be seen from Eq.~\eref{e:dbdz} with $\nu=\pm1$,
and from the fact that $D_-$ and $P_-$ take different values depending on $\zeta$.
However, everything is still consistent,
as $D_-$ and $P_-$ also satisfy the symmetries~\eref{e:rhopm=cdotsigma1},~\eref{e:rhopm=cdotsigma2},~\eref{e:rhopm=cdotsigma3} and~\eref{e:rhopm=cdotsigma4}.
Thus, again, we only focus on the first $\lambda$-sheet,
i.e., $\zeta\in(-\infty,-q_o]\cup[q_o,\infty)$.

\paragraph{Evolution equations for norming constants.}

Recall that $C_n$ is defined by Eq.~\eref{e:Res1}:
\[
C_n = \frac{b_n}{s_{1,1}'(\zeta_n)} = b_n \lim_{\zeta\to \zeta_n} \frac{\zeta-\zeta_n}{s_{1,1}(\zeta,z)}\,.
\label{e:dCdz01}
\]
In Appendix~\ref{A:evolution} we show that for $n = 1,\dots,N$,
\[
\partialderiv{C_n}z = -i\,R_{-,1,1}(\zeta_n,z)\,C_n\,,
\label{e:dCdz}
\]
where $R_{-,1,1}$ is given in Eq.~\eref{e:Rij1}.
Hereafter, one can use Eqs.~\eref{e:Csymm} and~\eref{e:symmnormconstdef} to obtain the propagation for $C_n$ with $n = N+1,\dots,2N$.
Like the propagation of the reflection coefficient, 
the propagation equation for the norming constant is completely determined 
by the asymptotic values as $t\to-\infty$, as it should be.
Also, in the ODE~\eref{e:dCdz}, the quantity $R_{-,1,1}$ is evaluated at $\zeta = \zeta_n\in \Gamma^+$,
i.e., in the first $\lambda$-sheet.
Therefore, norming constants $C_n$ are uniquely determined without any ambiguities.

\section{Inverse problem}
\label{s:inverseprob}

\subsection{Riemann-Hilbert problem and reconstruction formula}
\label{s:rhp}

The development of the inverse problem begins from the scattering relation~\eref{e:scattering}.
One can rewrite this scattering relation as 
$(\mu_{+,1},\mu_{+,2}) = (\mu_{-,1},\mu_{-,2})\,\e^{i\gamma t\sigma_3}S\,\e^{-i\gamma t\sigma_3}$.
Equivalently, one can write
\[
\label{e:scatt2}
\mu_{+,1}/s_{1,1} = \mu_{-,1} + b\,\e^{-2i\gamma t}\mu_{-,2}\,,
\qquad
\mu_{+,2}/s_{2,2} = \~b\,\e^{2i\gamma t}\mu_{-,1} + \mu_{-,2}\,,\qquad \forall \zeta\in\Sigma\,,
\]
where the $(t,z,\zeta)$-dependence is omitted for brevity and the reflection coefficients $b$ and $\~b$ are defined in Eq.~\eref{e:reflcoeffdef}.
One can view Eq. \eref{e:scatt2} as a relation between eigenfunctions analytic in $\Gamma^+$ and those in $\Gamma^-$.
Thus, we introduce the meromorphic matrices
\[
M^+(t,z,\zeta) = (\mu_{+,1}/s_{1,1},\,\mu_{-,2})\,,\qquad
M^-(t,z,\zeta) = (\mu_{-,1},\,\mu_{+,2}/s_{2,2})\,.
\label{e:Mpmdef}
\]
As mentioned in Section~\ref{s:jost}, subscripts $\pm$ indicate normalization as $t\to\pm\infty$, while superscripts $\pm$ distinguish between
analyticity in $\Gamma^+$ and $\Gamma^-$, respectively.
We then have,
\[
M^-(t,z,\zeta) = M^+(t,z,\zeta)\,(I-G(t,z,\zeta))\,,\qquad \forall \zeta\in\Sigma\,,
\label{e:RHP}
\]
where the jump matrix $G$ is
\[
G(t,z,\zeta) = \begin{pmatrix} 0 & -\e^{2i\gamma(\zeta)t} \~b(\zeta,z) \\
  \e^{-2i\gamma(\zeta)t}b(\zeta,z) & b(\zeta,z)\~b(\zeta,z)
  \end{pmatrix}\,.
\label{e:jump}
\]
Equation~\eref{e:RHP} is the jump condition for the desired RHP.
This RHP is solved explicitly in Appendix~\ref{A:RHPsoln}, where it is shown that the solution is
\begin{multline}
M(t,z,\zeta) = I + \frac{i}{\zeta}\sigma_3Q_- 
  + \sum_{n=1}^{2N}\bigg(\frac{\Res_{\zeta_n}M^+}{\zeta-\zeta_n} + \frac{\Res_{\zeta_n^*}M^-}{\zeta-\zeta_n^*}\bigg)
  + \frac1{2\pi i}\int_\Sigma \frac{M^+(t,z,\zeta')}{\zeta'-\zeta}\,G(t,z,\zeta')\,\d\zeta'\,,
\label{e:rhpsoln}
\end{multline}
with $\zeta_{N+n} = - q_o^2/\zeta_n^*$ for $n = 1,\dots,N$.
Note that the expressions for $M^+$ and $M^-$ are formally identical,
except for the fact that $M$ is evaluated in different regions of the complex plane.

With the relabeling in definition~\eref{e:symmnormconstdef}, 
the residue relations~\eref{e:Res1},~\eref{e:Res3a} and~\eref{e:Res3b} imply that 
only the first column of $M^+$ has a pole at $\zeta=\zeta_n$ with $n = 1,\dots,2N$, and its residue is proportional
to the second column of $M^+$ at that point, explicitly:
\begin{equation}
\label{e:rhpresidues1}
\Res_{\zeta = \zeta_n}M^+ 
    = \big( C_n(z)\,\e^{-2i\gamma(\zeta_n)t}\mu_{-,2}(t,z,\zeta_n) \,, 0 \big)
    = \big( C_n(z)\,\e^{-2i\gamma(\zeta_n)t}M_{2}^+(t,z,\zeta_n) \,, 0 \big) \,.
\end{equation}
Similarly,
\begin{equation}
\label{e:rhpresidues2}
\Res_{\zeta = \zeta_n^*}M^- 
    = \big( 0 \,, \~C_n(z)\,\e^{2i\gamma(\zeta_n^*)t}\mu_{-,1}(t,z,\zeta_n^*) \big)
    = \big( 0 \,, \~C_n(z)\,\e^{2i\gamma(\zeta_n^*)t}M_{1}^-(t,z,\zeta_n^*) \big) \,.
\end{equation}
We can therefore evaluate the second column of the solution~\eref{e:rhpsoln} at $\zeta=\zeta_n$, 
obtaining:
\[
\label{e:rhpdiscrete1}
\mu_{-,2}(t,z,\zeta_n) = \begin{pmatrix} iq_-/\zeta_n \\ 1 \end{pmatrix} 
  + \sum_{j=1}^{2N}\frac{\~C_j(z)\,\e^{2i\gamma(\zeta_j^*)t}}{\zeta_n-\zeta_j^*}\,\mu_{-,1}(t,z,\zeta_j^*)
  + \frac1{2\pi i}\int_\Sigma \frac{M^+(t,z,\zeta')}{\zeta'-\zeta_n}\,G_2(t,z,\zeta')\,\d\zeta'\,,
\]
for $n=1,\dots,2N$,
and where $G_2$ denotes the second column of $G$.
Similarly, we can evaluate the first column of the solution~\eref{e:rhpsoln} at $\zeta=\zeta_n^*$, 
obtaining:
\[
\label{e:rhpdiscrete2}
\mu_{-,1}(t,z,\zeta_n^*) = \begin{pmatrix} 1 \\ iq_-^*/\zeta_n^* \end{pmatrix}
  + \sum_{j=1}^{2N}\frac{C_j(z)\,\e^{-2i\gamma(\zeta_j)t}}{\zeta_n^*-\zeta_j}\,\mu_{-,2}(t,z,\zeta_j)
  + \frac1{2\pi i}\int_\Sigma \frac{M^+(t,z,\zeta')}{\zeta'-\zeta_n^*}\,G_1(t,z,\zeta')\,\d\zeta'\,,
\]
again for $n=1,\dots,2N$, and where $G_1$ denotes its first column.
Finally, evaluating the solution \eref{e:rhpsoln} with $\zeta\in\Sigma$, 
we obtain together with equations~\eref{e:rhpdiscrete1} and~\eref{e:rhpdiscrete2}, 
a closed linear system of algebraic-integral equations for the solution of the RHP.

The last remaining task in the inverse problem is to reconstruct the potential and the density matrix from the solution~\eref{e:rhpsoln} of the RHP.
It is easy to compute the asymptotic behavior of $M(t,z,\zeta)$ 
\begin{multline}
M(t,z,\zeta) = I + \frac1\zeta\bigg[i\sigma_3Q_- 
  + \sum_{n=1}^{2N} \big( \Res_{\zeta_n}M^+ + \Res_{\zeta_n^*}M^- \big)
\\[-1ex]
  - \frac1{2\pi i}\int_\Sigma M^+(t,z,\zeta')\,G(t,z,\zeta')\,\d\zeta'\,\bigg]
 + O(1/\zeta^2)\,,\qquad \zeta\to\infty\,,
\label{e:rhpsolnasymp}
\end{multline}
where the residues are given by Eqs.~\eref{e:rhpresidues1} and~\eref{e:rhpresidues2}.
Taking $M=M^+$ and comparing the $(1,2)$-element of Eq.~\eref{e:rhpsolnasymp} to the one of Eq.~\eref{e:muasymp3},
we then obtain the reconstruction formula for the potential:
\[
q(t,z) = q_- - i \sum_{n=1}^{2N}\~C_n(z)\,\e^{2i\gamma(\zeta_n^*)t}\,\mu_{-,1,1}(t,z,\zeta_n^*)
  + \frac1{2\pi}\int_\Sigma (M^+G)_{1,2}(t,z,\zeta')\,\d\zeta'\,,
\label{e:reconstruction}
\]
where $\mu_{-,2}(t,z,\zeta_n^*)$ and $M^+(t,z,\zeta)$ are obtained from the simultaneous solution~\eref{e:rhpsoln} for $\zeta\in\Sigma$ 
and Eqs.~\eref{e:rhpdiscrete1} and~\eref{e:rhpdiscrete2}.

To reconstruct the density matrix, one also starts from the solution~\eref{e:rhpsoln} and Eqs.~\eref{e:rhpdiscrete1} and~\eref{e:rhpdiscrete2}, 
then computes $M^-(t,z,\zeta)$ for $\zeta\in\Sigma$.
Recalling definition of $M^\pm$~\eref{e:Mpmdef}, we then have
\[
\nonumber
\mu_-(t,z,\zeta) = (M^-_1(t,z,\zeta),M^+_2(t,z,\zeta))\,,\qquad \forall \zeta\in\Sigma\,.
\]
In turn, from the definition~\eref{e:mudef} of $\mu$ and relation~\eref{e:rhophipmrel}, 
we obtain the density matrix as
\[
\rho(t,z,\zeta) = \mu_-(t,z,\zeta)\,\e^{i\gamma t\sigma_3}\rho_-(\zeta,z)\,\e^{-i\gamma t\sigma_3}\mu_-^{-1}(t,z,\zeta)\,,\qquad \forall \zeta\in\Sigma\,.
\label{e:densityreconstruction}
\]

Similarly to the NLS equation, it is also possible to derive a ``trace'' formula, which allows one to recover the analytic scattering coefficient
from the knowledge of the reflection coefficient and discrete eigenvalues, 
and also a so-called ``theta'' condition, which yields the 
asymptotic phase difference of the potential in terms of the discrete eigenvalues:
\begin{gather}
s_{1,1}(\zeta,z) = 
\exp\bigg(-\frac1{2\pi i}\int_\Sigma \frac{\log[1+|b(\zeta',z)|^2]}{\zeta-\zeta'}\d\zeta'\bigg)
\prod_{n=1}^N\frac{(\zeta-\zeta_n)(\zeta+q_o^2/\zeta_n^*)}{(\zeta-\zeta_n^*)(\zeta+q_o^2/\zeta_n)}\,,
\label{e:trace1}
\\
s_{2,2}(\zeta,z) = 
\exp\bigg(\frac1{2\pi i}\int_\Sigma \frac{\log[1+|b(\zeta')|^2]}{\zeta-\zeta'}\d\zeta'\bigg)
\prod_{n=1}^N\frac{(\zeta-\zeta_n^*)(\zeta+q_o^2/\zeta_n)}{(\zeta-\zeta_n)(\zeta+q_o^2/\zeta_n^*)}\,,
\label{e:trace2}
\end{gather}
as well as
\[
\arg(q_-/q_+) = 4\sum_{n=1}^N \arg \zeta_n+\frac1{2\pi i}
\int_\Sigma \log[1+|b(\zeta',z)|^2]\frac{\d\zeta'}{\zeta'}\,.
\label{e:theta}
\]
See Appendix~\ref{s:trace} for details.
In fact, the trace formulae~\eref{e:trace1} and~\eref{e:trace2} and the ``theta'' condition~\eref{e:theta} are exactly the same as the ones obtained in focusing NLS with NZBC~\cite{JMP55p031506}.

\subsection{Reflectionless potentials}
\label{s:solitons}

We now look at a special case of potentials $Q(t,z)$ for which the reflection coefficient $b(\zeta,z)$ vanishes identically.
Note that in order for the solution to remain reflectionless for $z>0$ it is not sufficient that $b(\zeta,0)=0$.
Indeed, the propagation equation~\eref{e:dbdz} for the reflection coefficient $b(\zeta,z)$ shows that $b(\zeta,z) = 0$ for $z>0$ only if $\rho_{-,o}(\zeta,z)=0$ $\forall z>0$.
We therefore assume that this is the case for the rest of this section.
Recalling the general asymptotic state~\eref{e:rhopm=cdotsigma1},~\eref{e:rhopm=cdotsigma2},~\eref{e:rhopm=cdotsigma3} and~\eref{e:rhopm=cdotsigma4} we then define the initial state for the medium as follows
\[
\label{e:solitonBC1}
\rho_-(\zeta,z) = \nu\, h_-\,\sigma_3\,,
\] 
where $\nu = 1$ for $\zeta\in(-\infty,-q_o]\cup[q_o,\infty)$, 
$\nu = -1$ for $\zeta\in(-q_o,q_o)$, and where $h_-=\pm1$ is related to the initial population inversion via Eq.~\eref{e:DPsoliton}.
Notice that the sign discontinuities introduced by $\nu$ are necessary,
because of the symmetry~\eref{e:rhopmsymmetry}.
Also recall that $h_- = \pm1$ indicates whether, initially, more atoms are in the excited state than in the ground state or vice versa,
respectively (cf.~the discussion at the end of Section~\ref{s:boundaryvalues}).
Again, we emphasize the fact, shown in Section~\ref{s:boundaryvalues}, that, due to the presence of background radiation,
the medium is always polarized, and 
as a result, pure states do not exist in general, 
which differs from the case of ZBG.
Moreover, without calculating the density matrix $\rho$ from the inverse problem explicitly,
we know that $D\to \lambda h_-/\gamma$ and $P \to -i h_- q_-/\gamma$ as $t\to-\infty$ from Eqs.~\eref{e:DP=DPpm1} and~\eref{e:DP=DPpm2}.
In other words, 
the initial state of the medium is independent of discrete eigenvalues and soliton types (discussed later), but is dependent on $\lambda$.

As usual, for reflectionless potentials there is no jump from $M^+$ to $M^-$ across the continuous spectrum, 
and the RHP therefore reduces to an algebraic system,
whose solution yields the soliton solutions of the nonlinear system.

It is convenient to introduce scalar functions
\[
\nonumber
c_j(\zeta,z) = \frac{C_j(z)}{\zeta-\zeta_j}\,\e^{-2i\gamma(\zeta_j)t},\qquad
j=1,\dots,2N.
\]
Also recall the symmetry~\eref{e:symmnormconstdef}, and $\gamma(\zeta^*)= \gamma^*(\zeta)$.
The algebraic system obtained from the RHP can then be expressed as
\begin{gather}
\mu_{-,2}(t,z,\zeta_j) = \begin{pmatrix}iq_-/\zeta_j\\1\end{pmatrix} - \sum_{l=1}^{2N} c_l^*(\zeta_j^*,z)\mu_{-,1}(t,z,\zeta_l^*)\,,
\qquad j=1,\dots,2N\,,
\label{e:linalgsyst1}
\\
\mu_{-,1}(t,z,\zeta_n^*) = \begin{pmatrix}1\\iq_-^*/\zeta_n^*\end{pmatrix}  
+ \sum_{j=1}^{2N} c_j(\zeta_n^*,z)\mu_{-,2}(t,z,\zeta_j)\,,\qquad
n=1,\dots,2N\,.
\label{e:linalgsyst2}
\end{gather}
Substituting Eq.~\eref{e:linalgsyst1} into Eq.~\eref{e:linalgsyst2} yields, for all $n=1,\dots,2N$,
\[
\mu_{-,1}(t,z,\zeta_n^*) = \begin{pmatrix}1\\iq_-^*/\zeta_n^*\end{pmatrix} 
  + \sum_{j=1}^{2N} c_j(\zeta_n^*,z) \begin{pmatrix}iq_-/\zeta_j\\1\end{pmatrix}
  - \sum_{j=1}^{2N}\sum_{l=1}^{2N} c_j(\zeta_n^*,z)c_l^*(\zeta_j^*,z)\,\mu_{-,1}(t,z,\zeta_l^*)\,.
\label{e:linalgsyst3}
\]
Note that only the first component of the above eigenfunctions is needed
in the reconstruction formula~\eref{e:reconstruction} for the potential.
Let us write the resulting system in matrix form.
Let $\@X= (X_1,\dots,X_{2N})^T$ and $\@B = (B_1,\dots,B_{2N})^T$, where
\[
\nonumber
X_n= \mu_{-,1,1}(t,z,\zeta_n^*)\,,\qquad
B_n = 1 + iq_- \sum_{j=1}^{2N}c_j(\zeta_n^*,z)/\zeta_j\,,\qquad
n=1,\dots,2N\,.
\]
Also, let us define the $2N\times2N$ matrix $A= (A_{n,l})$, where
\[
\nonumber
A_{n,l} = \sum_{j=1}^{2N} c_j(\zeta_n^*,z)c_l^*(\zeta_j^*,z)\,,
\qquad
n,l= 1,\dots,2N\,.
\]
The system~\eref{e:linalgsyst3} then becomes simply
$K\,\@X = \@B$, 
where 
\[
K = I + A= (\@K_1,\dots,\@K_{2N})\,.
\label{e:linsystMdef}
\]
The solution of the system is $X_n= \det\,K_n/\det K$
for $n=1,\dots,2N$, where the $2N\times2N$ matrices $K_1,\dots,K_{2N}$ are 
$K_n = (\@K_1,\dots,\@K_{n-1},\@B,\@K_{n+1},\dots,\@K_{2N})$.
Finally, upon substituting $X_1,\dots,X_{2N}$ into the reconstruction formula~\eref{e:reconstruction}, 
one can write the result for the potential compactly as
\[
q(t,z) = \frac{\det K^\mathrm{aug}}{\det K}\,, \qquad
K^\mathrm{aug} = \begin{pmatrix}q_- & i\,\@D^T \\ \@B & K \end{pmatrix}\,,
\label{e:Nsolitonsolution}
\]
with $\@D= (D_1,\dots,D_{2N})^T$ and $D_n = \~C_n(z)\,\e^{2i\gamma(\zeta_n^*)t}$ for $n=1,\dots,2N$.

The reconstruction formula for the density matrix also takes on a simpler form in the reflectionless case.
In this case we also need the second component of the eigenfunctions in the linear system~\eref{e:linalgsyst1} and~\eref{e:linalgsyst2}.
But it is trivial to see that, from Eq.~\eref{e:linalgsyst3}, one can obtain them from the solution of
\[
\nonumber
K\,\@Y = \@L\,,
\]
where $\@Y= (Y_1,\dots,Y_{2N})^T$ are the unknowns and $\@L = (L_1,\dots,L_{2N})^T$, with
\[
\nonumber
Y_n= \mu_{-,2,1}(t,z,\zeta_n^*)\,,\qquad
L_n = iq_-^*/\zeta_n^* + \sum_{j=1}^{2N}c_j(\zeta_n^*,z)\,,\qquad
n=1,\dots,2N\,,
\]
and where the coefficient matrix $K$ is still given by definition~\eref{e:linsystMdef}.
Once $\@X$ and $\@Y$ have been obtained, from Eqs.~\eref{e:linalgsyst1} and~\eref{e:linalgsyst2} we have
\[
\nonumber
\mu_{-,1}(t,z,\zeta_n^*) = (X_n,Y_n)^T\,,\quad
\mu_{-,2}(t,z,\zeta_n) = (\~X_n,\~Y_n)^T\,,\qquad
n=1,\dots,2N\,,
\]
which are related by Eq.~\eref{e:linalgsyst1}, i.e.,
\[
\nonumber
\begin{pmatrix}\~X_n\\\~Y_n\end{pmatrix} = 
\begin{pmatrix}iq_-/\zeta_n\\1\end{pmatrix} - \sum_{j=1}^{2N} c_j^*(\zeta_n^*,z)
\begin{pmatrix}X_j\\Y_j\end{pmatrix}\,,\qquad
n=1,\dots,2N\,. 
\]
Recalling that in the reflectionless case there is no jump, we then have 
$\mu_-(t,z,\zeta) = \left(M_{-,1},M_{+,2}\right)$ $\forall \zeta\in\Sigma$, and therefore, 
from Eq.~\eref{e:rhpsoln} we find
\[
\mu_-(t,z,\zeta) = I + \frac i\zeta\,\sigma_3Q_- 
  + \sum_{n=1}^{2N}\frac{C_n\,\e^{-2i\gamma(\zeta_n)t}}{\zeta-\zeta_n}\begin{pmatrix}\~X_n &0\\ \~Y_n &0\end{pmatrix}
  + \sum_{n=1}^{2N}\frac{\~C_n\,\e^{2i\gamma(\zeta_n^*)t}}{\zeta-\zeta_n^*}\begin{pmatrix}0 &X_n\\ 0 &Y_n\end{pmatrix}\,.
\label{e:mu-reflectionless}
\]
Recall the relationship between the density matrix $\rho$ and the eigenfunction matrix $\mu$ is given by relation~\eref{e:densityreconstruction}.
In the reflectionless situation, it simplifies to
\[
\rho(t,z,\zeta) = \nu h_- \mu_-(t,z,\zeta)\,\sigma_3\,\mu_-^{-1}(t,z,\zeta)\,,\qquad \forall \zeta\in\Sigma\,.
\label{e:solitonrho}
\]

\section{Exact solutions and their behavior}
\label{s:solutions}

\begin{figure}[t!]
    \centering
    \includegraphics[width=0.28\textwidth]{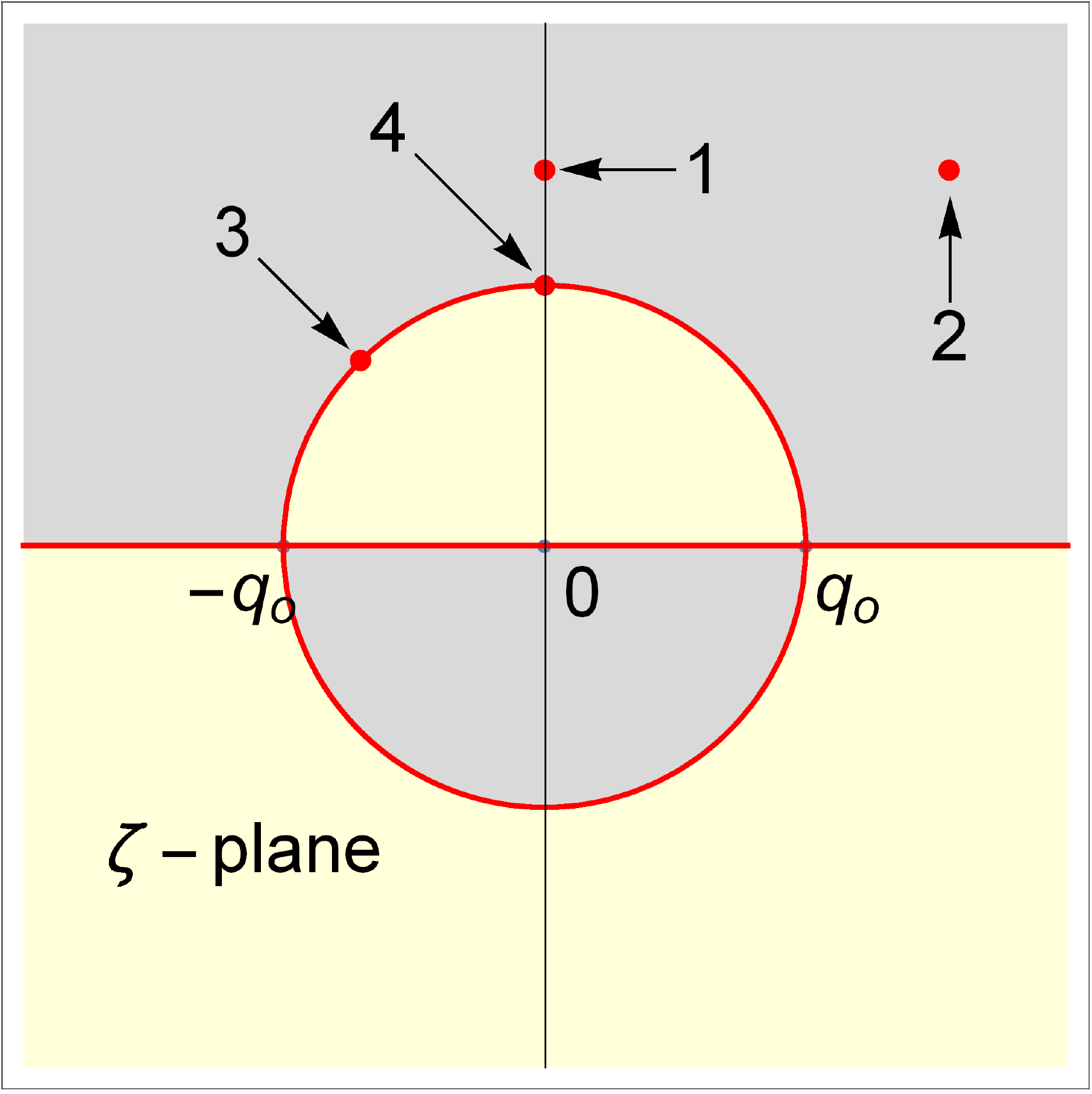}\hspace{4ex}
    \includegraphics[width=0.28\textwidth]{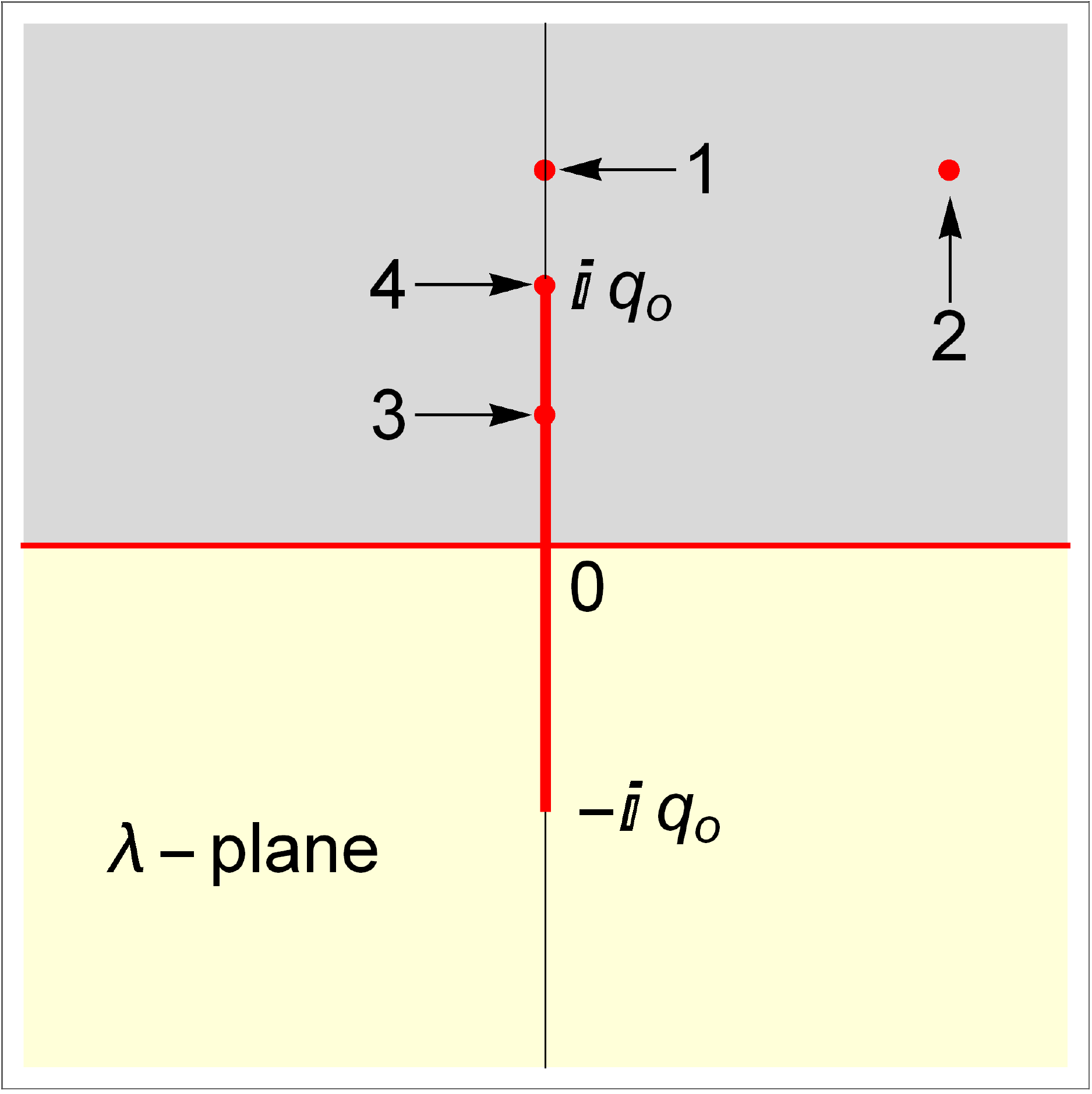}\hspace{4ex}
    \includegraphics[width=0.28\textwidth]{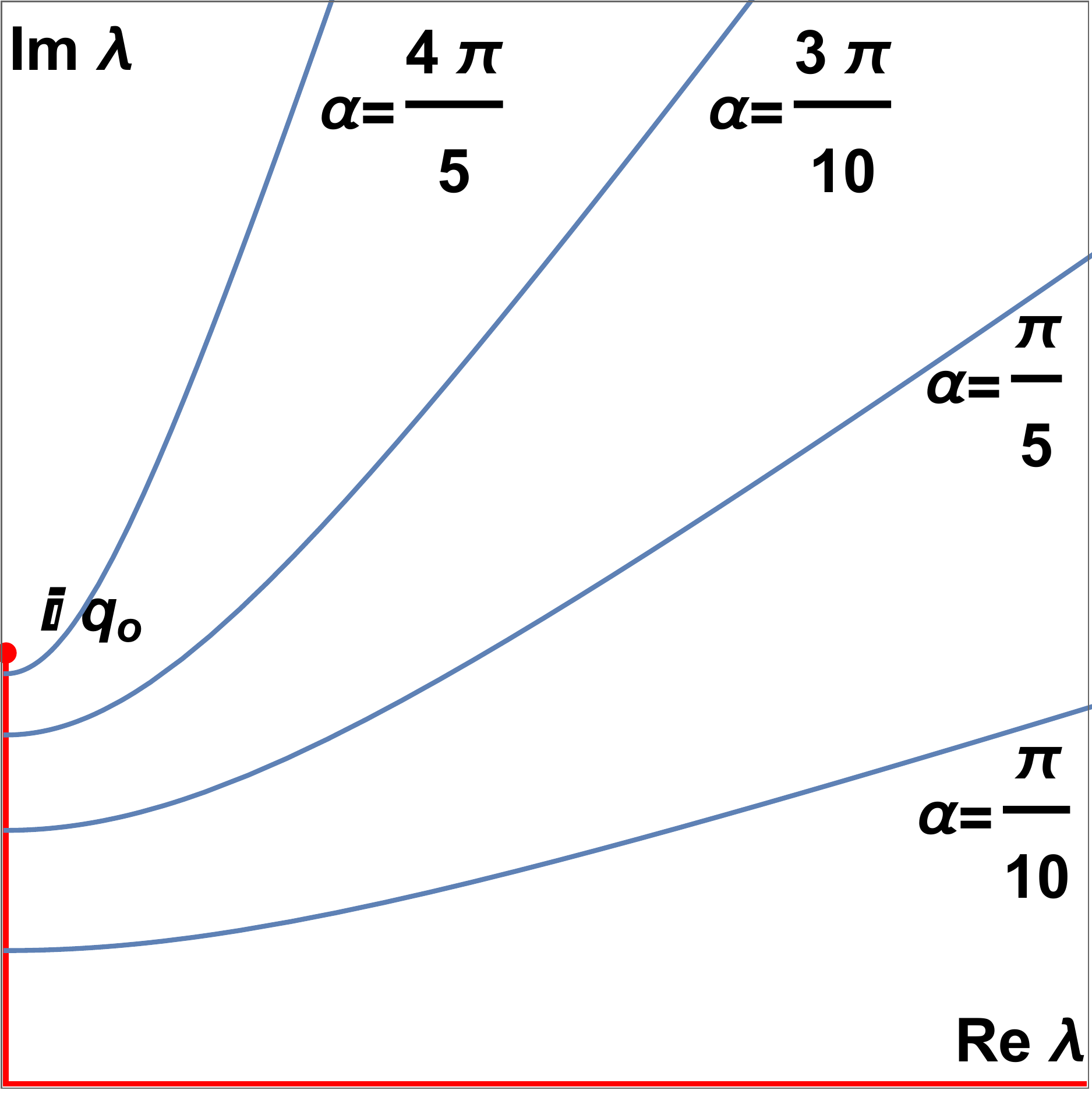}
    \caption{Left and center: four types of one-soliton solutions:
        Type~1, Traveling-wave soliton with a purely imaginary discrete eigenvalue;
        Type~2, Oscillatory soliton with a generic discrete eigenvalue;
        Type~3, Periodic solution with a discrete eigenvalue on the branch cut;
        Type~4, Rational solution with a discrete eigenvalue at the branch point $iq_o$.
        Red contours denote the continuous spectrum.
        Right: how the discrete eigenvalue~\eref{e:1solitonC1} changes as a function of $\eta$ with various values of $\alpha$ in the $\lambda$-plane.
    }
    \label{f:solitontype}
\end{figure}

In this section we use the IST formalism developed in previous sections to obtain a variety of exact solutions of the MBEs with NZBG.
In particular, we will focus on variation of one-soliton solutions with a single discrete eigenvalue $\zeta_1$.
According to the location of this discrete eigenvalue, 
there exist four types of solutions, 
shown in Fig.~\ref{f:solitontype}.
We will discuss all types separately in detail.
In addition, we will also discuss the stability of soliton solutions of the MBEs.

\subsection{One-soliton solutions: representation formulae}
\label{s:1soliton}

Without loss of generality, 
we take the input background to be $q_-(0) = q_o$ (owing to the phase invariance of the MBEs),
so that the background propagates as $q_-(z) = q_o \e^{2i W_-(z)}$ with $W_-(z)\in\Real$ defined in Eq.~\eref{e:q-timeevolution}. 
We parametrize the discrete eigenvalue $\zeta_1$ and the norming constant $C_1$ respectively as
\[
\label{e:1solitonC1}
\zeta_1= q_o\eta\,\e^{i\alpha}\,,\qquad
C_1(z)= \exp\big[\xi(z)+i\varphi(z)\big]\,,
\]
with $\eta>1$, $\alpha\in(0,\pi)$ and both $\xi(z)$ and $\varphi(z)$ real.
For later use, we also defined quantities
\[
\label{e:Deltad}
\Delta_\pm= \eta \pm 1/\eta\,,\qquad 
d_\pm = \eta ^2\pm 1/\eta ^2\,.
\]
Recall that, for all the soliton solutions, 
the initial state for the medium is given by Eq.~\eref{e:solitonBC1} with $h_- = \pm 1$. 
We further assume $h_-$ is independent of $z$ and $\zeta$.
We take $\rho_-$ to be diagonal because only in this case a reflectionless solution remains reflectionless upon propagation.
The formula for optical field will be given explicitly below.
However, due to complexity of the expressions,
we will only give explicit formulas for the modified eigenfunction matrix $\mu$ instead of the density matrix $\rho$.
One can then simply use the relationship~\eref{e:solitonrho} to reconstruct $\rho(t,z,\zeta)$.
Moreover, it turns out that $\mu$ satisfies the symmetries $\mu_{2,1} = -\mu_{1,2}^*$ and $\mu_{2,2} = \mu_{1,1}^*$,
so it is sufficient to give only the $(1,1)$ and $(1,2)$ components of $\mu$.

\paragraph{Type 1. Traveling-wave soliton: solutions with purely imaginary eigenvalues.}
Let us start from the simplest case $\alpha = \pi/2$ in Eq.~\eref{e:1solitonC1}. 
The theta condition~\eref{e:theta} implies that the corresponding asymptotic phase difference is $2\pi$
(i.e., no phase difference).
The general soliton solution~\eref{e:Nsolitonsolution} and~\eref{e:mu-reflectionless} reduces to
\begin{gather}
\label{e:staticsoliton}
q(t,z) = q_-(\Delta_+\,\cosh\chi + d_+\sin s +  i d_- \cos s)/(\Delta_+\,\cosh\chi + 2\,\sin s)\,,\\
\mu_{1,1} = 
(- d_- \zeta  q_o \sinh\chi + i\Delta_- q_o^2 Y + 2 i \gamma  \zeta  X)/(\zeta X Z)\,,\\
\mu_{1,2} = q_- (-i d_- q_o \sinh\chi + \Delta_- \zeta Y + 2 \gamma  X)/(\zeta  X Z^* )\,,
\end{gather}
with
\begin{gather}
X = \Delta_+\cosh\chi + 2\sin s\,,\qquad
Y = \Delta_- \sin s + i\Delta_+ \cos s\,,\qquad
Z = \Delta_- q_o + 2 i \gamma\,,\\
\chi(t,z) = q_o\Delta_- t + \Delta_0 + \xi(z)\,,\qquad
s(z) = \varphi(z) + 2W_-(z)\,,\qquad
\Delta_0 = \log[\Delta_+/(2q_o\Delta_-\eta)]\,.
\label{e:Deltao}
\end{gather}
Recall $W_-(z)$ was defined in Eq.~\eref{e:q-timeevolution} and all quantities in Eq.~\eref{e:Deltad}.
Note that $\Delta_+>2$, so the above soliton solution is always non-singular.
Equation~\eref{e:staticsoliton} is the analogue of the Kuznetsov-Ma solution of the focusing NLS equation with NZBC \cite{Kuznetsov,Ma}.
Indeed, 
the calculations to obtain the soliton solution~\eref{e:staticsoliton} are similar to those for the focusing NLS equation (e.g., see \cite{JMP55p031506} for details),
but the resulting solutions exhibit quite different behavior upon propagation.

Importantly, 
the $z$ dependence of the solution is entirely determined by the norming constant
via $\xi(z)$, $\varphi(z)$ and the background $q_-(z)$.
We will discuss the $z$ dependence later.

\paragraph{Type 2. Oscillatory soliton: solutions with eigenvalues in general position.}
A more general expression of the one-soliton solution for a general discrete eigenvalue can also be found, 
similarly to that for the focusing NLS equation.
(Again, see \cite{JMP55p031506} for details.)
Recalling the parametrization~\eref{e:1solitonC1} for the discrete eigenvalue and norming constant,
after tedious but straightforward calculations, we obtain 
\begin{align}
\label{e:oscillatorysoliton}
q(t,z) = & \e^{-2 i \alpha } q_-
[\cosh (\chi-2i \alpha) + d_+ \kappa_s + i d_- \kappa_c]/(\cosh\chi - 2 \kappa_s)\,,\\
\nonumber
\mu_{1,1}(t,z,\zeta) = & \big[q_o^2 \cosh (\chi-2i\alpha) + \Delta_- \zeta  q_o \cosh(\chi - i\alpha) - \zeta ^2 \cosh\chi\big]/(X Y)\\
& +i \sin\alpha\big[E_o \e^{-i s}|\zeta -\zeta_1|^2 - \e^{i s} E_o^* |\zeta -\hat\zeta_1| ^2\big]\big/(|E_o| \Delta_+ X Y) \,,\\
\nonumber
\mu_{1,2}(t,z,\zeta) = & - i \e^{-2i\alpha} q_-\big[\zeta \cosh (\chi - 2i \alpha) - \Delta_-  q_o \cosh(\chi - i\alpha) + \hat{\zeta} \cosh\chi\big]\big/
(X Y^*)\\
& + \e^{-2i\alpha}  q_-\sin\alpha\big[\e^{-i s} \eta^4 E_o |\zeta - \hat\zeta_1| ^2-\e^{i s} E_o^* |\zeta -\zeta_1|^2\big]\big/
(|E_o| \Delta_+ \zeta  \eta^2 X Y^*)\,,
\end{align}
where 
\begin{gather}
X = \cosh\chi - 2 \kappa_s\,,\qquad
Y = \e^{2 i \alpha } q_o^2 + \e^{i \alpha } \Delta_- \zeta  q_o-\zeta ^2\,,\\
\kappa_s(t,z) = E_1(\eta ^2 \sin (s-2 \alpha )+\sin s)\,,\qquad
\kappa_c(t,z) = E_1(\eta ^2 \cos (s-2 \alpha )+\cos s)\,,\\
\chi(t,z) =q_o \Delta_- t\sin\alpha+\xi(z) +\ln \big[\Delta_+/\big(2 q_o |E_o|\sin\alpha\big)\big]\,,\\
s(t,z) = -q_o \Delta_+ t\cos\alpha+\varphi(z)+2 W_-(z)\,,\qquad
E_o = 1 + \e^{2 i \alpha } \eta^2\,,\qquad
E_1 = \sin\alpha/(|E_o|\Delta_+)\,,
\end{gather}
and $\Delta_\pm$, $d_\pm$ are defined in Eq.~\eref{e:staticsoliton} and $W_-(z)$ is defined in Eq.~\eref{e:q-timeevolution}.
Note that the term $\cosh(\chi-2i\alpha)$ can be easily expressed in terms of real-valued functions using addition formulae.
Equation~\eref{e:oscillatorysoliton} is the analogue of the Tajiri-Watanabe solutions of the focusing NLS equation with NZBC \cite{TajiriWatanabe},
but exhibits different spatial-temporal behavior.

The same as for the traveling-wave soliton solutions, 
the propagation of the above solution is also determined by the three quantities $\xi(z)$, $\varphi(z)$ and $q_-(z)$ from Section~\ref{s:propagation},
which will be discussed later.
The names ``traveling-wave soliton" and ``oscillatory soliton" will become clear after we compute the $z$ propagation explicitly.

\paragraph{Soliton amplitude.}

We define the instantaneous amplitude compared to the background as 
\[
\label{e:generalamplitude}
A(z) = \max_{t\in\Real}|q(t,z)| - q_o\,.
\]
It is easy to see that the soliton achieves its maximum when $\chi = 0$ in both cases~\eref{e:staticsoliton} and~\eref{e:oscillatorysoliton}.
For the oscillatory soliton solution~\eref{e:oscillatorysoliton}, 
the explicit formula~\eref{e:generalamplitude} is complicated and does not simplify in general, 
so it is omitted for brevity.
However, for the traveling-wave soliton~\eref{e:staticsoliton},
simple calculations show that the amplitude~\eref{e:generalamplitude} becomes
\[
\label{e:amplitude}
A(z) = q_o\sqrt{\frac{(\Delta_2^2+1)\Delta_1+2\sin s}{\Delta_1+2\sin s}}-q_o\,.
\]
The maximal possible amplitude is attained when $s(z) =(2n +3/2)\pi$ with $n\in\Integer$, and is
\[
\nonumber
A_\mathrm{max} = q_o (1+\eta^2)/\eta\,.
\]
Notice that $A_\mathrm{max}$ is an increasing function of $\eta$.
That is, traveling-wave soliton solutions have a larger maximal possible amplitude when the discrete eigenvalues move farther away from the continuous spectrum.

Importantly, however, note that, for some solutions, the above theoretical maximum may not be achieved in practice. 
This is because, as we will show in Section~\ref{s:IBsoliton}, for discrete eigenvalues on the imaginary axis, 
$s(z)$ will turn out to be constant,
and therefore the condition $s(z) =(2n +3/2)\pi$ may not be achieved for any value of $z$ or $n$.
In such cases, the actual maximum of the solution 
depends on the phase of the norming constant.
In fact, 
the entire shape of the traveling-wave soliton solution depends on the phase of the norming constant.
This is unlike the case of ZBG and also unlike the solutions of the focusing NLS equation.

\paragraph{Soliton velocity.}

By inspecting soliton solutions~\eref{e:staticsoliton} or~\eref{e:oscillatorysoliton}, 
it is evident that solitons are localized along the curve $\chi(t,z) = y$.
Our assumption that the initial state for the medium~\eref{e:solitonBC1} is independent of $z$
implies that $\xi(z)$ from Eqs.~\eref{e:staticsoliton} and~\eref{e:oscillatorysoliton} has the form $\xi(z) = \xi_1\, z + \xi(0)$, 
where $\xi_1$, $\xi(0)\in\Real$ and are independent of~$z$.
Then, provided that $\xi_1\ne0$,
the soliton is located along the line $z-Vt = z_o$, and travels with velocity~$V$ in the $(t,z)$ frame, where
\[
\label{e:solitonvelocity}
V = -q_o \Delta_-\sin\alpha/\xi_1\,.
\] 
Now recall that the MBEs~\eref{e:MBEqscalar} and~\eref{e:MBErhoentries} are written in a comoving frame of reference.
Straightforward algebra shows that the physical soliton velocity in the laboratory frame is, 
\[
\nonumber
V_{\mathrm{lab}} = \frac{V}{1+V/c}\,,
\]
where $c$ is the speed of light in vacuum.
That is, in laboratory coordinates, the soliton is located along the line $z-V_\mathrm{lab}t = z_o$.
Conversely, $V = V_\mathrm{lab}/(1-V_\mathrm{lab}/c)$.
Therefore, the requirement $V_\mathrm{lab}\le c$ of physical signals translates into the requirement $V\ge0$.
Moreover, $V_\mathrm{lab}\to0$ implies $V\to 0$ and $V\to\infty$ as $V_\mathrm{lab} \to c$. 


\begin{figure}[t!]
\vglue-\bigskipamount
\centering
\begin{tabular}[b]{ccc}
\includegraphics[scale=0.21]{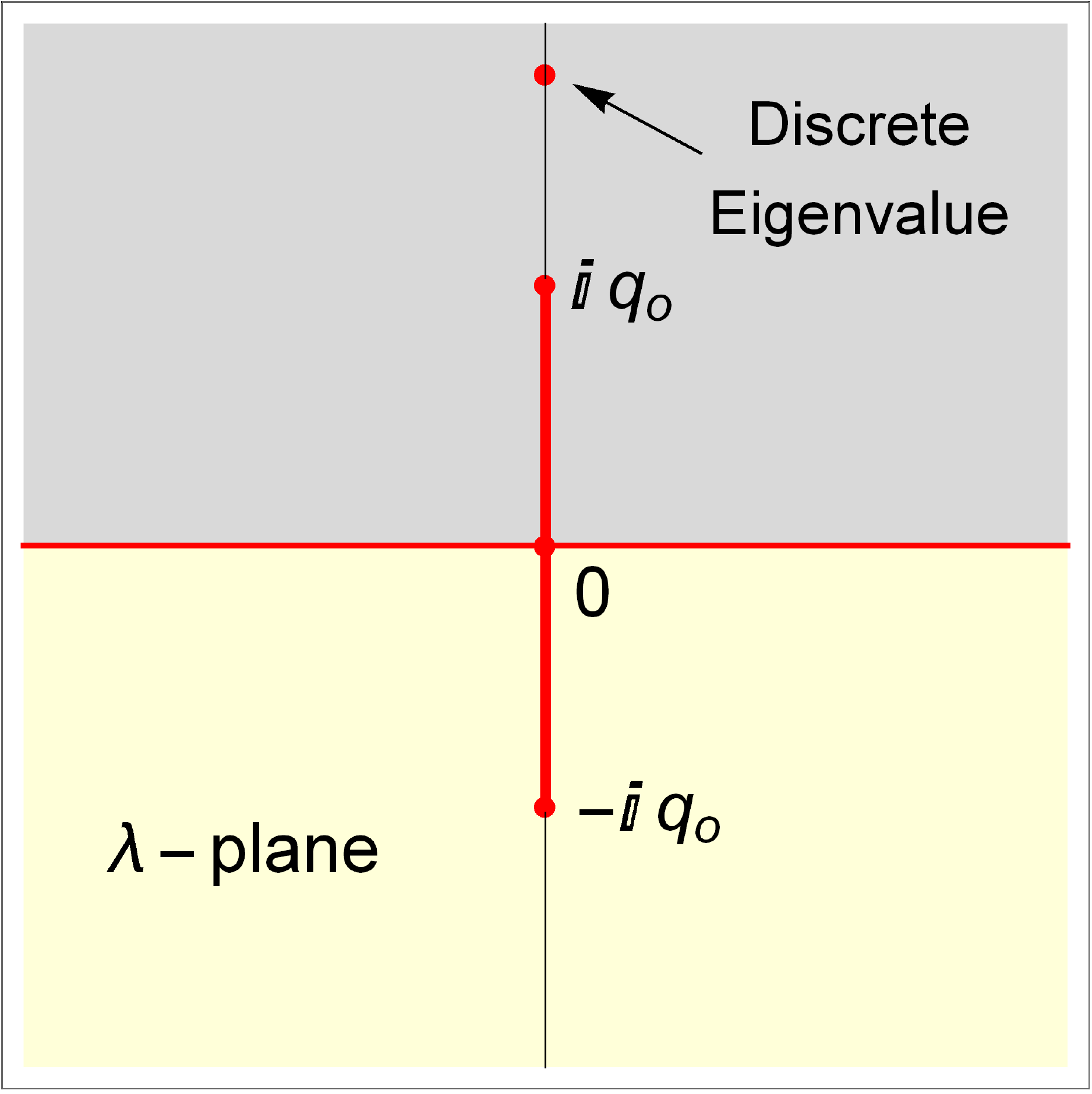}&
\includegraphics[scale=0.27]{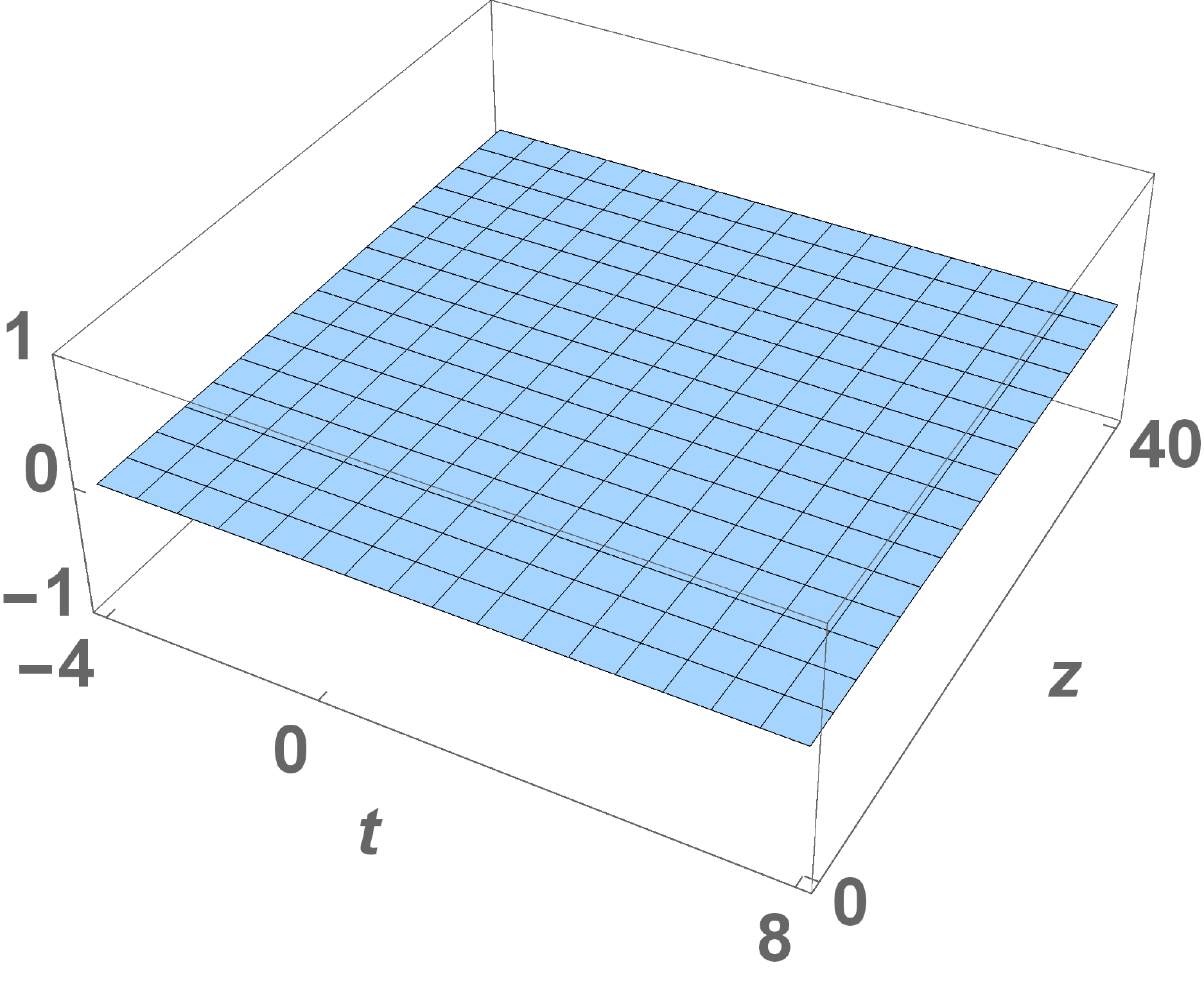}&
\includegraphics[scale=0.27]{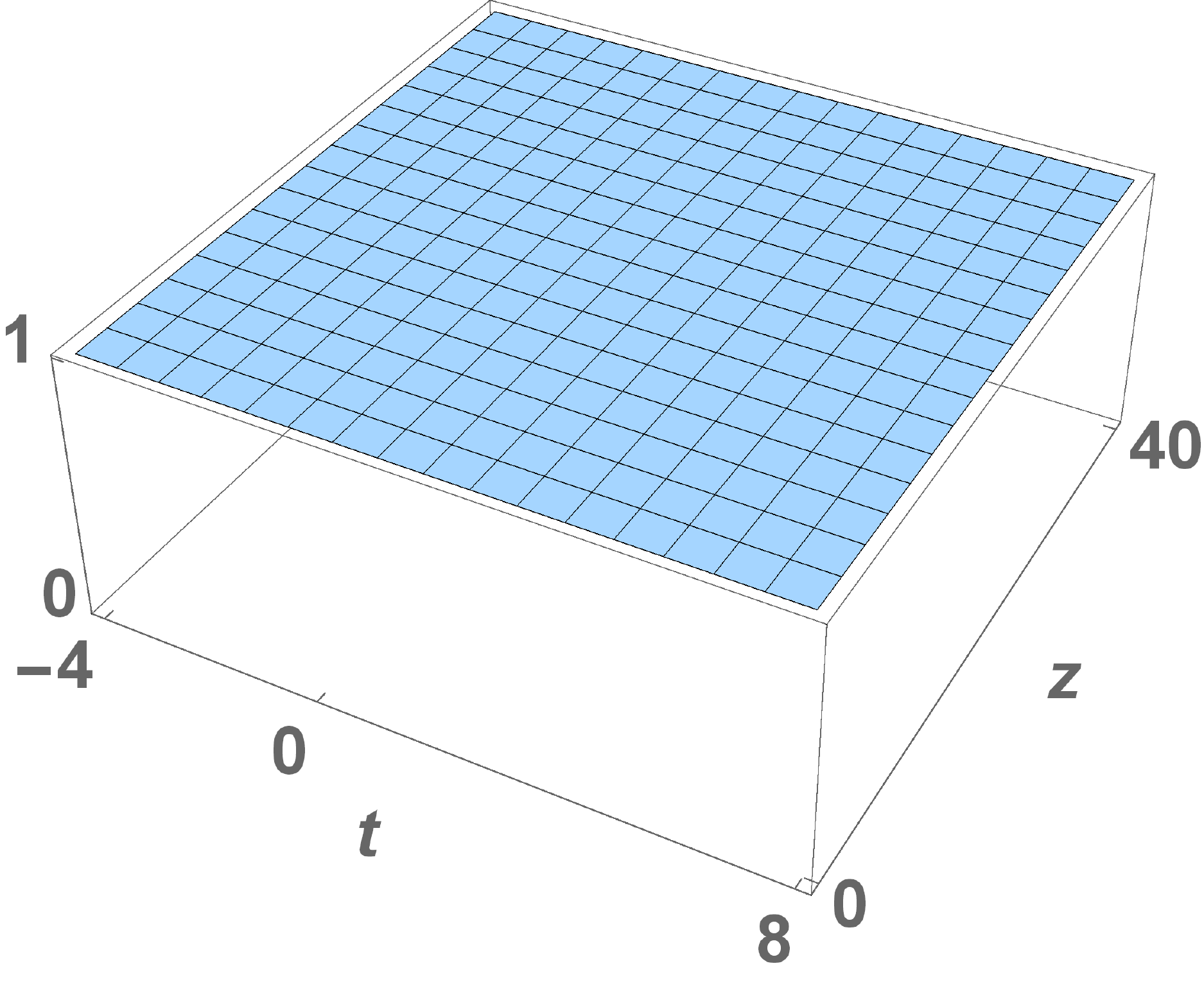}\\
\includegraphics[scale=0.27]{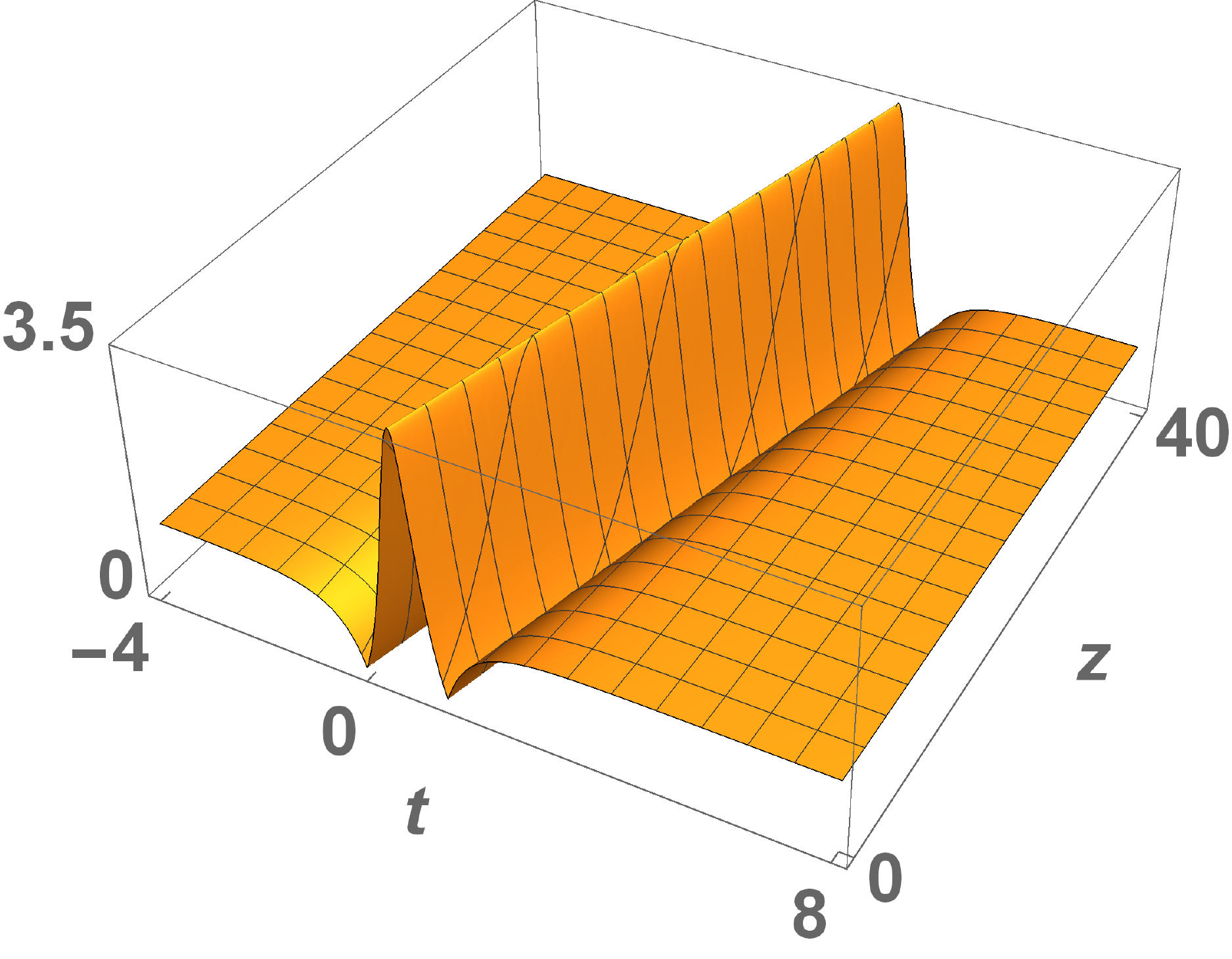}&
\includegraphics[scale=0.27]{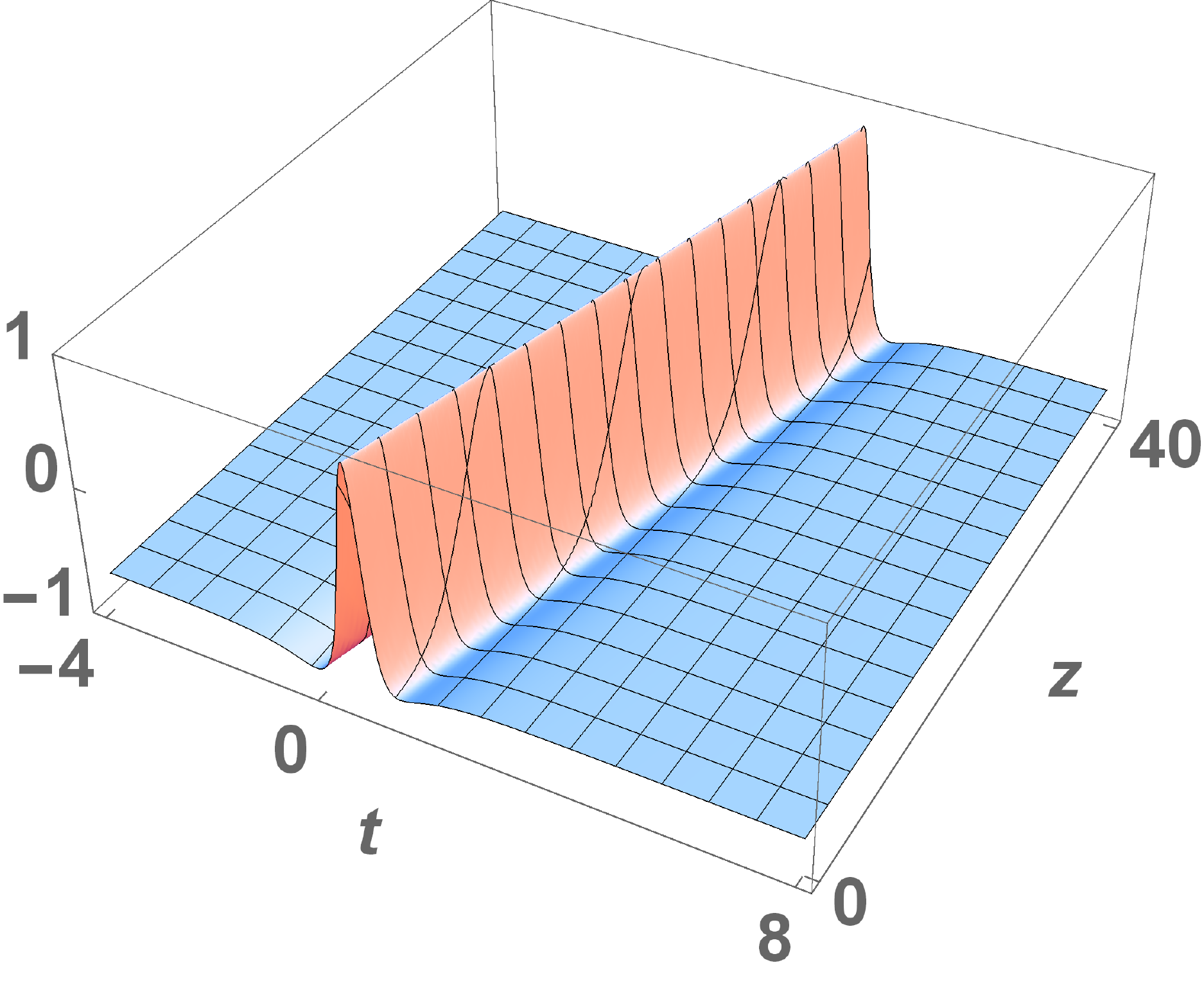}&
\includegraphics[scale=0.27]{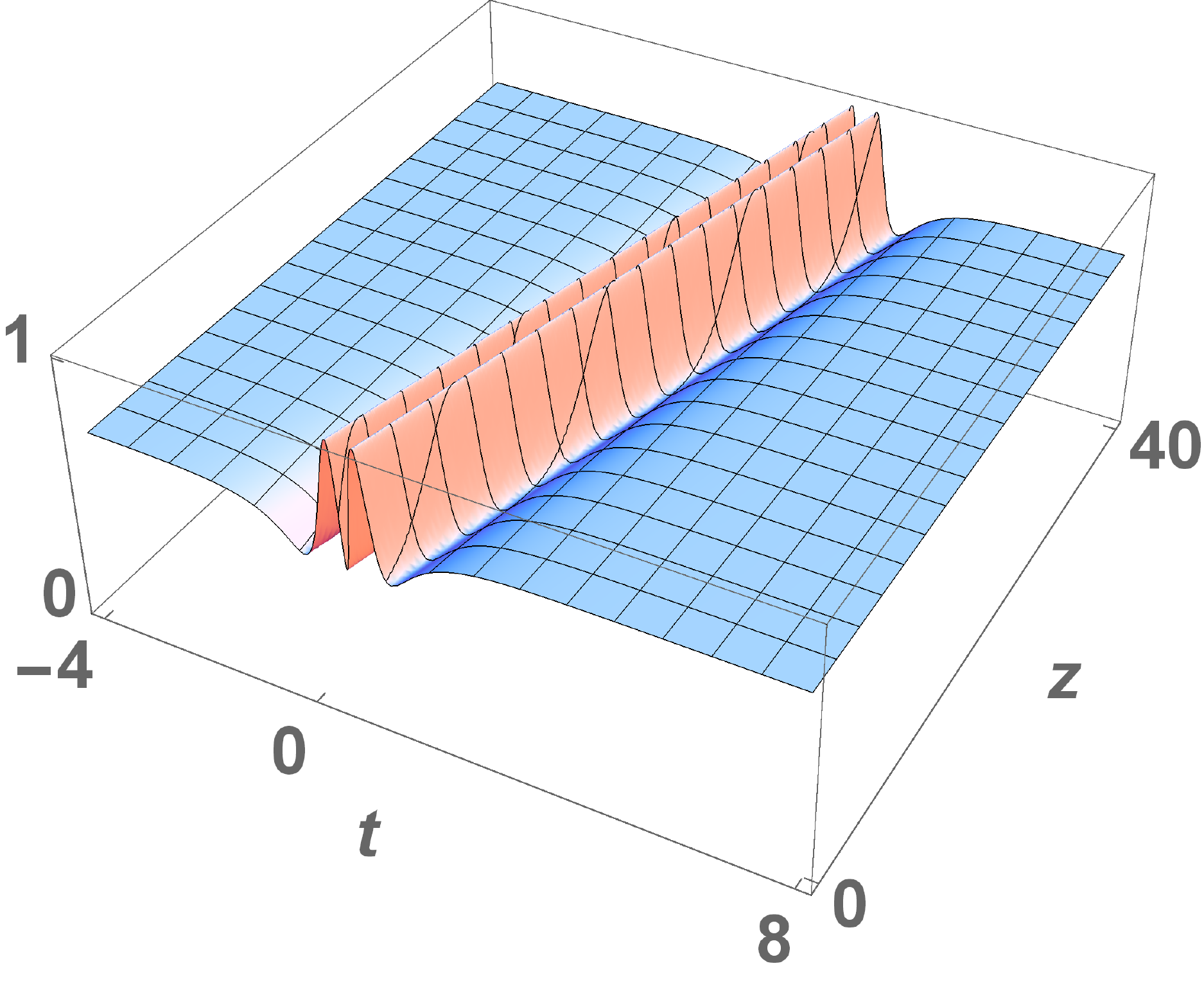}
\end{tabular}
\caption{One traveling-wave soliton solution~\eref{e:staticsoliton} with the initial state for the medium~\eref{e:solitonBC1}, 
	inhomogeneous broadening and a discrete eigenvalue $\zeta_1 = 2i$ on the imaginary axis:
	$h_- = -1$ (i.e., more atoms are initially in the ground state), 
    $q_o = 1$, $\epsilon = 2$, 
	$\xi(0) = 0$ and $\varphi(0) = -\pi/2$.
    Top left: the discrete eigenvalue in the $\lambda$-plane.
	Bottom left: the optical field $|q(t,z)|$.
	Center and right: components of density matrix $\rho(t,z,\lambda)$.
	Top center: $D(t,z,0^+)$.
	Top right: $|P(t,z,0^+)|$.
	Bottom center: $D(t,z,q_o)$.
	Bottom right: $|P(t,z,q_o)|$.
	}
\label{f:IB1solitontypeI}
\vskip2\bigskipamount
\centering
\begin{tabular}[b]{ccc}
\includegraphics[scale=0.21]{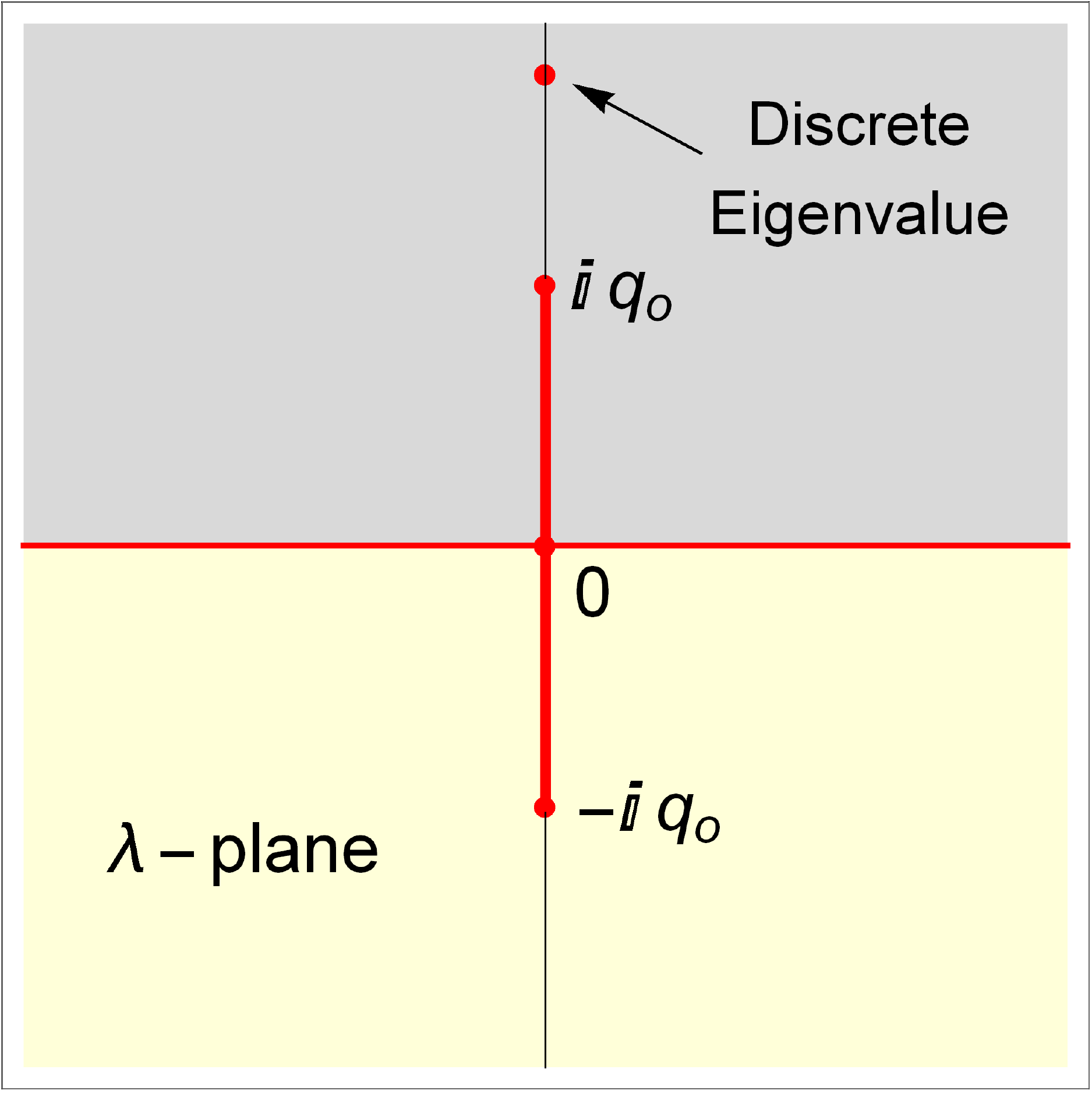}&
\includegraphics[scale=0.26]{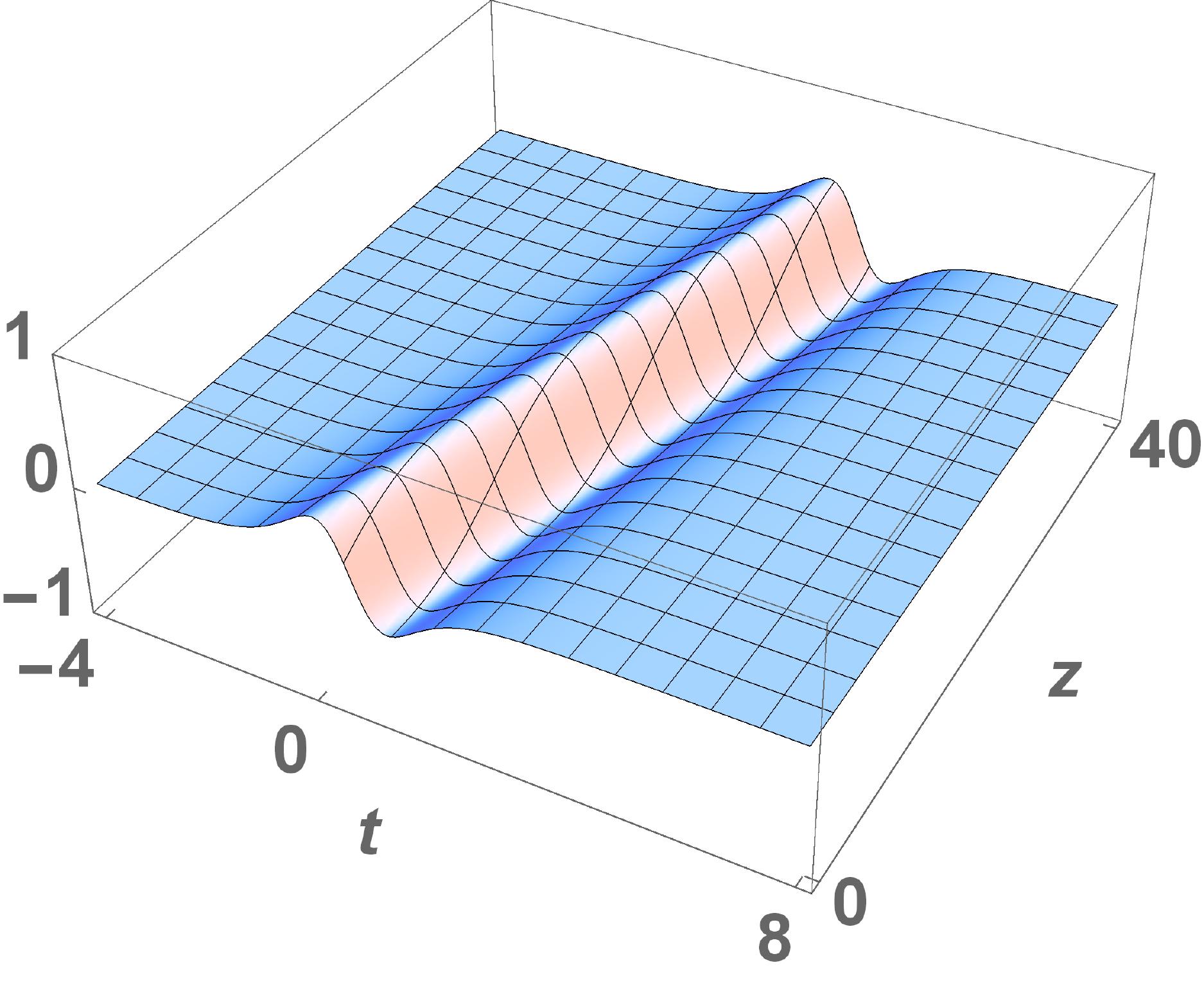}&
\includegraphics[scale=0.26]{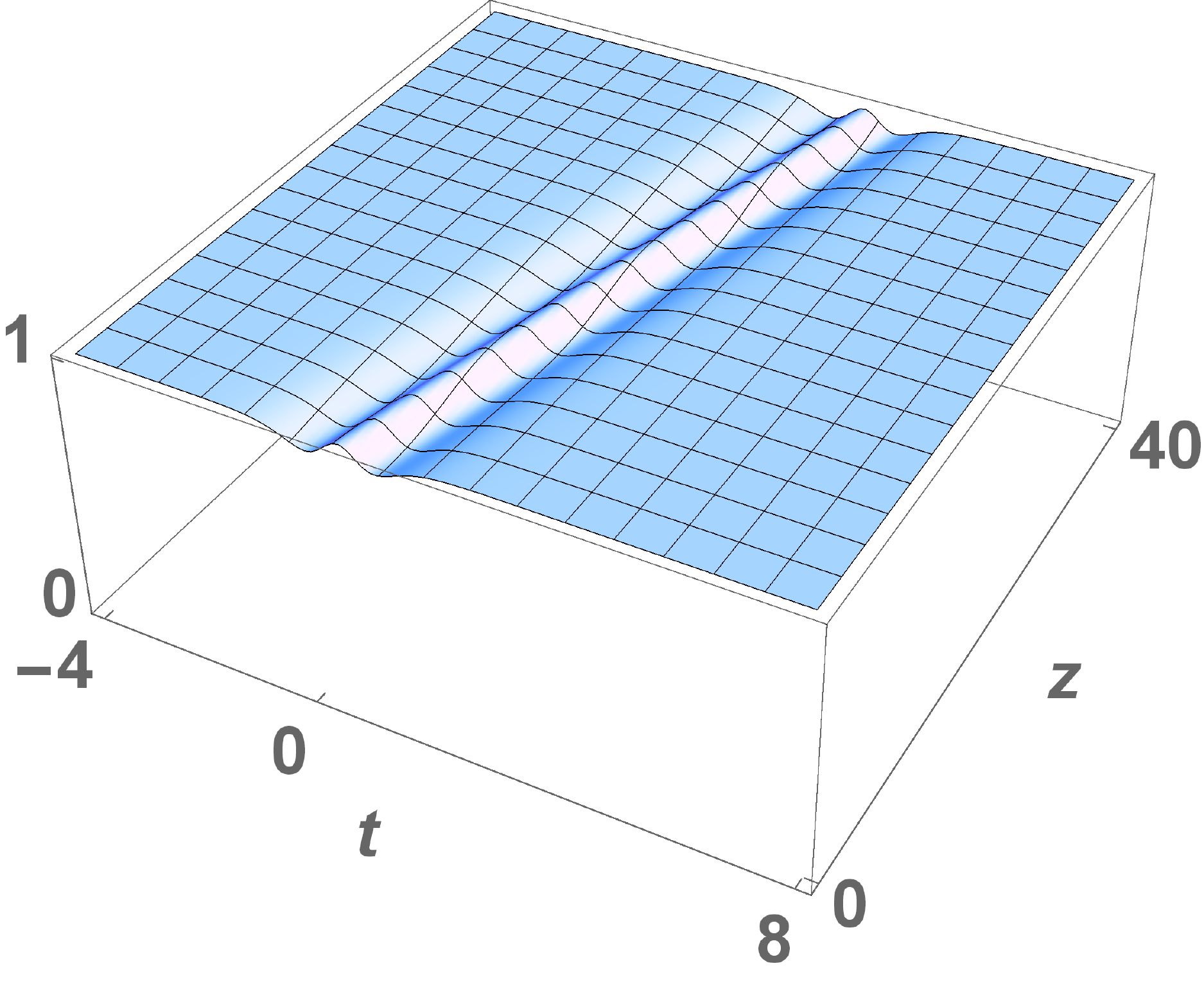}\\
\includegraphics[scale=0.26]{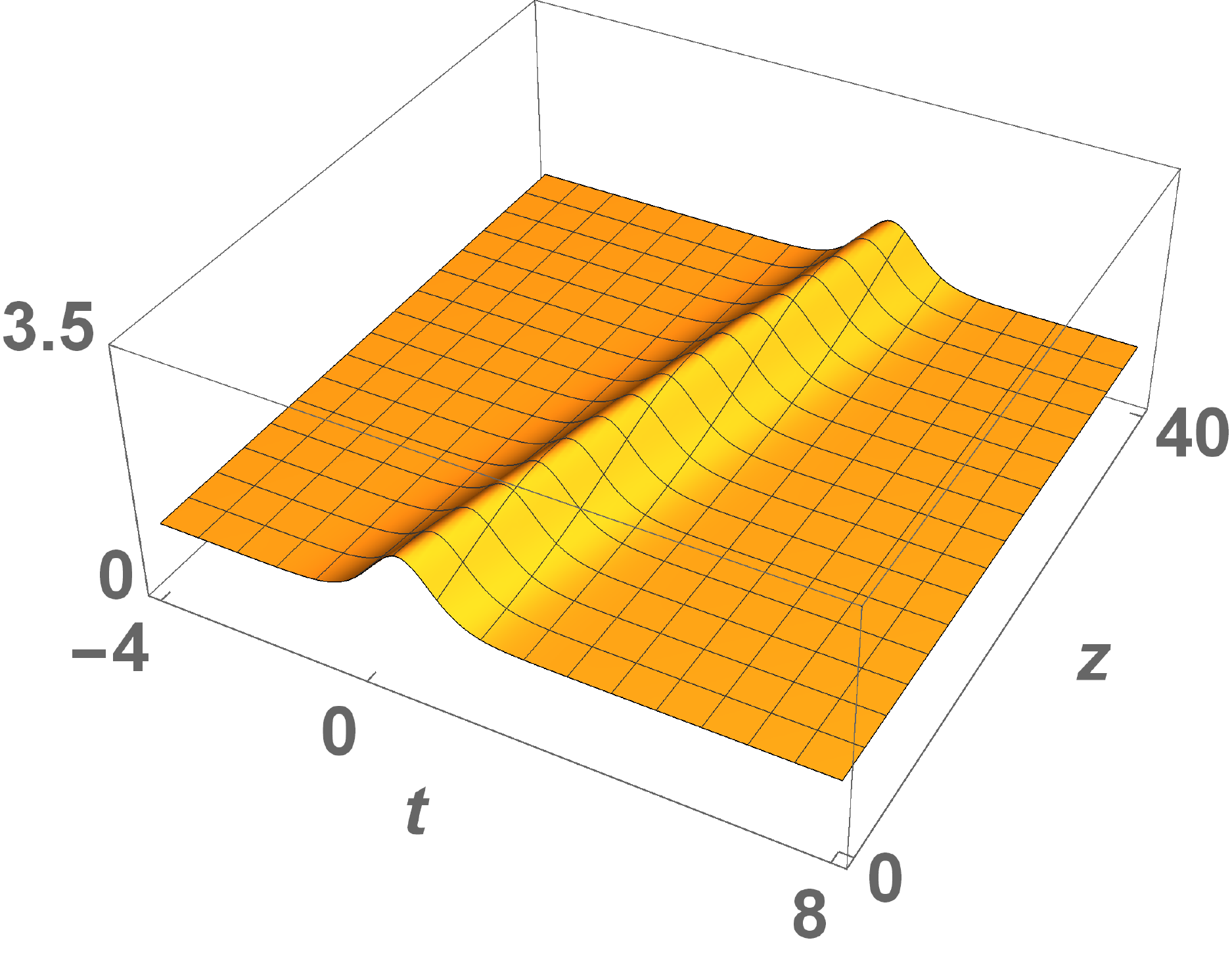}&
\includegraphics[scale=0.26]{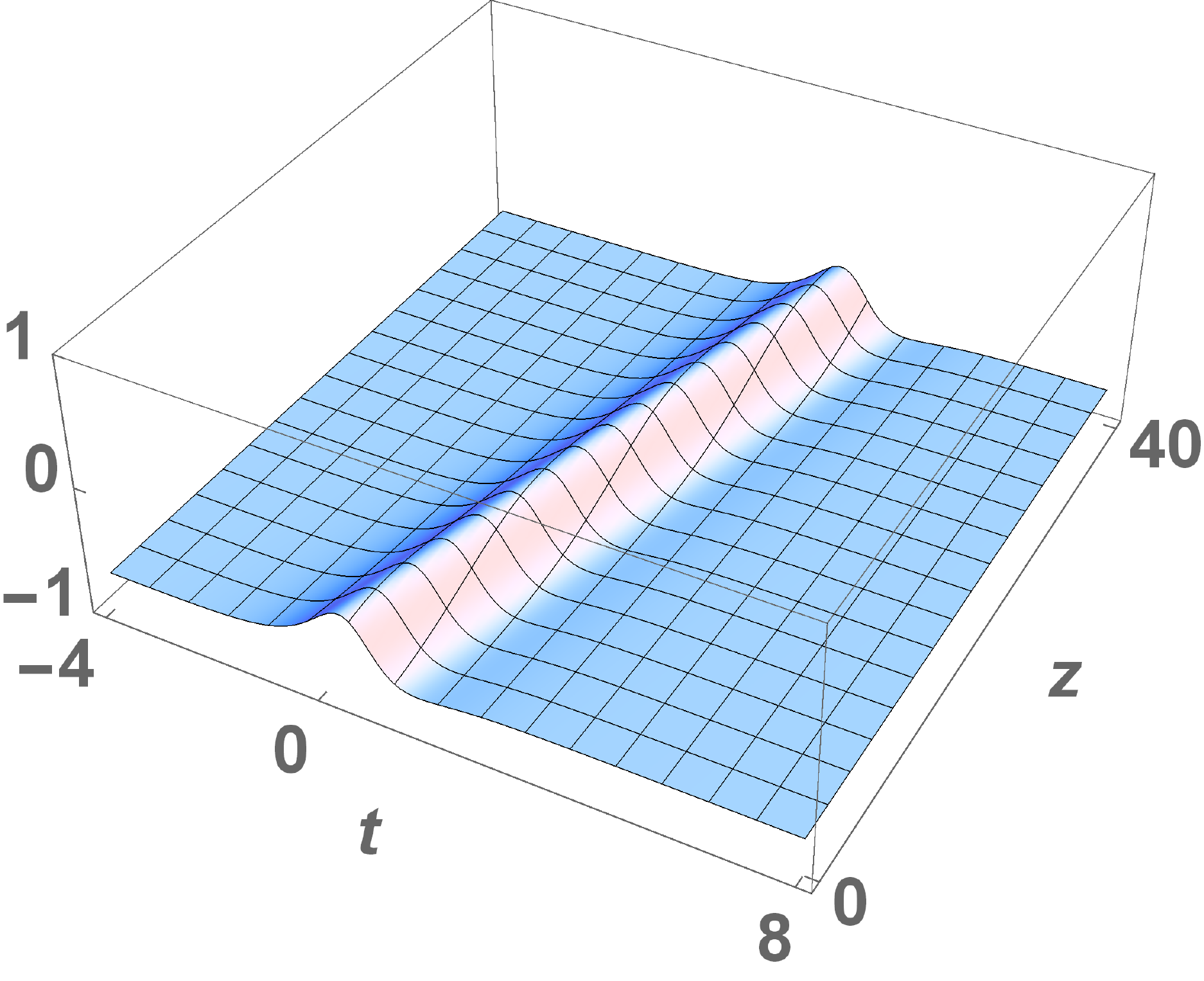}&
\includegraphics[scale=0.26]{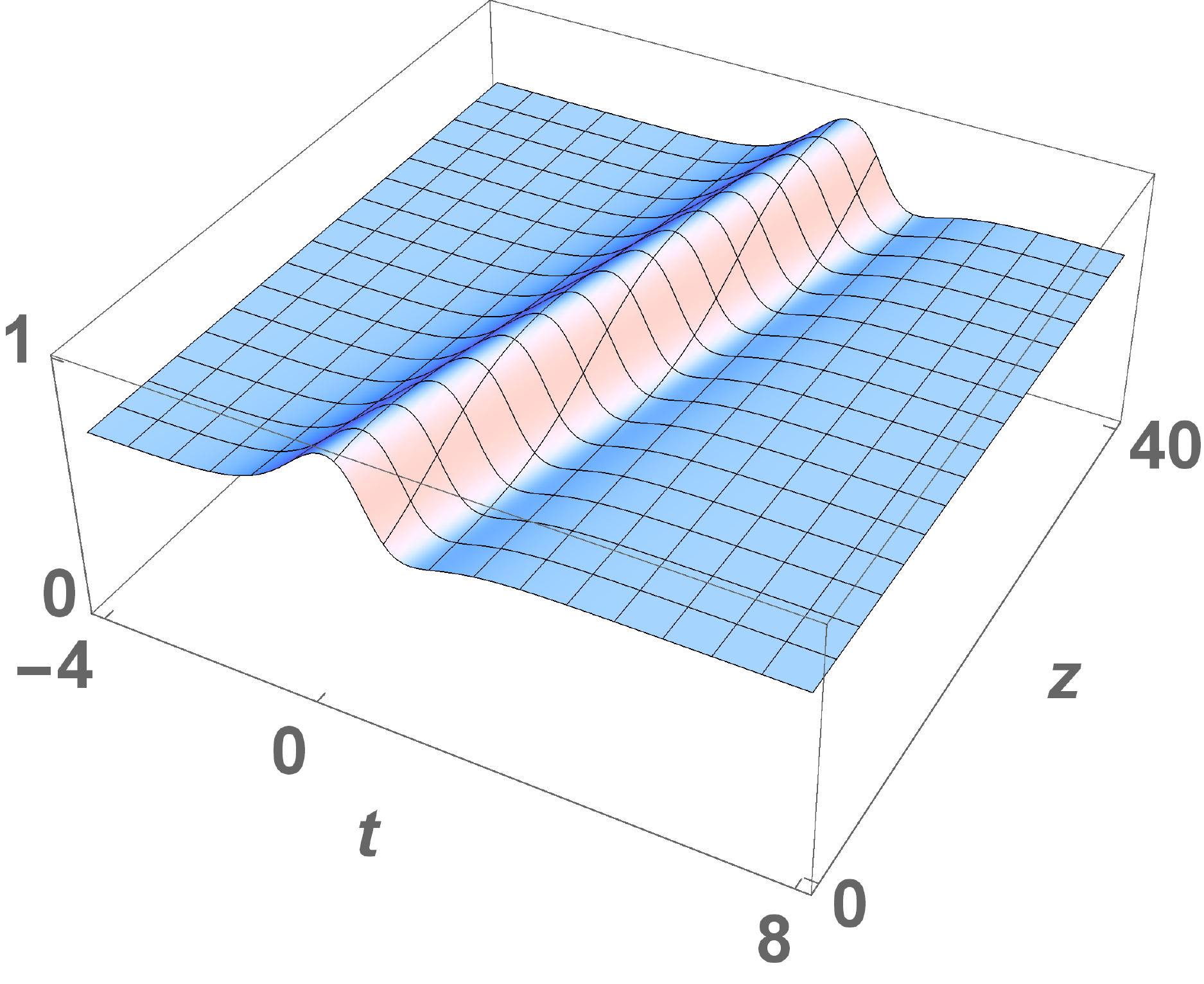}
\end{tabular}
\caption{Same as Fig.~\ref{f:IB1solitontypeI}, 
	but with a different choice of norming constant: $\varphi(0) = 0$.}
\label{f:IB1solitontypeI2}
\end{figure}

\subsection{Soliton solutions with inhomogeneous broadening: propagation}
\label{s:IBsoliton}

As usual, when studying the case of inhomogeneous broadening we take the spectral-line shape to be a Lorentzian, 
namely:
\[
g(\lambda) = \frac{\epsilon}{\pi}\frac{1}{\epsilon^2+\lambda^2}\,,\qquad
\epsilon>0\,.
\label{e:Lorentzian}
\]
The propagation of the optical background $q_-(z)$ is governed by Eq.~\eref{e:q-timeevolution}.
But because $g(\lambda)$ and $\gamma(\lambda)$ are respectively an even and an odd function of $\lambda$, 
the integral in the definition~\eref{e:wpmdef} is zero, i.e., $w_-(z)=0$.
Thus, the initial state for $q(t,z)$ is independent of $z$, 
namely $q_-(z) = q_o$.
We point out that a more general scenario, 
i.e., solutions with a shifted Lorentzian, 
is discussed in Appendix~\ref{s:soliton_shift_lorentzian}.

The propagation of the norming constant $C_1$ is governed by ODE~\eref{e:dCdz}.
Recall that we parameterized the norming constant in Eq.~\eref{e:1solitonC1}.
In order to obtain $\xi(z)$ and $\varphi(z)$ one must compute $R_{\pm,d}$ explicitly by using Eq.~\eref{e:R_d_axis}.
In this case the calculations are rather involved, and are presented in Appendix~\ref{A:Rpm_explicit}. 
Since we will substitute the discrete eigenvalue into $R_{-,d}(\zeta,z)$, 
we know $\zeta\in \Gamma^+$, and therefore we consider the case with $\zeta\in\Complex/\Real$,
i.e., $\lambda\in\Complex/\Real$.
At the end, we write the matrix in the following form
\[
\label{e:Rd-explicit_NZBG}
R_{-,d}(\zeta) =  h_- g(\lambda) \big[\Theta(\lambda) - \gamma\Theta(i\epsilon)/ (q_o^2-\epsilon^2)^{1/2}\big]\,\sigma_3 \,,\qquad \zeta\in \Complex/\Real\,,
\]
with $\Theta(\lambda)$ defined as
\[
\label{e:Theta}
\Theta(\lambda) = 2\,\arcsech(-i\lambda/q_o)\,.
\]
Importantly, in Appendix~\ref{A:Rpm_explicit}, 
we also show that $R_{-,d}$ has a different form on the continuous spectrum $\Sigma$ as follows,
\[
\label{e:R-dreal}
R_{-,d}(\zeta,z) = 
\sigma_3\,\frac{\epsilon h_-}{\pi(\epsilon^2+\lambda^2)}
\big[\, 2\arccsch(-\lambda/q_o) - \gamma\Theta(i\epsilon)\big/(q_o^2-\epsilon^2)^{1/2}\,\big]\,,\qquad \zeta\in\Real\,.
\]
Therefore, the matrix $R_{-,d}$ is discontinuous across the real $\zeta$-axis, i.e., the real $\lambda$-axis.  
(This is the same as in the case of ZBG, and follows immediately from its definition as a principal value integral.)
Clearly, the value of $R_{-,d}$~\eref{e:Rd-explicit_NZBG} depends on whether $\epsilon<q_o$ or $\epsilon>q_o$.

Using the expression above, we can now compute the propagation for the norming constants.
Recalling the parametrization~\eref{e:1solitonC1}, 
we have that
\[
\label{e:xiverphi_unshiftedLorentzian}
\xi(z) = \xi(0) + \Im(R_{-,1,1}(\zeta_1))z\,,\qquad
\varphi(z) = \varphi(0) - \Re(R_{-,1,1}(\zeta_1))z\,,
\]
Explicitly, from Eq.~\eref{e:dCdz} and~\eref{e:R-dreal} we have 
\begin{gather}
\label{e:xiphi1}
\xi(z) = h_- \Im\{g(\lambda) \big[\Theta(\lambda) - \gamma\Theta(i\epsilon)/ (q_o^2-\epsilon^2)^{1/2}\big]\}z + \xi(0)\,,\\
\label{e:xiphi2}
\varphi(z) = - h_-\Re\{g(\lambda) \big[\Theta(\lambda) - \gamma\Theta(i\epsilon)/ (q_o^2-\epsilon^2)^{1/2}\big]\}z+ \varphi(0)\,.
\end{gather}
It is easy to show that the one-soliton solution~\eref{e:oscillatorysoliton} oscillates with temporal frequency
\[
\nonumber
\omega = q_o|\Delta_-\sin\alpha\,\cot\arg R(\zeta_1) - \Delta_+ \cos\alpha|/(2\pi)\,.
\]
In particular, if $\alpha = \pi/2$ (traveling-wave soliton), simple calculations show that $\omega = 0$,
i.e., there are no oscillations.
This is why we name the solitons as traveling-wave or oscillatory solitons.

\begin{figure}[t!]
\vglue-\bigskipamount
\centering
\begin{tabular}[b]{ccc}
\includegraphics[scale=0.21]{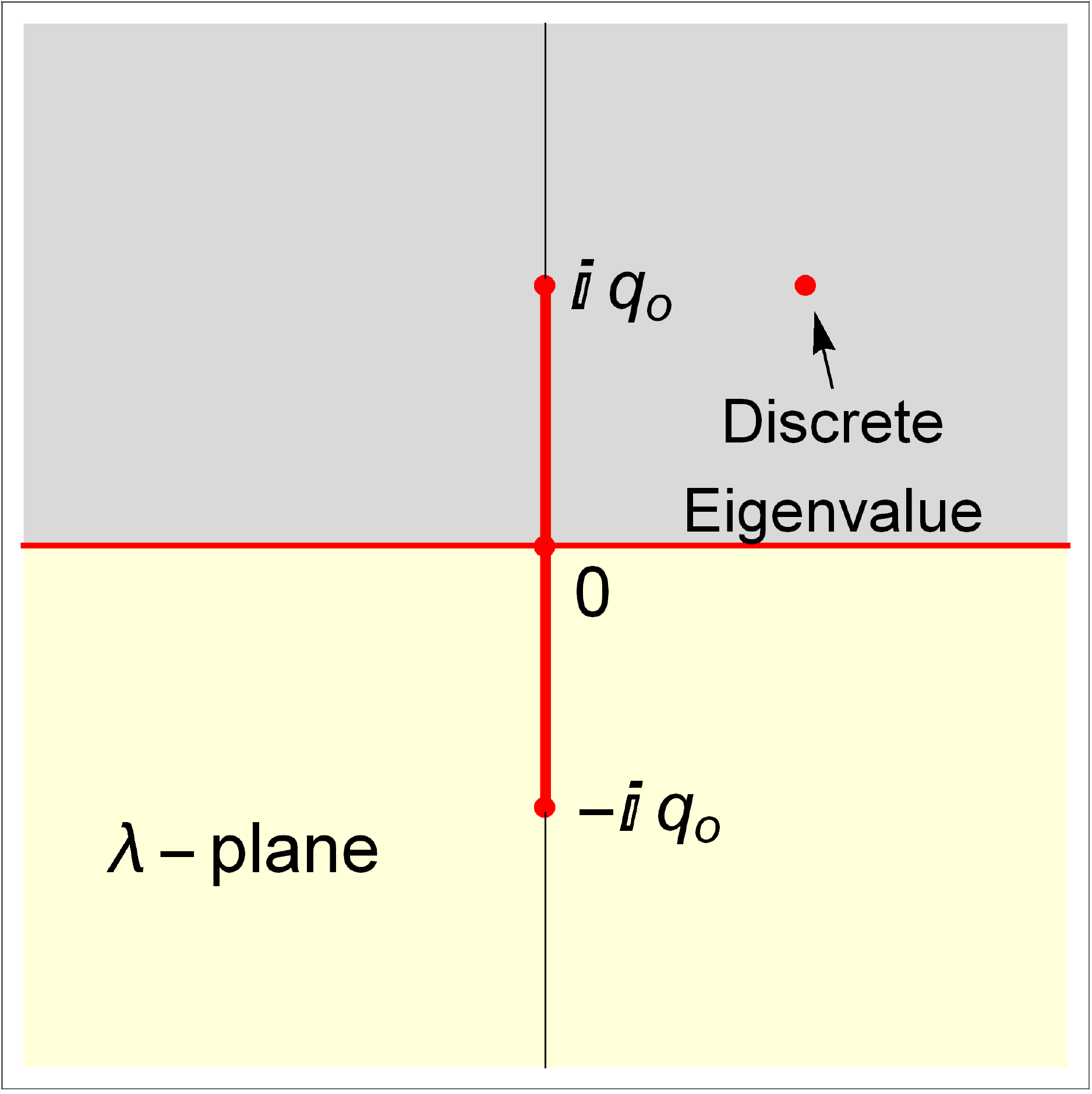}&
\includegraphics[scale=0.27]{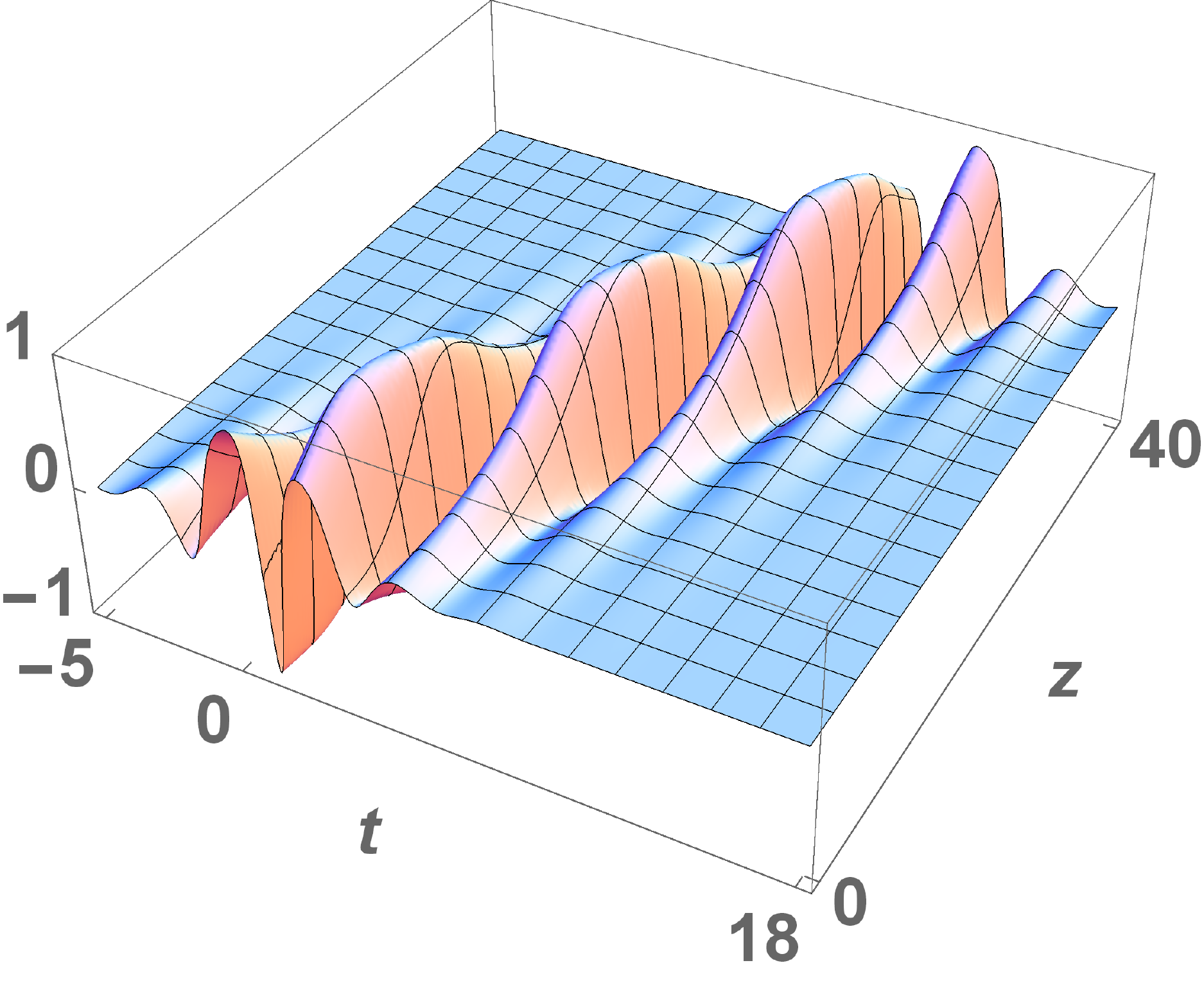}&
\includegraphics[scale=0.27]{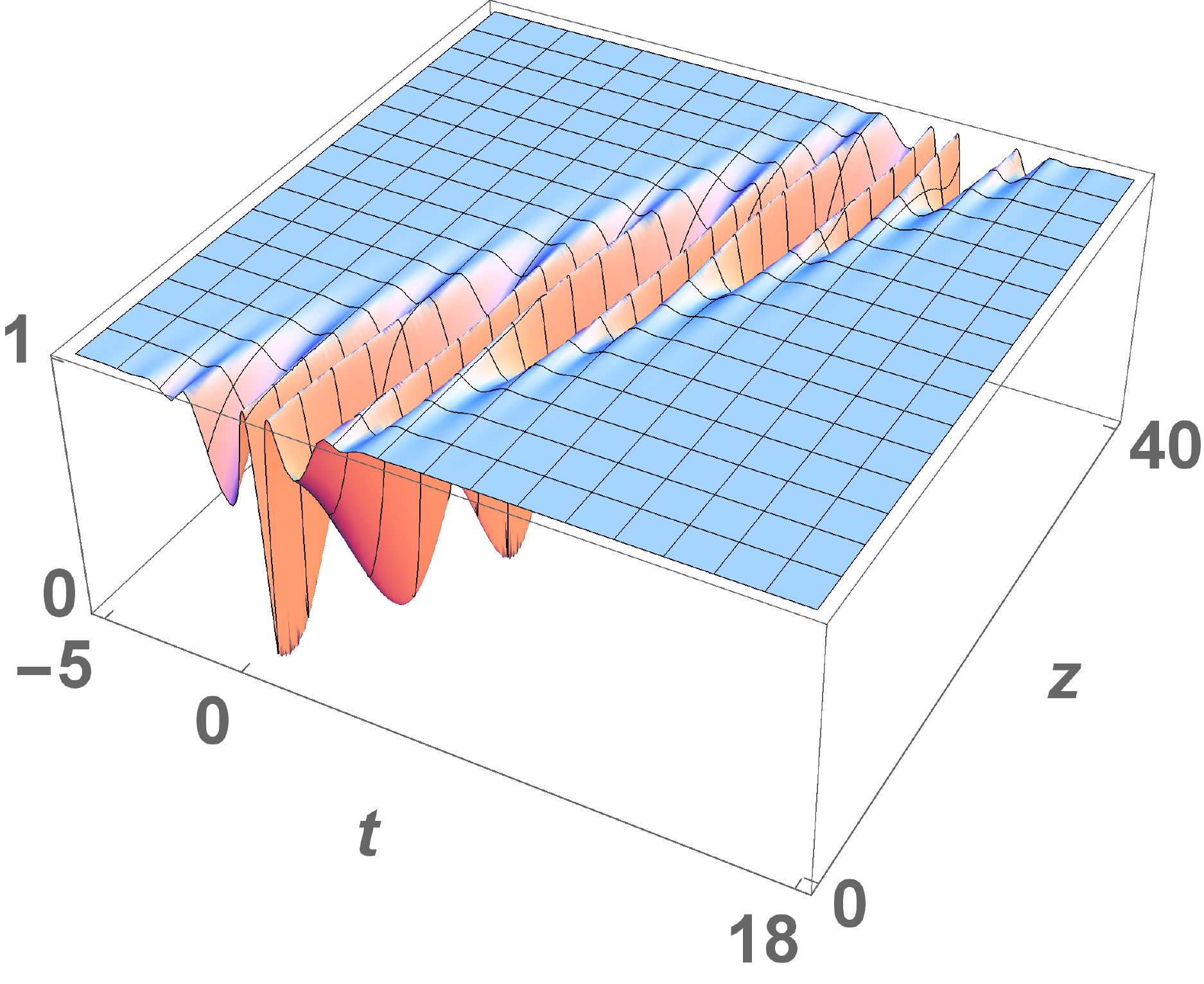}\\
\includegraphics[scale=0.27]{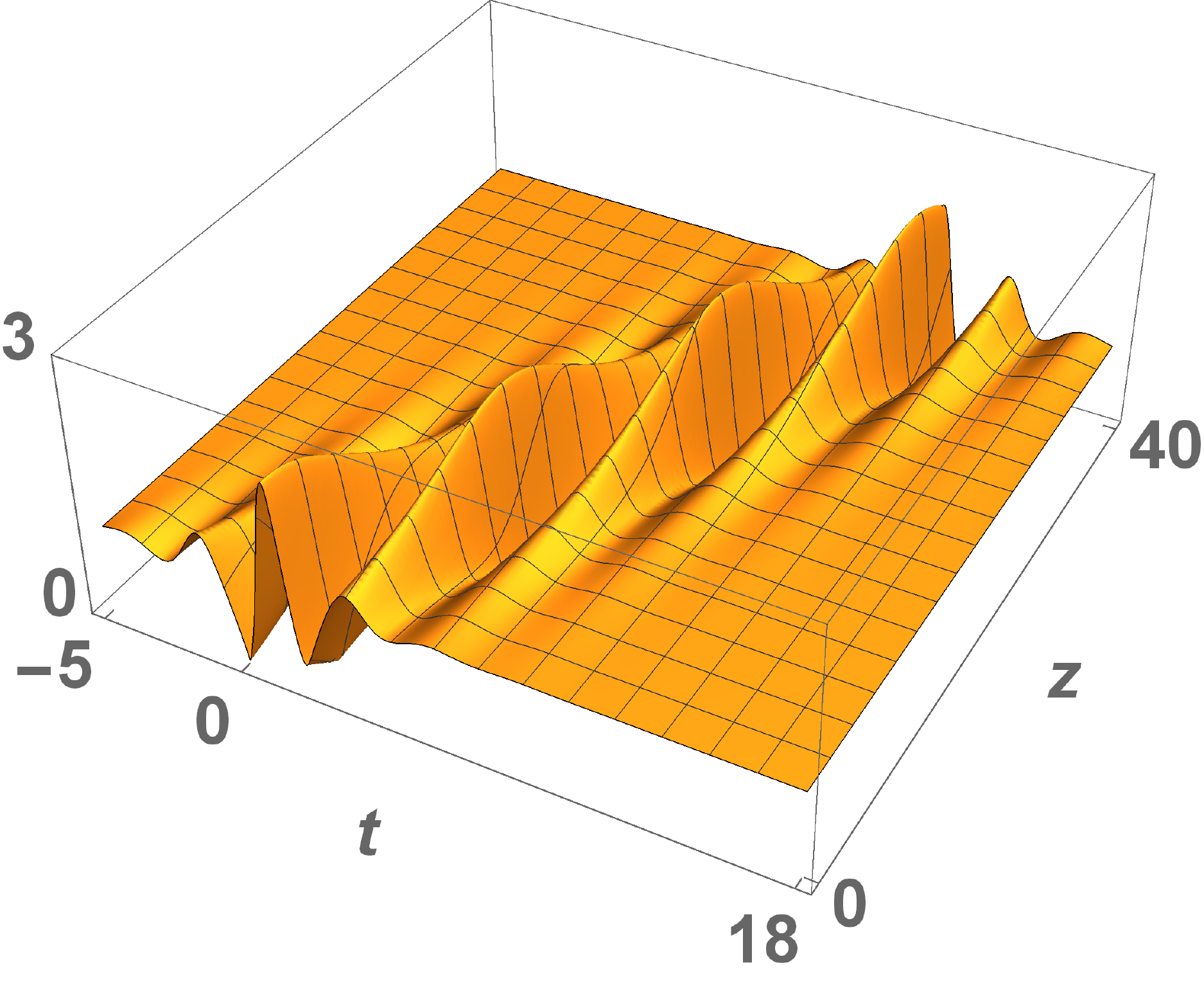}&
\includegraphics[scale=0.27]{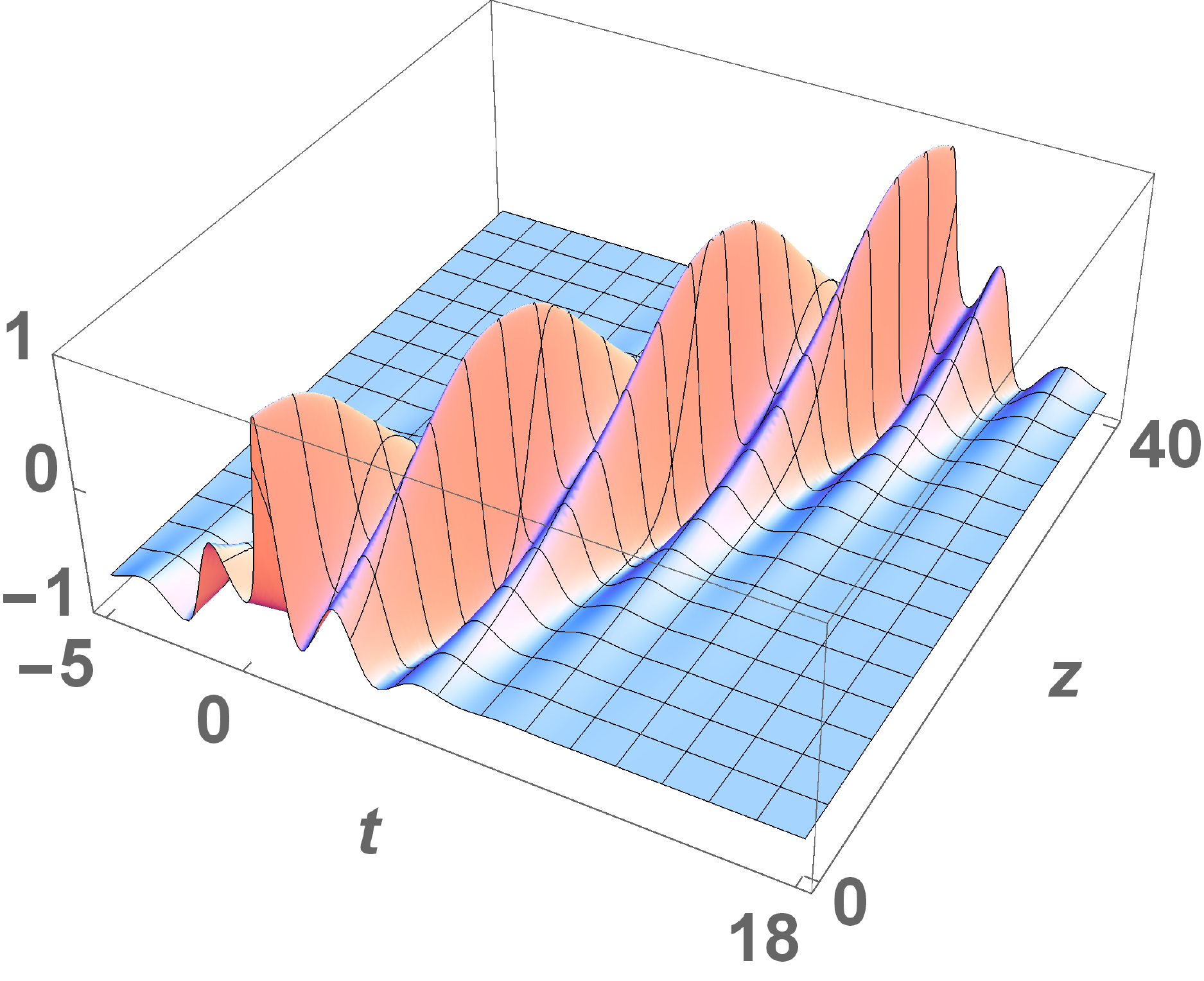}&
\includegraphics[scale=0.27]{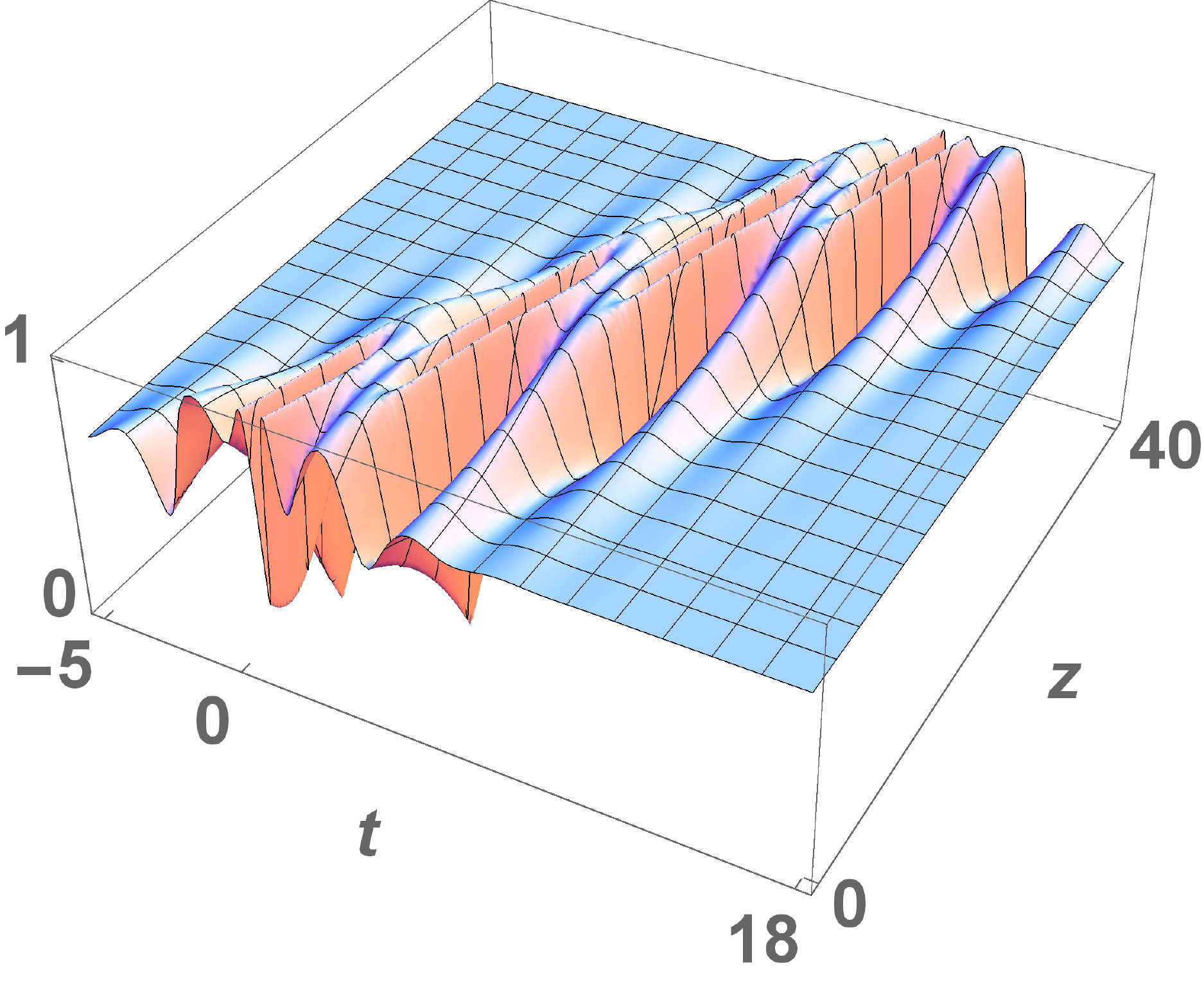}
\end{tabular}
\caption{Similar to Fig.~\ref{f:IB1solitontypeI}, 
	but for an oscillatory soliton solution~\eref{e:oscillatorysoliton} with discrete eigenvalue $\zeta_1 = \sqrt3+i$.
	All other parameters are the same as those in Fig.~\ref{f:IB1solitontypeI}.}
\label{f:generalsoliton}
\vskip2\bigskipamount
\centering
\includegraphics[scale=0.27]{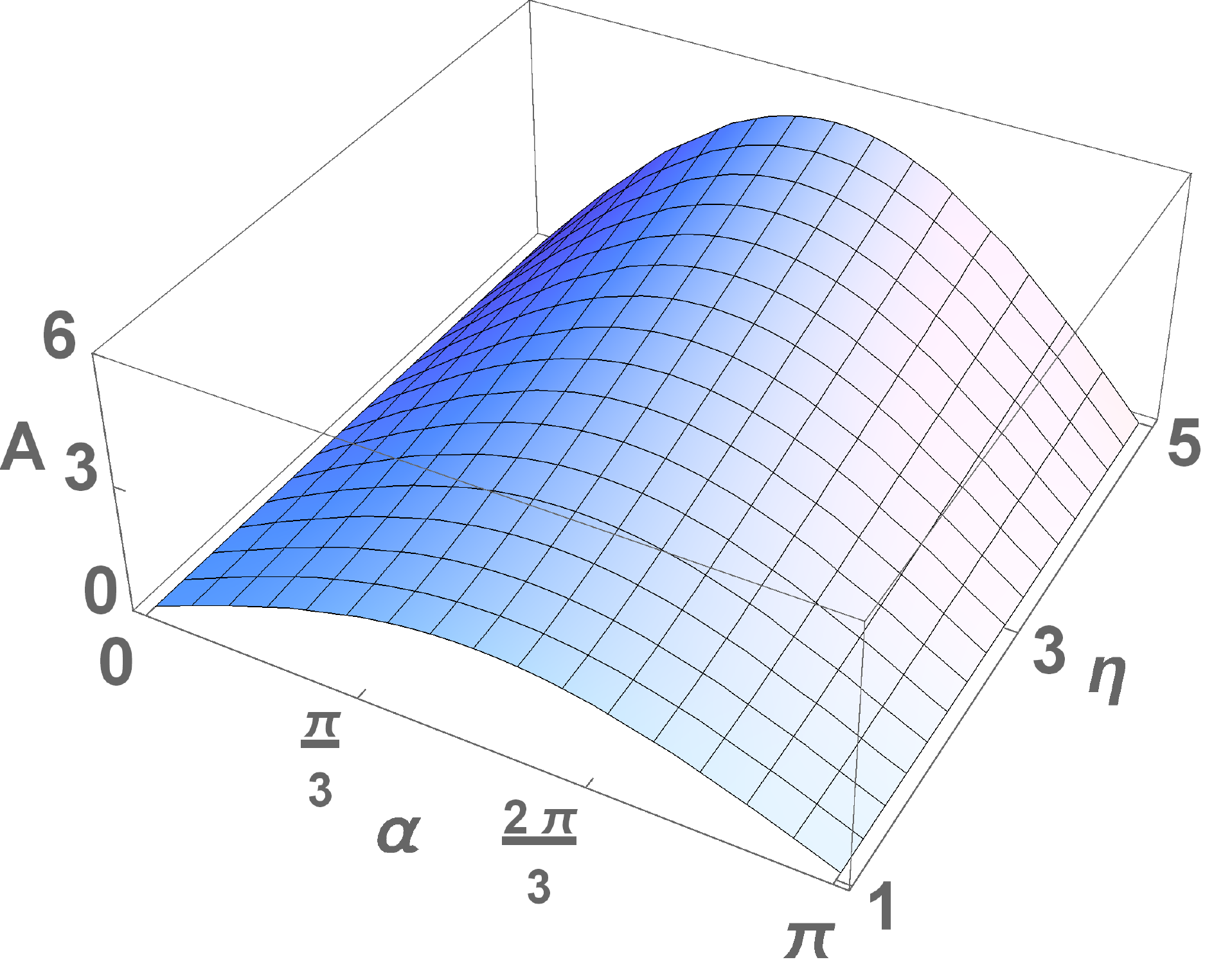}\quad
\includegraphics[scale=0.27]{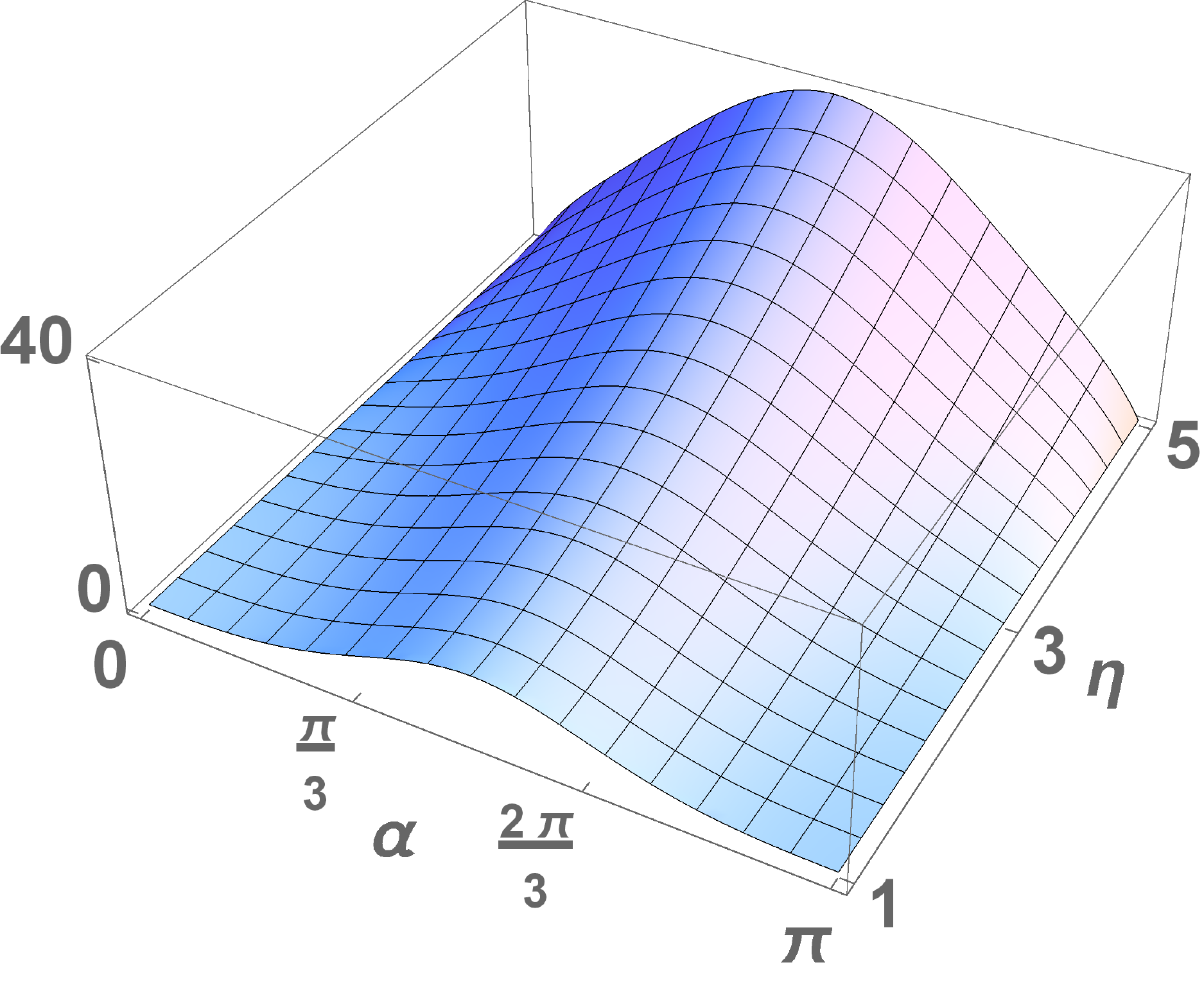}\quad
\includegraphics[scale=0.27]{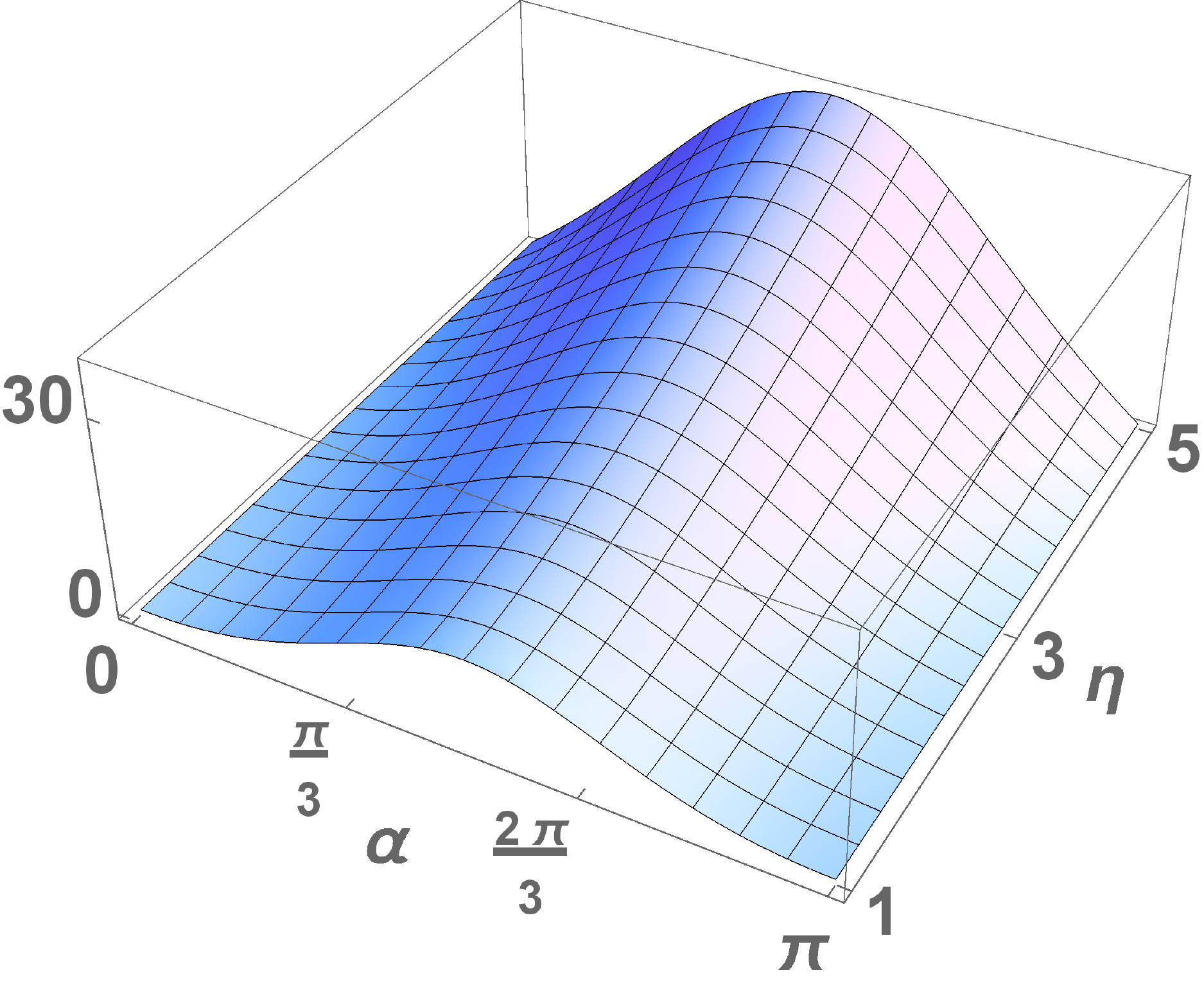}
\caption{
Soliton amplitude and velocity with inhomogeneous broadening 
corresponding to more atoms initially being in the ground than in the excited state, 
i.e., $h_- = -1$ (see text for further details).
Left: the maximum amplitude of a soliton with $q_o = 1$ as a function of $\eta$ and $\alpha$.
Center: the velocity of the soliton with $\epsilon = 0.5$.
Right: the velocity of the soliton with $\epsilon = 2$.
}
\label{f:AV}
\end{figure}

Two traveling-wave soliton solutions are shown in Figs.~\ref{f:IB1solitontypeI} and~\ref{f:IB1solitontypeI2},
whereas an oscillatory soliton solution is shown in Fig.~\ref{f:generalsoliton}.
Notice that, since $\gamma(\lambda)$ is discontinuous at $\lambda =0$,
the density matrix $\rho(t,z,\lambda)$ is also discontinuous there.
Thus, in the above figures as well as the subsequent ones, we plot $\rho(t,z,0^+)$.
Moreover, we take $\lambda$ on the first sheet, i.e., we take $\nu = 1$ in the initial state~\eref{e:solitonBC1}.
Notice also that oscillatory behavior can be easily observed from Fig.~\ref{f:generalsoliton}
just like for the focusing NLS equation with NZBC.
Moreover, one can easily observe from all the figures that $\lim_{t\to-\infty}D$ does not equal $h_-$,
and is a function of $\lambda$ even though $h_-$ is not.
This is consistent with our discussion in Section~\ref{s:boundaryvalues} and with Eqs.~\eref{e:DP=DPpm1} and~\eref{e:DP=DPpm2}.

Now that we have derived the propagation of the norming constant,
we can explicitly compute the amplitude and velocity of the soliton.
In Fig.~\ref{f:AV}(left) we show the maximal possible amplitude of the soliton, 
$A_\mathrm{max}(z)$, as a function of $\eta$ and $\alpha$, 
which is computed using Eq.~\eref{e:generalamplitude}.
Also, taking $\xi(z)$ from Eq.~\eref{e:xiphi1}, 
if $\Im\{g(\lambda) \big[\Theta(\lambda) - \gamma\Theta(i\epsilon)/ (q_o^2-\epsilon^2)^{1/2}\big]\}\ne0$ we can obtain the soliton velocity~\eref{e:solitonvelocity} as 
\[
\label{e:IBv}
V = -q_o h_-\Delta_-\sin\alpha/\Im\{g(\lambda) \big[\Theta(\lambda) - \gamma\Theta(i\epsilon)/ (q_o^2-\epsilon^2)^{1/2}\big]\}\,,
\]
where we use the notation from Section~\ref{s:1soliton}.
In Fig.~\ref{f:AV} (center and right), we show the dependence of the soliton velocity 
on the parameters $\eta$ and $\alpha$,
in the cases $\epsilon< q_o$ and $\epsilon> q_o$, respectively.
It is apparent that the velocity is an increasing function as $\eta\to\infty$. 
Thus, as the discrete eigenvalue moves closer to the real line,
the corresponding soliton travels slower. 

Note that, just like in the case of ZBG, the soliton velocity is symmetric with respect to the imaginary $\lambda$-axis.
That is, solitons corresponding to discrete eigenvalues $\lambda_n$ and $-\lambda_n^*$ (i.e., with the same imaginary part and opposite real part)
have the same velocity.  
(This is unlike what happens to the NLS equation, where such solitons would have opposite velocities.)

Importantly, note also that $\Im\{g(\lambda) \big[\Theta(\lambda) - \gamma\Theta(i\epsilon)/ (q_o^2-\epsilon^2)^{1/2}\big]\}>0$ for all $|\zeta|>q_o$.
Thus, $V\gl0$ (solitons are subluminal or superluminal) when $h_-=\mp1$ 
(i.e., more atoms initially are in the ground than in the excited state or vise versa),
respectively,
similarly to the case with ZBG.


\subsection{Soliton solutions in the sharp-line and zero-background limits}
\label{s:SLtype1}

We next discuss the two limits $\epsilon\to0$ and $q_o\to0$. 
The former is called sharp-line limit,
and the latter is called zero-background limit.
Importantly, in the development of the IST, 
one could take the limit $q_o\to0$ throughout to recover the formalism in the case of ZBG.
However, we will see shortly that the behavior of solutions differs depending on whether one lets $q_o\to0$
before or after taking the limit $\epsilon\to0$.
In other words, the limits $q_o\to0$ and $\epsilon\to0$ \textit{do not commute}.
In practice, what this means is that one must know a priori whether the physical scenario involves $\epsilon<q_o$ or $q_o<\epsilon$.

\paragraph{Case~1: Taking the limit $q_o\to0$ first.}
Taking $q_o\to 0$ in Eq.~\eref{e:q-timeevolution} we immediately obtain $q_- = 0$ for all $z>0$.
Also, as it was shown in~\cite{JMP55p031506} that, in the limit $q_o\to0$, 
the solution profile~\eref{e:staticsoliton} reduces to a hyperbolic secant.
Moreover, taking the limit $q_o\to0$ in Eq.~\eref{e:Rd-explicit_NZBG}, we obtain
\[
\nonumber
R_{-,d} = \frac{i h_-}{\epsilon-i\lambda}\sigma_3\,.
\]
Note that this expression is identical to the one in the case of ZBG~\cite{ablowitz74}.
Also, the expression~\eref{e:Rd-explicit_NZBG} was derived on sheet I. 
The same applies to $\rho(t,z,\lambda)$.
Thus, the entire solution reduces to that in the case of ZBG.
One obtains a similar result while computing everything on sheet II.

If we now take the limit $\epsilon\to0$, we recover the sharp-line limit in the case of ZBG
(in particular, $R_- = (h_-/\lambda)\,\sigma_3$).

\paragraph{Case 2: Taking the limit $\epsilon\to0$ first.}
We now take the limit $\epsilon\to0$ first.
Using Eq.~\eref{e:Rd-explicit_NZBG}, 
it is easy to show that the matrix $R_{-,d}(\zeta_1)$ is identically zero in this limit.
Therefore, the soliton solution is constant with respect to the spatial variable $z$.

If we now take the limit $q_o\to0$, we preserve a soliton solution that is constant with respect to~$z$, 
which is \textit{not} the same as the soliton solution of the case with ZBG in the sharp-line limit. 
Thus, as anticipated, the two limits $\epsilon\to0$ and $q_o\to0$ do not commute.
This is consistent with the fact that the limit $\epsilon\to0$ is subtle, and must be carried out carefully 
(e.g., see \cite{bygako1}).

\paragraph{Direct sharp-line limit.}

We can also compute the sharp-line limit directly by using the Dirac delta with the initial state for the medium~\eref{e:solitonBC1},
meaning that we take the spectral-line shape as follows
\[
g(\lambda)=\delta(\lambda)\,.
\label{e:gDirac}
\]
Again, the calculation requires evaluating the integral in Eq.~\eref{e:R_d_axis} on either $\lambda$-sheet.
Importantly, with our definition~\eref{e:defgamma} of $\gamma(\lambda)$,
both $\zeta(\lambda)$ and $\gamma(\lambda)$ are discontinuous at $\lambda=0$.
Therefore, we need to use identity~\eref{e:DiracdeltaII} in the calculation. 
We then obtain $q_-(z) = q_o$ from the propagation equations~\eref{e:wpmdef} and~\eref{e:q-timeevolution}.
Calculating the integral in Eq.~\eref{e:R_d_axis} we obtain
$R_{-,d} = 0$.
Thus, the norming constant $C_1$ satisfies
${\partial C_1}/{\partial z} = 0$ from the ODE~\eref{e:dCdz},
implying
\[
\nonumber
\xi(z) = \xi(0)\,,\qquad 
\varphi(z) = \varphi(0)\,.
\]
In other words, the soliton solution~\eref{e:staticsoliton} is constant in $z$,
and therefore the soliton travels at the speed of light. 
From these results, one can also reconstruct the density matrix $\rho(t,z,\lambda)$.
This result is precisely the same as that obtained in case~2 above, namely by taking the limit $\epsilon\to0$ first.

\subsection{Type 3. Periodic solutions}
\label{s:periodic}

Here, we discuss the nontrivial limiting case of oscillatory soliton solutions~\eref{e:oscillatorysoliton},
when the discrete eigenvalues tend to the branch cut of $\gamma$.
More precisely, 
we take the limit as $\eta\to1^+$ of the discrete eigenvalue $\zeta_1 = q_o \eta\e^{i\alpha}$, 
as in the parameterization~\eref{e:1solitonC1},
with the phase $\alpha\in(0,\pi/2)\cup(\pi/2,\pi)$, and define
$q_\mathrm{P}(t,z) = \lim_{\eta\to1}q(t,z)$. 
[Several traces of the discrete eigenvalue as $\eta\to1^+$ with fixed values of $\alpha$ are shown in Fig.~\ref{f:solitontype}(right).]
After simple calculations, this solution is found to be
\begin{gather}
\label{e:periodicq}
q_\mathrm{P}(t,z) = 
\e^{-2i\alpha} q_- \frac{\cosh(\chi - 2i\alpha)+ |\tan\alpha|\sin(s-\alpha) \cos\alpha}
{\cosh\chi - |\tan\alpha|\sin(s-\alpha) \cos\alpha}\,,\\
\mu_{1,1}(t,z,\zeta) = \big[\zeta^2 X - q_o^2 \cosh(\chi - 2 i\alpha) + q_o^2\kappa+i \zeta q_o \tilde{\kappa}\big]
\big/\big[(\zeta ^2-\zeta_1^2)X \big]\,,\\
\mu_{1,2}(t,z,\zeta) = i q_- \big[q_o^2 X - \zeta^2 \cosh (\chi - 2i\alpha) + \zeta ^2\kappa - i \zeta q_o \tilde{\kappa}\big]
\big/\big[\zeta (q_o^2 - \e^{2 i \alpha } \zeta ^2)X\big]\,,
\end{gather}
where
\begin{gather}
\tilde\kappa = \sin (2 \alpha ) \cos(s-\alpha )\,,\qquad
X = \cosh\chi - \sin(s-\alpha)\sin\alpha\,,\\
\chi(z) = \xi(z)-\ln(q_o|\sin(2\alpha)|)\,,\qquad
s(t,z) = -2q_ot \cos \alpha + \varphi(z) + 2W_-(z)\,,
\end{gather}
and $\xi(z)$ and $\varphi(z)$ will be discussed next.
Notice that the $t$-dependence [in $s(t,z)$] of the solution~\eref{e:periodicq} only appears in trigonometric functions.
As a result, this solution is periodic in both $t$ and localized along the curve $\chi(z) = y$.
We should point out that this periodic solution does not satisfy our initial data: 
$\lim\nolimits_{t\to\pm\infty}q(t,z) = q_-(z)$.
This, however, is similar to what happens when the same limiting process is applied to the focusing NLS equation \cite{JMP55p031506}.
Indeed, solution~\eref{e:periodicq} is the analogue of the so-called Akhmediev breathers \cite{Akhmediev} of the focusing NLS equation, 
which are spatially periodic.
However, the Akhmediev breathers are not temporally periodic, but the solutions of the MBEs are.
Of course, in our case the propagation depends on the choice of $\rho_{-,d}$ and the spectral-line shape $g(\lambda)$.
We next compute the $z$ dependence explicitly with inhomogeneous broadening.

\begin{figure}[t!]
\centering
\begin{tabular}[b]{ccc}
\includegraphics[scale=0.21]{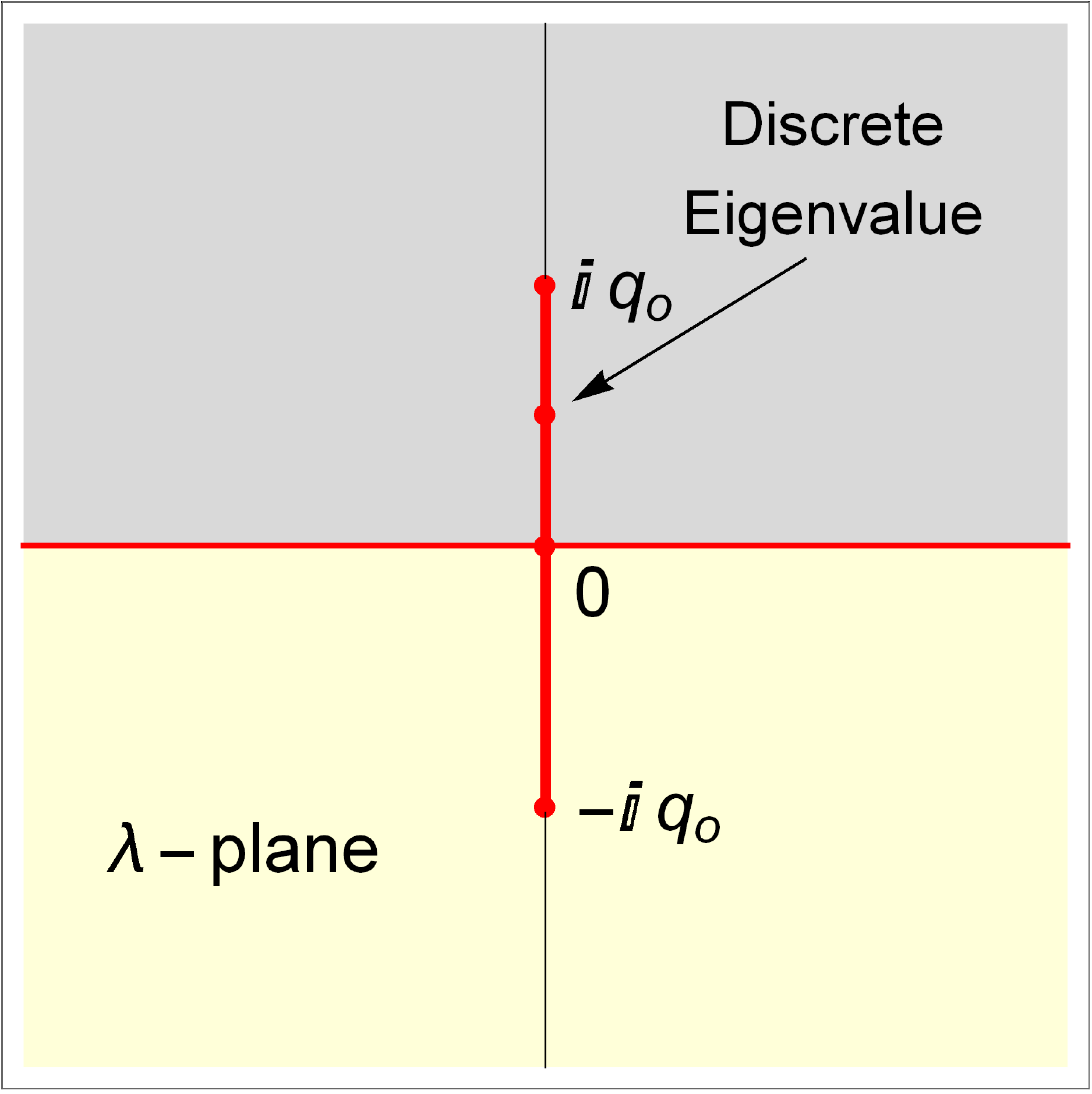}&
\includegraphics[scale=0.27]{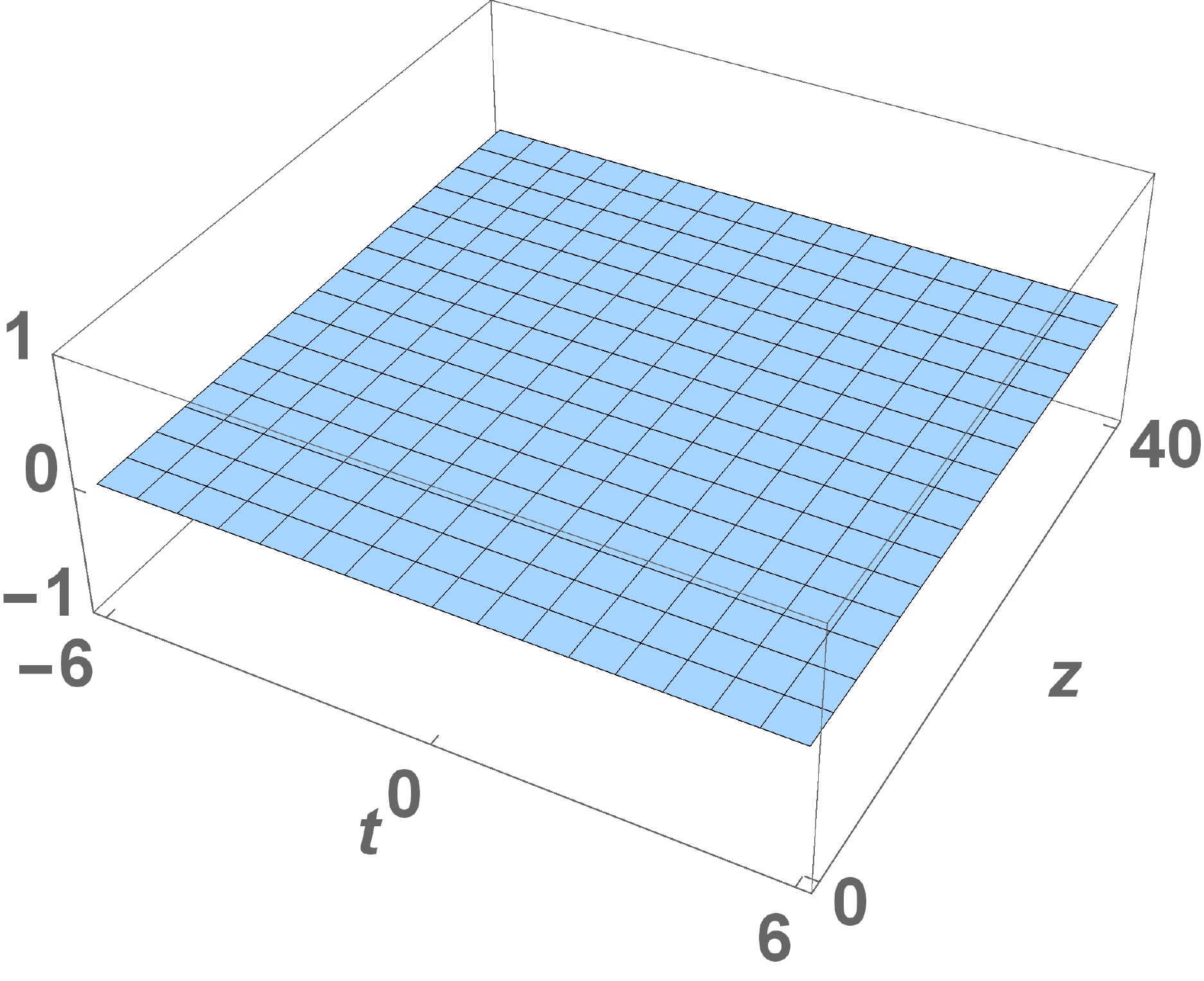}&
\includegraphics[scale=0.27]{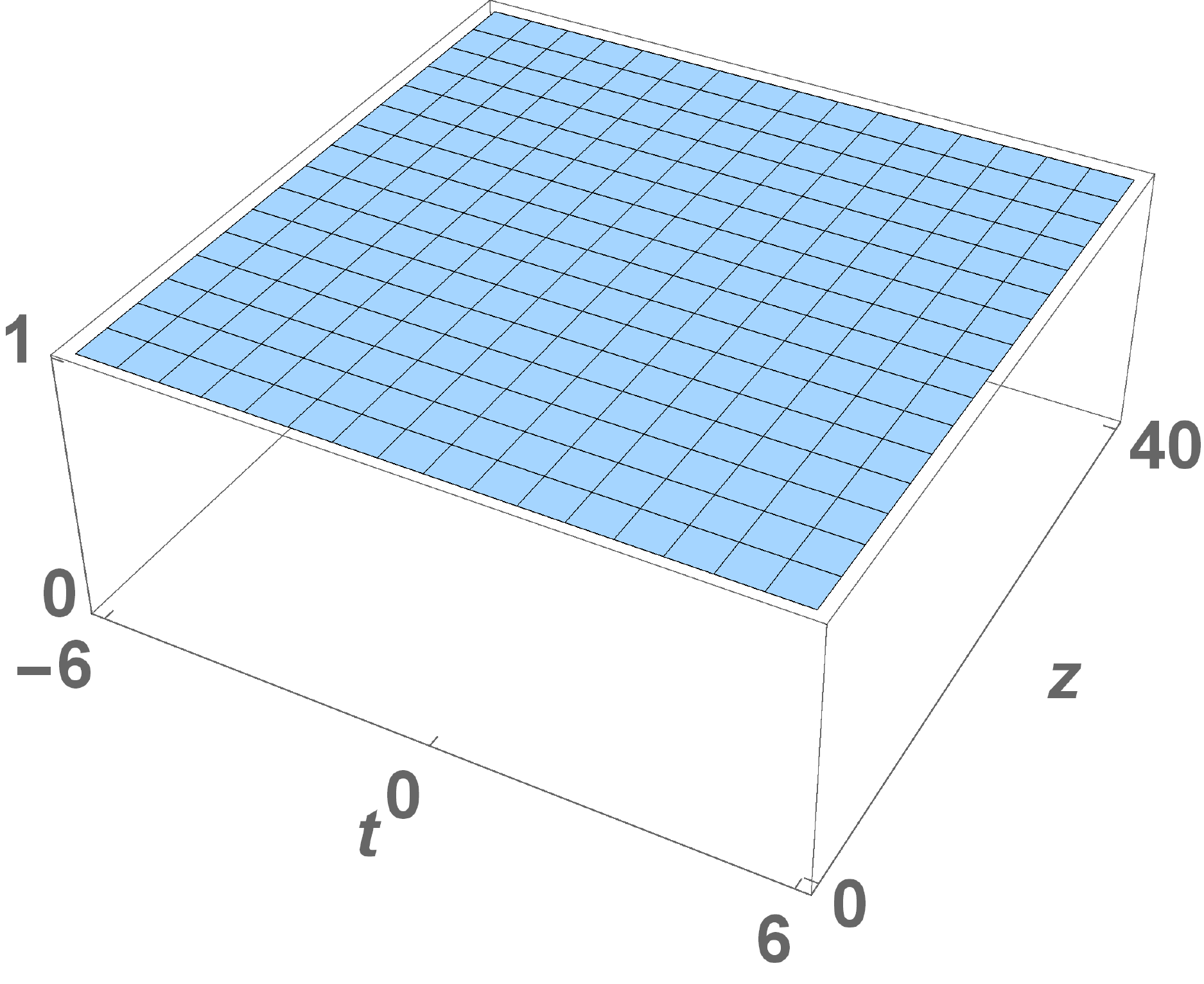}\\
\includegraphics[scale=0.27]{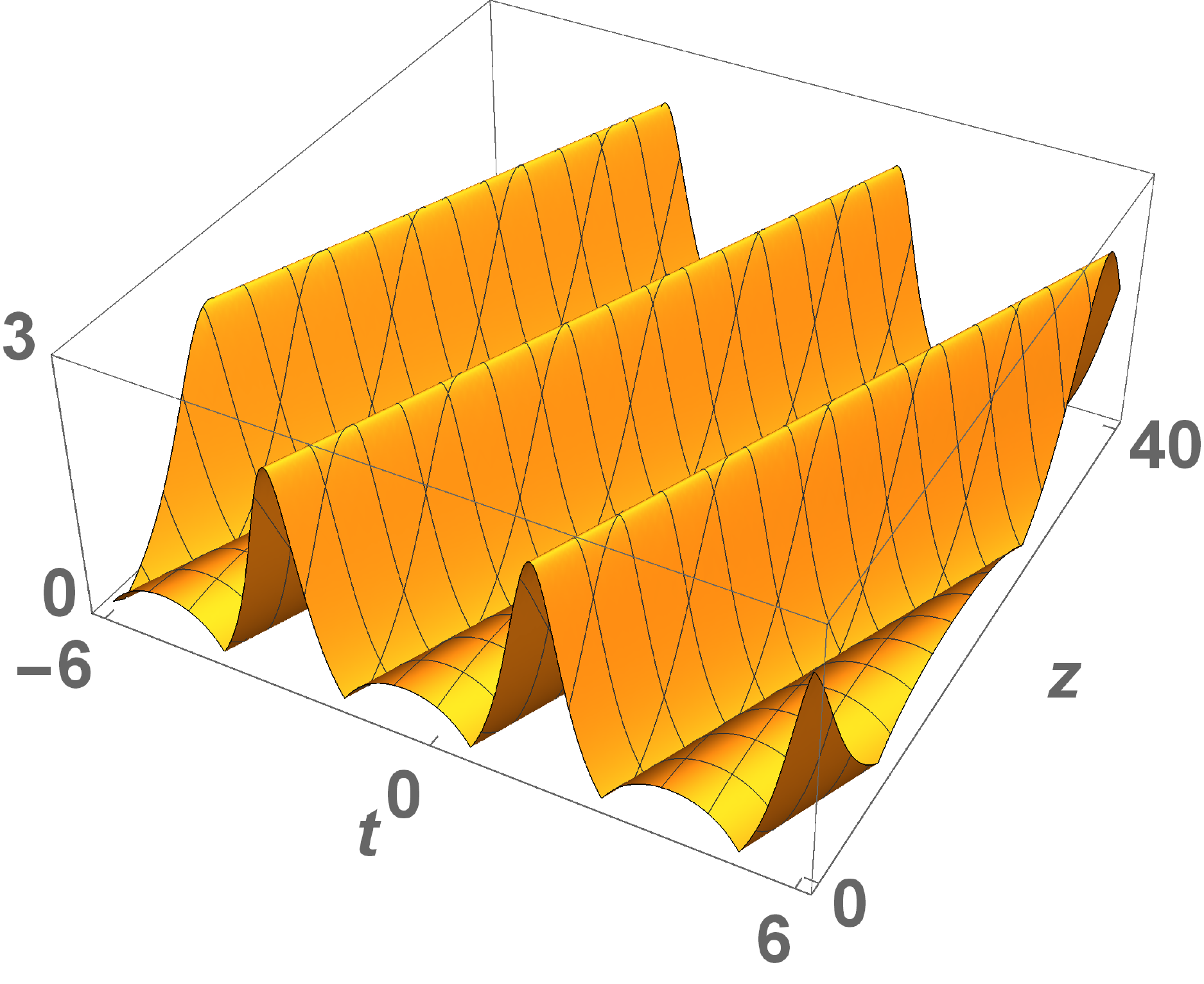}&
\includegraphics[scale=0.27]{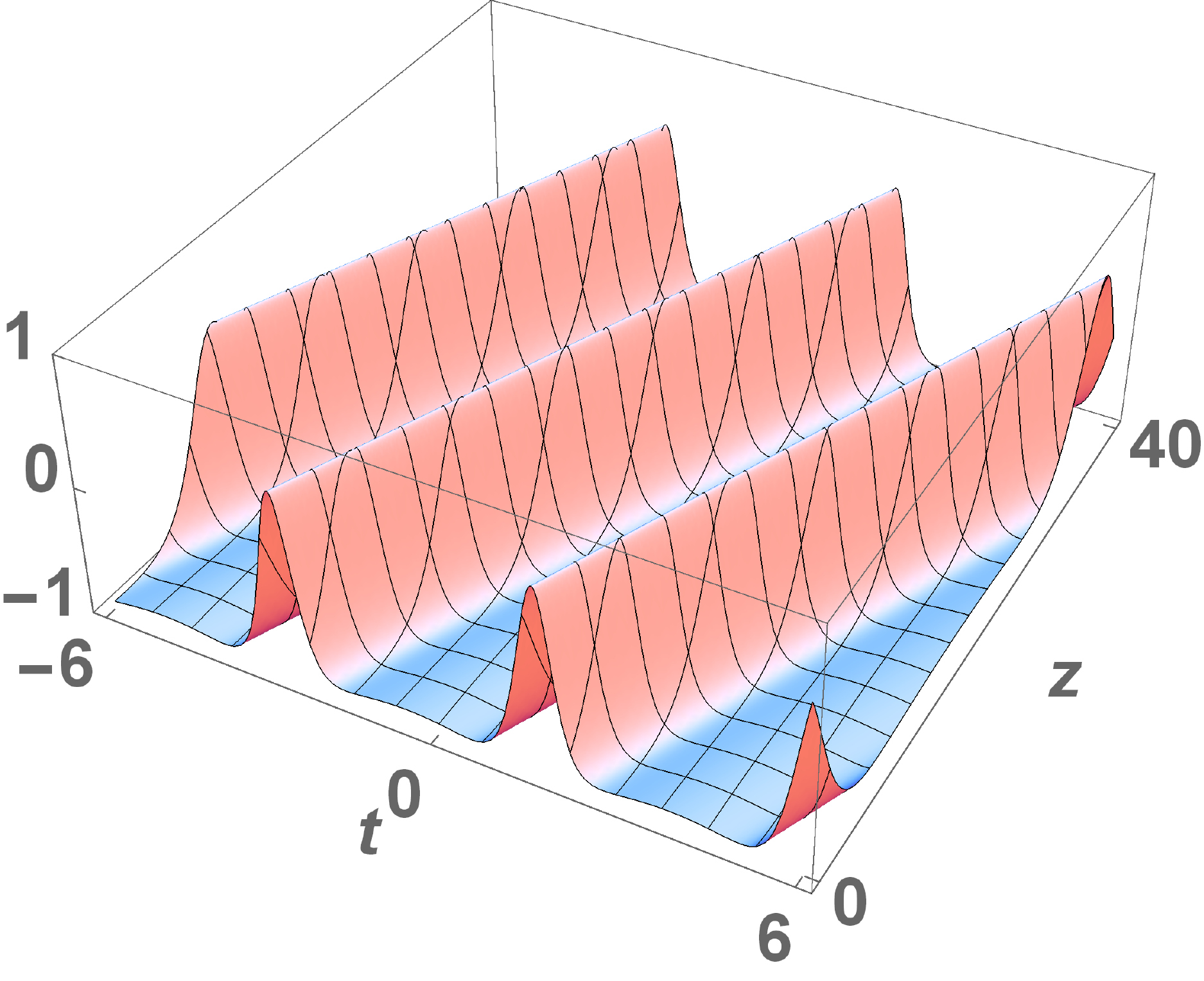}&
\includegraphics[scale=0.27]{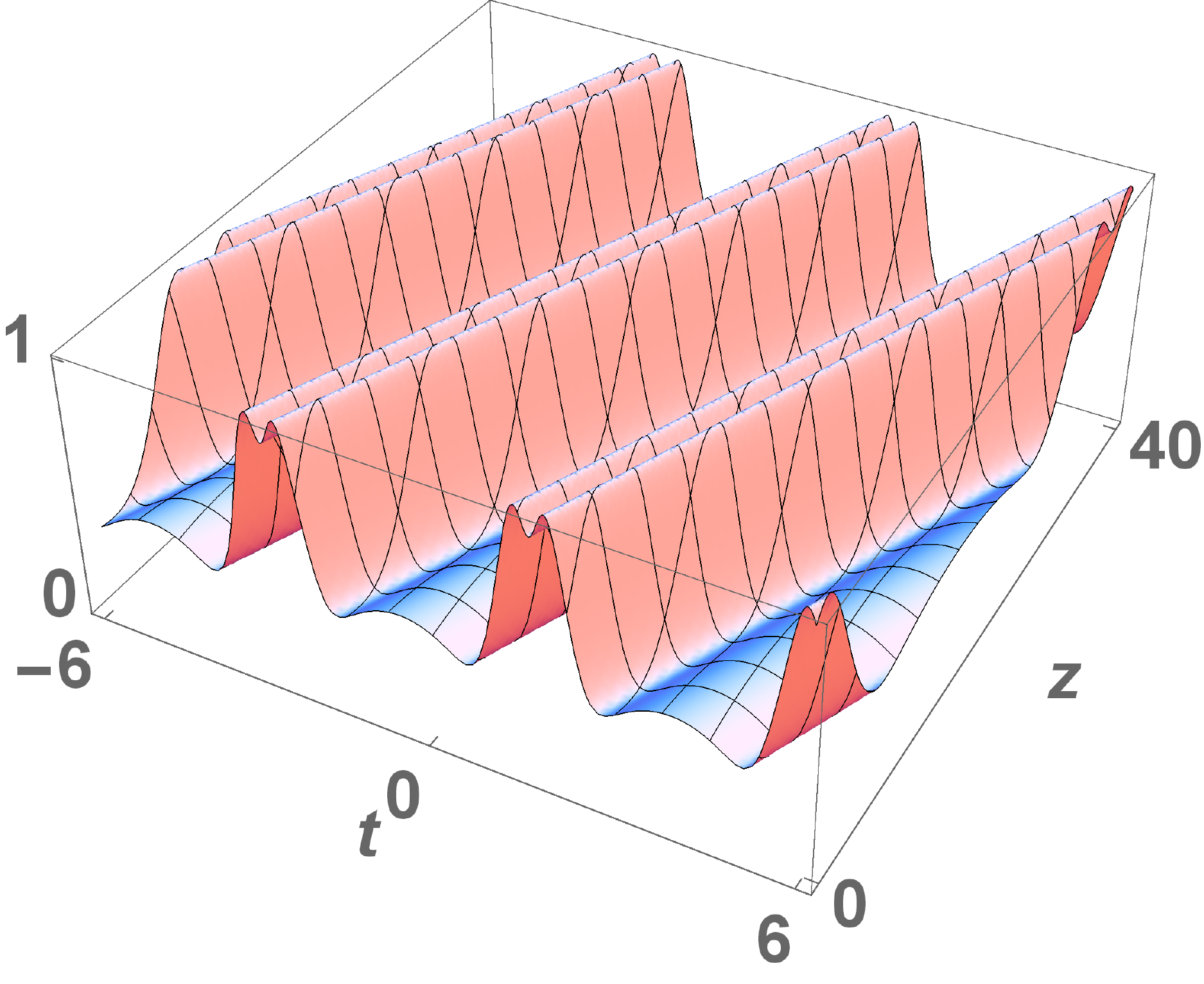}
\end{tabular}
\caption{
    One periodic solution~\eref{e:periodicq} with the initial state for the medium~\eref{e:solitonBC1}, 
    inhomogeneous broadening: $\alpha = \pi/4$,
    $h_- = -1$ (i.e., more atoms are  initially in the ground state than in the excited state), 
    $q_o = 1$, $\epsilon = 2$, 
    $\xi(0) = 0$ and $\varphi(0) = -\pi/2$.
    Top left: the discrete eigenvalue in the $\lambda$-plane.
    Bottom left: the optical field $|q(t,z)|$.
    Center and right: components of density matrix $\rho(t,z,\lambda)$.
    Top center: $D(t,z,0^+)$.
    Top right: $|P(t,z,0^+)|$.
    Bottom center: $D(t,z,q_o)$.
    Bottom right: $|P(t,z,q_o)|$.}
\label{f:IBperiodic1}
\end{figure}
%

\paragraph{Periodic solution with inhomogeneous broadening.}

Following the discussion in Section~\ref{s:IBsoliton}, 
taking the limit of Eq.~\eref{e:Rd-explicit_NZBG} as $\eta\to1$, we obtain
the following quantities appearing in the soliton solution
\[
\label{e:periodicparameter}
W_- = 0\,,\qquad
R_{-,d} = h_- R_\epsilon(\alpha)\,\sigma_3\,,
\]
with 
\[
R_\epsilon = g(iq_o\sin\alpha)
\bigg(2\,\mathrm{arcsech}(\sin\alpha) - \frac{q_o \Theta(i\epsilon)\cos\alpha}{\sqrt{\smash{q_o}^2-\epsilon^2}}\bigg)\,,
\label{e:Repsilondef}
\]
Therefore, we have the following two quantities from Eqs.~\eref{e:xiphi1} and~\eref{e:xiphi2}, 
which determine the norming constant,
\[
\nonumber
\xi(z) = \xi(0)\,,\qquad
\varphi(z) = -h_- R_\epsilon\,z + \varphi(0)\,.
\]
In this particular case, we can rewrite the periodic solution~\eref{e:periodicq} explicitly as
\[
\label{e:periodicIB}
q_\epsilon(t,z) = \e^{-2i\alpha} q_o 
\frac{\cosh\big[\ln(q_o|\sin2\alpha|) + 2i\alpha - \xi(0)\big] - \sin(s+\alpha)\sin\alpha}
{\cosh\big[\ln(q_o|\sin2\alpha|) - \xi(0)\big] + \sin(s+\alpha)\sin\alpha}\,,
\]
where 
\[
s(t,z) = 2q_ot \cos \alpha + h_- R_\epsilon(\alpha)\,z - \varphi(0)\,,
\]
with $\xi(0)$ and $\varphi(0)$ arbitrary real constants and $R_\epsilon(\alpha)$ given by Eq.~\eref{e:Repsilondef}.
The modified eigenfunction matrix $\mu$ can be computed in a similar way as in the previous cases and is omitted for brevity.

Due to the fact that both variables $t$ and $z$ appear in trigonometric functions, 
this solution $q_\epsilon(t,z)$ is periodic in both $t$ and $z$ with frequencies 
$\omega_t = q_o\cos\alpha/\pi$
and $\omega_z = R_\epsilon/(2\pi)$.
Moreover, each peak travels with velocity $V = -2q_o h_- \cos\alpha/R_\epsilon$.
One such solution is shown in Fig.~\ref{f:IBperiodic1}.

It is easy to compute the sharp-line limit, i.e., $\epsilon \to 0$, of the periodic solution.
By Eq.~\eref{e:periodicparameter}, we find that $R_\epsilon \to 0$ as $\epsilon\to0$.
Thus, the solution becomes
\[
\nonumber
q_\mathrm{P}(t) = \e^{-2i\alpha} q_o 
\frac{\cosh\big[\ln(q_o|\sin2\alpha|) + 2i\alpha - \xi(0)\big] - \sin(s+\alpha)\sin\alpha}
{\cosh\big[\ln(q_o|\sin2\alpha|) - \xi(0)\big] + \sin(s+\alpha)\sin\alpha}\,,
\]
with 
\[
\nonumber
s(t) = 2q_ot \cos \alpha - \varphi(0)\,.
\]
Notice that this sharp-line limit solution is independent of $z$.
Again, notice that all the solutions discussed in this section are the Maxwell-Bloch analogue of the so-called ``Akhmediev breathers'' of the focusing NLS equation.

\subsection{Type 4. Rational solutions}
\label{s:rational}

Recall that, for the focusing NLS with NZBC, one can obtain the Peregrine solution as a suitable limit of its one-soliton solutions~\cite{Peregrine}.
We next show that it is possible to obtain rational solutions of the MBEs with NZBG in a similar way.
We compute such a solution by taking the limit of the traveling-wave soliton solution~\eref{e:staticsoliton} 
as the discrete eigenvalue tends to the branch point $iq_o$, i.e., $\eta\to1$.
(Recall that the discrete eigenvalue is $\zeta_1 = iq_o\eta$.)
Notice that this limit is different from the one in the previous section due to different phases of the discrete eigenvalue.
To continue our calculation, we first expand the norming constant $C_1$ around $\eta=1$. 
Recall that we parameterized the norming constant $C_1$ in the parameterization~\eref{e:1solitonC1}. 
Therefore, we have the following expansions
\begin{gather}
\label{e:xisexpansion1}
\xi(\eta,z) = \xi_o(z) + \xi_1(z)(\eta - 1) + O(\eta - 1)^2\,,\qquad
\eta\to1\,,\\
\label{e:xisexpansion2}
s(\eta,z) = s_o(z) + s_1(z)(\eta - 1) + O(\eta - 1)^2\,,\qquad
\eta \to 1\,,
\end{gather}
where
\[
\xi_o(z) = \lim_{\eta\to1} \xi(\eta,z)\,,\qquad 
\xi_1(z) = \lim_{\eta\to1} \big[\xi(\eta,z) - \xi_o(z)\big]/(\eta - 1)\,,
\nonumber
\]
and similarly for $s_o(z)$ and $s_1(z)$.
In order to write down these two Taylor expansions explicitly, 
we need to compute the expansion of the matrix $R_{-,d}(\zeta_1,z)$ as $\eta\to1$, namely:
\vspace*{-0.4ex}
\[
\label{e:R-dexpansion}
R_{-,d}(iq_o\eta,z) = R^{(0)}(z) + R^{(1)}(z)(\eta - 1) + O(\eta - 1)^2\,.
\]
Using Eq.~\eref{e:R_d_axis}, straightforward calculations show that the coefficients in the above expression are
\vspace*{-0.4ex}
\[
\label{e:RoR1}
R^{(0)}(z) = 2w_-(z)\sigma_3\,,\qquad
R^{(1)}(z) = i q_o \int \frac{\rho_{-,d} g(\lambda)}{\gamma(\lambda-i q_o)}\d\lambda\,,
\]
where $w_-(z)$ is defined in Eq.~\eref{e:wpmdef}.
Notice that $R^{(1)}(z)$ is nonzero in general, and can be computed on either $\lambda$-sheet.
Let us recall the propagation equation for the norming constant~\eref{e:dCdz} and the definition of $W_-(z)$~\eref{e:q-timeevolution}.
The norming constant in this limit is given by
$\lim_{\eta\to1} C_1(i q_o \eta,z) = \e^{-2iW_-(z)} \lim_{\eta\to1} C_1(i q_o \eta,0)$.
Comparing this expression with Eq.~\eref{e:1solitonC1},
we find 
\[
\nonumber
\xi_o(z) = \lim_{\eta\to1} \xi(i q_o \eta,0)\,,\qquad
s_o(z) =\lim_{\eta\to1}\varphi(i q_o \eta,0)\,.
\]
Therefore, both $\xi_o(z)$ and $s_o(z)$ are independent of~$z$.  
Correspondingly, hereafter we will simply write them as $\xi_o$ and $s_o$.

Importantly, if one takes the limit $\eta\to1$ with an arbitrary choice of ICs for $\xi_o$ and $s_o$, 
the soliton solution~\eref{e:staticsoliton} reduces to the background solution (i.e., a trivial solution).
On the other hand, choosing the ICs to be 
$\xi_o = -\Delta_o$ [recall $\Delta_o$ is defined in Eq.~\eref{e:Deltao}] and $s_o = -\pi/2$,
the limit of the solution~\eref{e:staticsoliton} as $\eta\to1$ yields
a solution of the MBEs with rational dependence on $t$, namely: 
\[
\label{e:rationalq}
q_R(t,z) = \lim_{\eta\to1} q(t,z)
= q_-(z)\frac{\big[ 2q_o t+\xi_1(z)\big]^2 + s_1(z)^2 + 4i s_1(z) - 3}
{\big[ 2q_o t+\xi_1(z)\big]^2 + s_1^2(z) + 1}\,.
\]
The center of the soliton and its velocity are given by the constraint $2q_o t+\xi_1(z) = 0$.
Furthermore, the density matrix is also a rational function of $t$.  
Indeed, computing the limit of the eigenfunction $\mu_{-}(t,z,\zeta)$ in Eq.~\eref{e:mu-reflectionless} 
using the same methods as above, we find
\begin{multline}
\label{e:rationalmu}
\mu_R(t,z,\zeta) = 
\lim_{\eta\to1}\mu_{-}(t,z,\zeta) = \\
\big[
(Y\gamma\zeta-2q_o^2) I + 2iq_o(q_o s_1 + X\zeta) \sigma_3 + i(Y \gamma - 2 \zeta)\sigma_3 Q_- +2(q_o X-\zeta s_1) Q_-
\big]/(\gamma  \zeta  Y)\,,
\end{multline}
with $I$ the $2\times2$ identity matrix and
\[
q_- = q_o \e^{2i W_-(z)} \,,\qquad
X = 2q_o t+\xi_1\,,\qquad
Y = X^2 + s_1^2+1\,.
\]
The $z$-dependence of all the entries was omitted for brevity.
One can now use~\eref{e:solitonrho} to obtain the density matrix $\rho(t,z,\lambda)$.
The final result is omitted for brevity due to its complexity.
We refer to the solution~\eref{e:rationalq} 
and the density matrix generated by Eqs.~\eref{e:rationalmu} and~\eref{e:solitonrho}  as the rational solution of the MBEs with NZBG.

In the limit $q_o\to0$ of the above solution, 
$q_\mathrm{R}(t,z)\to0$ and $\mu_\mathrm{R}(t,z,\zeta)\to I$, implying $\rho_\mathrm{R}(t,z,\zeta) \to \rho_-(z,\zeta)$. 
Therefore, the rational solution reduces to the trivial solution of the MBEs with ZBG.

\paragraph{Rational solution with inhomogeneous broadening.}

Recall that the propagation of the above solution~\eref{e:rationalq} is determined by three quantities $W_-(z)$, $\xi_1(z)$ and $s_1(z)$. 
The latter two are obtained from the second term $R_1(z)$ in Eq.~\eref{e:RoR1}.
We turn to calculating them next.

\begin{figure}[t!]
\vglue-2\bigskipamount
\centering
\begin{tabular}[b]{ccc}
\includegraphics[scale=0.21]{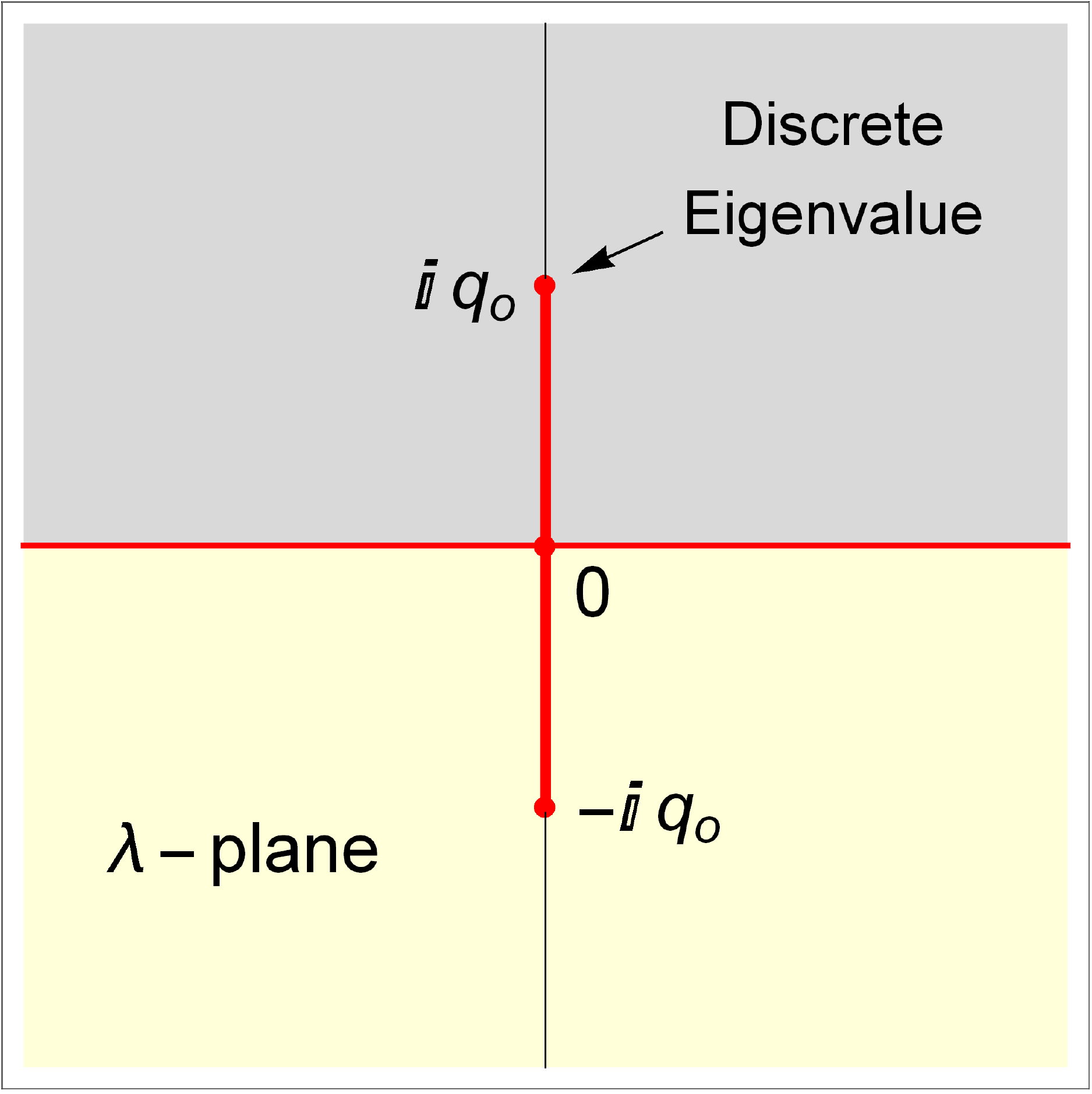}&
\includegraphics[scale=0.27]{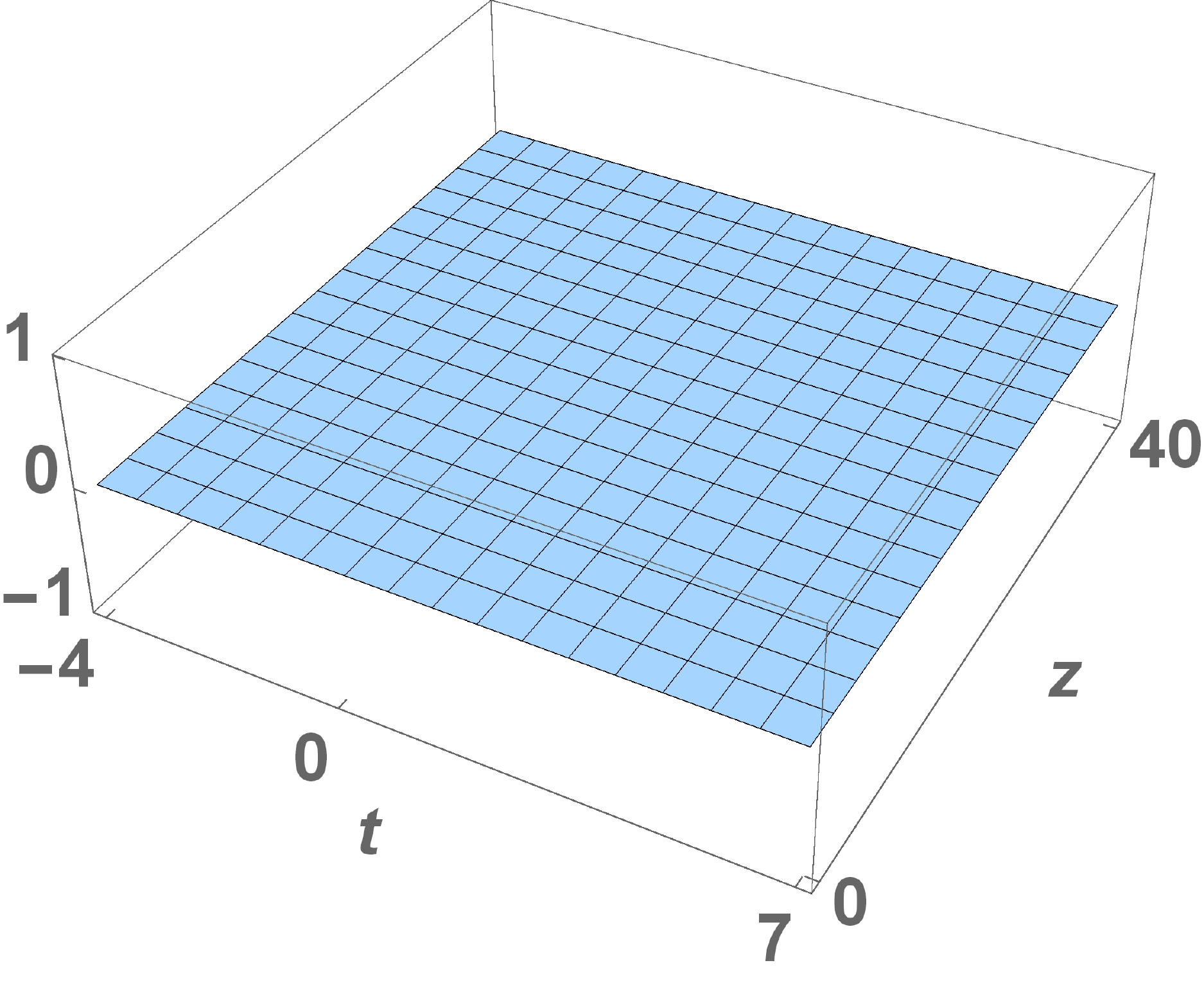}&
\includegraphics[scale=0.27]{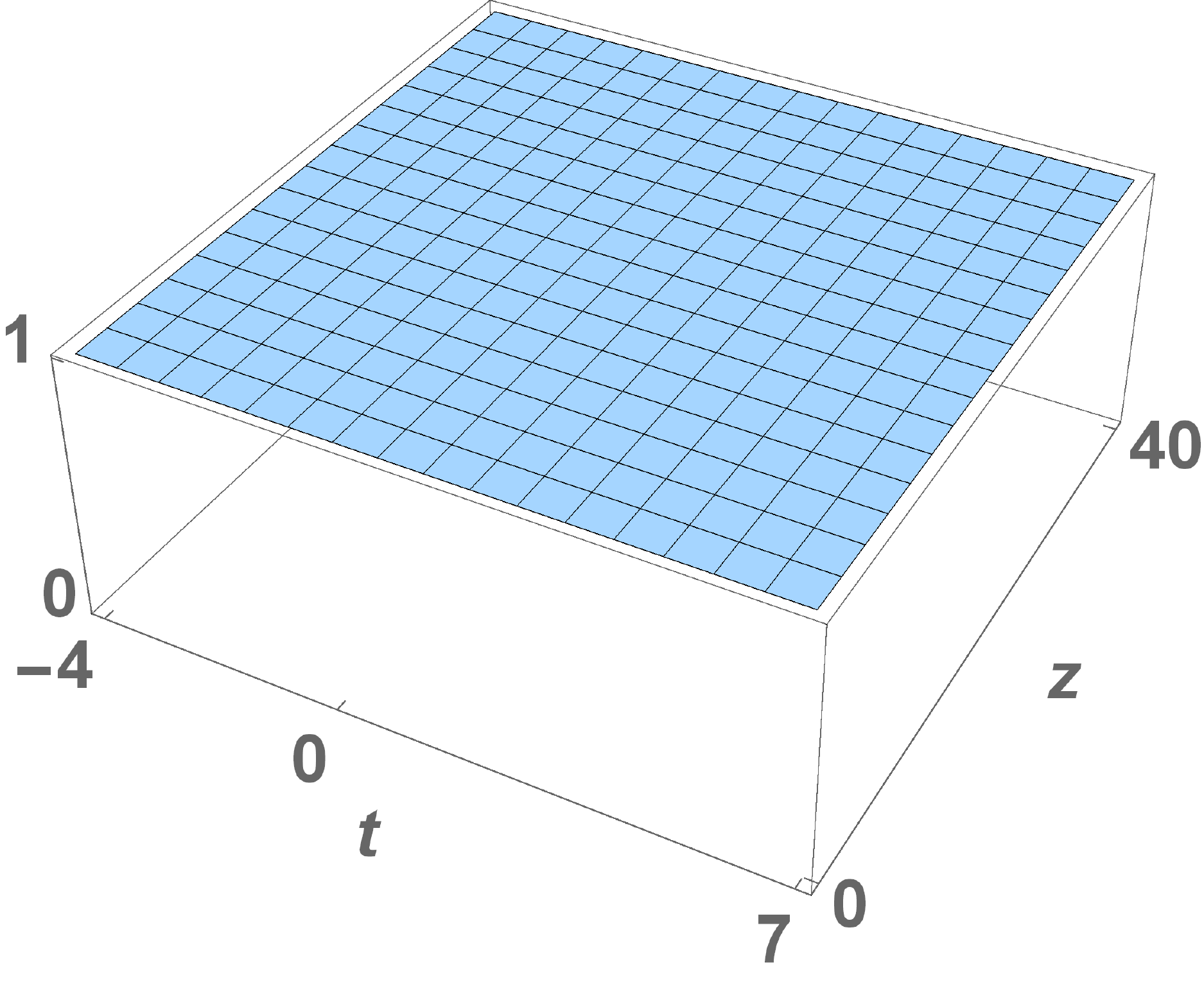}\\
\includegraphics[scale=0.27]{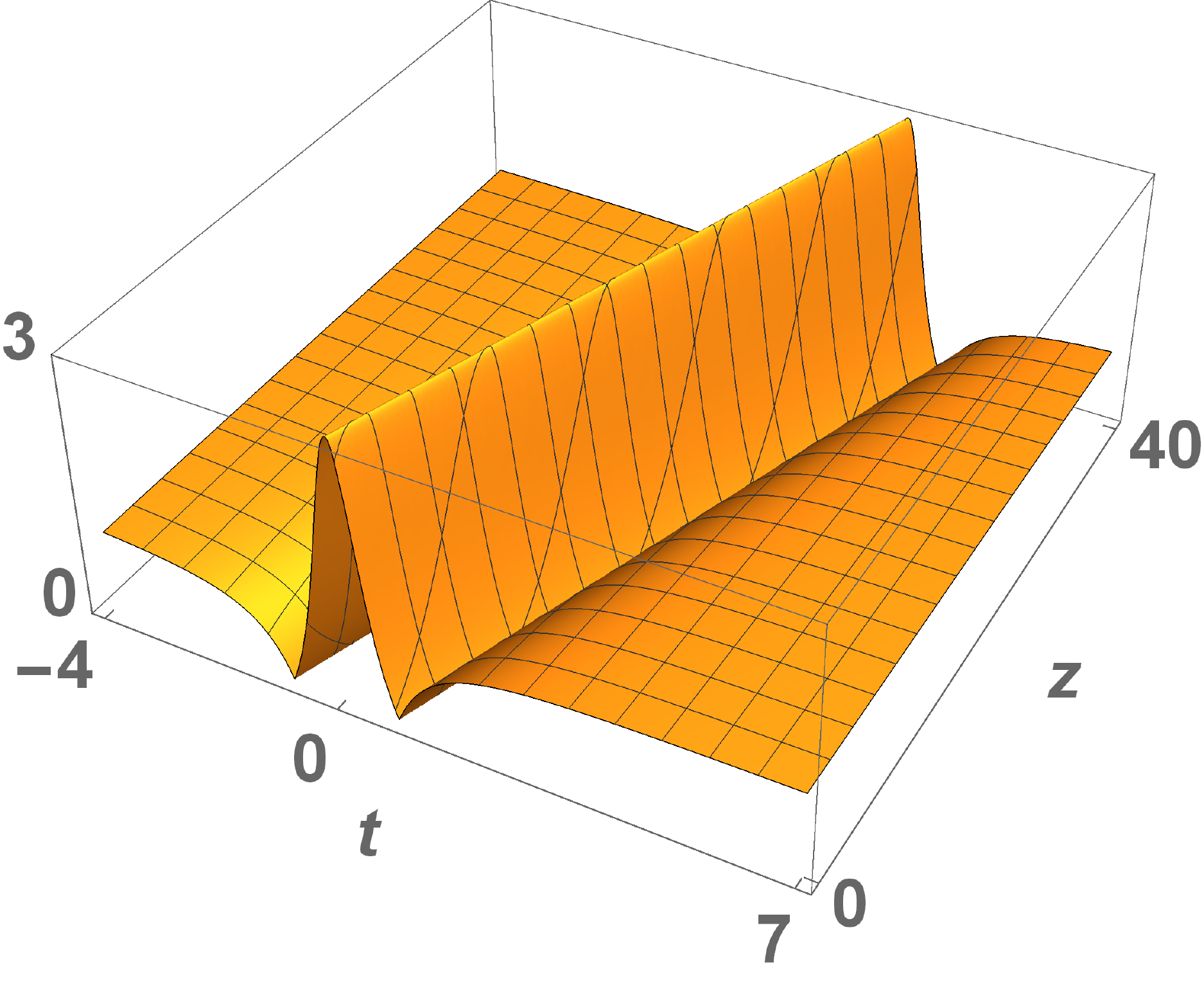}&
\includegraphics[scale=0.27]{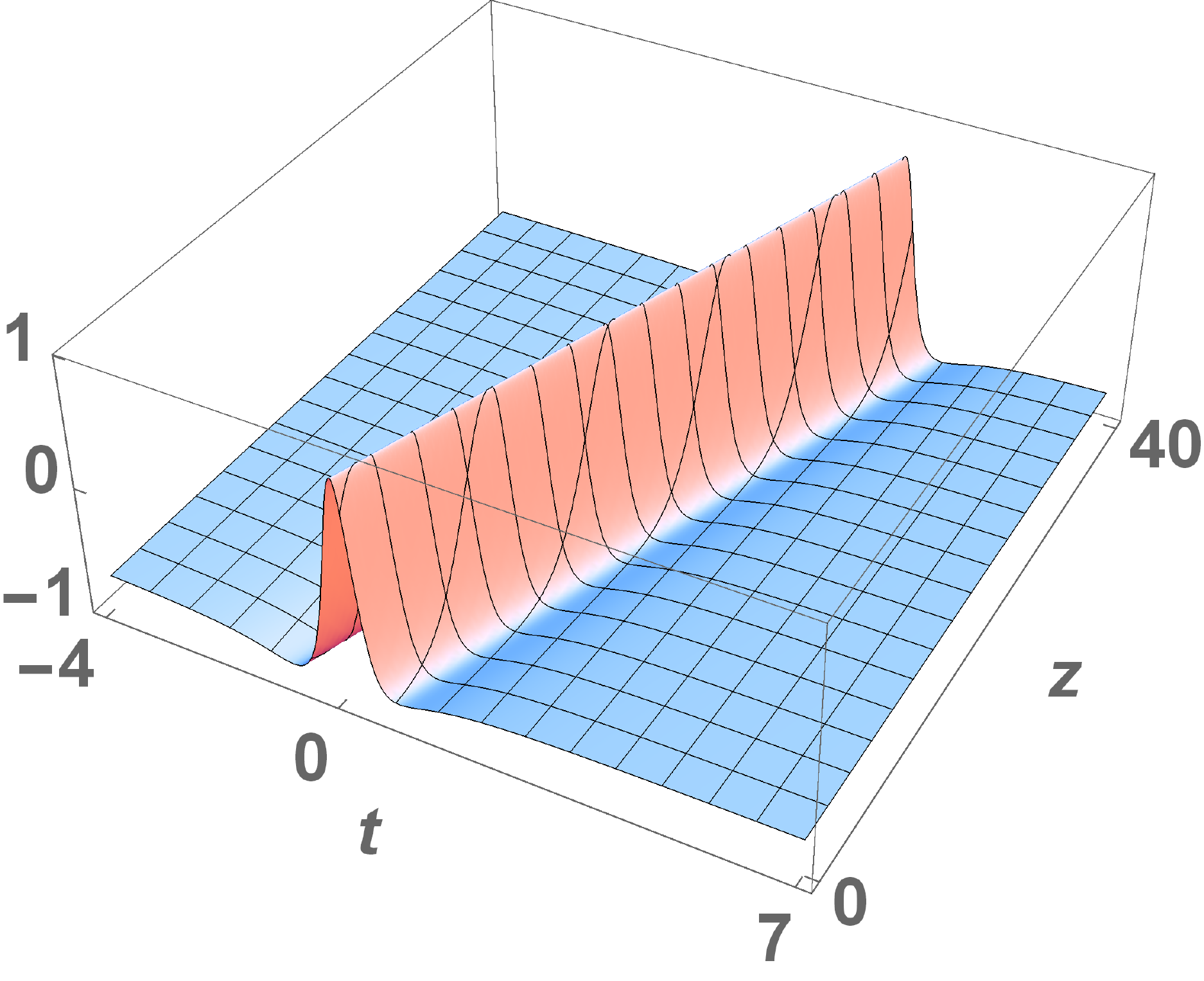}&
\includegraphics[scale=0.27]{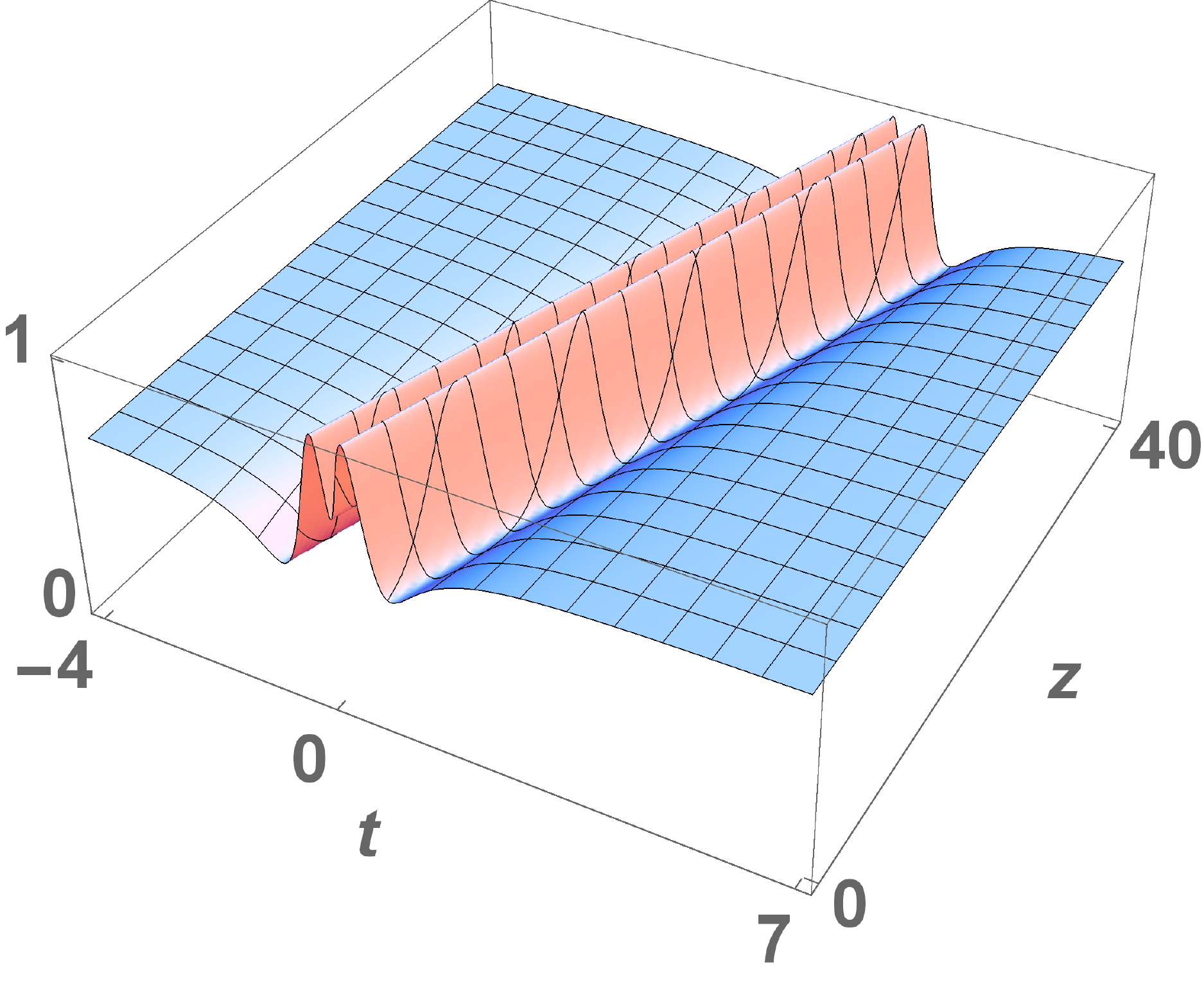}
\end{tabular}
\caption{
    One rational solution~\eref{e:rationalq} with the initial state for the medium~\eref{e:solitonBC1}, 
    inhomogeneous broadening:
    $h_- = -1$ (i.e., more atoms are  initially in the ground state than in the excited state), 
    $q_o = 1$, $\epsilon = 2$, 
    $\~\xi_1 = \~s_1 = 0$.
    Top left: the discrete eigenvalue in the $\lambda$-plane.
    Bottom left: the optical field $|q(t,z)|$.
    Center and right: components of density matrix $\rho(t,z,\lambda)$.
    Top center: $D(t,z,0^+)$.
    Top right: $|P(t,z,0^+)|$.
    Bottom center: $D(t,z,q_o)$.
    Bottom right: $|P(t,z,q_o)|$.}
\label{f:IBrational1}
\end{figure}

We again use the Lorentzian spectral-line shape~\eref{e:Lorentzian}.  Recall from the discussion in Section~\ref{s:IBsoliton} that in this case $w_-(z) = 0$.
Then from Eq.~\eref{e:RoR1}, we obtain the explicit expansion~\eref{e:R-dexpansion} for the matrix $R_{-,d}$ as $\eta\to1$, with
\vspace*{-1ex}
\[
\nonumber
R^{(0)}(z) = 0\,,\qquad R^{(1)}(z) = i h_- g_o \~R_1\, \sigma_3\,,
\]
which again, can be computed on either sheet. 
Notice that $h_- = \pm1$ denotes the initial state of the medium as in Eq.~\eref{e:solitonBC1}, 
and 
\vspace*{-1ex}
\[
g_o = g(i q_o) = \frac{\epsilon}{\pi(\epsilon^2-q_o^2)}\,,\qquad
\~R_1 = 2 - q_o\Theta(i\epsilon)\big/\sqrt{q_o^2-\epsilon^2}\,,
\label{e:rationalparameters}
\]
where $\Theta(\lambda)$ is defined in Eq.~\eref{e:Theta}.
Notice that $g_o$ and $\~R_1$ are both real, and both $R^{(0)}(z)$ and $R^{(1)}(z)$ are independent of $z$. 
By using Eqs.~\eref{e:dCdz},~\eref{e:xisexpansion1} and~\eref{e:xisexpansion2}, we then find
\vspace*{-1ex}
\begin{gather*}
\xi_1(z) = h_- g_o\~R_1\, z+ \~\xi_1\,,\qquad s_1(z) = \~s_1\,,\qquad
\\[-0.4ex]
\noalign{\noindent where}
\~\xi_1 = \lim_{\eta\to1} \big[\xi(\eta,0)-\xi_o\big]/(\eta-1)\,,\qquad
\~s_1 = \lim_{\eta\to1} \big[\varphi(\eta,0)-s_o\big]/(\eta-1)\,.
\end{gather*}
Substituting all the components into Eq.~\eref{e:rationalq},
we then obtain the fully explicit rational solution
\vspace*{-1ex}
\[
\label{e:rationalexplicit}
q_R (t,z) = q_o \frac{\big( 2q_o t + h_- g_o\~R_1 z + \~\xi_1 \big)^2 + \~s_1^2 + 4i \~s_1 - 3}
{\big( 2q_o t + h_- g_o\~R_1 z + \~\xi_1 \big)^2 + \~s_1^2 + 1}\,,
\]
where $\~\xi_1$ and $\~s_1$ are arbitrary real constants, 
and both $g_o$ and $\~R_1$ are given in Eq.~\eref{e:rationalparameters}.
Notice that this solution is a traveling wave.
In particular, if $\~s_1 = 0$, solution~\eref{e:rationalexplicit} is purely real.
The parameter $\~\xi_1$ determines a spatial/temporal displacement, while $\~s_1$ determines a ``phase shift''.
The density matrix $\rho(t,z,\zeta)$ can also be calculated from Eq.~\eref{e:rationalmu},
but is omitted for brevity due to its complexity.
One such solution is shown in Fig.~\ref{f:IBrational1}.

\begin{figure}[t!]
\centering
\begin{tabular}[b]{ccc}
\includegraphics[scale=0.21]{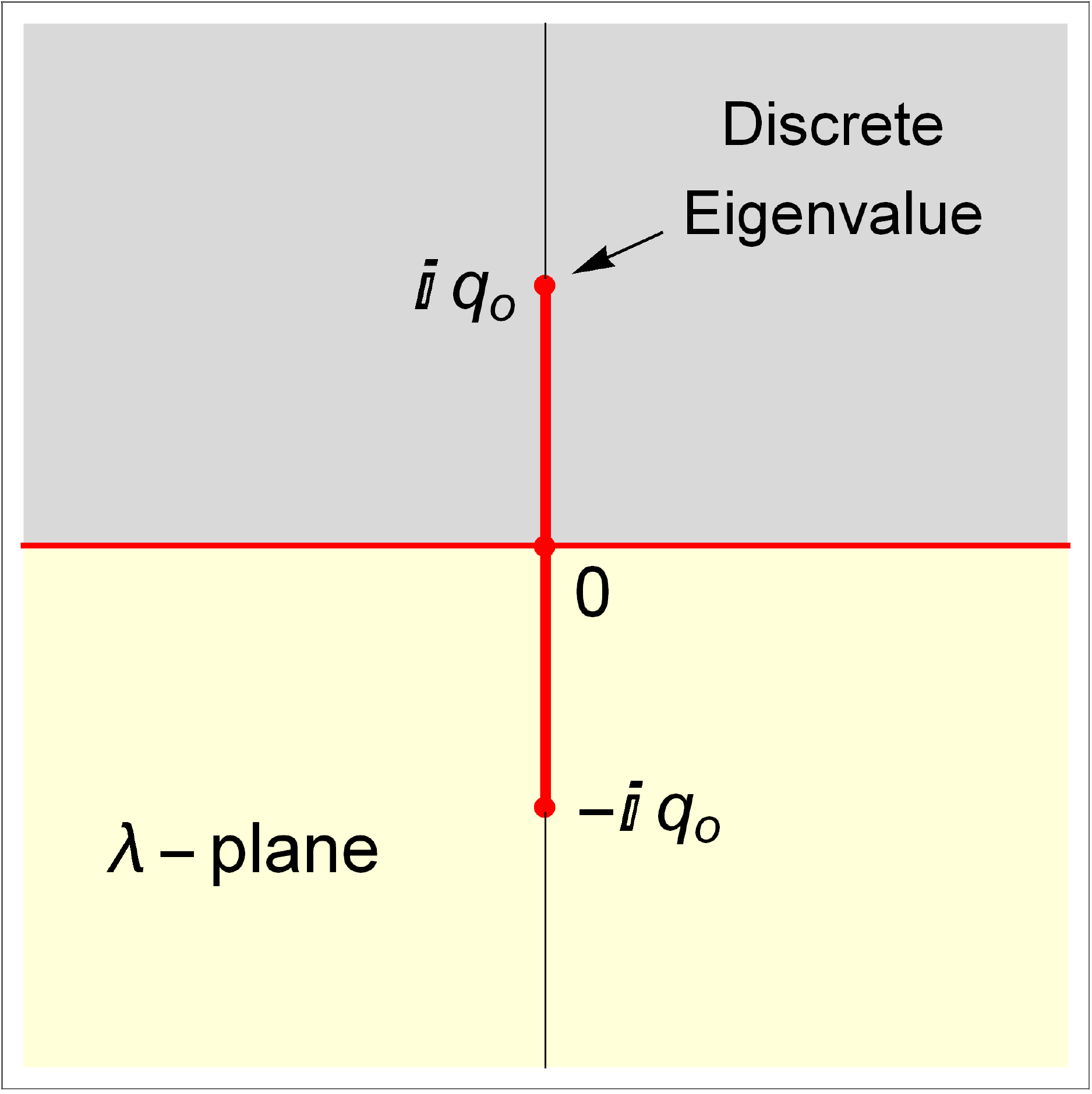}&
\includegraphics[scale=0.27]{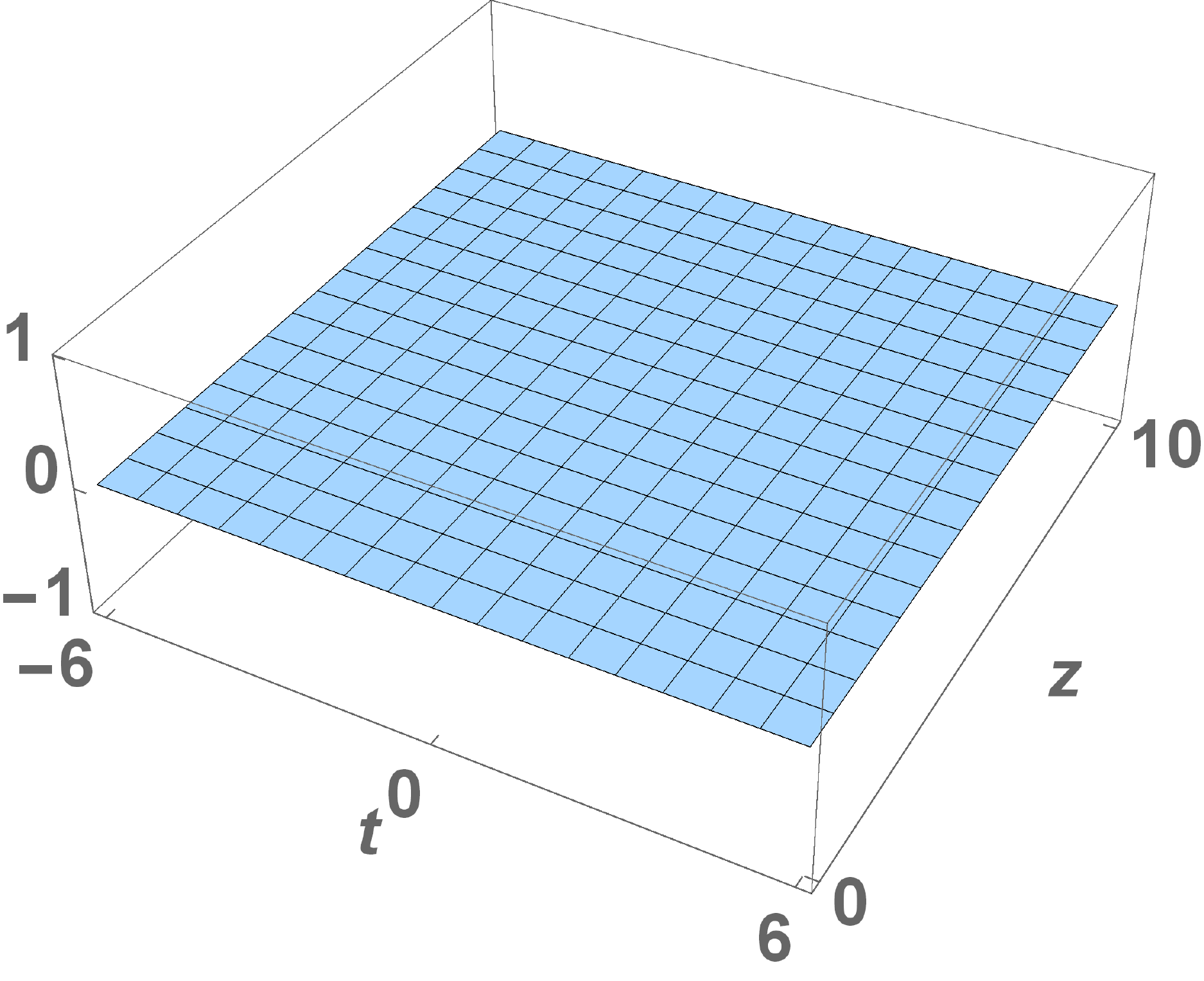}&
\includegraphics[scale=0.27]{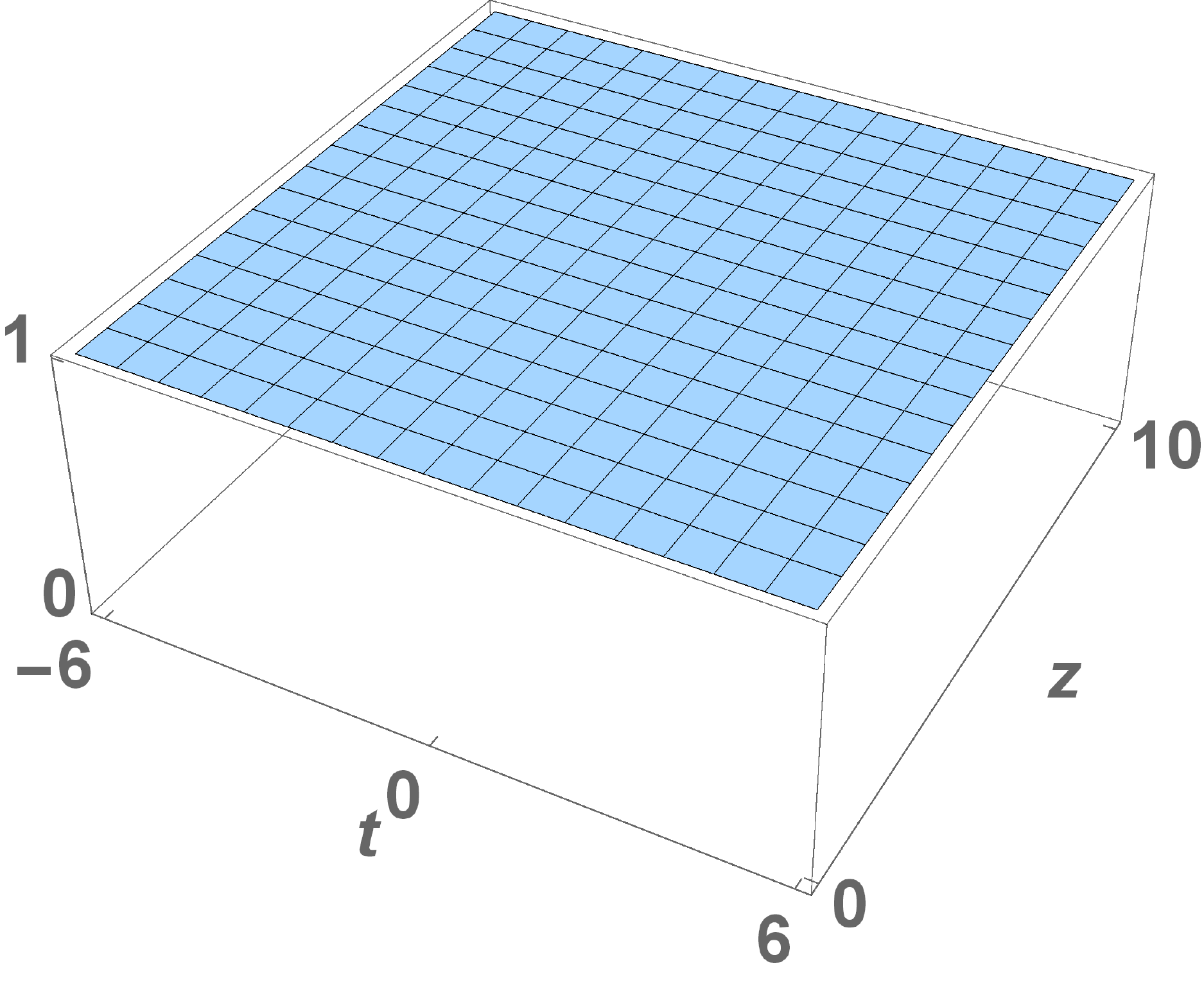}\\
\includegraphics[scale=0.27]{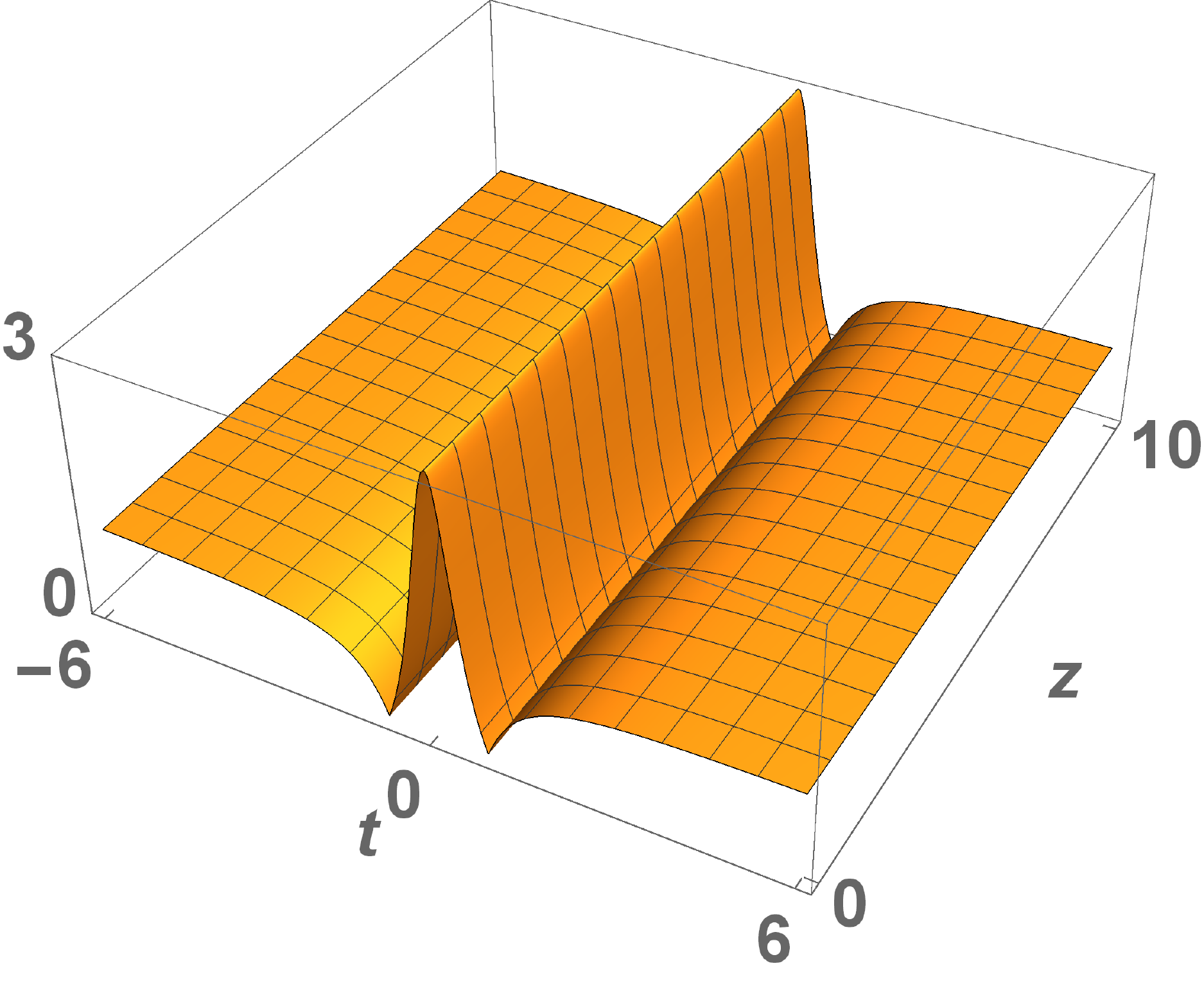}&
\includegraphics[scale=0.27]{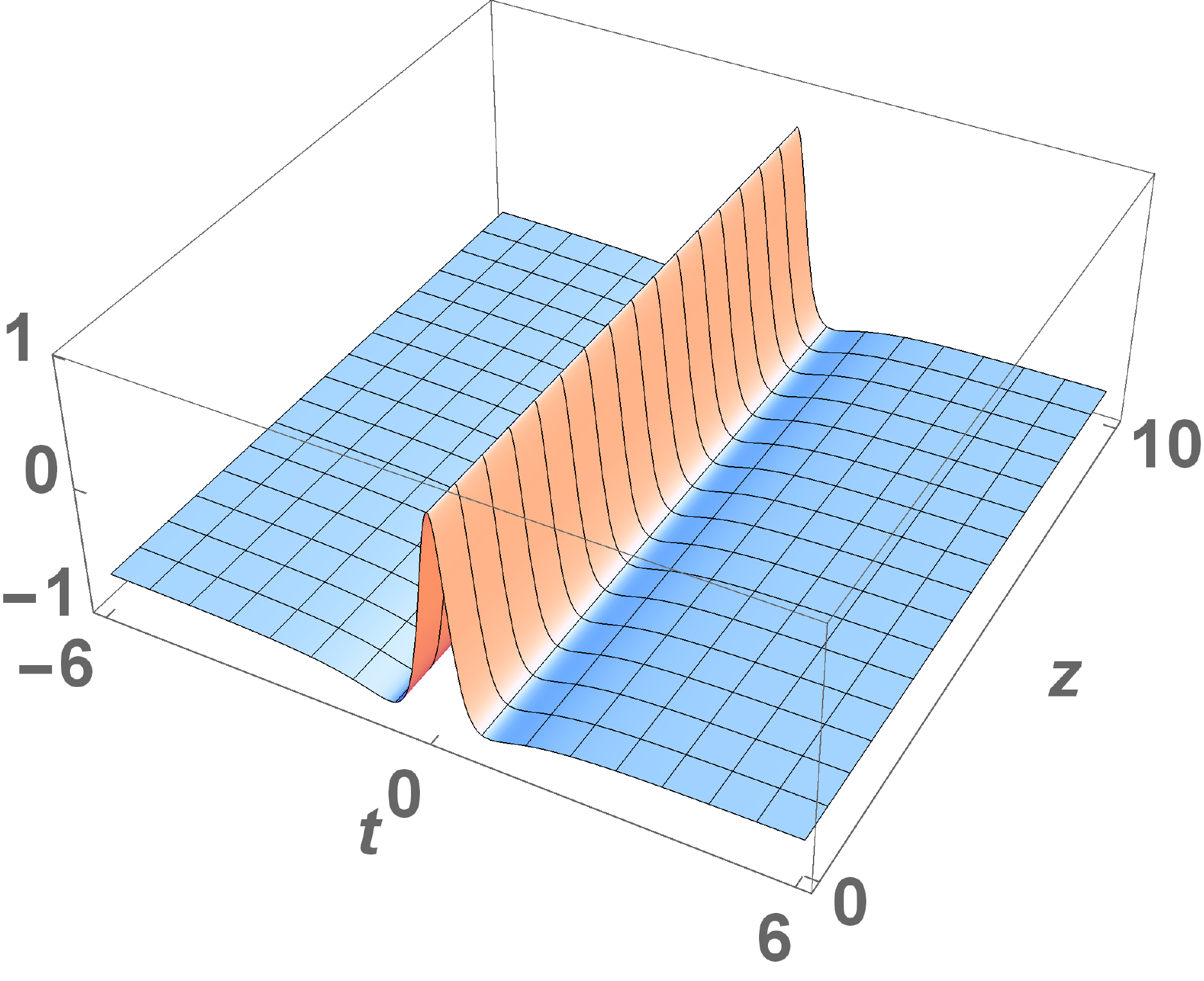}&
\includegraphics[scale=0.27]{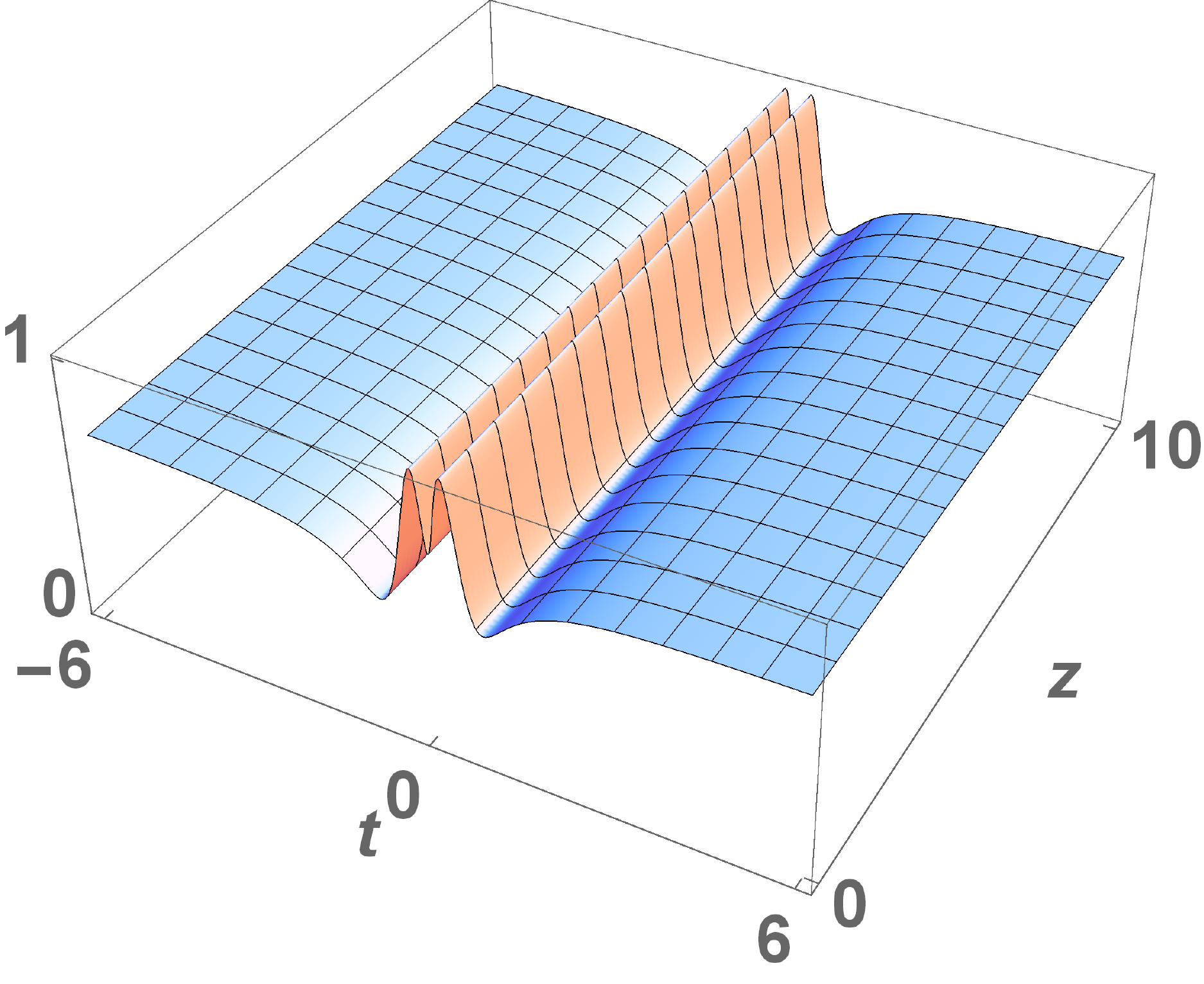}
\end{tabular}
\caption{Similarly to Fig.~\ref{f:IBrational1}, but for a rational solution in the sharp-line limit:
$h_- = -1$, $q_o = 1$, $\~\xi_1 = 0$ and $\~s_1 = 0$.
}
\label{f:SLrational1}
\vskip2\medskipamount
\centering
\begin{tabular}[b]{ccc}
\includegraphics[scale=0.21]{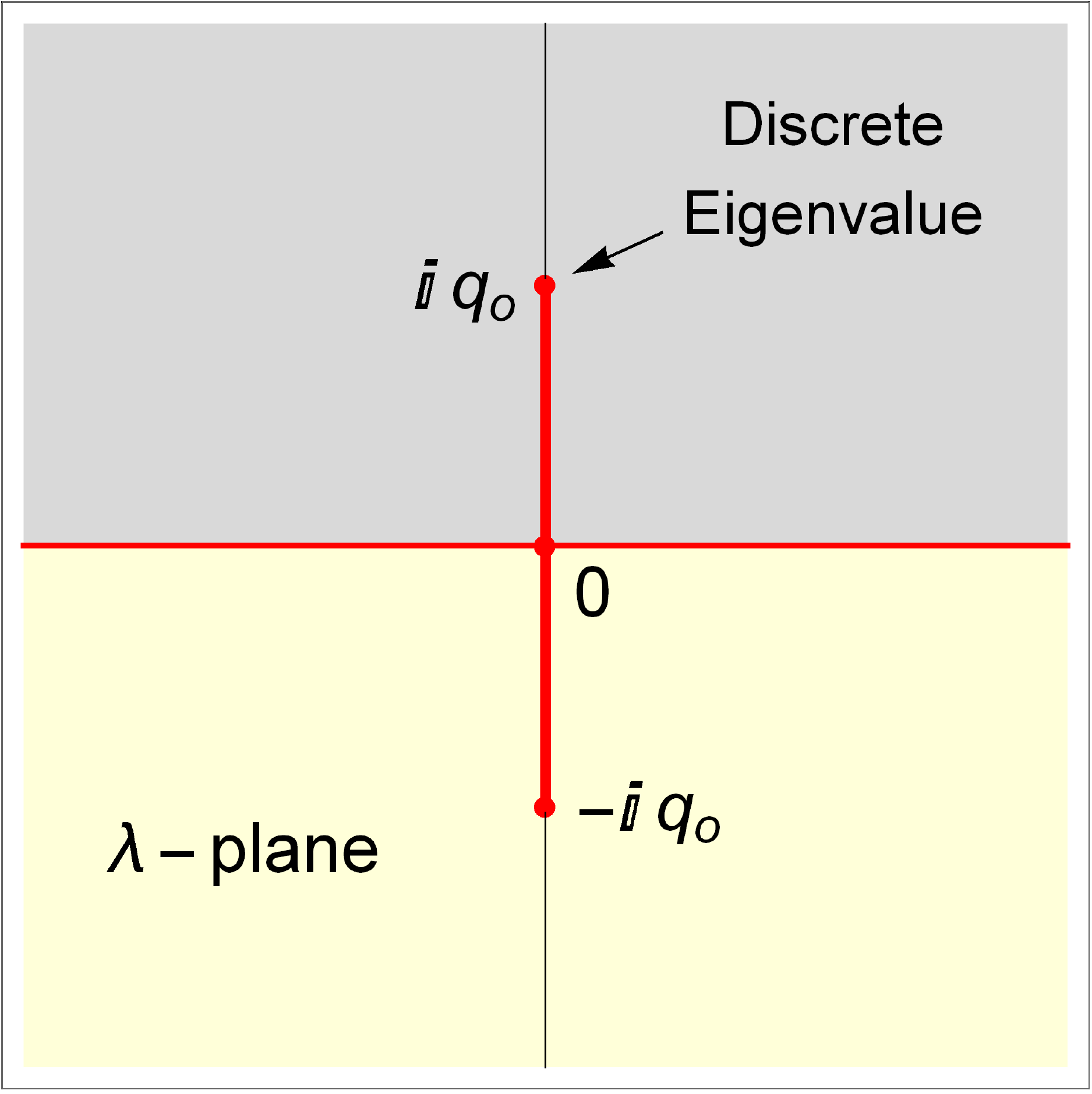}&
\includegraphics[scale=0.27]{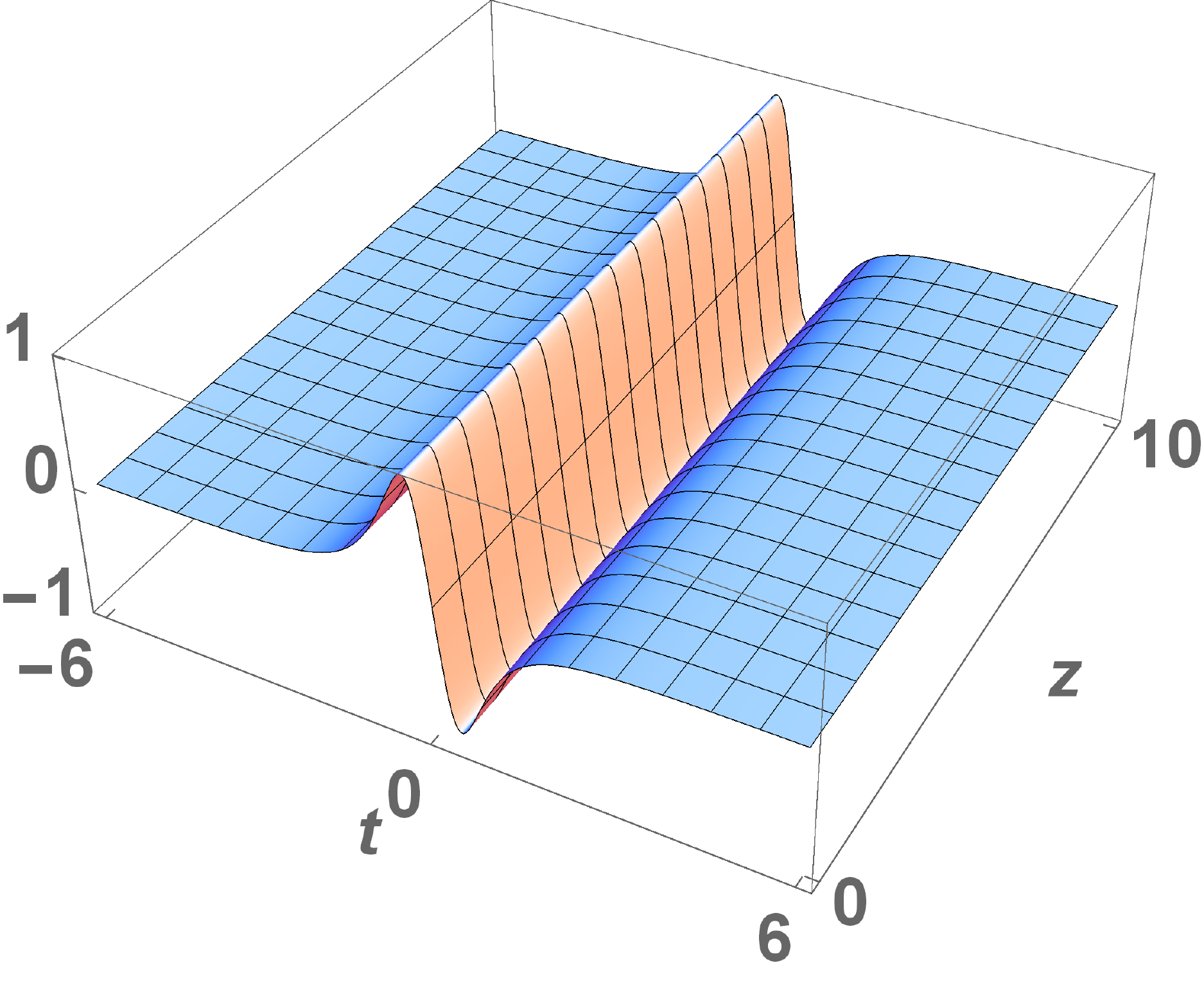}&
\includegraphics[scale=0.27]{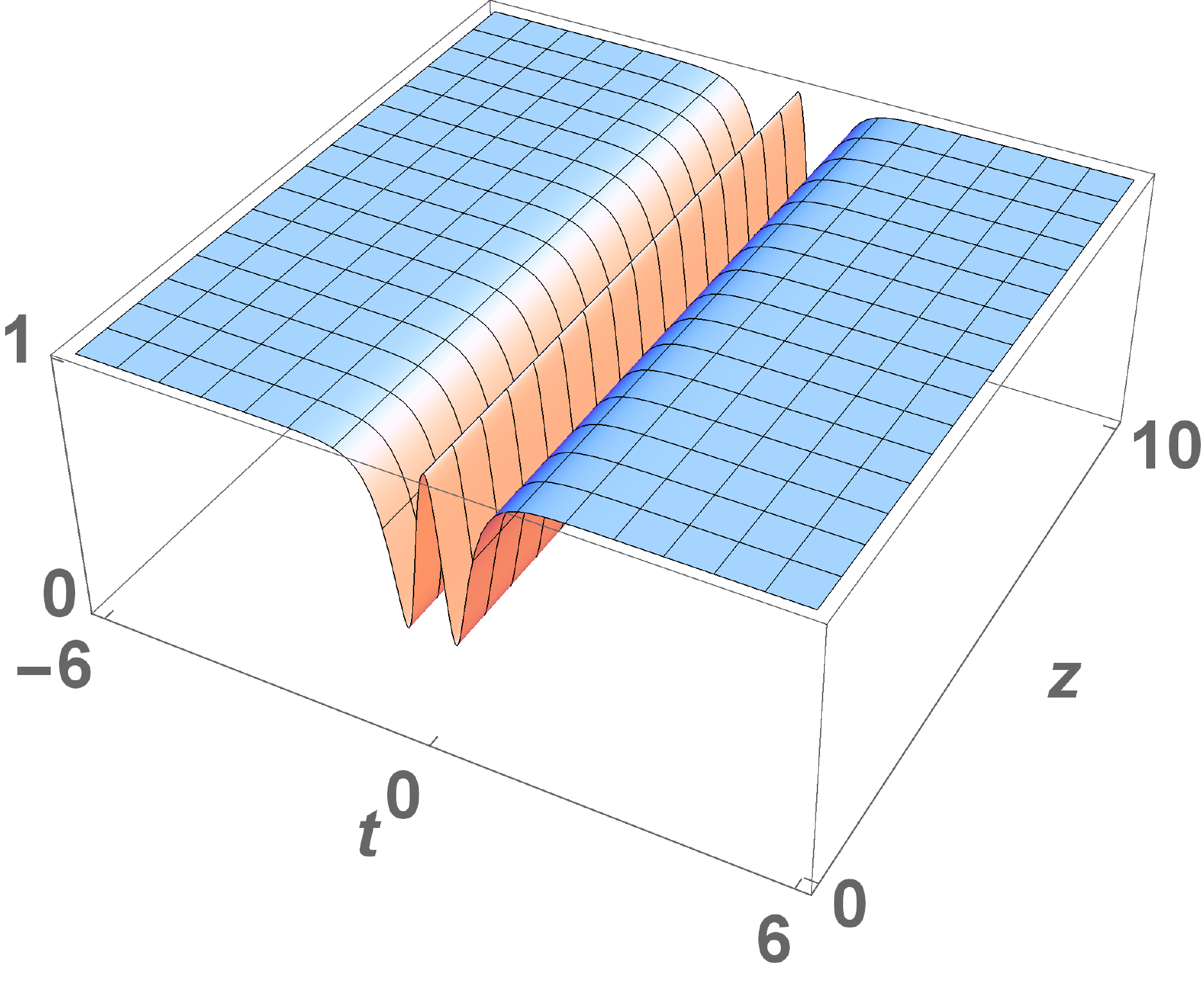}\\
\includegraphics[scale=0.27]{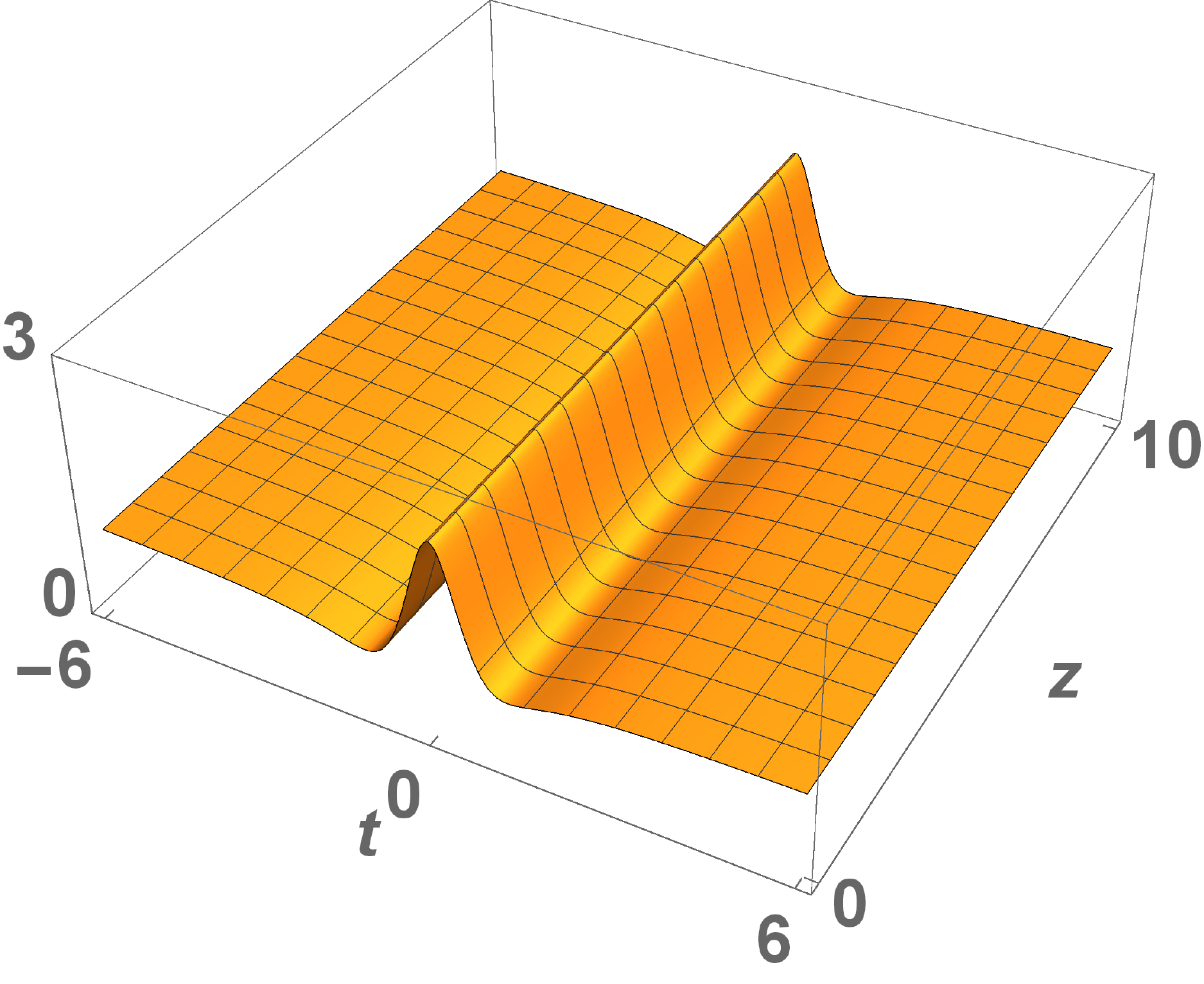}&
\includegraphics[scale=0.27]{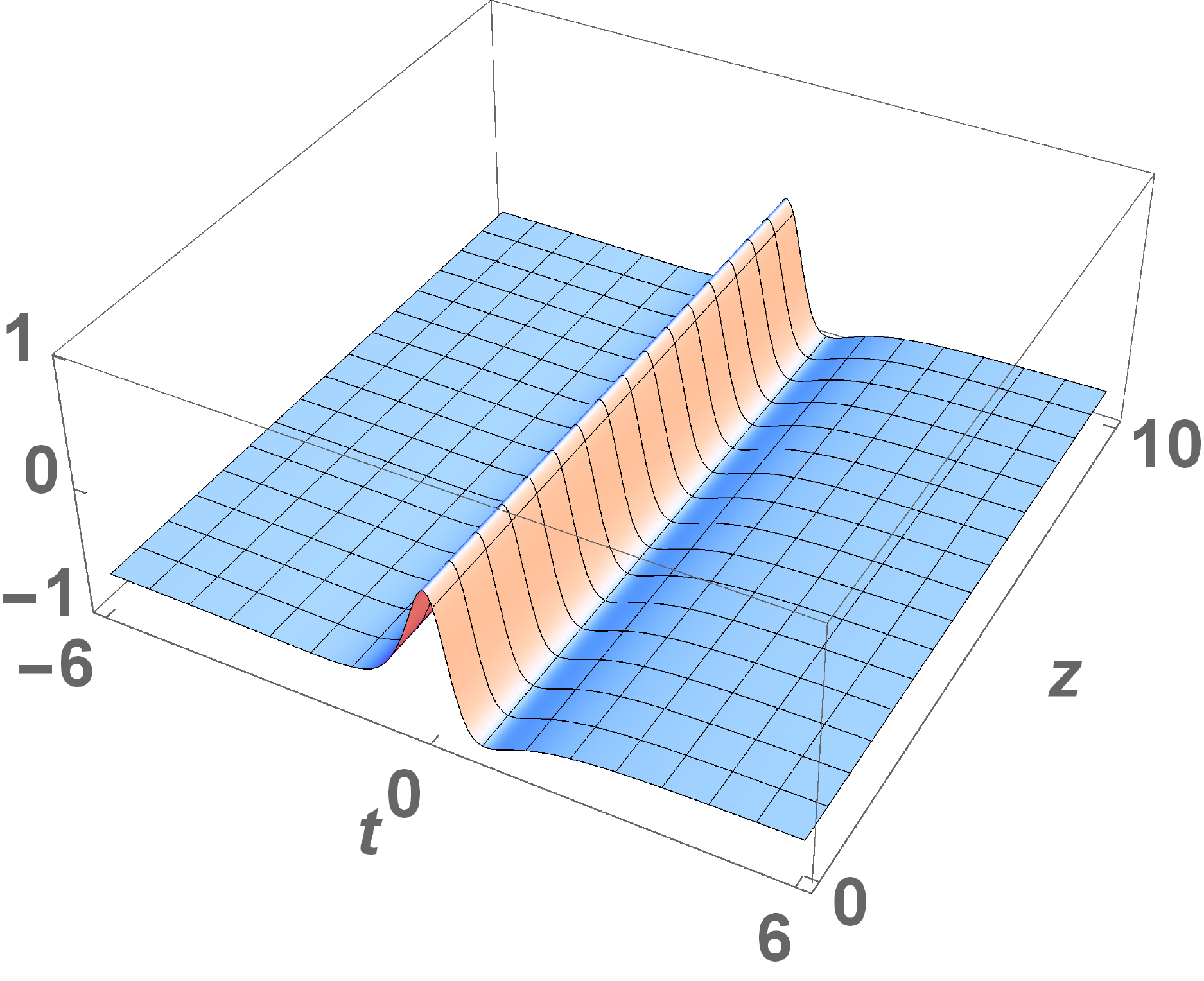}&
\includegraphics[scale=0.27]{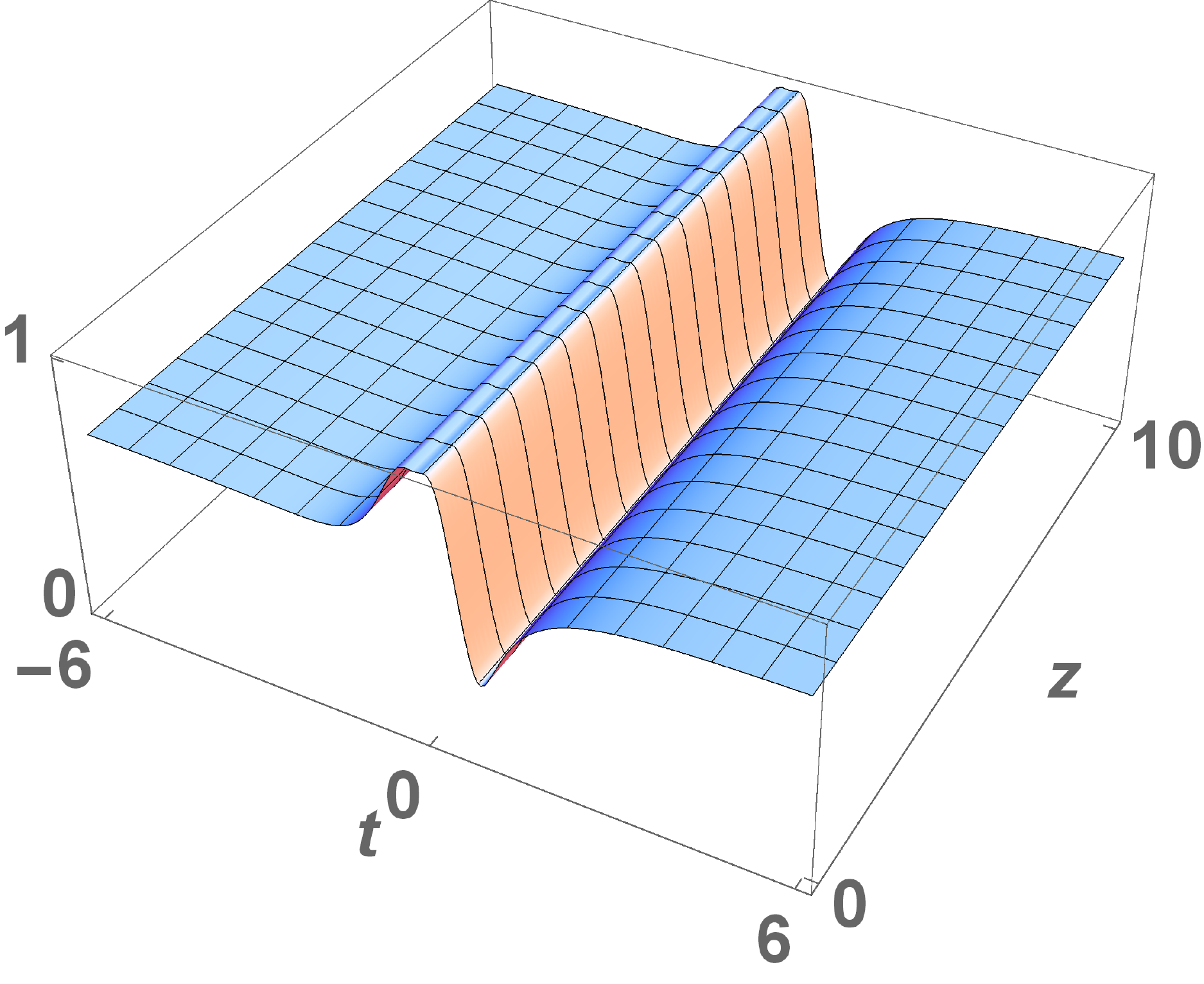}
\end{tabular}
\caption{Same as Fig.~\ref{f:SLrational1}, but with $\~s_1 = 1$.}
\label{f:SLrational2}
\end{figure}

The amplitude compared to the background and the velocity of such a solution are, respectively,
\[
\nonumber
A = \frac{q_o}{1+\~s_1^2}\sqrt{\~s_1^4+ 10 \~s_1^2+9}-q_o\,,\qquad
V = - 2h_-q_o\big/\big(g_o\~R_1\big)\,.
\vspace*{-1ex}
\]
The maximal possible value for $A$ is $A_\mathrm{max} = 2q_o$ when $\~s_1 = 0$.

It is easy to compute the sharp-line limit of the above rational solution.
In the limit $\epsilon\to0$, we have $g_o\~R_1\to0$, while all other parts of the solution remain the same 
as the ones in the inhomogeneous broadening case, implying
\[
q_{\mathrm{R},0} (t,z) = q_o \big[\big( 2q_o t + \~\xi_1 \big)^2 + \~s_1^2 + 4i \~s_1 - 3\big]\big/
\big[\big( 2q_o t + \~\xi_1 \big)^2 + \~s_1^2 + 1\big]\,.
\]
Therefore, the solution travels at the speed of light in vacuum.
Also, we can use relation~\eref{e:solitonrho} to write down the density matrix with $\~\xi_1 = 0$ explicitly
\vspace*{-1ex}
\begin{multline*}
\rho(t,z,\lambda) = \frac{h_-}{\gamma ^3 \~y^2}\big[
-4q_o( 2 q_o^2 \~s_1 + 4 \lambda  q_o^2 t - \gamma ^2 \~s_1 \~y)\sigma_1
\\[-0.4ex]
+(16 q_o^5 t^2-4 q_o^3 \~s_1^2-16 \lambda  q_o^3 \~s_1 t+4 q_o^3+\gamma ^2 q_o \~y^2-4 \gamma ^2 q_o \~y)\sigma_2\\ 
+(16 q_o^4 \~s_1 t+32 \lambda  q_o^4 t^2-4  \lambda  q_o^2 \~y+ \gamma ^2 \lambda  \~y^2)\sigma_3
\big]\,,
\end{multline*}
where $\~y = 4q_o^2 t^2 +\~s_1^2+1$ and $\sigma_j$ are defined in Eq.~\eref{e:Paulidef}.
Two such solutions are shown in Figs.~\ref{f:SLrational1} and~\ref{f:SLrational2}.

\subsection{On the stability of the soliton solutions}
\label{s:stability1}

We now discuss the stability of general soliton solutions.
To do so, we use a general form for the initial state $\rho_-$ consistent with Eqs.~\eref{e:rhopm=cdotsigma1},~\eref{e:rhopm=cdotsigma2},~\eref{e:rhopm=cdotsigma3} and~\eref{e:rhopm=cdotsigma4}, 
which we rewrite in a convenient form as
\vspace*{-1ex}
\[
\rho_-(\zeta,z) = \begin{pmatrix}
D_- & P_- \\ P^*_- & -D_-
\end{pmatrix}
 = \begin{pmatrix}
\nu\,\cos\beta(z) & \e^{i\phi(\zeta,z)}\sin\beta(z)\\ 
\e^{-i\phi(\zeta,z)}\sin\beta(z) & -\nu\,\cos\beta(z)
\end{pmatrix}\,,
\label{e:rho-BC}
\]
where $\nu = 1$ for $\zeta\in(-\infty,-q_o]\cup[q_o,\infty)$, $\nu = -1$ for $\zeta\in(-q_o,q_o)$.
Functions $\beta(z)$ and $\phi(\zeta,z)$ are real-valued, 
and ensure that $\det\rho_-(z) = -1$.
Moreover, $\phi(-q_o^2/\zeta,z) = -\phi(\zeta,z) + 4W_-(z)$ 
with $W_-(z)$ defined in Eq.~\eref{e:q-timeevolution} and $\zeta\in(-\infty,-q_o]\cup[q_o,\infty)$,
because of the enforced symmetries~\eref{e:rhopm=cdotsigma1},~\eref{e:rhopm=cdotsigma2},~\eref{e:rhopm=cdotsigma3} and~\eref{e:rhopm=cdotsigma4}.
Recall the discussion in Section~\ref{s:boundaryvalues},
$\beta(z)$ relates to the initial amount of population inversion of this system.
If $0\le \beta(z)<\pi/2$,
initially, there are more atoms in the excited state than in the ground state,
whereas if $\pi/2<\beta(z)\le\pi$,
initially, there are more atoms in the ground state than in the excited state.
The special values $\beta(z) = 0$ and $\beta(z) = \pi$ 
identify the situation where the largest proportion of the atoms is initially in the excited and ground states, respectively.
Moreover, we consider the inhomogeneous broadening case, i.e., we take $g(\lambda)$ as in Eq.~\eref{e:Lorentzian}.

We first point out that the stability is characterized by the reflection coefficient $b(\zeta,z)$ appearing in the jump condition of the RHP~\eref{e:RHP}.
Therefore, we need to analyze the quantity $b(\zeta,z)$ whose dependence on $z$ is governed by the ODE~\eref{e:dbdz}.
Since the coefficients of this ODE contain the entries of $R_{-,d}$,
we consider the matrix $R_{-,d}$, as defined in Eq.~\eref{e:R_d_axis}. 
Because $\rho_{-,d}(\zeta,z) = \nu\,\cos\beta(z)\sigma_3$ from Eq.~\eref{e:rho-BC},
$R_{-,d}$ can be calculated in the same way as in Appendix~\ref{A:Rpm_explicit}.
We thus obtain Eqs.~\eref{e:Rd-explicit_NZBG} and~\eref{e:R-dreal} for $\lambda\in\Complex\backslash\Real$ and $\lambda\in\Real$, respectively, 
on either $\lambda$-sheet, with $h_- = \cos\beta(z)$.
It is then convenient to write the resulting expression as a function of $\zeta$ in the form of
\[
R_{-,1,1}(\zeta,z) = (\bar R_\re +i \bar R_\im)\cos\beta(z) \,,\qquad \zeta\in\Sigma\,,
\label{e:R-dreim}
\]
with $\bar R_\re$ and $\bar R_\im$ both real.
Comparing Eqs.~\eref{e:Rij1},~\eref{e:Rij2} and~\eref{e:Ypmdef}, we then obtain
\[
\nonumber
A = [\bar R_\re + i(\bar R_\im+\nu_o\pi g(\lambda))]\cos\beta(z) \,,
\]
where $\nu_o = 1$ if $\zeta\in\Real$ and $\nu_o = 0$ if $\zeta\in C_o$.
Using the propagation equation~\eref{e:dbdz} for the reflection coefficient,
we find the following explicit expression
\begin{equation}
\label{e:bformula}
b(\zeta,z) =  \Omega(z,0,\zeta)\,b(\zeta,0) - 
    \nu_o\pi g(\lambda)\int_0^z \Omega(z,z',\zeta)\,\e^{- i\phi(\zeta,z')}\sin\beta(z')\d z'\,,
\end{equation}
where the primes denote the integration variable (not differentiation) as before, 
and where $\nu_o = 1$ if $\zeta\in\Real$, $\nu_o = 0$ if $\zeta\in C_o$, and
\begin{equation} 
\label{e:Omega}
\Omega(z,z',\zeta) = \exp\big[\big(\bar R_\im + \nu_o \pi g(\lambda) 
    -i \bar R_\re\big)\int\nolimits_{z'}^z \cos\beta(s)\d s\big]\,.
\end{equation}
Note that Eq.~\eref{e:bformula} on the real axis is formally the same as the one in the case of ZBG.
If $b(\zeta,0)$ (i.e., the reflection coefficient at $z=0$) and $\sin\beta(z)$ are both zero,
then $b(\zeta,z)$ is identically zero (i.e., there is no radiation).
We assume this is not the case.

Notice that the reflection coefficient~\eref{e:bformula} derived from the inhomogeneous ODE contains a homogeneous and an inhomogeneous portion, 
and $\Omega(z,z',\zeta)$ appears in both of them.
As a result, the long-distance behavior of the reflection coefficient $b(\zeta,z)$ depends on the absolute value of $\Omega(z,z',\zeta)$.
The situation is further complicated by the fact that both $\beta(z)$ and $\phi(\zeta,z)$ are in principle arbitrary.
For simplicity, and by analogy with the case with ZBG, 
here we will study the simpler situation in which $\beta(z)$ and $\phi(\zeta,z)$ are constants with respect to $z$
and in which $g(\lambda)$ is the Lorentzian function~\eref{e:Lorentzian}.
In this case, the expression \eref{e:bformula} simplifies to:
\[
\label{e:bformula2}
b(\zeta,z) =  \bigg[
    b(\zeta,0) - \frac{\nu_o\pi g(\lambda) \e^{-i\phi(\zeta)}\tan\beta}{\bar R_\im + \nu_o\pi g(\lambda) -i \bar R_\re}
  \bigg]\e^{(\bar R_\im+\nu_o\pi g(\lambda) -i \bar R_\re)z\cos\beta} 
  +\frac{\nu_o\pi g(\lambda) \e^{-i\phi(\zeta)}\tan\beta}{\bar R_\im + \nu_o\pi g(\lambda) -i \bar R_\re}\,,
\]
where again, $\nu_o = 1$ if $\zeta\in\Real$ and $\nu_o = 0$ if $\zeta\in C_o$,
$\bar R_\re$ and $\bar R_\im$ given by Eq.~\eref{e:R-dreim} as before.
In this situation, the growth/decay of $b(\zeta,z)$ depends on the modulus of 
$\exp\big[(\bar R_\im + \nu_o\pi g(\lambda)-i\bar R_\re)\cos\beta\,z\big]$,
so the most important part by now is the unknown quantity $\bar R_\im$.
Since any complex phase of the exponential produces inessential oscillations,
the quantity $\bar R_\re$ is therefore not crucial for determining the long-distance behavior of the reflection coefficient.
By this discussion, 
we next turn to characterize the imaginary part of the matrix $R_{-,d}$ given by 
Eqs.~\eref{e:Rd-explicit_NZBG} and~\eref{e:R-dreal}, 
depending on the location of $\zeta$.
It is necessary to discuss separately two cases depending on whether $q_o>\epsilon$ or $q_o<\epsilon$.

\paragraph{Case I: $0\le q_o<\epsilon$.}

Firstly, we discuss the case where the amplitude of the background field is small relative to the width of the spectral-line shape.
In this case $\Theta(i\epsilon) = 2\arcsech(\epsilon/q_o)$ is pure imaginary.
We can then rewrite $R_{-,d}$ from Eqs.~\eref{e:Rd-explicit_NZBG} and~\eref{e:R-dreal} as
\[
\label{e:R-d1}
R_{-,d}(\zeta,z) = 
2g(\lambda)\big[C(\lambda)
    - \gamma(\epsilon^2-q_o^2)^{-1/2}\arccos(q_o/\epsilon)\big]\sigma_3\cos\beta\,,\qquad
    \zeta\in\Sigma\,,
\]
where 
\[
\label{e:Clambda}
C(\lambda) = \begin{cases}
\displaystyle 
\arccsch(-\lambda/q_o) & \qquad \zeta\in\Real\,,\\
\displaystyle 
\arcsech(-i\lambda/q_o) & \qquad\zeta\in C_o\,,
\end{cases}
\]
$C_o$ is the complex portion of the continuous spectrum
and $g(\lambda)$ is given in Eq.~\eref{e:Lorentzian} as before.
Notice that the function $C(\lambda)$ is always real on the continuous spectrum,
because $\zeta\in\Real$ implies $\lambda\in\Real$ and $\zeta\in C_o$ implies $\lambda\in i[-q_o,q_o]$.
Also $\gamma$ is real for all $\zeta\in\Sigma$.
We then conclude that the quantity $R_{-,d}(\zeta,z)$ is always real. 
In other words, $\bar R_\im = 0$, for all $\zeta\in\Sigma$.

\paragraph{Case II: $0<\epsilon<q_o$.}

Next, we consider the case in which the width of the spectral-line shape is small compared to the amplitude of the background optical field.
In this case, $\Theta(i\epsilon) = 2\arcsech(\epsilon/q_o)$ is real.
Similarly to the previous case, 
we rewrite the matrix from Eqs.~\eref{e:Rd-explicit_NZBG} and~\eref{e:R-dreal} into the form of Eq.~\eref{e:R-dreim}. 
We obtain
\[
\label{e:R-d2}
R_{-,d}(\zeta,z) = 
2g(\lambda)\big[C(\lambda)
- \gamma(q_o^2 - \epsilon^2)^{-1/2}\arcsech(\epsilon/q_o)\big]\sigma_3\cos\beta\,,\qquad
\zeta\in\Sigma\,,
\]
where $C(\lambda)$ is defined in Eq.~\eref{e:Clambda}.
Following a similar discussion as in case I, we know that the above expression is purely real, 
implying $\bar R_\im = 0$ for all $\zeta\in\Sigma$.

\paragraph{Implications for stability.}

In summary, we have that 
$\bar R_\im = 0$ on the continuous spectrum, no matter what the values of $\epsilon$ and $q_o$ are.
Therefore, the solution~\eref{e:bformula2} further reduces to
\[
\nonumber
b(\zeta,z) =  \begin{cases}
\displaystyle 
   \bigg[b(\zeta,0) - \frac{\pi g(\lambda) \e^{-i\phi(\zeta)}\tan\beta}{\pi g(\lambda) -i \bar R_\re}\bigg]\e^{[\pi g(\lambda) -i \bar R_\re]z\cos\beta} 
       +\frac{\pi g(\lambda)}{\pi g(\lambda) -i \bar R_\re} \e^{-i\phi(\zeta)}\tan\beta\,, & \zeta\in\Real\,,\\
\displaystyle 
   b(\zeta,0)\e^{-i \bar R_\re z\cos\beta}\,, & \zeta\in C_o\,,
\end{cases}
\]
where $\bar R_\re$ can be computed from Eq.~\eref{e:R-d1} or~\eref{e:R-d2} and is omitted here due to its lack of importance.
Obviously, the reflection coefficient $b(\zeta,z)$ is always bounded on the portion $\zeta\in C_o$ for $z>0$, 
so we next turn to discuss the case $\zeta\in\Real$.

It is easy to see that $b(\zeta,z)$ (and therefore the radiation) grows exponentially as $z\to\infty$ if $\cos\beta>0$,
whereas $b(\zeta,z)$ (and therefore the radiation) decays exponentially as $z\to\infty$ if $\cos\beta<0$.
Therefore, the stability of general solutions with NZBG is identical to the case of ZBG.  
Namely,
\textit{radiation decays (implying that the background and the solitons are stable) 
when initially the atoms in the ground state dominate (population inversion is negative);
radiation grows (implying that the background and the solitons are unstable) 
when initially the atoms in the excited state dominate (population inversion is positive).}

In particular, for a purely diagonal initial state (i.e., $\sin\beta = 0$) without any discrete eigenvalues, 
the solution of MBE with NZBG tends to the background value 
when the medium initially has the lowest population inversion.
That is, $q(t,z) = q_-(z) + o(1)$ as $z\to\infty$ if $\rho_- = -\nu\sigma_3$ and if no solitons are present.

\section{Soliton solutions with a shifted Lorentzian}
\label{s:soliton_shift_lorentzian}

In this section we present one-soliton solutions~\eref{e:oscillatorysoliton} obtained when 
the shape of spectral line is a shifted Lorentzian, namely 
\[
\label{e:shiftedLorentzian}
g(\lambda) = \frac{\epsilon}{\pi}\frac{1}{\epsilon^2 + (\lambda-\lambda_o)^2}\,,\qquad
\epsilon > 0\,,\qquad
\lambda_o\in\Real\,,
\]
where $\lambda_o$ is the center of the Lorentzian.
In the sharp-line limit ($\epsilon\to0$),
Eq.~\eref{e:shiftedLorentzian} becomes a shifted Dirac delta $g(\lambda) = \delta(\lambda-\lambda_o)$.

Note that changing the shape of spectral line only affects the $z$-dependence of solutions.
Therefore, all the calculations for the pure soliton solutions in Sections~\ref{s:solitons} and~\ref{s:1soliton}
remain valid.
On the other hand, since the function~\eref{e:shiftedLorentzian} is not even,
the calculations are more complex than with the standard Lorentzian~\eref{e:Lorentzian}.
As a result, here we present the calculations in more detail than in Section~\ref{s:IBsoliton}.
More importantly, we show that the modified $z$-dependence results in drastically different dynamical behavior of the soliton solutions.

\paragraph{Asymptotic background.}

In the case of a standard Lorentzian, $q_-(z)$ is constant with respect to $z$ (cf.\ Section~\ref{s:IBsoliton}). 
This will not be true anymore in our case, however.
Recall the relevant equations are~\eref{e:dQpmdx},~\eref{e:wpmdef} and~\eref{e:q-timeevolution}.
Moreover, recall that the choice of $\rho_-$ resulting in pure soliton solutions is~\eref{e:solitonBC1}.
In other words, on the first $\lambda$-sheet, 
$D_- = h_-$ with $h_- = \pm1$ and $\lambda\in\Real$.
For this value of $D_-$, Eq.~\eref{e:wpmdef} becomes
\[
\label{e:w_-_temp0}
w_- = \frac{h_-}{2}\int_\Real g(\lambda)/\gamma\d\lambda\,.
\]
We calculate this $w_-$ as follows [where we recall $\gamma = \sign(\lambda)\sqrt{\lambda^2 + q_o^2}$ for $\lambda\in\Real$]
\begin{align*}
w_- & = \frac{\epsilon h_-}{2\pi}\int_\Real \frac{\sign(\lambda)}{\sqrt{\lambda^2+q_o^2}[\epsilon^2 + (\lambda-\lambda_o)^2]}\d\lambda\\
& = \frac{2\epsilon \lambda_o h_-}{\pi}
\int_0^\infty\frac{\lambda}{\sqrt{\lambda^2+q_o^2}}\,\frac{1}{[\lambda^2 - (\lambda_o - i\epsilon)^2][\lambda^2 - (\lambda_o + i\epsilon)^2]}\d\lambda\,.
\end{align*}
Performing the change of variable $u = \sqrt{\lambda^2 + q_o^2}$, we then obtain
\begin{multline}
\label{e:w-1}
w_- = \frac{i h_-}{4\pi} \bigg[
\frac{1}{\sqrt{q_o^2 + (\lambda_o +i\epsilon)^2}}\ln\frac{q_o - \sqrt{q_o^2 + (\lambda_o +i\epsilon)^2}}{q_o + \sqrt{q_o^2 + (\lambda_o +i\epsilon)^2}}\\
- \frac{1}{\sqrt{q_o^2 + (\lambda_o -i\epsilon)^2}}\ln\frac{q_o - \sqrt{q_o^2 + (\lambda_o -i\epsilon)^2}}{q_o + \sqrt{q_o^2 + (\lambda_o -i\epsilon)^2}}
\bigg]
\end{multline}
Importantly, we point out that Eq~\eref{e:w_-_temp0} implies $w_-\in\Real$ for all values of $q_o$, $h_1$, $\epsilon$ and $\lambda_o$,
even though this is not trivially seen from Eq.~\eref{e:w-1}.
If we consider the special case $\lambda_o = 0$ (an unshifted Lorentzian), the two terms in Eq.~\eref{e:w-1} cancel out exactly,
and we recover the result $w_-=0$ of Section~\ref{s:IBsoliton}.
We next consider the limit of Eq.~\eref{e:w-1} as $\epsilon\to0$.
After a careful calculation which takes into account
the branch cut of the complex logarithmic function, we obtain the limit of Eq.~\eref{e:w-1} as
\[
\nonumber
w_- \to \frac{h_-}{2\sqrt{q_o^2 + \lambda_o^2}}\sign(\lambda_o) = \frac{h_-}{2\gamma(\lambda_o)}\,,\qquad \epsilon\to0\,.
\]
This result coincides with direct calculations of the sharp-line limit where 
$g(\lambda) = \delta(\lambda- \lambda_o)$.
Thus, the $z$-dependence of the background is given by 
Eq.~\eref{e:q-(z)} with $w_-$ given by Eq.~\eref{e:w-1}.
Thus, with a shifted Lorentzian, the background value is not independent of the propagation variable $z$.

\paragraph{The auxiliary matrix and propagation of the norming constant.}

We next discuss the $z$-dependence of the norming constant,
which is governed by Eq.~\eref{e:dCdz}.
As a result, we need to compute the auxiliary matrix $R$ first.
In particular, we next calculate $R_{-,1,1}$ from Eq.~\eref{e:Rij1}.
Notice that in Eq.~\eref{e:Rij1}, the quantity $R_{-,1,1}$ is expressed as a principal value integral,
but later on is evaluated at $\zeta=\zeta_n$, i.e., at that discrete spectrum. 
In other words, eventually, the quantity $R_{-,1,1}$ is evaluated off the real line.
Thus, the principal value is not needed for soliton solutions.
Equation~\eref{e:Rij1} becomes
\begin{align}
\label{e:R-11_temp1}
R_{-,1,1}(\lambda) & = \int_\Real \frac{\gamma+\lambda'-\lambda}{\gamma'(\lambda'-\lambda)}D_-g(\lambda')\d\lambda'
= \frac{\gamma h_-\epsilon}{\pi}\int_\Real \frac{1}{\gamma'(\lambda'-\lambda)[\epsilon^2 + (\lambda' - \lambda_o)^2]}\d\lambda' + 2 w_-\,,
\end{align}
where $D_- = h_- = \pm1$.
We have already calculated the quantity $w_-$,
so we only need to consider the integral in the first term.
We name this integral $I_1$ and we can rewrite it as
\begin{align*}
I_1 = & \int_\Real \frac{1}{\gamma'(\lambda'-\lambda)[\epsilon^2 + (\lambda' - \lambda_o)^2]}\d\lambda'\\
= & \int_0^\infty \frac{2u}{\sqrt{u^2 + q_o^2}} 
\frac{\epsilon^2 + u^2 + \lambda_o^2 + 2\lambda\lambda_o}{(u^2 - \lambda^2)[u^2 - (\lambda_o + i\epsilon)^2][u^2 - (\lambda_o - i\epsilon)^2]}
\d u\,,
\end{align*}
where we rename the integration variable $\lambda'$ to $u$ for clarity.
With a change of variable $x = \sqrt{u^2 + q_o^2}$,
the above integral becomes
\begin{align*}
I_1 = 2\int_{q_o}^\infty  
\frac{\epsilon^2 + x^2 - q_o^2 + \lambda_o^2 + 2\lambda\lambda_o}{(x^2 - q_o^2 - \lambda^2)[x^2 - q_o^2 - (\lambda_o + i\epsilon)^2][x^2 - q_o^2 - (\lambda_o - i\epsilon)^2]}
\d x
\end{align*}
Decomposing this integral into partial fractions, using standard integration formulae, and combining all pieces, we then find
\begin{align}
\nonumber
R_{-,1,1}(\lambda) & = \frac{\gamma h_-\epsilon}{\pi}I_1 + 2w_-\\
\nonumber
& = - h_-g(\lambda;\lambda_o)\ln\frac{q_o - \sqrt{\lambda^2 + q_o^2}}{q_o + \sqrt{\lambda^2 + q_o^2}}\\
\nonumber
& - \frac{ih_-}{2\pi}\frac{q_o^2/\zeta + \lambda_o + i\epsilon}{\sqrt{q_o^2 +(\lambda_o + i\epsilon)^2}(\lambda - \lambda_o - i\epsilon)}
\ln\frac{q_o - \sqrt{q_o^2 + (\lambda_o + i\epsilon)^2}}{q_o + \sqrt{q_o^2 + (\lambda_o + i\epsilon)^2}}\\
\label{e:R-11_shiftedLorentzian}
& + \frac{ih_-}{2\pi}\frac{q_o^2/\zeta + \lambda_o - i\epsilon}{\sqrt{q_o^2 +(\lambda_o - i\epsilon)^2}(\lambda - \lambda_o + i\epsilon)}
\ln\frac{q_o - \sqrt{q_o^2 + (\lambda_o - i\epsilon)^2}}{q_o + \sqrt{q_o^2 + (\lambda_o - i\epsilon)^2}}\,.
\end{align}
Note that, in the limit $\lambda_o\to0$ (an unshifted Lorentzian),
we have
\begin{align*}
R_{-,1,1}(\lambda) & \to - h_-g(\lambda;0)\ln\frac{q_o - \sqrt{\lambda^2 + q_o^2}}{q_o + \sqrt{\lambda^2 + q_o^2}}\\
& + \frac{ih_-}{2\pi\sqrt{q_o^2 - \epsilon^2}} 
\bigg[-\frac{q_o^2/\zeta + i\epsilon}{\lambda - i\epsilon}
+ \frac{q_o^2/\zeta - i\epsilon}{\lambda + i\epsilon}\bigg]
\ln\frac{q_o - \sqrt{q_o^2 - \epsilon^2}}{q_o + \sqrt{q_o^2 - \epsilon^2}}\\
& = - h_-g(\lambda;0)\bigg[\ln\frac{q_o - \sqrt{\lambda^2 + q_o^2}}{q_o + \sqrt{\lambda^2 + q_o^2}} 
- \frac{\gamma}{\sqrt{q_o^2 - \epsilon^2}}
\ln\frac{q_o - \sqrt{q_o^2 - \epsilon^2}}{q_o + \sqrt{q_o^2 - \epsilon^2}}\bigg]\,.
\end{align*}
This expression is equivalent to Eq.~\eref{e:Rd-explicit_NZBG}.
We now have all the components needed to determine the $z$-dependence of the soliton solutions.
Recall that there are four types of one-soliton solutions depending on the location of the discrete eigenvalue
(cf. Fig.~\ref{f:solitontype} left).
Next we discuss all of them, and we show how they differ from the ones in the main text.

\paragraph{Soliton solutions of Types~1--3.}

\begin{figure}[t!]
    \centering
    \includegraphics[width=0.32\textwidth]{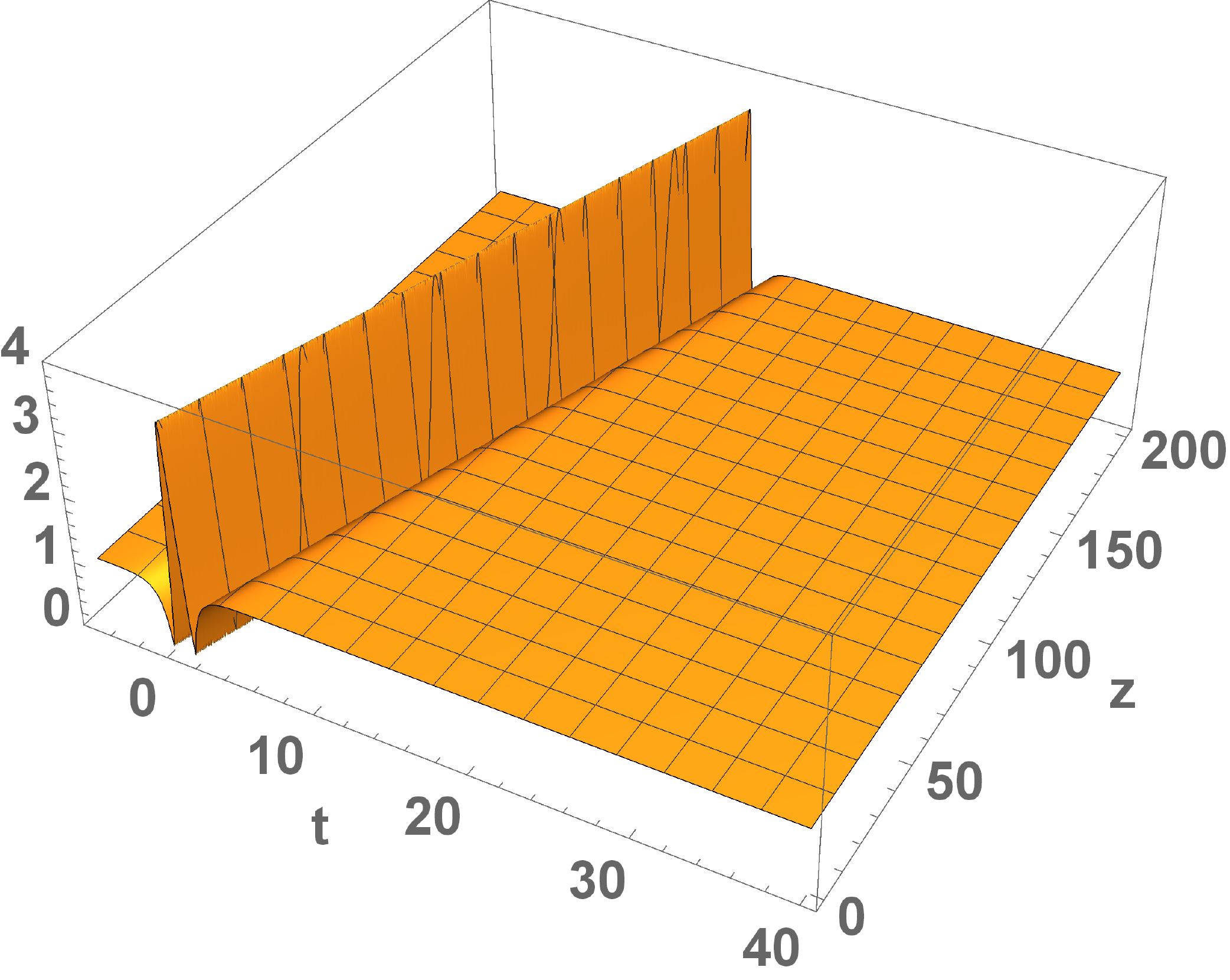}
    \includegraphics[width=0.32\textwidth]{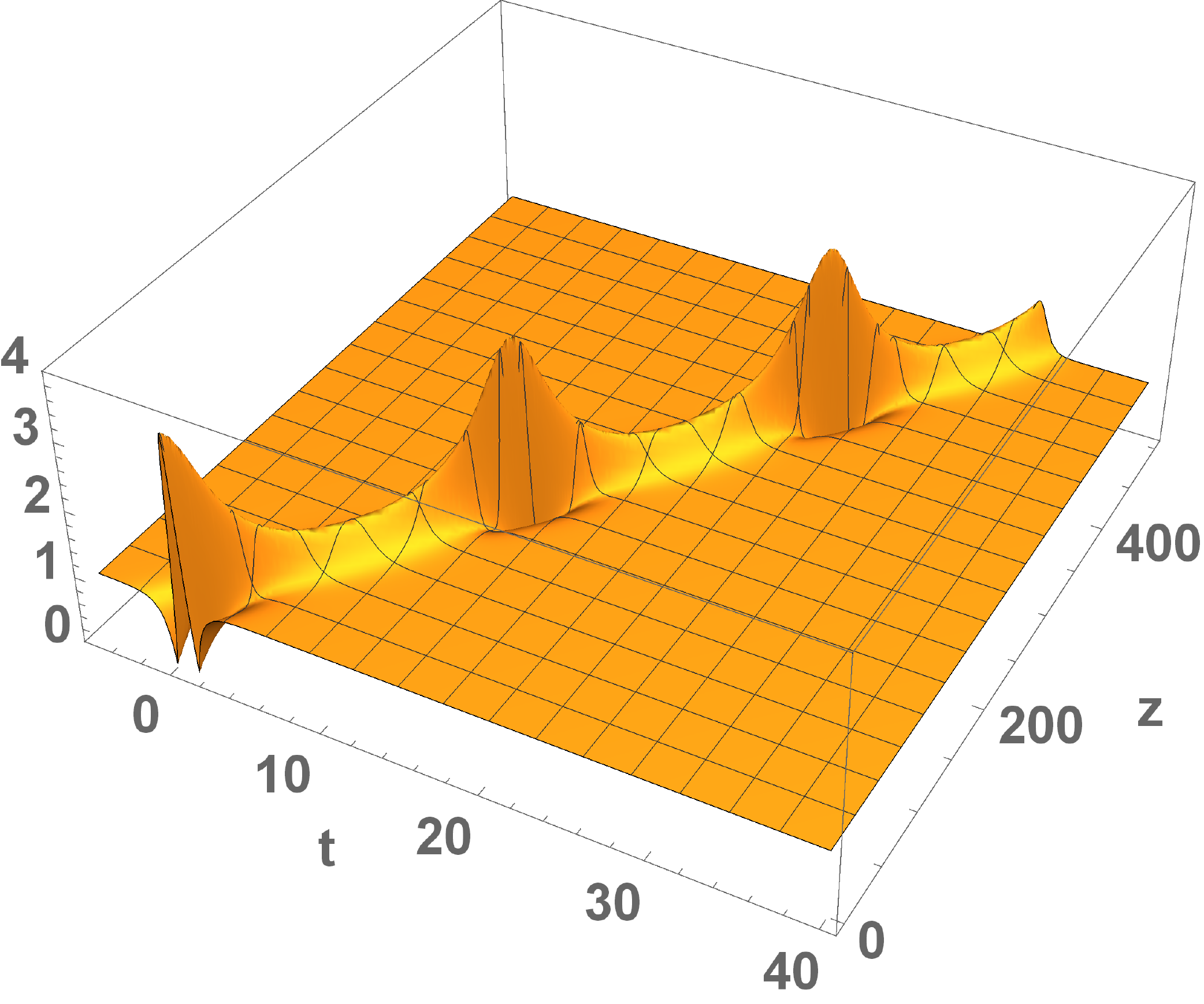}
    \includegraphics[width=0.32\textwidth]{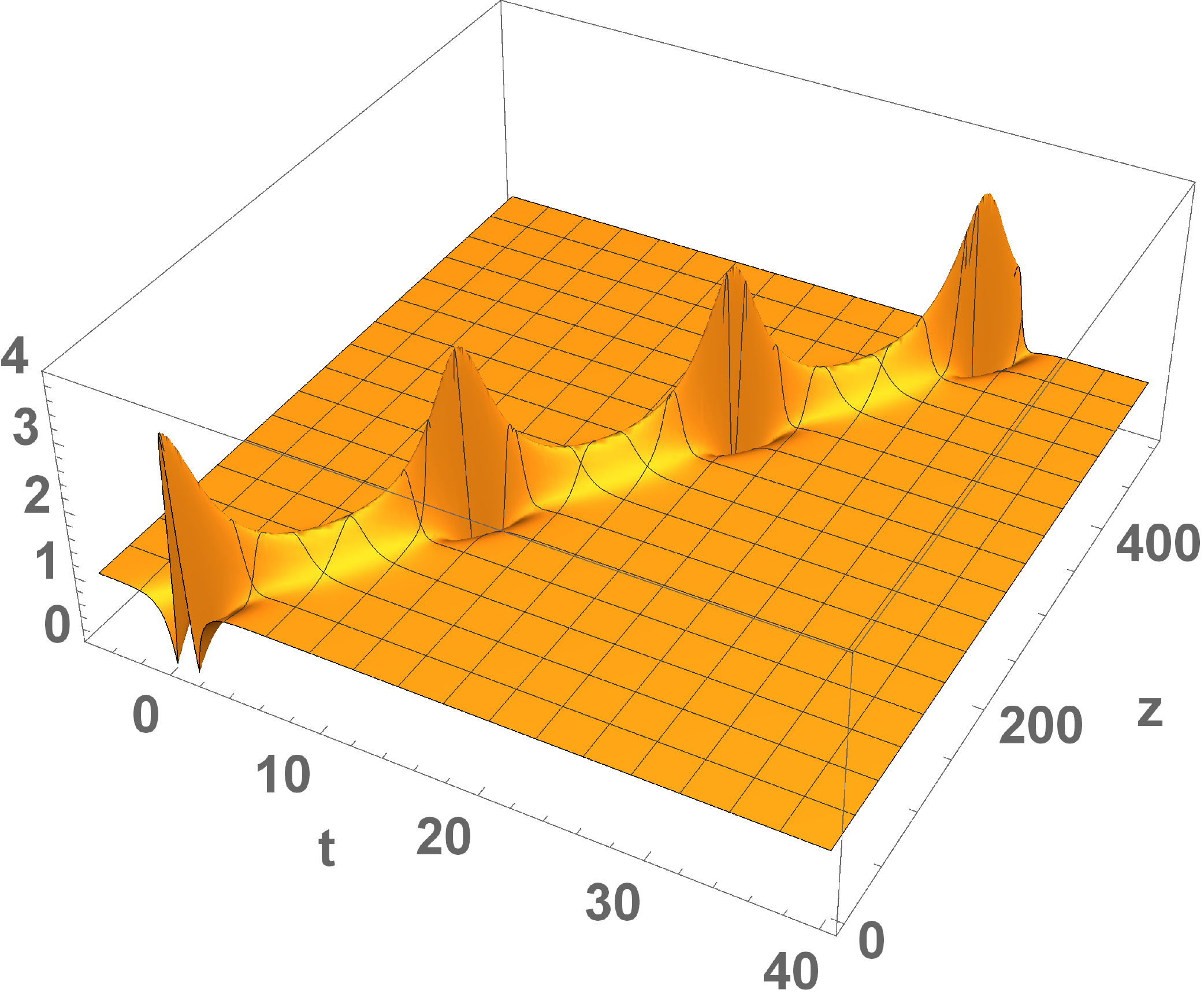}\\
    \includegraphics[width=0.32\textwidth]{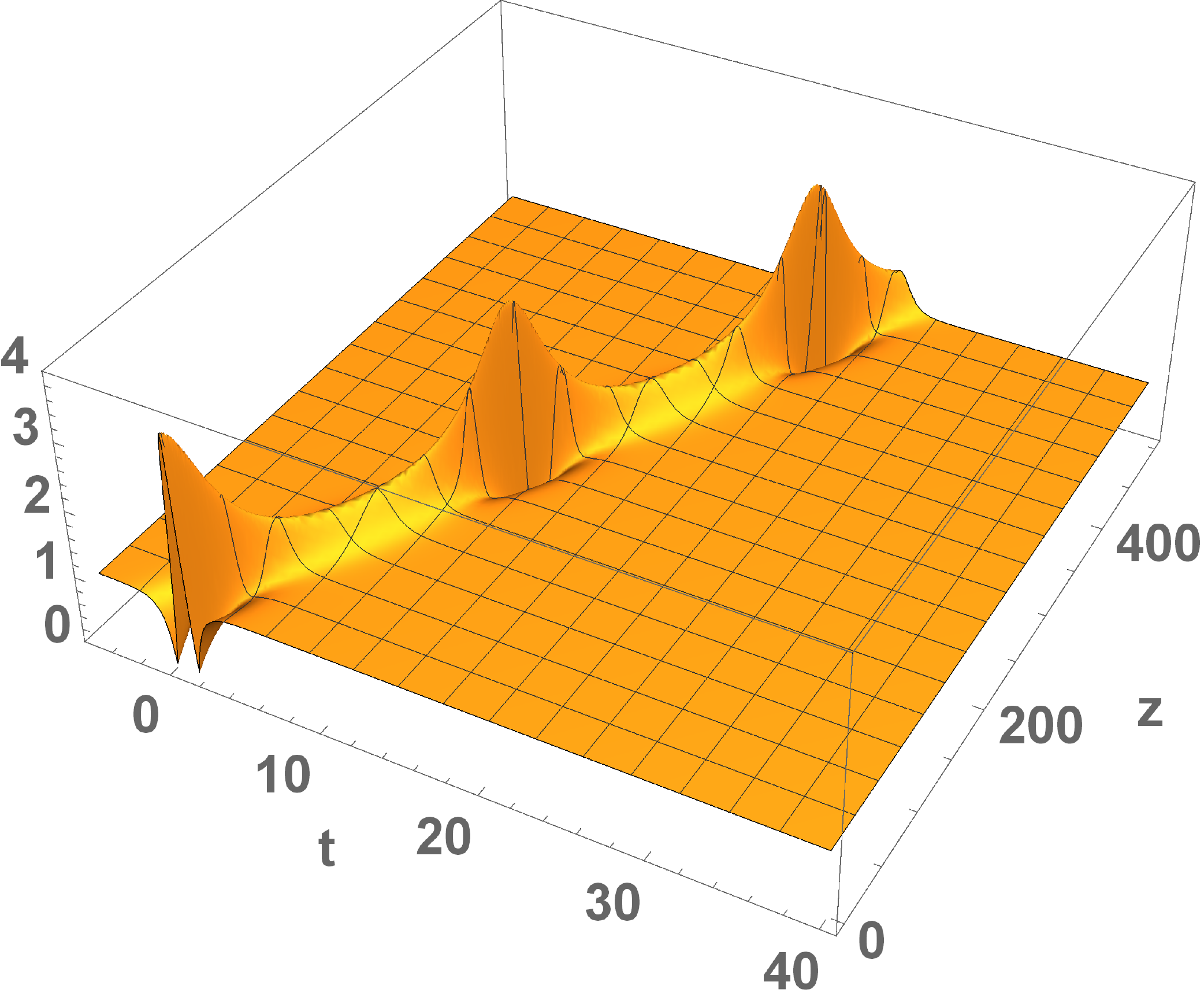}
    \includegraphics[width=0.32\textwidth]{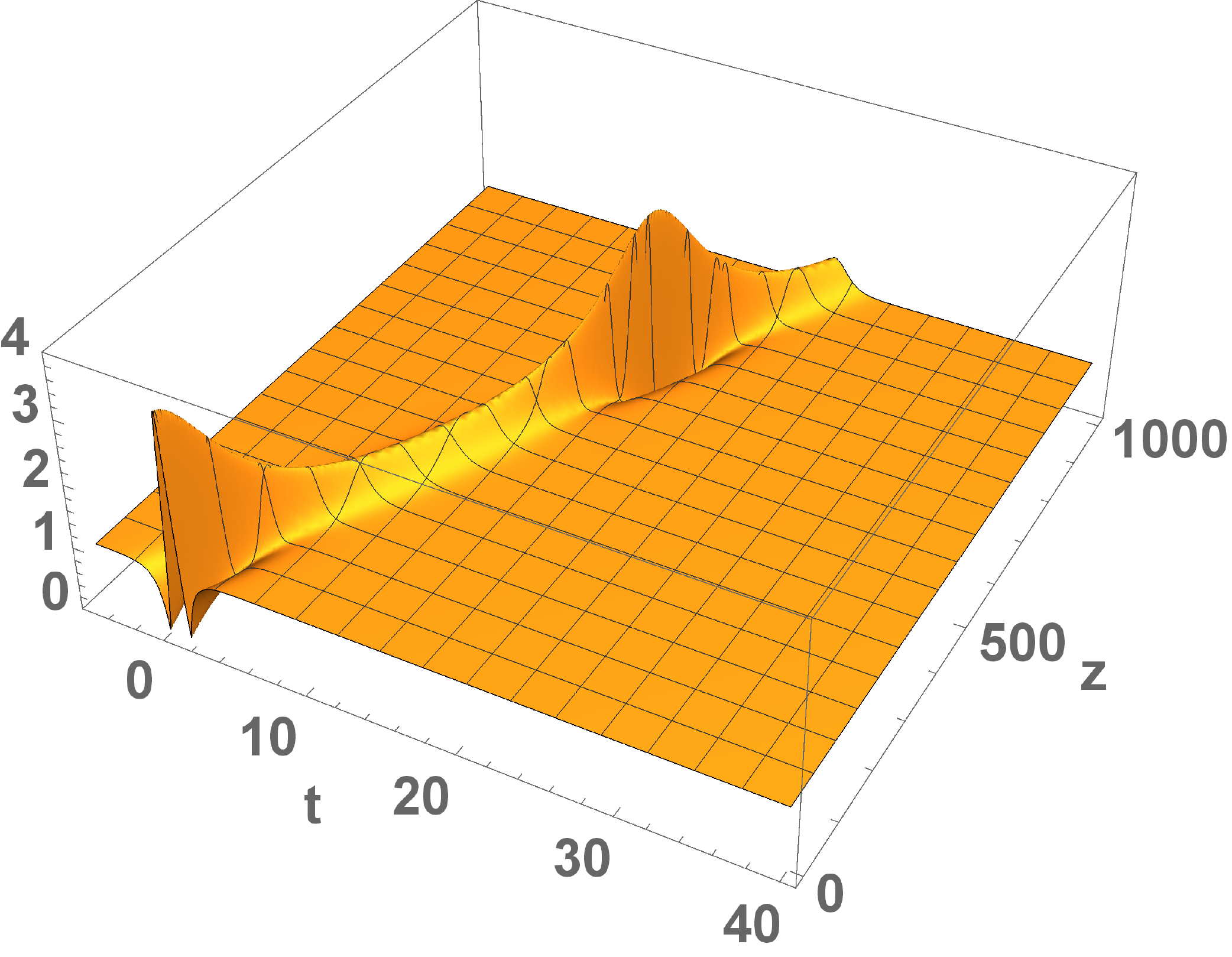}
    \includegraphics[width=0.32\textwidth]{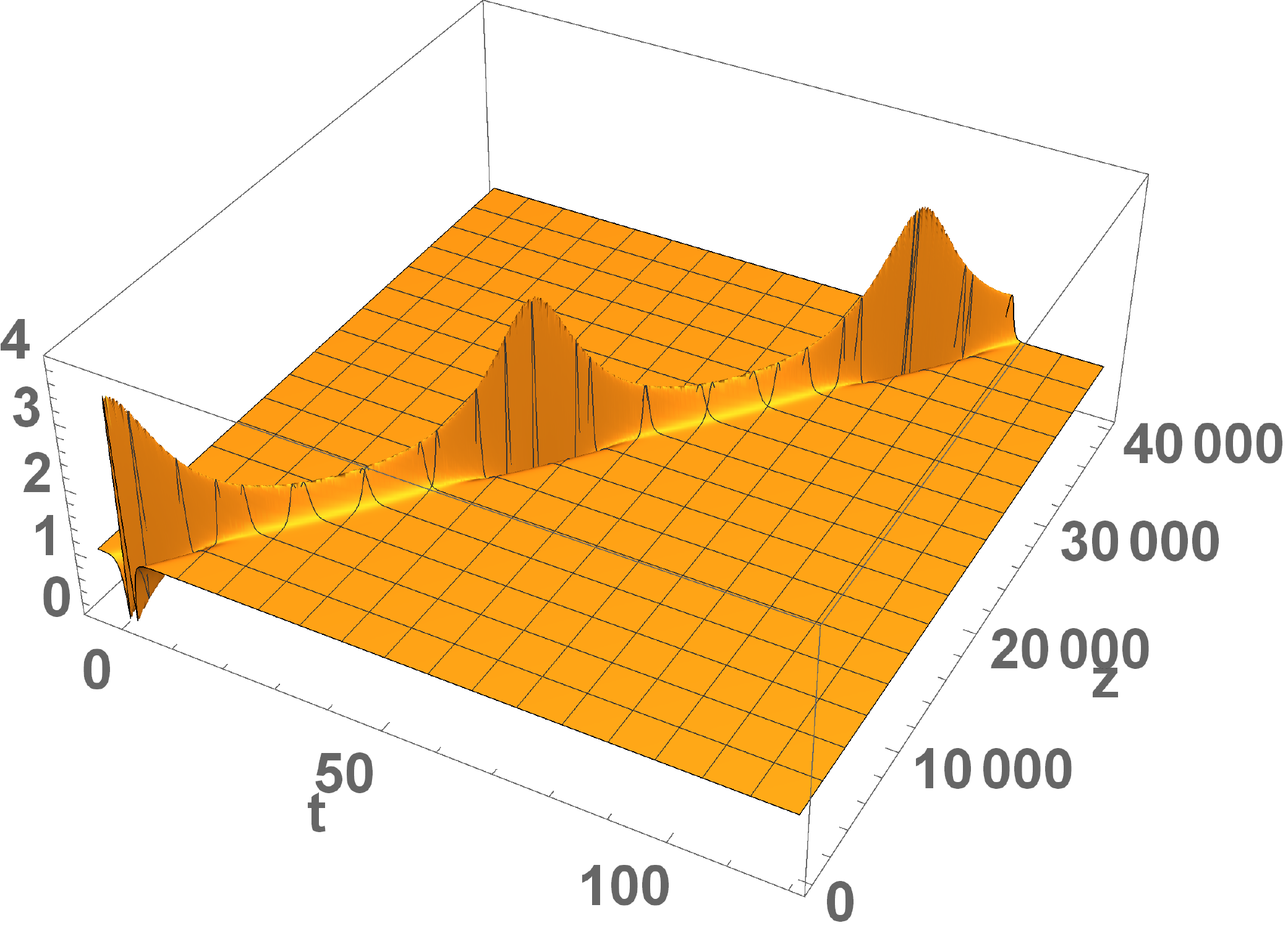}
    \caption{Amplitudes $|q(t,z)|$ of the Type~1 solitons~\eref{e:staticsoliton} with a shifted Lorentzian, 
        a discrete eigenvalue $\zeta_1 = 2i$,
        and $D_- = h_- = -1$, $q_o = 1$, $\epsilon = 2$, $\xi(0) = 0$ and $\varphi(0) = -\pi/2$.
        From left to right, from top to bottom,
        $\lambda_o = 0, 1, 1.5, 3, 6, 20$.
                Note the different scale in each plot, corresponding to the different spatial period.
                }
    \label{f:travel}
\vspace{3ex}
    \centering
    \includegraphics[width=0.32\textwidth]{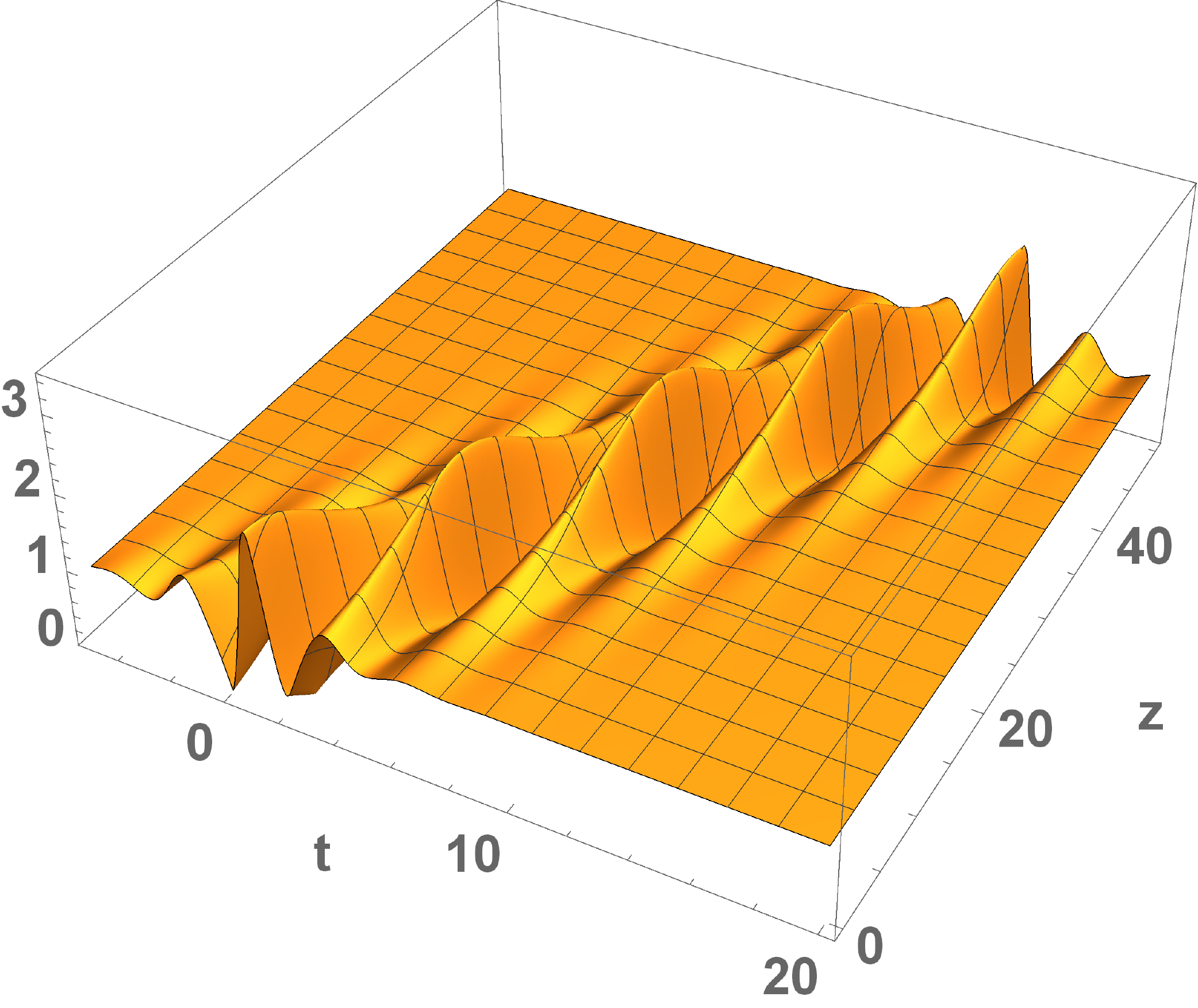}
    \includegraphics[width=0.32\textwidth]{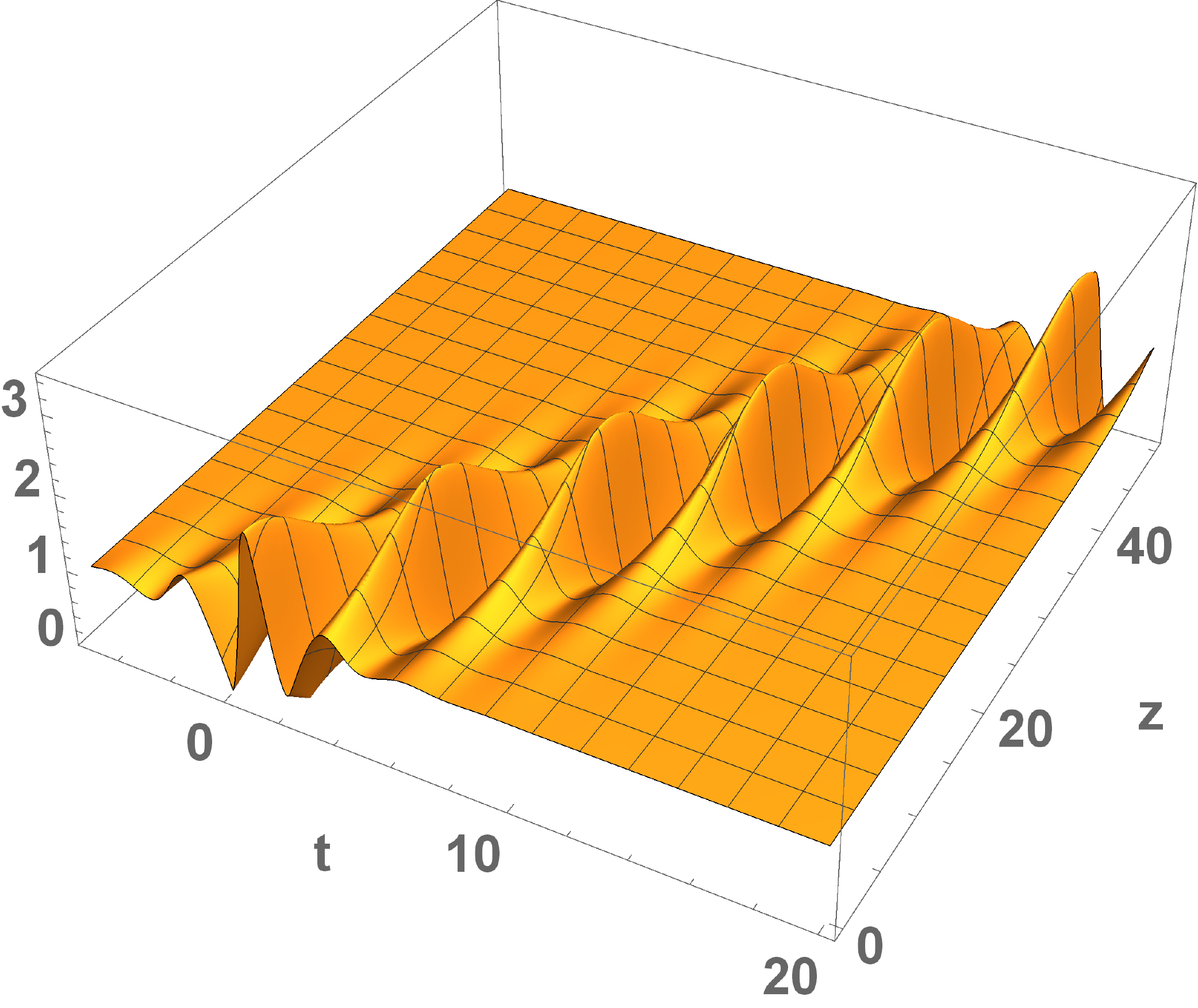}
    \includegraphics[width=0.32\textwidth]{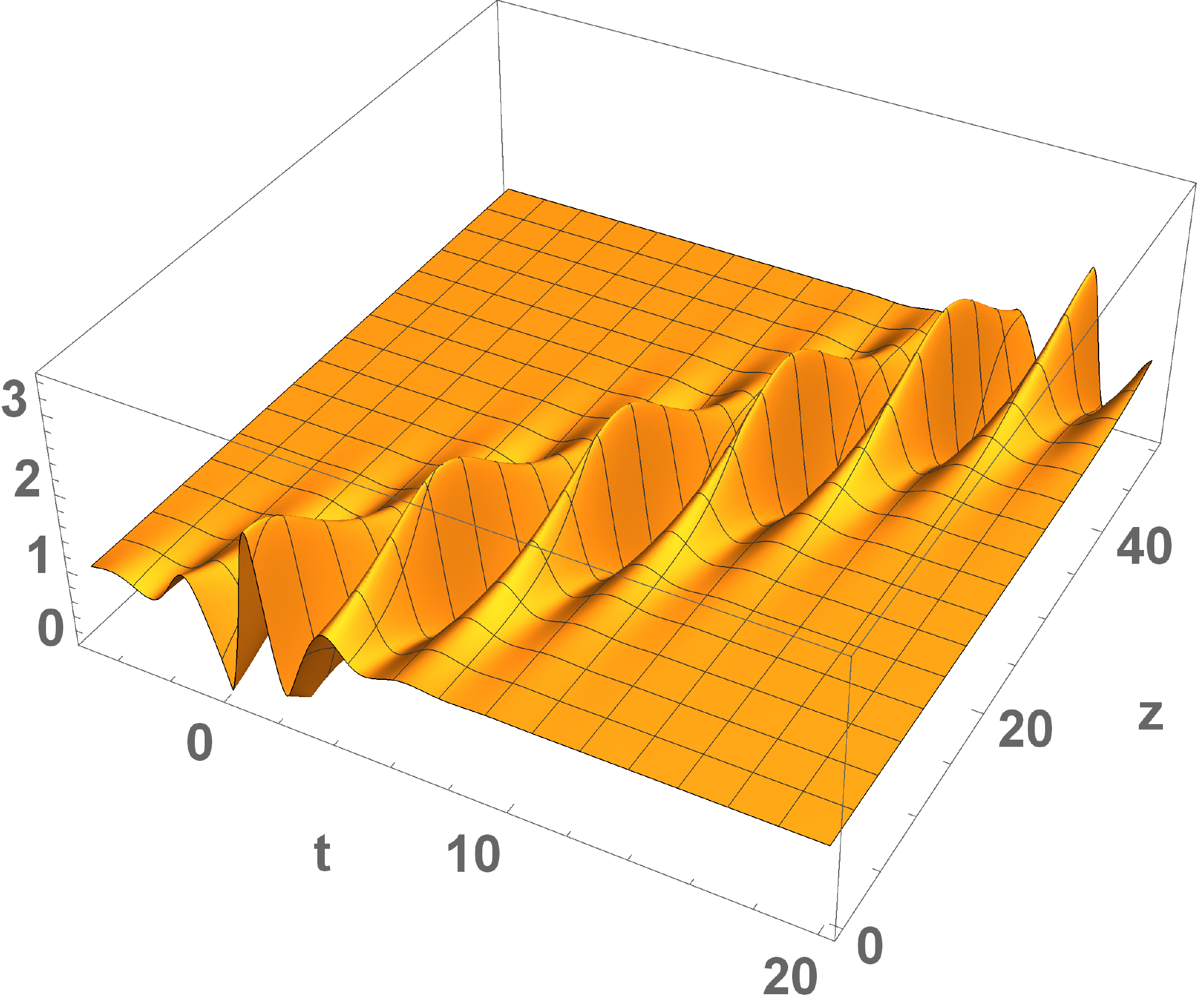}\\
    \includegraphics[width=0.32\textwidth]{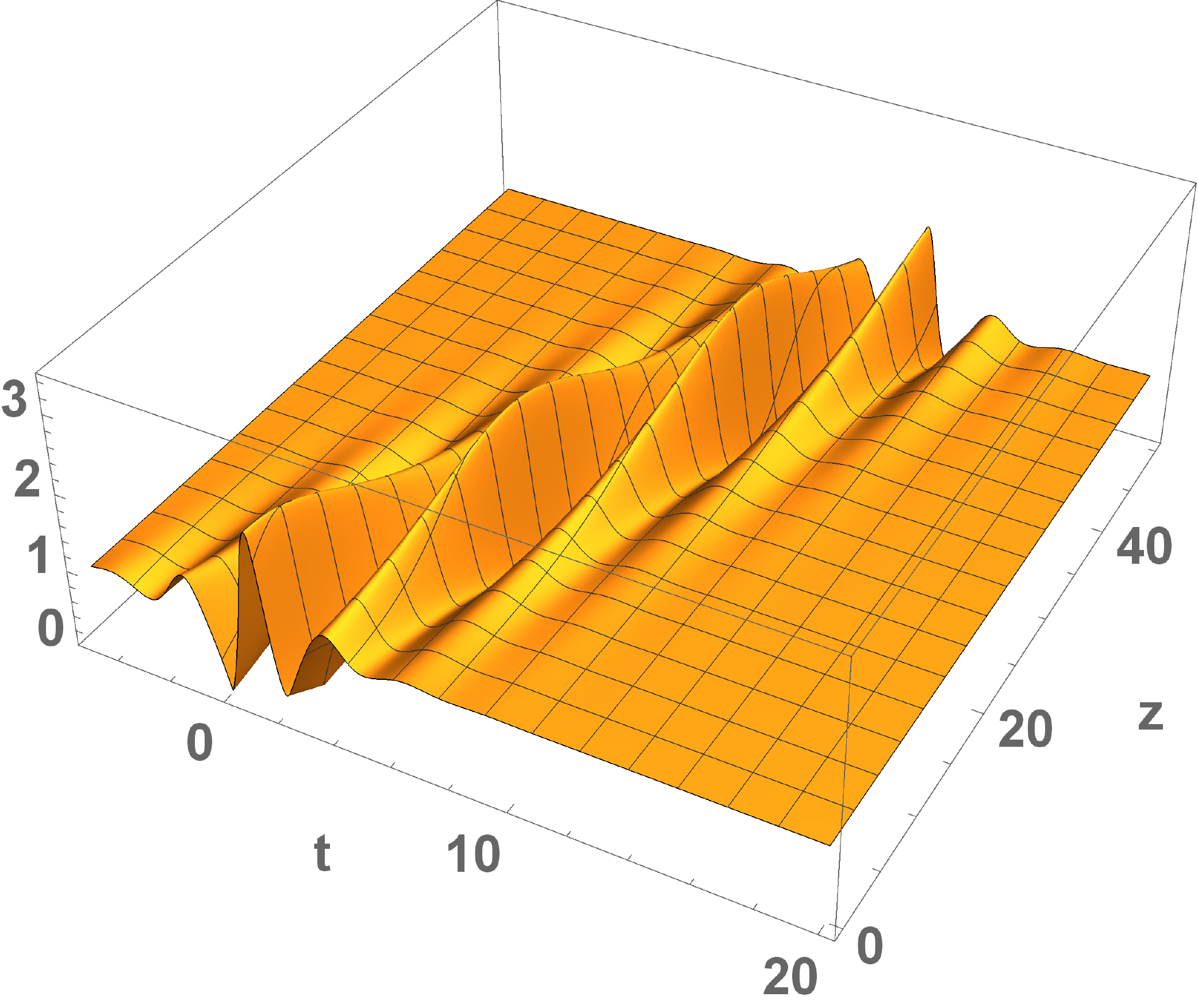}
    \includegraphics[width=0.32\textwidth]{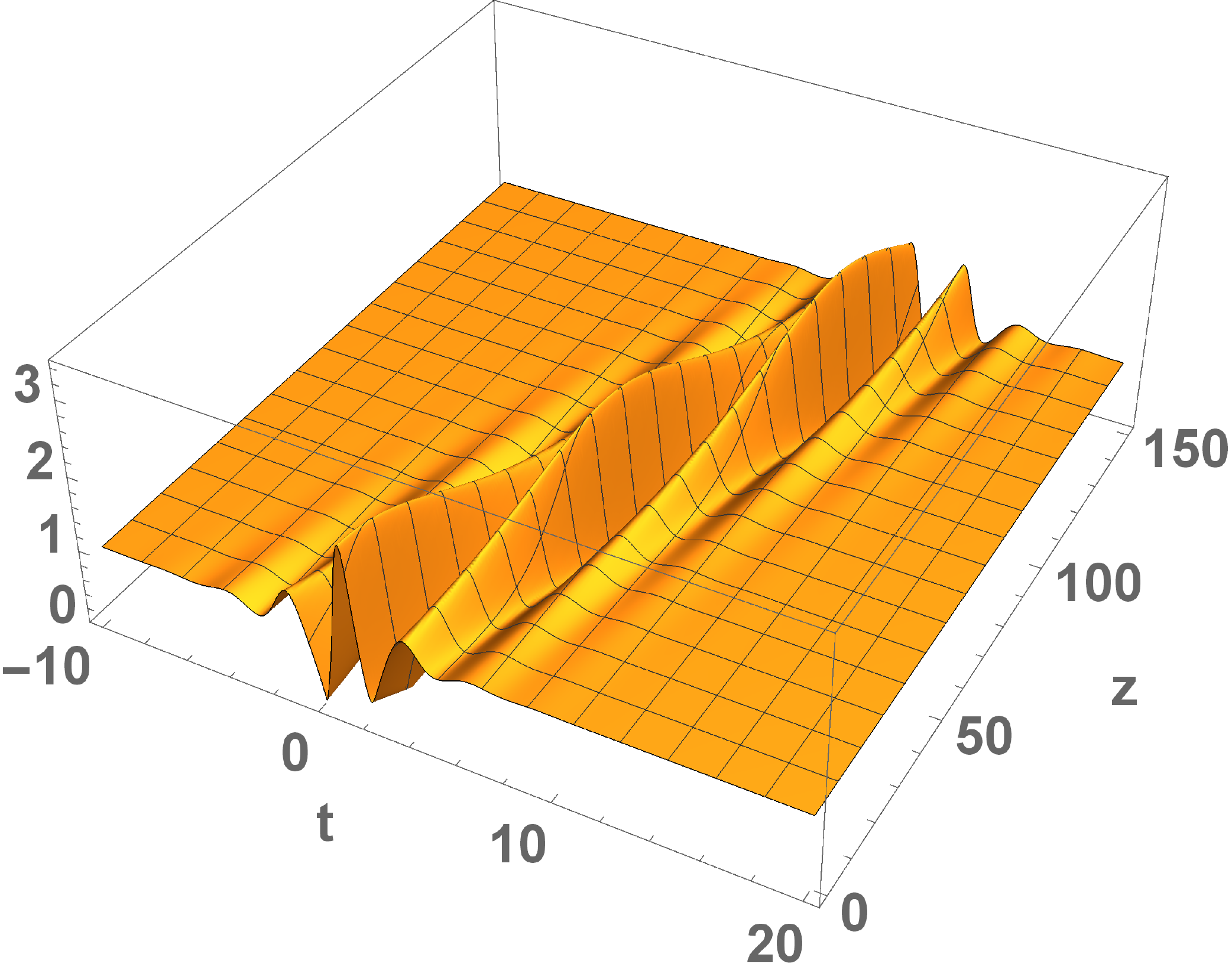}
    \includegraphics[width=0.32\textwidth]{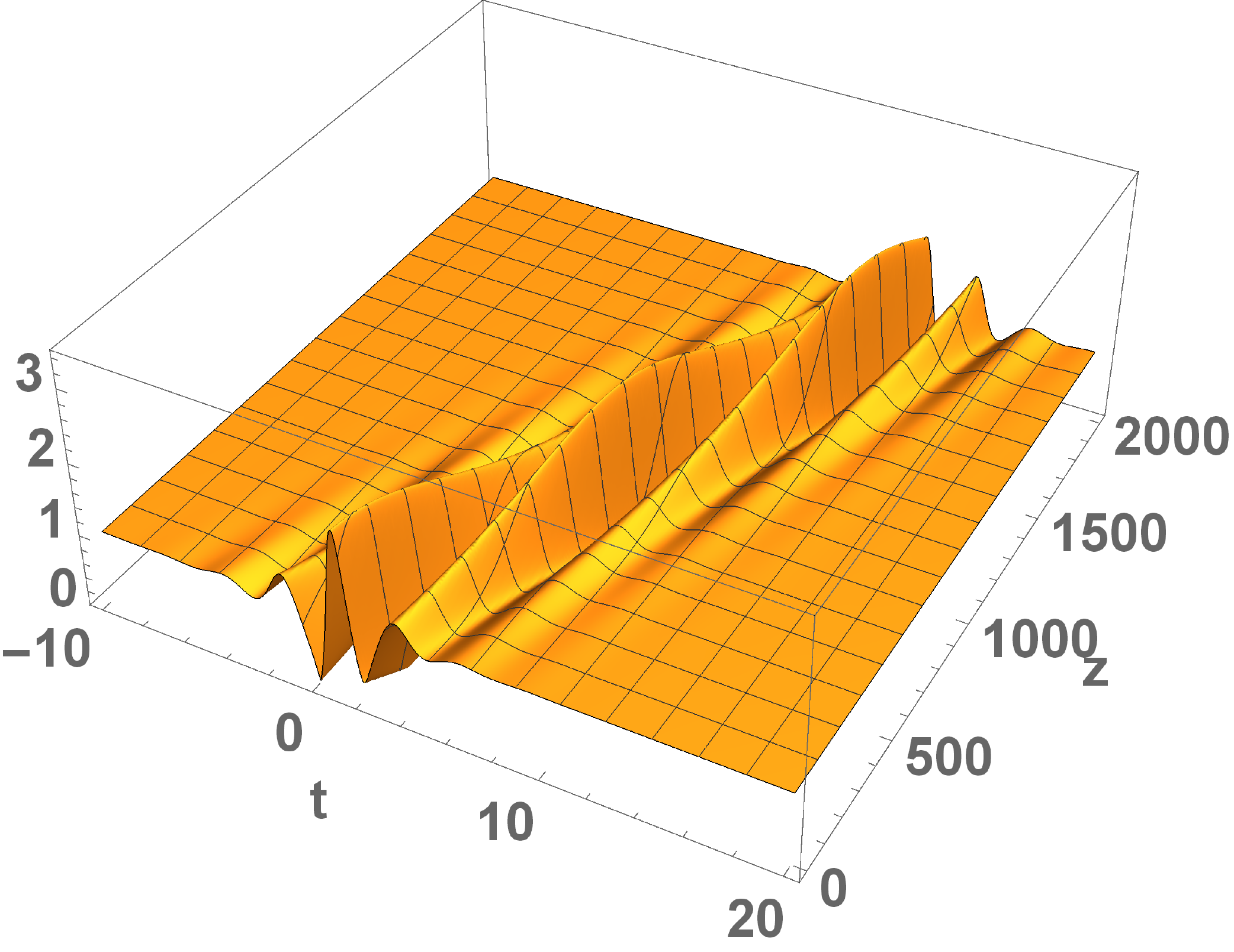}
    \caption{Similarly to Fig.~\ref{f:travel},
        but with amplitudes $|q(t,z)|$ of the Type~2 solitons~\eref{e:oscillatorysoliton} with a shifted Lorentzian,
        and a discrete eigenvalue $\zeta_1 = \sqrt3+i$.
        All other parameters remain the same.
                Note that the spatial scale is different in each plot.
    }
    \label{f:oscillatory}
\kern-3\bigskipamount
\end{figure}

\begin{figure}[t!]
\centering
\includegraphics[width=0.32\textwidth]{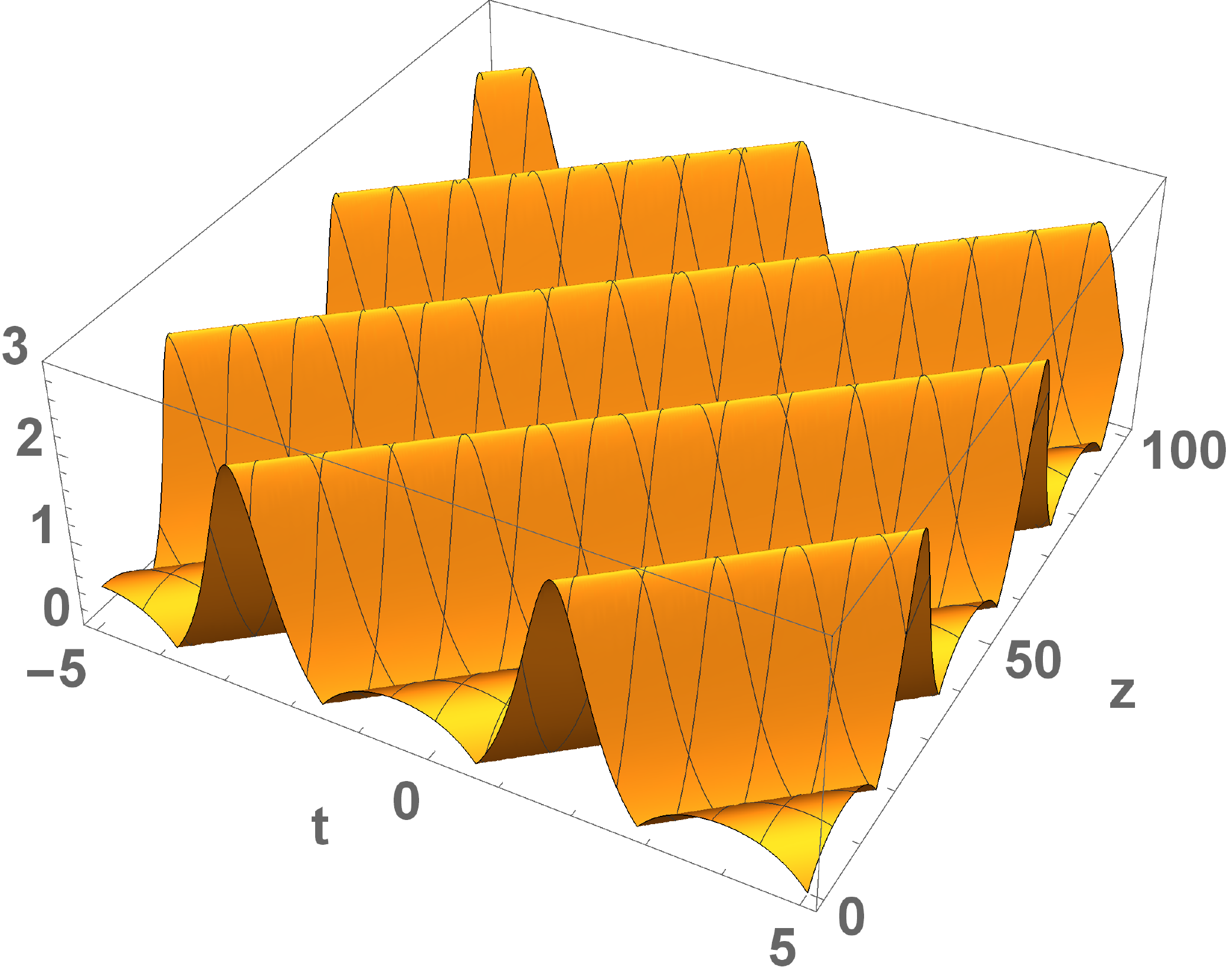}
\includegraphics[width=0.32\textwidth]{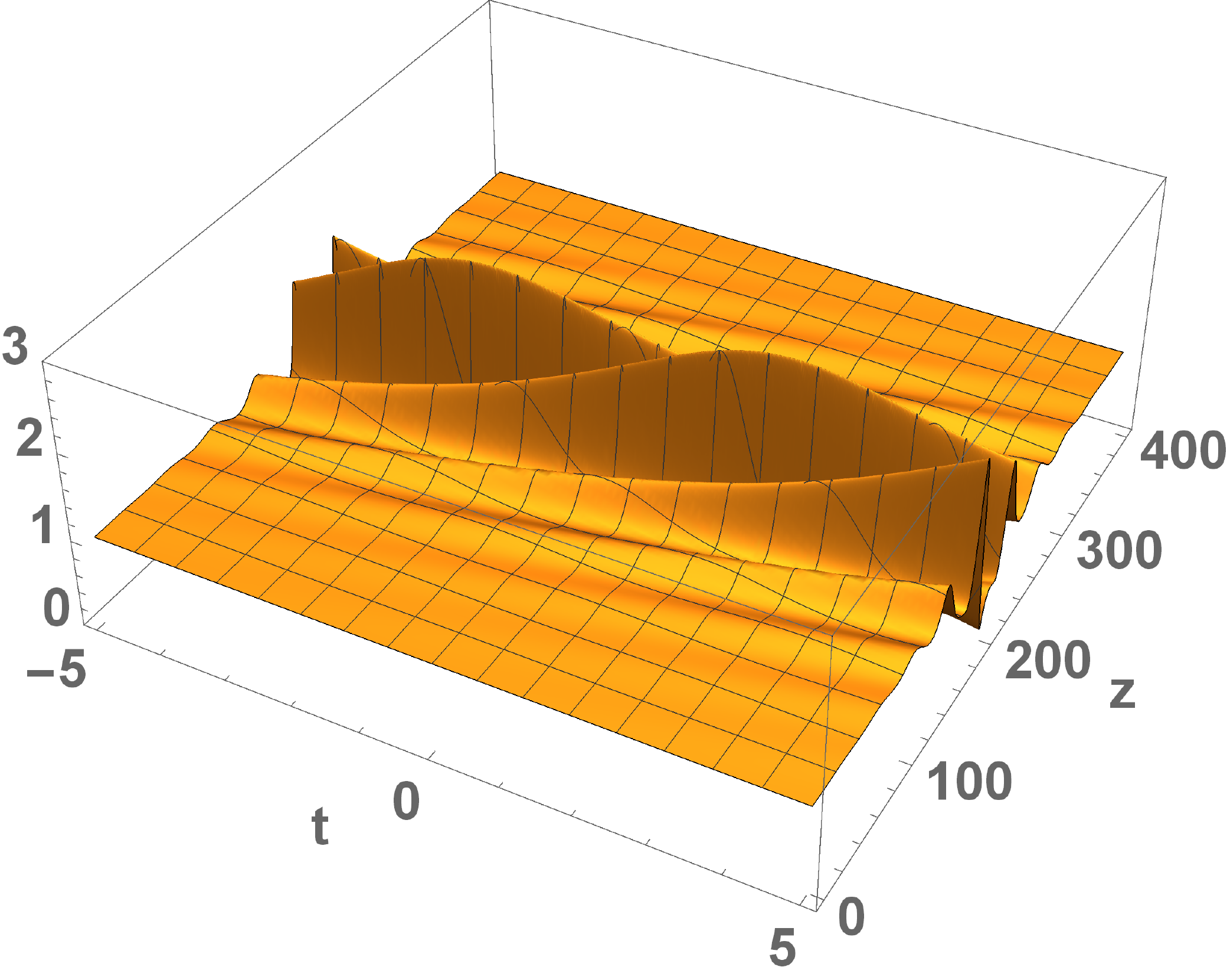}
\includegraphics[width=0.32\textwidth]{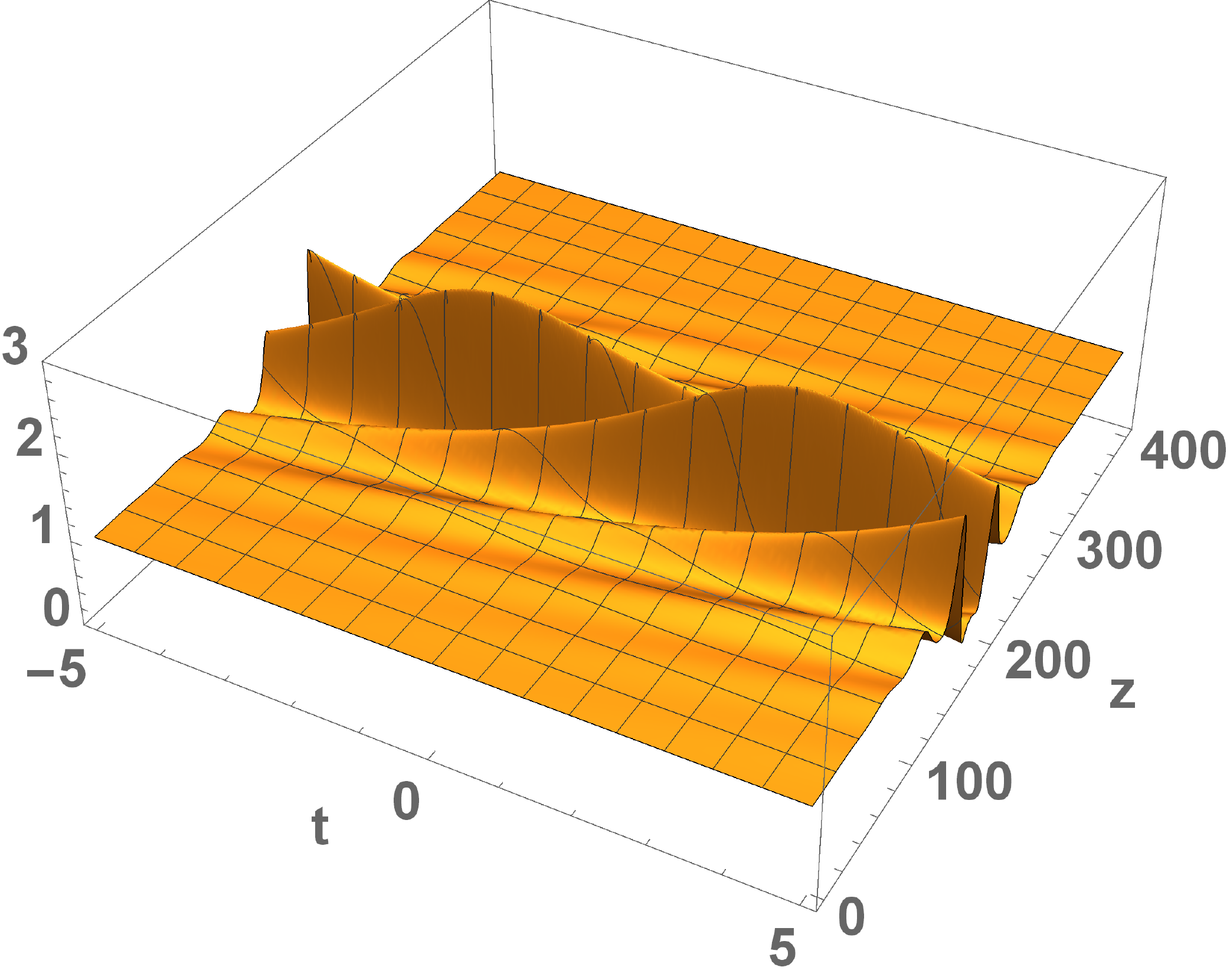}\\
\includegraphics[width=0.32\textwidth]{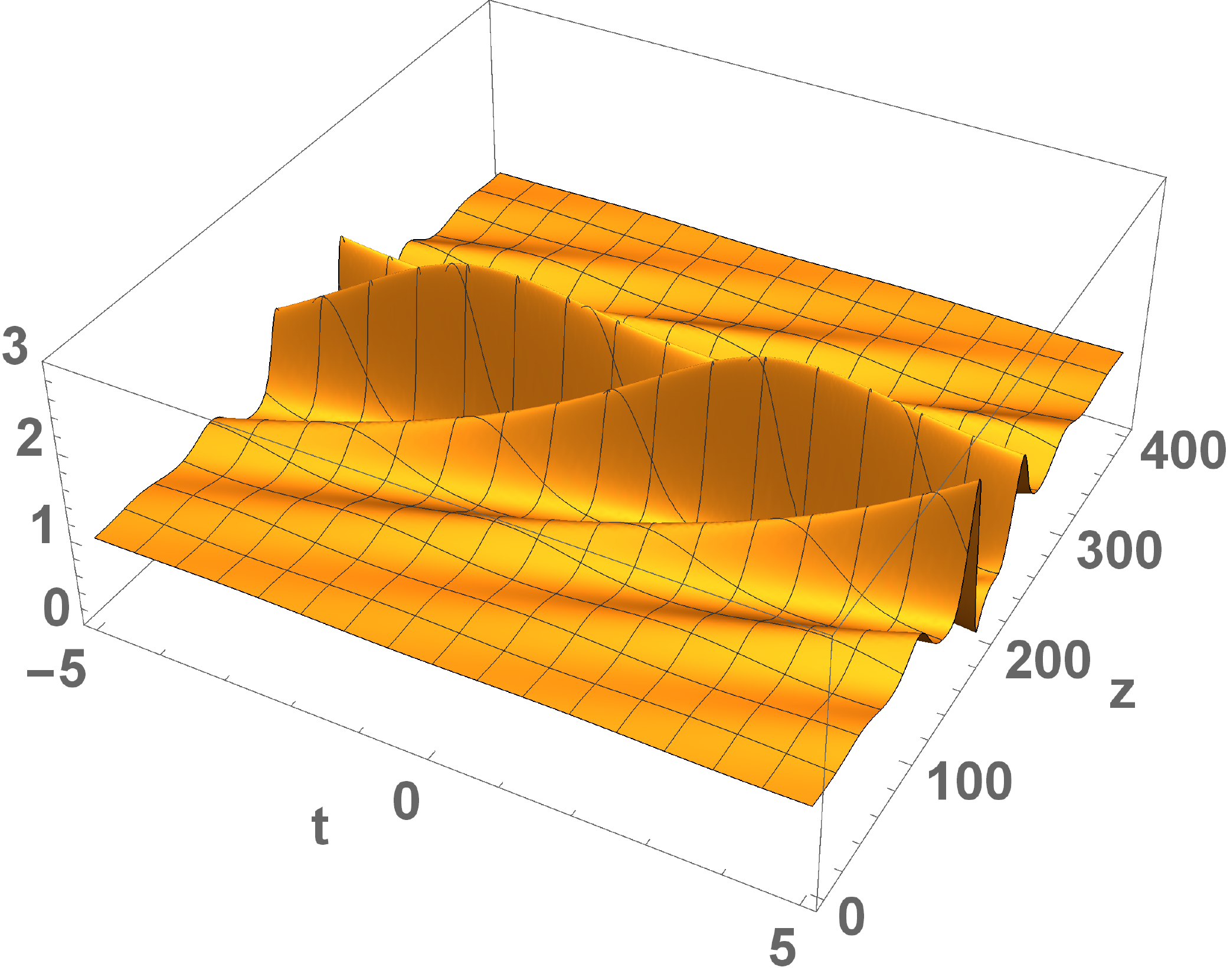}
\includegraphics[width=0.32\textwidth]{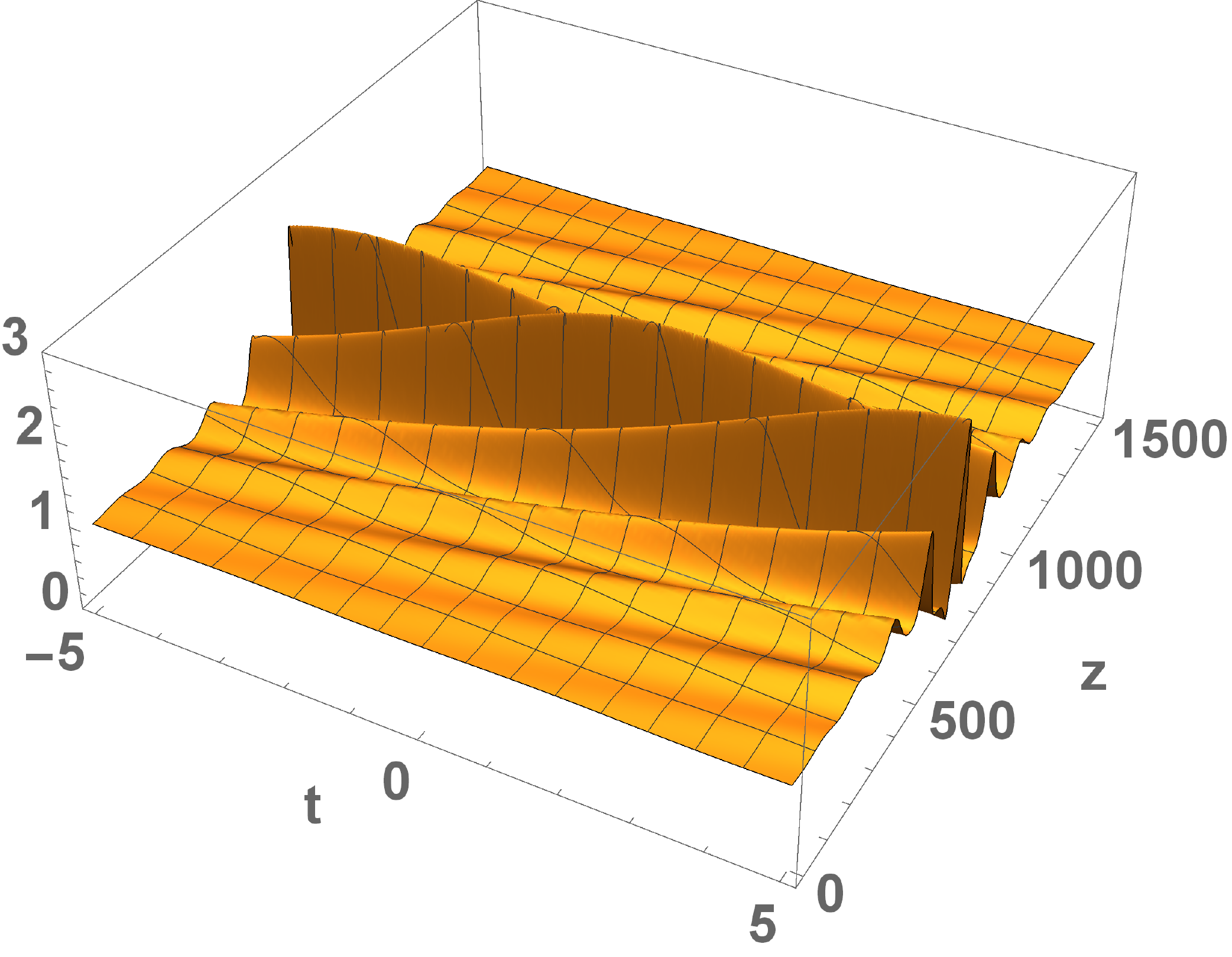}
\includegraphics[width=0.32\textwidth]{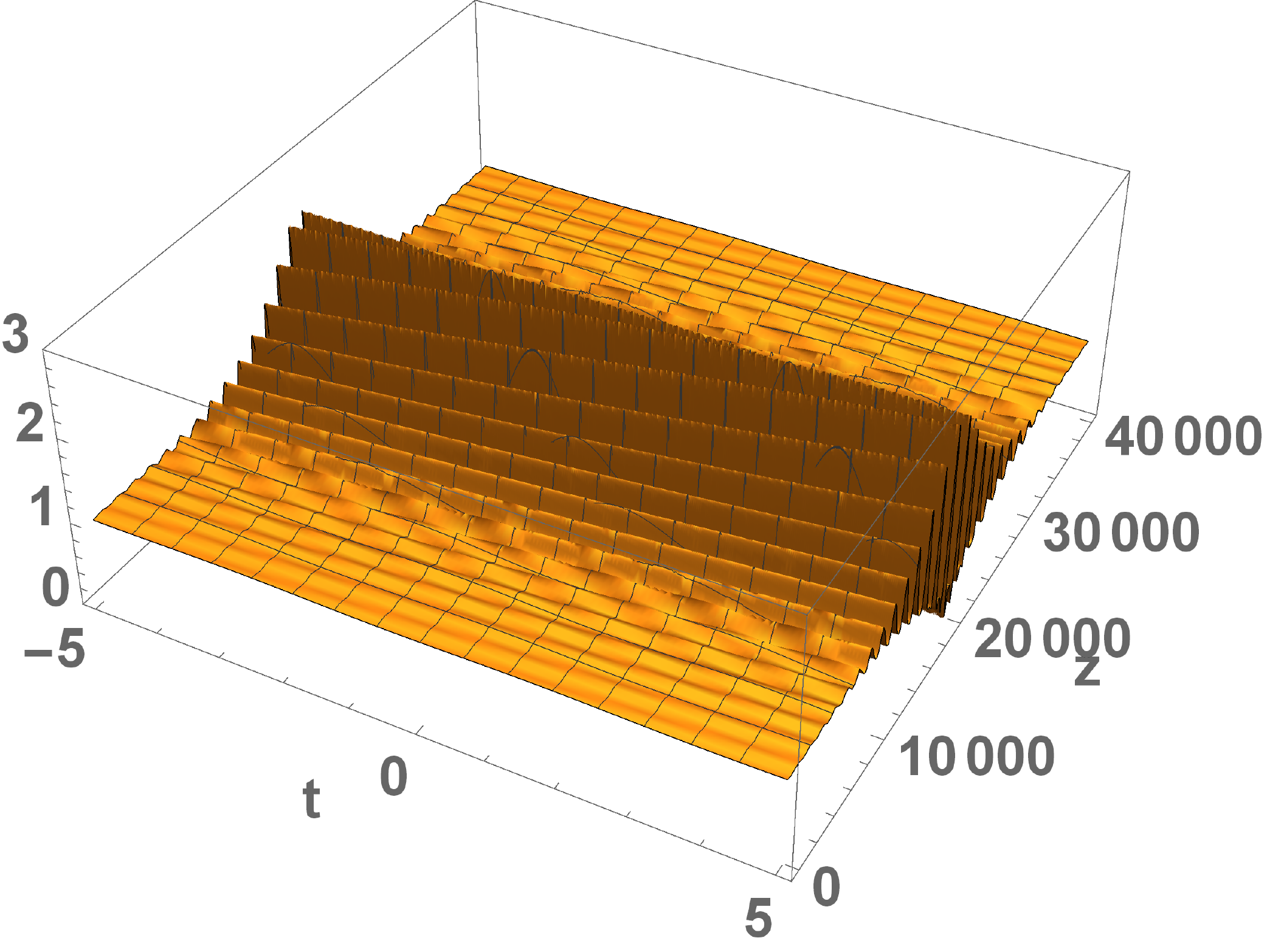}
\vspace*{-1ex}
\caption{Similarly to Fig.~\ref{f:travel},
    but with amplitudes $|q(t,z)|$ of the Type~3 solutions~\eref{e:periodicq} with a shifted Lorentzian
    and a discrete eigenvalue $\zeta_1 = (1+i)/\sqrt{2}$.
    All other parameters remain the same,
        except that in the second and third plots $\xi(0) = 7$,
    and in the fourth to the last plots $\xi(0) = 5$.
    Note that the spatial scale is different in each plot.
}
\label{f:periodic}
\vspace*{2ex}
    \centering
    \includegraphics[width=0.3\textwidth]{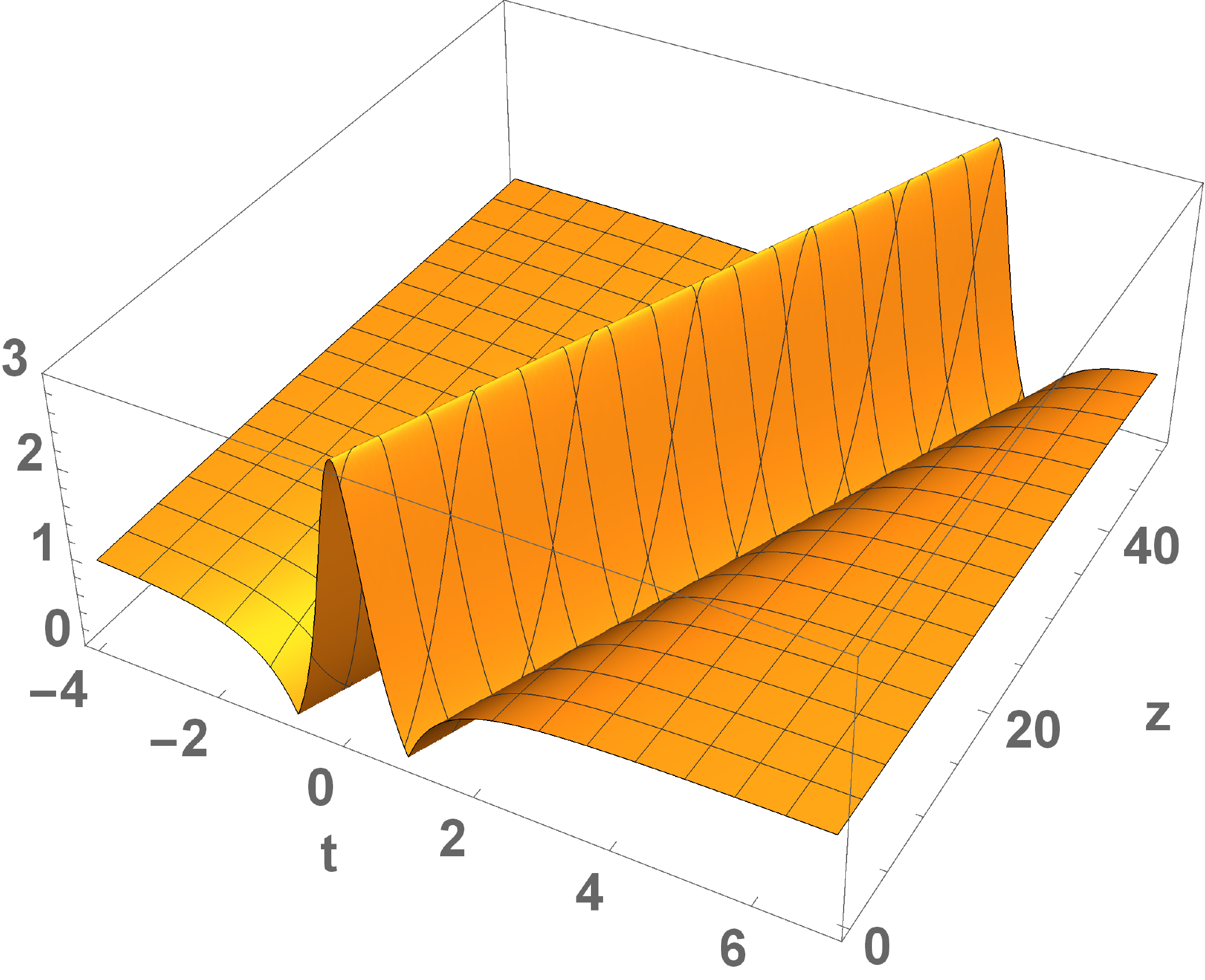}
    \includegraphics[width=0.3\textwidth]{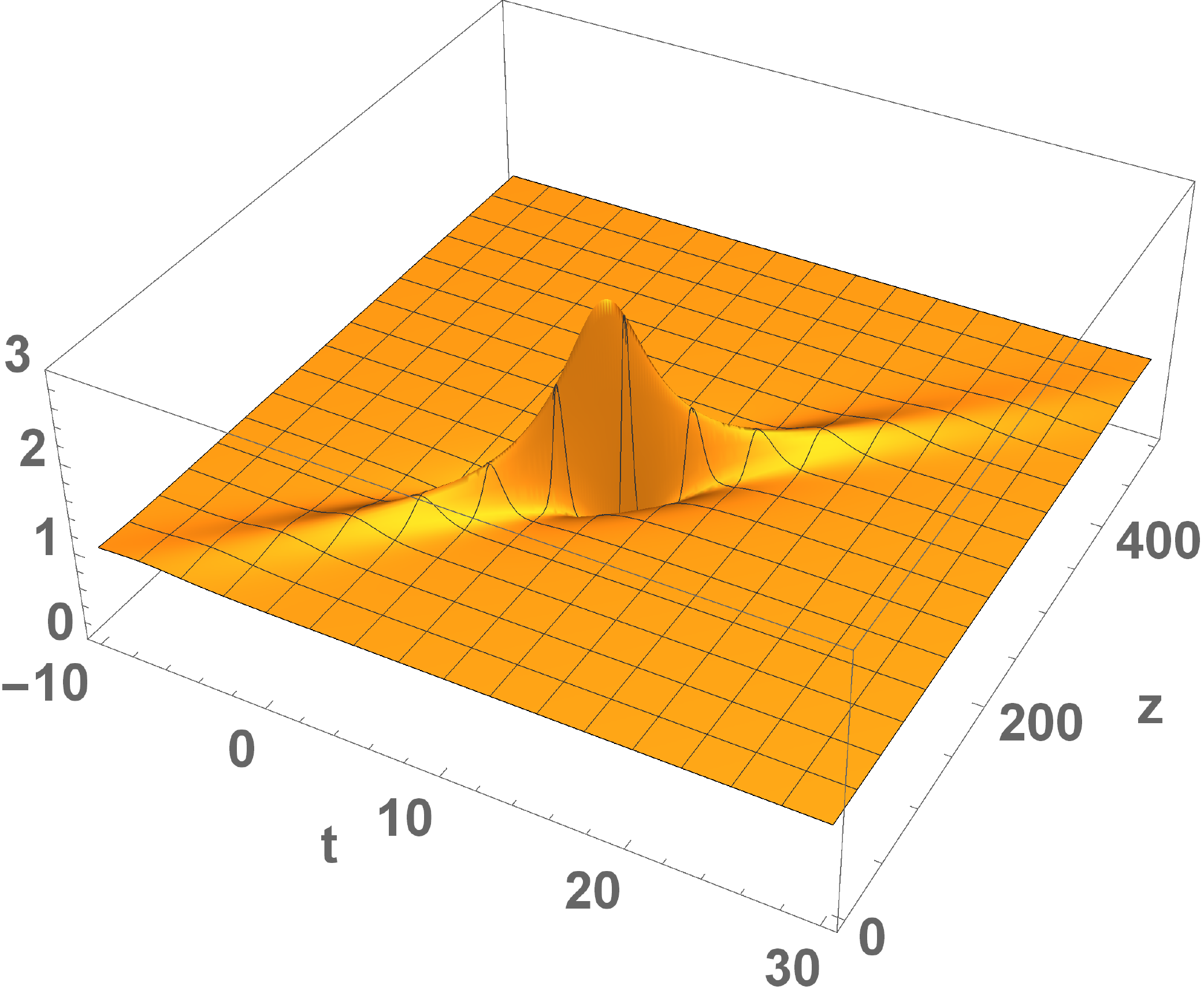}
    \includegraphics[width=0.3\textwidth]{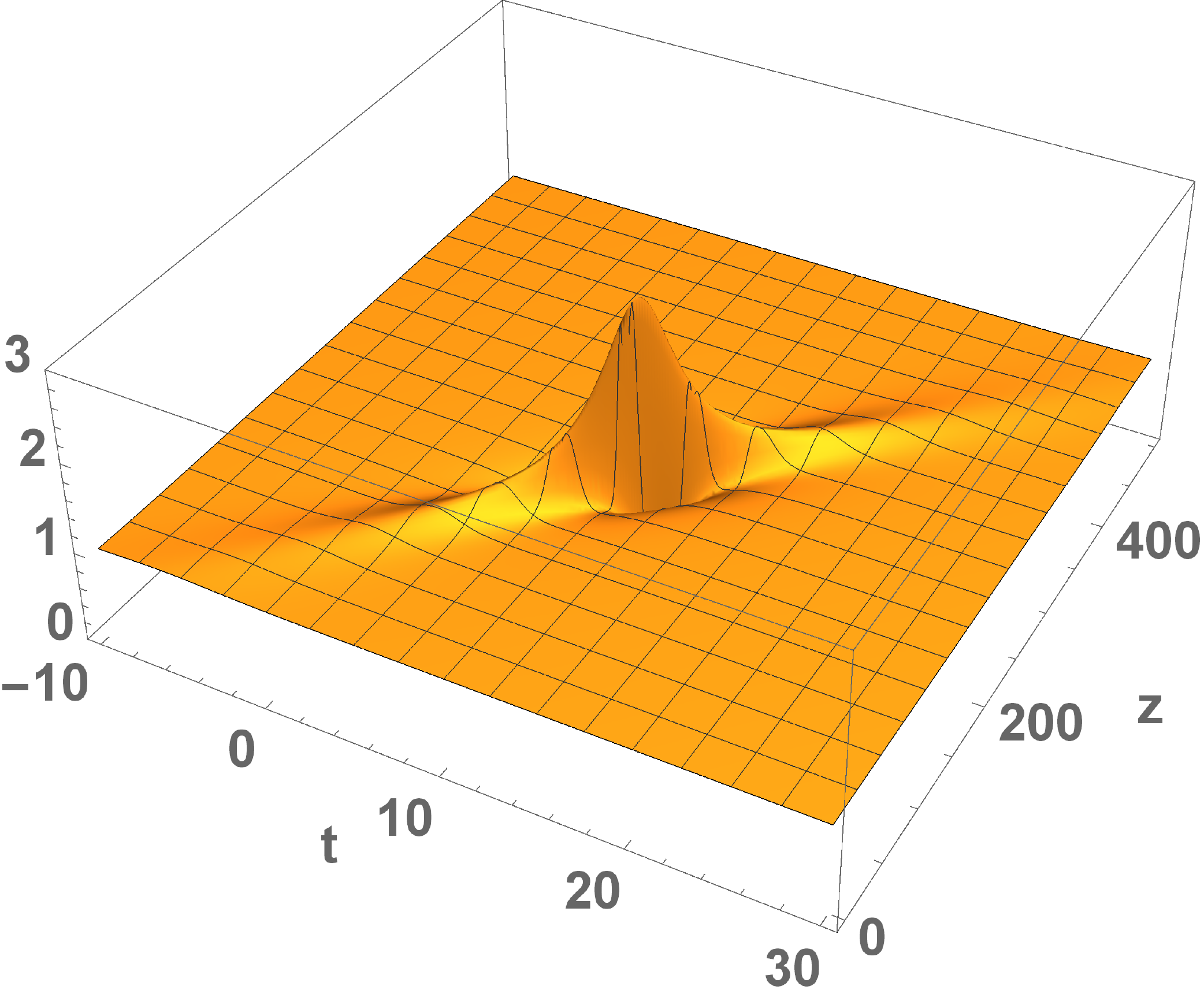}\\
    \includegraphics[width=0.3\textwidth]{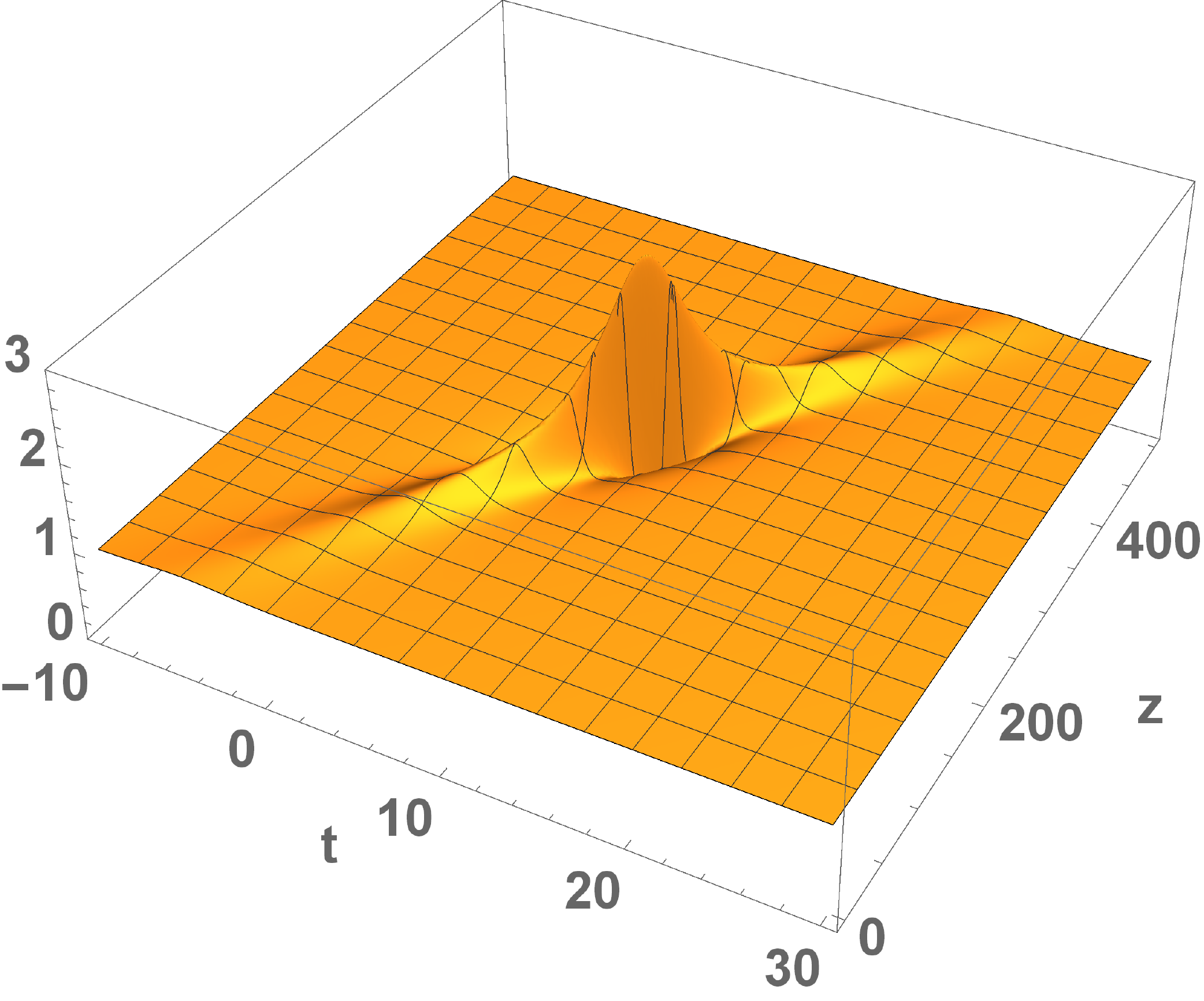}
    \includegraphics[width=0.3\textwidth]{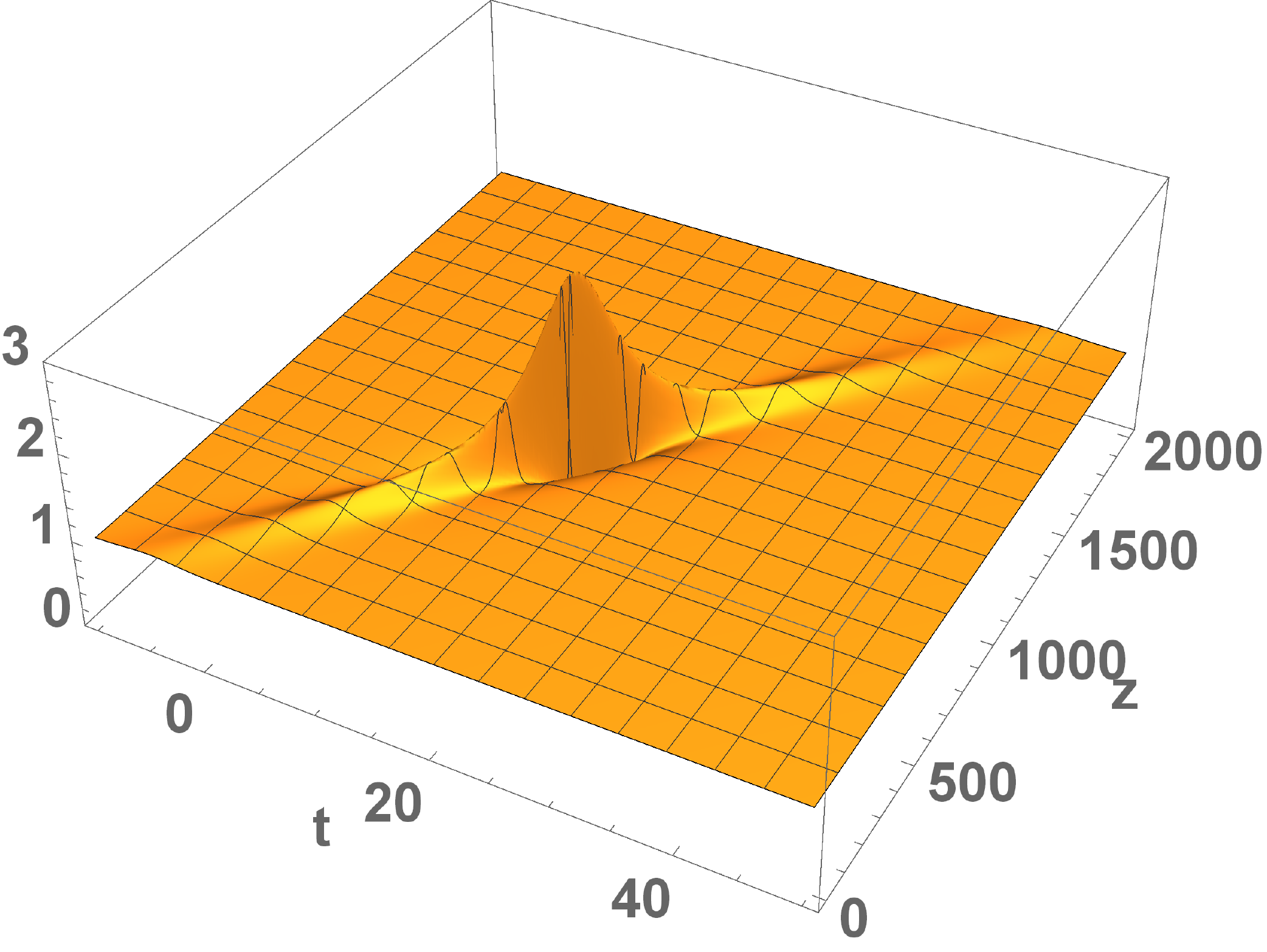}
    \includegraphics[width=0.3\textwidth]{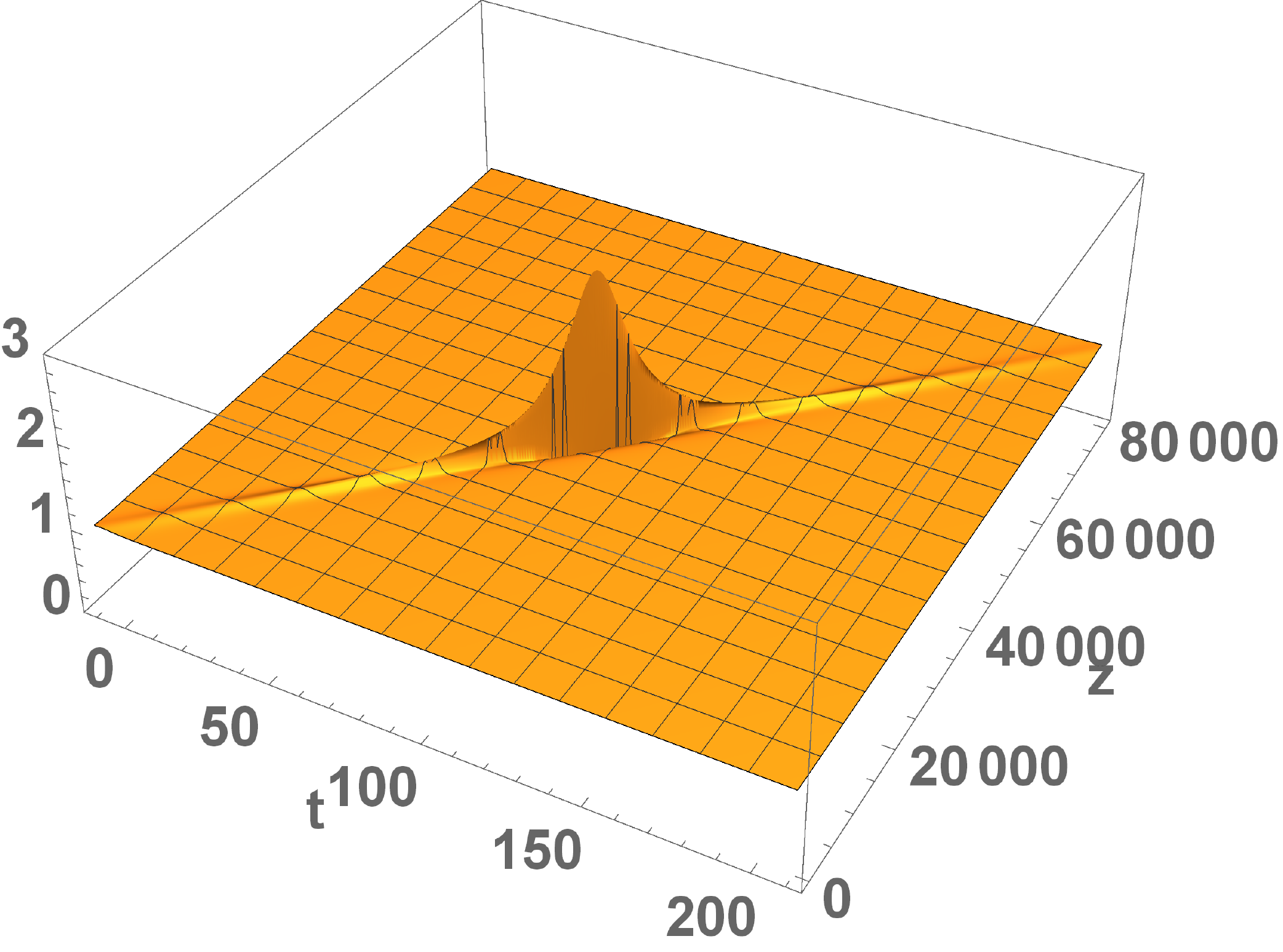}
    \caption{Amplitudes $|q(t,z)|$ of rational solutions~\eref{e:rational_shiftedlorentzian} with
        a shifted Lorentzian,
        $D_- = -1$, $q_o = 1$, $\epsilon = 2$. 
        From left to right, from top to bottom,
        $\lambda_o = 0$ with $\xi_o^{(1)} = \varphi_o^{(1)} = 0$;
        $\lambda = 1$ with $\xi_o^{(1)} = 15$ and $\varphi_o^{(1)} = 10$;
        $\lambda = 1.5$ with $\xi_o^{(1)} = \varphi_o^{(1)} = 12$;
        $\lambda = 3$ with $\xi_o^{(1)} = \varphi_o^{(1)} = 10$;
        $\lambda = 6$ with $\xi_o^{(1)} = \varphi_o^{(1)} = 10$;
        $\lambda = 20$ with $\xi_o^{(1)} = \varphi_o^{(1)} = 16$.
        One should notice the different variable ranges among all plots.
    }
    \label{f:rational}
\kern-3\bigskipamount
\end{figure}

Recall that the norming constant is parameterized by Eq.~\eref{e:xiverphi_unshiftedLorentzian}, 
where $\zeta_1$ is the discrete eigenvalue and $R_{-,1,1}(\zeta)$ is given by Eq.~\eref{e:R-11_shiftedLorentzian}.
The formulas for the first three types of soliton solutions all remain the same except for the $z$ dependence.
Expliticly, 
Eq.~\eref{e:staticsoliton} for Type~I,
Eq.~\eref{e:oscillatorysoliton} for Type~II and 
Eq.~\eref{e:periodicq} for Type~III
all remain valid except that one simply substitutes $w_-$ from Eq.~\eref{e:w-1}, 
and $\xi(z)$ and $\varphi(z)$ from Eq.~\eref{e:xiverphi_unshiftedLorentzian}
with $R_{-,1,1}$ from Eq.~\eref{e:R-11_shiftedLorentzian}.
The three types of solutions with a 
shifted Lorentzian are shown in Figs.~\ref{f:travel},~\ref{f:oscillatory} and~\ref{f:periodic},
respectively.


Figure~\ref{f:travel} shows Type-1 soliton solutions obtained with a few different choices of $\lambda_o$.
It is evident from the figure that,
as the peak of detuning function is shifted from the origin to infinity,
the solution becomes oscillatory.
As a result, this is not a simple traveling wave anymore,
but becomes a breather-like solution. 
Eventually, as $\lambda_o\to\infty$ the spatial period of the oscillation increases without bound.
Moreover, 
it appears that, as $\lambda_o$ gets large,
the soliton velocity tends to the speed of light.
This is to be expected:
as the atoms are detuned from the resonant frequency,
the nonlinear interactions between atoms and light become weaker,
so the medium becomes essentially transparent to the light pulse, 
which then becomes regular light traveling with speed $c$.

As shown in Fig.~\ref{f:oscillatory}, 
Type-2 solitons
retain the main characteristics of the corresponding solutions with an unshifted Lorentzian,
namely a localized, traveling breather-like solution with an internal oscillatory structure.
However, as the center of the Lorentzian approaches infinity,
the soliton velocity, i.e., the group velocity, and the phase velocity both tend to the speed of light.
Eventually, as $\lambda_o\to\infty$ this type of soliton solution becomes a localized traveling wave.

On the other hand, as shown in Fig.~\ref{f:periodic}, 
there is a dramatic change in the characteristics of the solution for Type-3 solitons.
Recall that, if $\lambda_o = 0$ (i.e., for an unshifted Lorentzian), these solutions are periodic with respect to both variables $t$ and $z$ 
(cf. Section~\ref{s:periodic}).
If $\lambda_o\ne0$, however, 
Type-3 soliton solutions are only periodic in $t$, and are not periodic in $z$ in general.
This is because when $\lambda_o\ne0$ and $\zeta_1 = \e^{i\alpha}$, the quantity $R_{-,1,1}(\zeta_1)$ is not purely real anymore.
So both $\xi(z)$ and $\varphi(z)$ acquire a nontrivial dependence on $z$.
Moreover, when $\lambda_o\ne0$, the solution decays to the background as $z\to\infty$,
i.e., the solution becomes localized in $z$.
As a result, the character of this solution becomes similar to that of the Akhmediev breathers of the NLS equation.

\paragraph{Type~4. rational solutions.}

The general expression~\eref{e:rationalq} for the rational solutions
still applies to the current case.
The difference from before is in the spatial dependence of the 
three quantities $q_-(z)$, $\xi_1(z)$ and $s_1(z)$.
As before, 
the background $q_-(z)$ is obtained from Eq.~\eref{e:q-timeevolution} with $w_-$ given by Eq.~\eref{e:w-1}.
The quantities $\xi_1(z)$ and $s_1(z)$ are obtained from the Taylor expansions~\eref{e:xisexpansion1} and~\eref{e:xisexpansion2}.
To calculate these quantities explicitly,
we first expand $R_{-,1,1}(iq_o\eta,z)$ similarly to Eq.~\eref{e:R-dexpansion} as
\[
\nonumber
R_{-,1,1}(iq_o\eta) = R_{-,1,1}^{(0)} + R_{-,1,1}^{(1)} (\eta-1) + O(\eta-1)^2\,,\qquad \eta\to1\,,
\]
where 
$R_{-,1,1}^{(0)} = 2w_-\in\Real$
and
\begin{multline}
R_{-,1,1}^{(1)} = 2ih_-g(iq_o) 
    + \frac{q_oh_-}{2\pi}\bigg[\frac{1}{iq_o - \lambda_o -i\epsilon}\frac{1}{\sqrt{q_o^2 + (\lambda_o + i\epsilon)^2}}\ln\frac{q_o - \sqrt{q_o^2 + (\lambda_o + i\epsilon)^2}}{q_o + \sqrt{q_o^2 + (\lambda_o + i\epsilon)^2}}\\
- \frac{1}{iq_o -\lambda_o + i\epsilon}\frac{1}{\sqrt{q_o^2 + (\lambda_o - i\epsilon)^2}}\ln\frac{q_o - \sqrt{q_o^2 + (\lambda_o - i\epsilon)^2}}{q_o + \sqrt{q_o^2 + (\lambda_o - i\epsilon)^2}}\bigg]\,.
\label{e:R-11O(1)rat}
\end{multline}
Then, from Eq.~\eref{e:xiverphi_unshiftedLorentzian}, we can expand $\xi$ and $\varphi$ as
\begin{gather*}
\xi(z) = \xi^{(0)}(z) + (\eta - 1)\xi^{(1)}(z) + O(\eta - 1)^2\,,\qquad
\varphi(z) = \varphi^{(0)}(z) + (\eta - 1)\varphi^{(1)}(z) + O(\eta - 1)^2\,,\qquad
\end{gather*}
as $\eta\to1$, with
\begin{gather}
\xi^{(0)} = \Im(R_{-,1,1}^{(0)})z + \xi_o^{(0)} = \xi_o^{(0)}\,,\qquad
\varphi^{(0)} = -\Re(R_{-,1,1}^{(0)})z + \varphi_o^{(0)} = -2w_- z + \varphi_o^{(0)}\,,
\label{e:xiphiO(1)shifted1}
\\
\xi^{(1)} = \Im(R_{-,1,1}^{(1)})z + \xi_o^{(1)}\,,\qquad
\varphi^{(1)} = -\Re(R_{-1,1,}^{(1)})z + \varphi_o^{(1)}\,,
\label{e:xiphiO(1)shifted2}
\end{gather}
and where $\xi_o^{(j)}$ and $\varphi_o^{(j)}$ for $j = 0,1$ are arbitrary real constants.
Following a similar procedure from Section~\ref{s:rational},
we can write down the rational solution as 
\[
\label{e:rational_shiftedlorentzian}
q(t,z) = q_o\e^{2iw_-z}\frac{\big[ 2q_o t+\xi^{(1)}(z)\big]^2 + [\varphi^{(1)}(z)]^2 + 4i \varphi^{(1)}(z) - 3}
{\big[ 2q_o t+\xi^{(1)}(z)\big]^2 + [\varphi^{(1)}(z)]^2 + 1}\,,
\]
where $w_-$ is given by Eq~\eref{e:w-1} 
and $\xi^{(1)}(z)$ and $\varphi^{(1)}(z)$ are given by Eqs.~\eref{e:xiphiO(1)shifted1} and~\eref{e:xiphiO(1)shifted2},
with $R_{-,1,1}^{(1)}$ given by Eq.~\eref{e:R-11O(1)rat}.
As before, a few examples of the resulting solutions are shown in Fig.~\ref{f:rational}
for a few different choices of $\lambda_o$.
Importantly, it should be clear from Eq.~\eref{e:rational_shiftedlorentzian} and Fig.~\ref{f:rational} that,
when $\lambda_o\ne0$,
Type-4 soliton solutions are not periodic traveling waves, but rather a localized structure in both $t$ and $z$,
and exhibits rogue-wave characteristics.
This is also evident from Fig.~\ref{f:periodic}.
In other words, rational solutions with $\lambda_o\ne0$ are similar to the Peregrine solutions of the focusing NLS equation.
Indeed, the very expression of the rational solution~\eref{e:rational_shiftedlorentzian} is essentially identical to that of the Peregrine solution
of the focusing NLS equation.

\section{Final remarks}
\label{s:conclusion}

In conclusion, we formulated the inverse scattering transform for the Maxwell-Bloch system of equations 
describing resonant interaction between light and active optical media in the case when the light intensity does not vanish in the distant past and future.  
We characterized the asymptotics of the density matrix.
We then used the formalism to compute explicitly the soliton solutions of the system. 
We obtained a representation for the soliton solutions in determinant form, and we 
wrote down explicitly the one-soliton solutions.
Afterwards, we also derived periodic solutions and rational solutions.
We then analyzed the properties of these solutions, 
we discussed the sharp-line and small-amplitude limits, 
and we show that the two limits do not commute.  
Finally, we investigated the behavior of radiation, 
showing that the background and solitons are stable
(i.e., radiative part of the solution is decaying)
when the initial population inversion is negative,
which is similar to the case with ZBG.

An important future step will be to generalize the IST theory of this work in order to study coupled Maxwell-Bloch systems 
(and in particular system describing the so-called ``lambda'' configuration) with NZBG, 
using the recent results of \cite{NLTY28p3101}.
The IST for such systems in the case with ZBG was done in \cite{gabitov85,bygako1}, 
and was further studied in \cite{Chakravarty201458}.
The study of the coupled MBEs with NZBG may be instrumental to develop a mathematical theory of the phenomenon of slow light
\cite{Hau99}.
In this respect, we note that a few solutions, obtained using direct methods were reported in \cite{RVB20051,RVB20052},
but a comprehensive description of the phenomenon is still missing.

\section*{Acknowledgments}

We thank Barbara Prinari for many interesting discussions on topics related to this work.
This work was partially supported by the National Science Foundation under grant numbers DMS-1615524 and DMS-1615859.

\section*{Appendix}
\addcontentsline{toc}{section}{Appendix}

\setcounter{section}1
\setcounter{subsection}0
\def\thesection{\Alph{section}}

\subsection{Notation}
\label{s:notations}

Let $\delta(\lambda)$ be a suitable representation of the Dirac delta.
If $f(\lambda)$ is continuous at $\lambda=0$,
\[
\int f(\lambda)\delta(\lambda)\,\d\lambda = f(0)\,.
\label{e:DiracdeltaI}
\]
On the other hand, if $f(\lambda)$ has a jump discontinuity at $\lambda=0$,
\[
\int f(\lambda) \delta(\lambda)\,\d\lambda = \half(f(0^-)+f(0^+))\,.
\label{e:DiracdeltaII}
\]
For any 2$\times$2 matrix $M$, one has
\[
\sigma_* M \sigma_*  = - (\det M)\,(M^{-1})^T\,,
\label{e:sigma1invT}
\]
where $\sigma_*$ is defined in Eq.~\eref{e:sigma*}.
\\
For reference, note that 
\[
\sigma_3Q_\pm = \begin{pmatrix}0 &q_\pm \\ q_\pm^* &0\end{pmatrix}\,,\qquad
(\sigma_3Q_\pm)^{-1} = \begin{pmatrix}0 &1/q_\pm^* \\ 1/q_\pm &0\end{pmatrix} = \frac1{q_o^2}\,\sigma_3Q_\pm\,.
\label{e:sQJs}
\]

\subsection{Symmetries}
\label{A:symmetries}

We consider the two symmetries separately.

1. $\zeta\mapsto \zeta^*$~ (upper/lower half plane), ~implying~ $(\lambda,\gamma)\mapsto(\lambda^*,\gamma^*)$~ (same sheet).
It can be shown that, if $\phi(t,z,\zeta)$ is a solution of the scattering problem~\eref{e:scatteringproblem}, 
so is 
\[
\nonumber
\tilde\phi(t,z,\zeta)= \sigma_*\,\phi^*(t,z,\zeta^*)\,.
\]
where $\sigma_*$ is defined in Eq.~\eref{e:sigma*}.
[That is because~ $\sigma_*^2= -I$,~ 
$\sigma_*\sigma_3\sigma_*= -\sigma_3$~ and~ $\sigma_*Q^*\sigma_*= Q^\dag= -Q$.]
By the same token, so is $\tilde\phi\,C$, where $C$ is any constant $2\times2$ matrix.
Now restrict our attention to $\zeta\in\Sigma$.
Let $\phi\equiv\phi_\pm$, and look at the asymptotic behavior of $\tilde\phi_\pm$ as $t\to\pm\infty$:
Recall that $\phi(t,z,\zeta)\sim Y_\pm(\zeta)\,\e^{i\gamma(\zeta)t\sigma_3}$ as $t\to\pm\infty$.
Note that $\gamma^*(\zeta^*)= \gamma(\zeta)$ and
$\sigma_*Y_\pm^*(\zeta^*) = \sigma_*(I - i\sigma_3Q_\pm^*/\zeta)= Y(\zeta)\sigma_*$.
Also, a little algebra shows that $\sigma_*\e^{ia\sigma_3}\sigma_*= -\e^{-ia\sigma_3}$. 
Therefore, $\tilde\phi_\pm(t,z,\zeta)\,\sigma_*\sim Y_\pm(\zeta)\,\e^{i\gamma(\zeta)t\sigma_3}$ as $t\to\pm\infty$.
But since the solution of the scattering problem with given BCs is unique, 
we must have $\tilde\phi_\pm\,\sigma_*= \phi_\pm$.
Hence $\forall \zeta\in\Sigma$ we have
\[
\phi_\pm(t,z,\zeta) = -\sigma_*\phi_\pm^*(t,z,\zeta^*)\,\sigma_*\,.
\label{e:phisymm1}
\]
Recalling Eq.~\eref{e:sigma1invT}, 
the desired results for the eigenfunctions are obtained by rewriting the symmetry \eref{e:phisymm1}.
Recalling the scattering relation~\eref{e:scattering} and using symmetry~\eref{e:phisymm1}, 
we have, $\forall \zeta\in\Sigma$,
\[
S^*(\zeta^*,z) = -\sigma_* S(\zeta,z)\sigma_*\,.
\label{e:S1symm}
\]
Using Eq. \eref{e:sigma1invT}, one can compute the corresponding symmetries for the scattering matrix.

2. $\zeta\mapsto -q_o^2/\zeta$~ (outside/inside $C_o$), ~implying~ $(\lambda,\gamma)\mapsto(\lambda,-\gamma)$~ (opposite sheets).
Since $\lambda(-q_o^2/\zeta)= \lambda(\zeta)$, it is easy to show that if $\phi(t,z,\zeta)$ is a solution of the scattering problem,
so is 
\[
\nonumber
\tilde\phi(t,z,\zeta)= \phi(t,z,-q_o^2/\zeta)\,.
\]
As before, this implies that $w\,C$ is also a solution, for any matrix $C$ independent of $t$.
With $\phi=\phi_\pm$, it is $\tilde\phi_\pm(t,z,\zeta)C\sim Y_\pm(-q_o^2/\zeta)\,\e^{-i\gamma(\zeta)\,t\sigma_3}C$
as $t\to\pm\infty$, because $\gamma(-q_o^2/\zeta)= -\gamma(\zeta)$.
Now note that~
$Y_\pm(-q_o^2/\zeta)\,\e^{-i\gamma t\sigma_3}\,\sigma_3Q_\pm = - i\zeta Y_\pm(\zeta)\,\e^{i\gamma t\sigma_3}$
[because $\e^{-ia\sigma_3}Q=Q\e^{ia\sigma_3}$ and $(\sigma_3Q)^2= q_o^2I$].
Thus, taking~ $C= (i/\zeta)\,\sigma_3Q$, 
we obtain the desired results for $\phi_\pm$ in this case.
For the scattering matrix, 
from the relation~\eref{e:scattering} and by using symmetry~\eref{e:phisymm2} and Eq.~\eref{e:sQJs}, 
we finally obtain symmetry~\eref{e:S2symm}.

1+2.  We can also combine the results of the
first two symmetries to get
\[
\nonumber
\phi_\pm^*(t,z,\zeta^*) = (1/i\zeta)\,\sigma_*\,\phi_\pm(t,z,-q_o^2/\zeta)\,\sigma_3Q_\pm\sigma_*\,.
\]
The corresponding symmetry for the columns follows as before.

\subsection{Asymptotics of the eigenfunctions as $\zeta\to\infty$ and $\zeta\to0$}
\label{A:asymptotics}

Here we will prove the asymptotic expansions~\eref{e:muasymp_zeta2infty1},~\eref{e:muasymp_zeta2infty2},~\eref{e:muasymp_zeta2zero1} and~\eref{e:muasymp_zeta2zero2}.

We start deriving the asymptotic behavior~\eref{e:muasymp_zeta2infty1} and~\eref{e:muasymp_zeta2infty2}.
The derivation proceeds by induction.
The result is trivially true for $\mu_d\O0$ and $\mu_o\O0$.
Moreover, using definition~\eref{e:Ydef} and separating the diagonal and off-diagonal parts of the expansion~\eref{e:muasymp}, 
we have
\begin{multline}
\mu_d\O{n+1} = \frac{\zeta}{2\gamma}\int_{-\infty}^t\! \big(\,
    \Delta Q_-(y,t)\mu_o\O{n}(y,t,\zeta) \,-\, \frac{i}{\zeta}\,\sigma_3Q_-\Delta Q_-(y,t)\,\mu_d\O{n}(y,t,\zeta) \,\big)\,\d y 
\\
  + \frac{i\sigma_3 Q_-}{2\gamma}\int_{-\infty}^t\! \e^{i\gamma(\zeta)(t-y)\sigma_3}
    \bigg[\Delta Q_-(y,t)\mu_d\O{n}(y,t,\zeta) \, -\, \frac{i}{\zeta}\,\sigma_3Q_-\Delta Q_-(y,t)\,\mu_o\O{n}(y,t,\zeta)\bigg]\e^{i\gamma(\zeta)(y-t)\sigma_3}\,\d y\,, 
\label{e:muasympd}
\end{multline}
\par\kern-2.4\bigskipamount
\begin{multline}
\mu_o\O{n+1} = \frac{i\sigma_3 Q_-}{2\gamma}\int_{-\infty}^t\! \big(\,
    \Delta Q_-(y,t)\mu_o\O{n}(y,t,\zeta) \,-\, \frac{i}{\zeta}\,\sigma_3Q_-\Delta Q_-(y,t)\,\mu_d\O{n}(y,t,\zeta) \,\big)\,\d y 
\\
  + \frac{\zeta}{2\gamma}\int_{-\infty}^t\! \e^{i\gamma(\zeta)(t-y)\sigma_3} \bigg[
    \xi \Delta Q_-(y,t)\mu_d\O{n}(y,t,\zeta) \,-\, \frac{i}{\zeta}\,\sigma_3Q_-\Delta Q_-(y,t)\,\mu_o\O{n}(y,t,\zeta)\bigg] \e^{i\gamma(\zeta)(y-t)\sigma_3}\,\d y\,.
\label{e:muasympo}
\end{multline}
As $\zeta\to\infty$, the four terms on the RHS of Eq.~\eref{e:muasympd} are, respectively:
$O(\mu_o\O{n})$, $O(\mu_d\O{n}/\zeta)$, $O(\mu_d\O{n}/\zeta^2)$ and $O(\mu_o\O{n}/\zeta^3)$.
The last two estimates are obtained using integration by parts, 
taking advantage of the fact that $\gamma(\zeta)= \zeta/2+O(1/\zeta)$ as $\zeta\to\infty$ from Eq.~\eref{e:unifinverse}.
Similarly, 
as $\zeta\to\infty$, the four terms on the RHS of Eq.~\eref{e:muasympo} are, respectively:
$O(\mu_o\O{n}/\zeta)$, $O(\mu_d\O{n}/\zeta^2)$, $O(\mu_d\O{n}/\zeta)$ and $O(\mu_o\O{n}/\zeta^2)$
(where again the last two estimates are obtained using integration by parts).
$\square$

Next we describe the asymptotic behavior~\eref{e:muasymp_zeta2zero1} and~\eref{e:muasymp_zeta2zero2}.
Again, the result is derived by induction.
The result is trivially true for $\mu_d\O0$ and $\mu_o\O0$.
Decomposing expansion~\eref{e:muasymp} into its diagonal and off-diagonal parts yields
Eqs. \eref{e:muasympd} and~\eref{e:muasympo} as before.
Finally, one shows that the four terms on the RHS of Eqs.~\eref{e:muasympd} and~\eref{e:muasympo} are, respectively:
for Eq. \eref{e:muasympd}: $O(\zeta^2\mu_o\O{n})$, $O(\zeta\,\mu_d\O{n})$, 
$O(\zeta^2\mu_d\O{n})$ and $O(\zeta\,\mu_o\O{n})$;
for Eq. \eref{e:muasympo}: $O(\zeta\mu_o\O{n})$, $O(\mu_d\O{n})$, 
$O(\zeta^3\mu_d\O{n})$ and $O(\zeta\,\mu_o\O{n})$.
Again, some estimates are obtained using integration by parts and $\gamma(\zeta) = O(1/\zeta)$ as $\zeta\to0$ from Eq.~\eref{e:unifinverse}.
$\square$

\subsection{Calculation of $R_\pm$}
\label{A:Rpm}

We first establish the following result, which will be needed in the calculation:
\begin{lemma}
\label{l:integral}
\[
\label{e:Ipmdef}
\lim_{t\to\pm\infty}\pvint_{-\infty}^\infty 
  \e^{\pm i(\gamma'-\gamma)t}f(\zeta,\zeta')\frac{\d\lambda'}{\lambda'-\lambda}  = 
\begin{cases}
\pm i\,\nu\pi f(\zeta,\zeta)\,, & \lambda\in\Real\,,\\
0\,, & \lambda\in i[-q_o,q_o]\,,
\end{cases}
\]
where $\nu = \pm1$ when $\lambda$ on sheet I or II, respectively,
and where we used the shorthand notation $\gamma=\gamma(\lambda)$ and $\gamma'= \gamma(\lambda')$.
\end{lemma}
Proof:
Notice that for both $\lambda'$ and $\lambda$, we have two choices: sheet I and sheet II.
Here, we compute the integral in the way that $\lambda'$ and $\lambda$ are always on the same sheet.
In this case, we can eliminate the mixed case: $\lambda'$ is on sheet I and $\lambda$ is on sheet II or vise versa.
We choose such a calculation because, 
later we will show that this version of IST naturally reduces to the ZBCs as $q_o\to0$.

Denote the LHS of the definition~\eref{e:Ipmdef} by $I_\pm(\zeta)$.
Note that, since $\gamma(\lambda)$ takes on different values on each sheet of the Riemann surface, so do $I_\pm$.
Performing the change of variable $\lambda'\mapsto\gamma'$ we have
\[
I_\pm(\zeta) = \nu\lim_{t\to\pm\infty}\pvint_L
  \e^{\pm i(\gamma'-\gamma)t}f(\zeta,\zeta')j(\zeta,\zeta')\frac{\d\gamma'}{\gamma'-\gamma}\,,
\label{e:Ipmgamma}
\]
where again $\nu = \pm1$ when both $\lambda$ and $\lambda'$ are on sheet I or sheet II, respectively,
$L=(-\infty,-q_o)\cup(q_o,\infty)$ and
\[
\nonumber
j(\zeta,\zeta')= \frac{\gamma'-\gamma}{\lambda'-\lambda}\,\deriv{\lambda'}{\gamma'}\,.
\]
Further letting $y=\gamma'-\gamma$ in Eq.~\eref{e:Ipmgamma}, we have
\[
I_\pm(\zeta) = \nu\lim_{t\to\pm\infty}\pvint_{L'}
  \e^{\pm iy't}f(\zeta,\zeta')j(\zeta,\zeta')\frac{\d y}y\,,
\label{e:Ipmy}
\]
where $L'= (-\infty,-q_o-\gamma)\cup(q_o-\gamma,\infty)$.
From here, we need to discuss two cases depending on whether $0\in L'$ or $0\not\in L'$.
\begin{itemize}[leftmargin=*]
\item 
If $\lambda\in\Real$, i.e., $\gamma \in(-\infty,-q_o]\cup[q_o,\infty)$, then we know $0\in L'$ and we obtain
\[
\nonumber
I_\pm(\zeta) = \pm i\nu\pi f(\zeta,\zeta)j(\zeta,\zeta)\,. 
\]
from the formula $\lim_{t\to\pm\infty}\pvint\nolimits_{-\infty}^\infty \e^{\pm iy t}F(y)\,\d y/y = \pm i \pi\,F(0)$.
Equation~\eref{e:Ipmdef} then follows by noting that 
$\lim_{\gamma'\to\gamma}(\lambda'-\lambda)/(\gamma'-\gamma) = d\lambda/d\gamma$,
and therefore $\lim_{\gamma'\to\gamma}j(\zeta,\zeta')=1$.
\item
If $\lambda\in i[-q_o,0)\cup i(0,q_o]$, i.e, $\zeta\in C_o$ and $-q_o < \gamma < q_o$, 
then $0\not\in L'$.
By using the Riemann-Lebesgue lemma, we know $I_\pm(\zeta) = 0$.
$\square$
\end{itemize}
Moreover, note that if the term $(\gamma'-\gamma)t$ in the exponent of Eq.~\eref{e:Ipmdef} 
were replaced by $\gamma'+\gamma$ or $\gamma't$,
one can show that the integral is zero using the Riemann-Lebesgue lemma.

We now proceed to compute $R_\pm$.
We rewrite Eq.~\eref{e:Rpmlim} for convenience as 
\[
\nonumber
R_\pm = -2i\lim_{t\to\infty}[\phi_\pm^{-1}V\phi_\pm - \phi_\pm^{-1}(\phi_\pm)_z]\,.
\]
We will first evaluate the two terms on the RHS separately.
Note first that, from Eq.~\eref{e:phipmevol} we get, as $t\to\pm\infty$,
\[
\nonumber
(\phi_\pm)_z = iw_\pm[\sigma_3,\phi_\pm] + o(1)\,,
\]
where $w_\pm$ was defined in Eq.~\eref{e:wpmdef},
implying $\phi_\pm^{-1}(\phi_\pm)_z = (\phi_\pm^{-1}(\phi_\pm)_z )_\eff + o(1)$, 
where $(\phi_\pm^{-1}(\phi_\pm)_z )_\eff$ denotes the leading order term in the limit given by
\[
\nonumber
(\phi_\pm^{-1}(\phi_\pm)_z )_\eff = -iw_\pm \sigma_3 + iw_\pm\,\e^{-i\gamma \sigma_3t}Y_\pm^{-1}\sigma_3 Y_\pm\,\e^{i\gamma \sigma_3t}\,.
\]
Moreover, recalling definitions~\eref{e:Ydef} and~\eref{e:YdetYinv}, we have
\[
(\phi_\pm^{-1}(\phi_\pm)_z )_\eff = -iw_\pm \sigma_3 + \frac{i\zeta w_\pm}{2\gamma}\bigg( \sigma_3
    + \frac i\zeta\,\e^{-i\gamma \sigma_3t}[\sigma_3,\sigma_3Q_\pm]\,\e^{i\gamma \sigma_3t} 
    + \frac1{\zeta^2}\,\sigma_3Q_\pm^2 \bigg)\,.
\nonumber
\]
Combining the first two terms on the RHS and recalling that $Q_\pm^2= -q_o^2I$ we then have
\[
(\phi_\pm^{-1}(\phi_\pm)_z)_\eff = 
  -i\frac{q_o^2w_\pm}{\gamma \zeta}\,\sigma_3  - \frac{w_\pm}{2\gamma}\,\e^{-i\gamma \sigma_3t}[\sigma_3,\sigma_3Q_\pm]\,\e^{i\gamma \sigma_3t}\,,
\label{e:phiinvdphidzeff}
\]
where $\sigma_1$ was defined in Eq.~\eref{e:Paulidef}.
Note that $(\phi_\pm^{-1}(\phi_\pm)_z)_\eff\to0$ as $q_o\to0$, as it should be,
since then the BCs for $\phi_\pm$ are independent of $z$.
On the other hand, the second term on the RHS of Eq.~\eref{e:phiinvdphidzeff} does not have a finite limit as $t\to\pm\infty$.
One would therefore expect that this term will be canceled by the first term in Eq.~\eref{e:Rpmlim}.
We will see that this is indeed the case.
The calculation of the first term of $R_\pm$ is considerably more involved, however.

To calculate the first term in $R_\pm$, we first note that, as $t\to\pm\infty$,
\[
\label{e:phiinvVphieff}
\phi_\pm^{-1}V\phi_\pm \sim \e^{-i\gamma \sigma_3t}Y_\pm^{-1}V Y_\pm\,\e^{i\gamma \sigma_3t}
  = (\phi_\pm^{-1}V\phi_\pm)_{\eff,d} + (\phi_\pm^{-1}V\phi_\pm)_{\eff,o}
\]
where $d$ and $o$ denote the diagonal and off-diagonal parts of a matrix, as before, and
\begin{align}
(\phi_\pm^{-1}V\phi_\pm)_{\eff,d} = & \frac \zeta{2\gamma} \,\bigg(
    V_d - \frac i\zeta [\sigma_3Q_\pm,V_o] + \frac1{\zeta^2} \sigma_3Q_\pm V_d \sigma_3Q_\pm \bigg)\,,
\\
(\phi_\pm^{-1}V\phi_\pm)_{\eff,o} = & \frac \zeta{2\gamma}\,
    \e^{-i\gamma \sigma_3t}\bigg( V_o -\frac i\zeta [\sigma_3Q_\pm,V_d] + \frac1{\zeta^2} \sigma_3Q_\pm V_o \sigma_3Q_\pm
     \bigg)\,\e^{i\gamma \sigma_3t}\,.
\label{e:phiinvphieffodef}
\end{align}
Now note that, as $t\to\pm\infty$,
\[
V\sim \frac i2\pvint_{-\infty}^\infty Y_\pm'\,e^{i\gamma'\sigma_3t}\rho_\pm'\e^{-i\gamma'\sigma_3t}(Y_\pm')^{-1}\frac{g(\lambda')\d\lambda'}{\lambda'-\lambda}\,,
\nonumber
\]
where as before we used primes to indicate functional dependence on $\lambda'$.
We therefore have
\begin{gather}
\label{e:Veff1}
V_d \sim \frac i2\pvint \frac{\zeta'}{2\gamma'}\bigg(
     \rho_{\pm,d}' + \frac i{\zeta'}[\sigma_3Q_\pm,\e^{i\gamma'\sigma_3t}\rho_{\pm,o}'\e^{-i\gamma'\sigma_3t}] + \frac1{\zeta'^2}\sigma_3Q_\pm\rho_{\pm,d}'\sigma_3Q_\pm
  \bigg)\,\frac{g(\lambda')\d\lambda'}{\lambda'-\lambda}\,,
\\
\label{e:Veff2}
V_o \sim \frac i2\pvint \frac{\zeta'}{2\gamma'}\bigg(
  \e^{i\gamma'\sigma_3t}\rho_{\pm,o}'\e^{-i\gamma'\sigma_3t} + \frac i{\zeta'}[\sigma_3Q_\pm,\rho_{\pm,d}'] + \frac1{\zeta'^2}\sigma_3Q_\pm\e^{i\gamma'\sigma_3t}\rho_{\pm,o}'\e^{-i\gamma'\sigma_3t}\sigma_3Q_\pm
  \bigg)\,\frac{g(\lambda')\d\lambda'}{\lambda'-\lambda}\,.
\end{gather}
Substituting these expressions in Eq.~\eref{e:phiinvVphieff}, and using the results we obtained earlier, 
it is straightforward to see that all the
terms in Eqs.~\eref{e:Veff1} and~\eref{e:Veff2} containing oscillating exponentials 
drop out of $(\phi_\pm^{-1}V\phi_\pm)_{\eff,d}$ in the limit $t\to\pm\infty$, 
leaving
\begin{multline*}
\lim_{t\to\pm\infty}(\phi_\pm^{-1}V\phi_\pm)_{\eff,d} = \frac i2 \pvint \frac{\zeta\zeta'}{4\gamma\gamma'}\bigg\{ 
  \rho_{\pm,d}' + \frac1{\zeta'^2}\,\sigma_3Q_\pm\rho_{\pm,d}'\sigma_3Q_\pm + 
\\
  \frac1{\zeta^2}\sigma_3Q_\pm\bigg( \rho_{\pm,d}' + \frac1{\zeta'^2}\,\sigma_3Q_\pm\rho_{\pm,d}'\sigma_3Q_\pm \bigg)\sigma_3Q_\pm
  + \frac1{\zeta\zeta'}[\sigma_3Q_\pm,[\sigma_3Q_\pm,\rho_{\pm,d}']]
  \bigg\}\frac{g(\lambda')\d\lambda'}{\lambda'-\lambda}\,.
\end{multline*}
We can simplify this expression, 
noting that $[\sigma_3Q,[\sigma_3Q,\rho_d]]= 2(q_o^2\rho_d - 2\sigma_3Q\rho_d\sigma_3Q)$.
After some algebra, we then obtain simply
\[
\lim_{t\to\pm\infty}(\phi_\pm^{-1}V\phi_\pm)_{\eff,d} = 
 i\frac\pi2\,\H_{\lambda}\bigg[ \frac{g(\lambda')}{4\gamma\gamma'\zeta\zeta'}\big\{
    (\zeta\zeta'+q_o^2)^2\rho_{\pm,d}' + (\zeta-\zeta')^2q_o^2\sigma_1\,\rho_{\pm,d}'\sigma_1 \big\} \bigg]\,,
\label{e:phiinvVphieffd}
\]
Next, since $\rho$ and $\rho_\pm$ are traceless, it is
$\sigma_1\rho_{\pm,d}\sigma_1 = - \rho_{\pm,d}$.
We can therefore combine the two terms in the curly brackets to obtain a simplified version (using the identity $\zeta^2-q_o^2 = 2\lambda\zeta$).
So, we obtain
\[
\label{e:Rpmdtmp}
R_{\pm,d}(\zeta,z) = 
 \pi\,\H_{\lambda}\bigg[ \frac{\lambda\lambda'+q_o^2}{\gamma\gamma'}\,\rho_{\pm,d}'\,g(\lambda')\bigg] + \frac{2q_o^2}{\gamma \zeta}w_\pm\,\sigma_3\,.
\]
Recall that the definition of $w_\pm$ is given in Eq.~\eref{e:wpmdef}.
The last step is to simplify Eq.~\eref{e:Rpmdtmp}.
The key idea is to rewrite $w_\pm$ as a Hilbert transform,
namely
\[
\nonumber
w_\pm = \frac{\pi}{2}\H_{\lambda}\bigg[D_\pm'g(\lambda')\frac{\lambda'-\lambda}{\gamma'}\bigg]\,.
\]
By using the identity $D_\pm\sigma_3 = \rho_{\pm,d}$, 
we can combine the two Hilbert transforms in Eq.~\eref{e:Rpmdtmp} and notice that $1/\zeta = (\gamma-\lambda)/q_o^2$.
Finally, we obtain the desired result~\eref{e:R_d_axis}.
Note that, as $q_o\to0$, Eq. \eref{e:phiinvVphieffd} reduces to the result with ZBC.
Also, one should notice that this formula holds for all $\zeta\in\Complex$.

Our next task is to compute Eq.~\eref{e:phiinvphieffodef}.
We first look at the terms coming from the $t$-independent parts of $V_d$ and $V_o$, of which there are four.
They yield:
\begin{align*}
(\phi_\pm^{-1}V\phi_\pm)_{\eff,o.1} = &
\frac i2\,\e^{-i\gamma \sigma_3t} \pvint \frac{\zeta\zeta'}{4\gamma\gamma'}\bigg\{ 
  \frac i{\zeta'}\bigg( [\sigma_3Q_\pm,\rho_{\pm,d}'] + \frac1{\zeta^2}\,\sigma_3Q_\pm[\sigma_3Q_\pm,\rho_{\pm,d}']\sigma_3Q_\pm \bigg)
\\
  & -\frac i\zeta\bigg( [\sigma_3Q_\pm,\rho_{\pm,d}'] + \frac1{\zeta'^2}\,[\sigma_3Q_\pm,\sigma_3Q_\pm\rho_{\pm,d}'\sigma_3Q_\pm] \bigg)
  \bigg\} \, \frac{g(\lambda')\d\lambda'}{\lambda'-\lambda} \,\e^{i\gamma \sigma_3t}\,.
\end{align*}
Now note that 
$\sigma_3Q[\sigma_3Q,\rho_d]\sigma_3Q= [\sigma_3Q,\sigma_3Q\rho_d \sigma_3Q]= -q_o^2[\sigma_3Q,\rho_d]$.
The terms in the curly brackets are then simply
$(2i/\zeta\zeta')(\lambda'-\lambda)[\rho_{\pm,d},\sigma_3Q_\pm]$.
We thus see that $(\phi_\pm^{-1}V\phi_\pm)_{\eff,o.1}$ 
cancels exactly the second term on the RHS of Eq.~\eref{e:phiinvdphidzeff}, as expected.

Our last task is to compute the terms in Eq.~\eref{e:phiinvphieffodef} coming from the
$t$-dependent parts of $V_d$ and $V_o$.
There are five terms, which yield:
\begin{align*}
(\phi_\pm^{-1}V\phi_\pm)_{\eff,o.2} = & \frac i2 \pvint \frac{\zeta\zeta'}{4\gamma\gamma'} \,\e^{-i\gamma \sigma_3t} \bigg\{ 
  \bigg( 1 + \frac{q_o^4}{\zeta^2\zeta'^2} \bigg) \rho_{\pm,o\sigma_3}'
\\
  & + \bigg( \frac1{\zeta^2} + \frac1{\zeta'^2} \bigg) \sigma_3Q_\pm \rho_{\pm,o\sigma_3}' \sigma_3Q_\pm
  + \frac1{\zeta\zeta'}[\sigma_3Q_\pm,[\sigma_3Q_\pm,\rho_{\pm,o\sigma_3}']] \bigg\} \,\e^{i\gamma \sigma_3t}\, \frac{g(\lambda')\d\lambda'}{\lambda'-\lambda} \,.
\end{align*}
where we used that $(\sigma_3Q_\pm)^2= q_o^2$, and where 
for brevity we denoted $\rho_{\pm,o\sigma_3}' = \e^{i\gamma'\sigma_3t}\rho_{\pm,o}'\e^{-i\gamma'\sigma_3t}$.
Now note that  $[\sigma_3Q_\pm,[\sigma_3Q,\rho_{\pm,o\sigma_3}']]= 2(q_o^2\rho_{\pm,o\sigma_3}' - \sigma_3Q_\pm \rho_{\pm,o\sigma_3}' \sigma_3Q_\pm)$, and recall
$\sigma_3Q\,\e^{i\theta \sigma_3} = \e^{-i\theta \sigma_3}\sigma_3Q$. 
Thus we can rewrite $(\phi_\pm^{-1}V\phi_\pm)_{\eff,o.2}$ as
\begin{align*}
(\phi_\pm^{-1}V\phi_\pm)_{\eff,o.2} = & \frac i2 \pvint \frac{\zeta\zeta'}{4\gamma\gamma'} \bigg\{ 
  \bigg( 1 + \frac{2q_o^2}{\zeta\zeta'} + \frac{q_o^4}{\zeta^2\zeta'^2} \bigg) \e^{i(\gamma'-\gamma)\sigma_3t} \rho_{\pm,o}' \e^{-i(\gamma'-\gamma)\sigma_3t}
\\
  & + \bigg( \frac1{\zeta^2} - \frac{2}{\zeta\zeta'} + \frac1{\zeta'^2}\bigg) \sigma_3Q_\pm \e^{i(\gamma'+\gamma)\sigma_3t} \rho_{\pm,o}' \e^{-i(\gamma'+\gamma)\sigma_3t} \sigma_3Q_\pm
  \bigg\} \,\, \frac{g(\lambda')\d\lambda'}{\lambda'-\lambda} \,.
\end{align*}
Recalling Lemma~\ref{l:integral}, we see that this integral yields
\[
\label{e:phiinvVphieffo}
\lim_{t\to\pm\infty}(\phi_\pm^{-1}V\phi_\pm)_{\eff,o.2} = 
\begin{cases}
  \displaystyle\pm\frac{\pi}{2}\, g(\lambda)\,\rho_{\pm,o}(\lambda,z) \sigma_3\,, & \zeta\in(-\infty,-q_o]\cup[q_o,\infty)\,,\\[0.8ex]
  \displaystyle\mp\frac{\pi}{2}\, g(\lambda)\,\rho_{\pm,o}(\lambda,z) \sigma_3\,, & \zeta\in(-q_o,q_o)\,,\\
 \displaystyle0\,, & \zeta\in C_o\,.
\end{cases}
\]
Finally, inserting Eq.~\eref{e:phiinvVphieffo} in Eq.~\eref{e:Rpmlim} we obtain Eq.~\eref{e:R_offd_axis}.
Note that the expression for $R_{\pm,o}$ is exactly the same as that for ZBC.

\subsection{Propagation of reflection coefficients and norming constants}
\label{A:evolution}

Here we derive the propagation equations for the reflection coefficients and norming constants.

We start to derive the ODE~\eref{e:dBdz2}.
It is obviously
\[
\partialderiv Bz = (S_o)_z S_o^{-1}\,B - B\,(S_d)_z\,S_d^{-1}\,.
\nonumber 
\]
Decomposing Eq.~\eref{e:dSdz} into its diagonal and off-diagonal parts, we then have
\[
-2i \partialderiv Bz = R_{-,o} + R_{-,d}B - B\,R_{-,d}
  -B\,R_{-,o}B - S_dR_{+,o}S_d^{-1} + BS_o\,R_{+,o}S_d^{-1}\,.
\label{e:dBdz}
\]
We need to:
(i) express the RHS in terms of the limiting values as $t\to-\infty$, and
(ii) linearize the resulting equations.
To accomplish these tasks we need to look at the last three terms on the RHS.

Recall that $\rho^+$ is expressed in terms of $\rho^-$ via relation~\eref{e:rhopmrel}.
Decomposing Eq.~\eref{e:rhopmrel} into its diagonal and off-diagonal parts 
(and recalling that $S^{-1} = S^\dag$) we have
\[
\nonumber
\rho_{+,o} = S_d^\dag\rho_{-,o}S_d + S_o^\dag\rho_{-,o}S_o + S_d^\dag\rho_{-,d}S_o + S_o^\dag\rho_{-,d}S_d\,.
\]
Finally, since $S^\dag S = S S^\dag = I$, it is
\[
S_d^\dag S_d + S_o^\dag S_o = S_d S_d^\dag + S_o S_o^\dag = I\,,\qquad
S_d^\dag S_o + S_o^\dag S_d = S_d S_o^\dag + S_o S_d^\dag = O\,.
\nonumber 
\]
Substituting all of these expressions into ODE~\eref{e:dBdz}, 
and after some tedious but straightforward algebra,
and using that $R_{+,o}$ is given by Eq.~\eref{e:R_offd_axis},
we finally get the desired ODE~\eref{e:dBdz2}.

We now derive the propagation equation for the norming constants~\eref{e:dCdz}.
Recalling definition~\eref{e:dCdz01} and taking the derivative 
(and assuming the limit and the derivative commute), we have
\[
\partialderiv{C_n}z = \partialderiv{b_n}z\,\frac1{s_{1,1}'(\zeta_n,z)} 
  - C_n \lim_{\zeta\to \zeta_n} \frac1{s_{1,1}(\zeta,z)}\partialderiv{s_{1,1}(\zeta,z)}z\,.
\label{e:dCdz1}
\]
We compute each of the two derivatives on the RHS separately.

Recall that, for $z\in\Sigma$, the propagation of the scattering matrix is governed by ODE~\eref{e:dSdz}.
Thanks to Eqs.~\eref{e:RpmUHP1} and~\eref{e:RpmUHP2} some elements of ODE~\eref{e:dSdz} can be extended into the UHP.
In particular:
\[
\partialderiv{s_{1,1}}z = \txtfrac i2(R_{-,1,1} - R_{+,1,1})\,s_{1,1}\,. 
\label{e:ds11dz}
\]
Also recall from definition~\eref{e:Rpmdef} that, $\forall z\in\Sigma$,
\[
\partialderiv{\phi_\pm}z = -\txtfrac i2\phi_\pm R_\pm + V\,\phi_\pm\,. 
\]
Again, thanks to Eqs.~\eref{e:RpmUHP1} and~\eref{e:RpmUHP2} some columns of the above equation can be extended into the UHP.
Explicitly,
\[
\partialderiv{\phi_{+,1}}z = -\txtfrac i2 R_{+,1,1}\phi_{+,1} + V\,\phi_{+,1}\,,\qquad  
\partialderiv{\phi_{-,2}}z = -\txtfrac i2 R_{+,2,2}\phi_{-,2} + V\,\phi_{-,2}\,. 
\label{e:dphidzUHP}
\]
Now recall that at $\zeta = \zeta_n$ it is $\phi_{+,1}(t,z,\zeta_n) = b_n\phi_{-,2}(t,z,\zeta_n)$.
Thus
\[
\bigg[ \partialderiv{b_n}z + \txtfrac i2(R_{+,1,1} - R_{-,2,2} )|_{\zeta=\zeta_n}\,b_n \bigg] \phi_{-,2} = 0\,.
\nonumber 
\]
Evaluating the above system in the limit $t\to-\infty$ we see that 
the terms in square bracket must vanish, yielding
\[
\partialderiv{b_n}z = - \txtfrac i2(R_{+,1,1} - R_{-,2,2} )|_{\zeta=\zeta_n}\,b_n\,.
\label{e:dcdz}
\]
Substituting Eqs.~\eref{e:ds11dz} and \eref{e:dcdz} into Eq.~\eref{e:dCdz1} we then finally obtain the desired ODE~\eref{e:dCdz}.

\subsection{Solution of the Riemann-Hilbert problem}
\label{A:RHPsoln}

As usual, in order to solve the RHP we need to take into account its normalization, namely, 
the asymptotic behavior of $M^\pm$ as $\zeta\to\infty$.
Recalling the asymptotic behavior of the Jost eigenfunctions and scattering coefficients obtained in Section~\ref{s:asymp},
it is easy to check that
\[
\nonumber
M^\pm = I + O(1/\zeta)\,,\qquad \zeta\to\infty\,.
\]
Moreover, from the asymptotic behavior of the Jost eigenfunctions and scattering data as $\zeta\to0$ 
we see that
\[
\nonumber
M^\pm = (i/\zeta)\sigma_3Q_- + O(1)\,,\qquad \zeta\to0\,.
\]
Thus, as in the case of the defocusing NLS with NZBC, in addition to the behavior at $\zeta=\infty$ and the poles from the discrete spectrum
one also needs to subtract the pole at $\zeta=0$ in order to obtain a regular RHP.
On the other hand, the asymptotic behavior of the off-diagonal scattering coefficients implies that 
the jump matrix $G(t,z,\zeta)$ from Eq.~\eref{e:jump} is $O(1/\zeta)$ as $\zeta\to\pm\infty$ and $O(\zeta)$ as $\zeta\to0$ along the real axis.

To solve, we subtract out the asymptotic behavior and the pole contributions.
Recall that discrete eigenvalues come in symmetric quartets: 
$\{\zeta_n,\zeta_n^*,-q_o^2/\zeta_n,-q_o^2/\zeta_n^*\}_{n=1}^N$.
More precisely, 
$s_{2,2}(\zeta_n,z)= s_{2,2}(-q_o^2/\zeta_n^*,z) = s_{1,1}(\zeta_n^*,z) = s_{1,1}(-q_o^2/\zeta_n,z) = 0$ 
for $n=1,\dots,N$.
We therefore rewrite the jump condition~\eref{e:RHP} as
\begin{multline*}
\kern-1em
M^- - I - (i/\zeta)\sigma_3Q_- 
  - \sum_{n=1}^N\bigg(\frac{\Res_{\zeta_n^*}M^-}{\zeta-\zeta_n^*} + \frac{\Res_{-q_o^2/\zeta_n}M^-}{\zeta+q_o^2/\zeta_n}\bigg)
  - \sum_{n=1}^N\bigg(\frac{\Res_{\zeta_n}M^+}{\zeta-\zeta_n} + \frac{\Res_{-q_o^2/\zeta_n^*}M^+}{\zeta+q_o^2/\zeta_n^*}\bigg)
\\
= M^+ - I - (i/\zeta)\sigma_3Q_- 
  - \sum_{n=1}^N\bigg(\frac{\Res_{\zeta_n}M^+}{\zeta-\zeta_n} + \frac{\Res_{-q_o^2/\zeta_n^*}M^+}{\zeta+q_o^2/\zeta_n^*}\bigg)
  - \sum_{n=1}^N\bigg(\frac{\Res_{\zeta_n^*}M^-}{\zeta-\zeta_n^*} + \frac{\Res_{-q_o^2/\zeta_n}M^-}{\zeta+q_o^2/\zeta_n}\bigg)
\\ \kern16em
  - M^+G\,, 
\end{multline*}
Recall we relabeled all the discrete eigenvalues at the end of Section~\ref{s:symmetries} as $\zeta_{N+n} = - q_o^2/\zeta_n^*$.
Therefore, the above equation reduces to
\begin{multline}
M^- - I - (i/\zeta)\sigma_3Q_- 
  - \sum_{n=1}^{2N}\bigg(\frac{\Res_{\zeta_n^*}M^-}{\zeta-\zeta_n^*} + \frac{\Res_{\zeta_n}M^+}{\zeta-\zeta_n}\bigg)
\\
= M^+ - I - (i/\zeta)\sigma_3Q_- 
  - \sum_{n=1}^{2N}\bigg(\frac{\Res_{\zeta_n}M^+}{\zeta-\zeta_n} + \frac{\Res_{\zeta_n^*}M^-}{\zeta-\zeta_n^*}\bigg)
  - M^+G\,, 
\label{e:RHP2}
\end{multline}
Now note that the LHS is analytic in $\Gamma^-$ and is $O(1/\zeta)$ as $\zeta\to\infty$ there,
while the sum of the first four terms of the RHS is analytic in $\Gamma^+$ and is $O(1/\zeta)$ as $\zeta\to\infty$ there.
We then introduce modified Cauchy projectors $\P_\pm$ by
\[
\label{e:Ppmdef}
\P_\pm[f](\zeta) = \frac1{2\pi i}\int_\Sigma\frac{f(\zeta')}{\zeta' - (\zeta\pm i0)}\,\d\zeta'\,,
\]
where $\int\nolimits_\Sigma$ denotes the integral along the oriented contour shown in Fig.~\ref{f:domains},
and the notation $z\pm i0$ indicates that when $z\in\Sigma$, the limit is taken from the left/right of it.
That is, $F(\zeta\pm i0)= \lim_{\epsilon\to0^+}F(\zeta\pm i\epsilon)$.
Recall Plemelj's formulae: if $f_\pm$ are analytic in the $\Complex^\pm$ and are $O(1/\zeta)$ as 
$\zeta$ tends to $\infty$ there, it is
\[
\P_+ f_+ = f_+\,,\qquad 
\P_-f_- = - f_-\,,\qquad 
\P_+f_-= \P_-f_+=0\,.
\nonumber 
\]
Applying $\P_+$ and $\P_-$ to the jump condition~\eref{e:RHP2} we then obtain the solution~\eref{e:rhpsoln} of the RHP.

\subsection{Trace formulae and ``theta'' condition}
\label{s:trace}

Recall that the components of the scattering matrix $s_{1,1}$ and $s_{2,2}$ are analytic respectively in $\Gamma^+$ and $\Gamma^-$ from Eq.~\eref{e:sanalytic}.
Also recall that the discrete spectrum is composed of quartets: 
$\zeta_n,\zeta_n^*,-q_o^2/\zeta_n,-q_o^2/\zeta_n^*$ $\forall n=1,\dots,N$.
Then define the quantities
\[
\beta^+(\zeta,z)= s_{1,1}(\zeta,z)\prod_{n=1}^N\frac{(\zeta-\zeta_n^*)(\zeta+q_o^2/\zeta_n)}{(\zeta-\zeta_n)(\zeta+q_o^2/\zeta_n^*)}\,,\quad 
\beta^-(\zeta)= s_{2,2}(\zeta,z)\prod_{n=1}^N\frac{(\zeta-\zeta_n)(\zeta+q_o^2/\zeta_n^*)}{(\zeta-\zeta_n^*)(\zeta+q_o^2/\zeta_n)}\,.
\label{e:betadef}
\]
The functions $\beta^\pm$ are analytic respectively in $\Gamma^\pm$, 
like $s_{1,1}$ and $s_{2,2}$.
But, unlike $s_{1,1}$ and $s_{2,2}$, they have no zeros.
Moreover, $\beta^\pm(\zeta,z)\to1$ as $\zeta\to\infty$ in the proper half planes.
Finally, note that
\[
\beta^+(\zeta,z)\beta^-(\zeta,z)= s_{1,1}(\zeta,z)s_{2,2}(\zeta,z)\,.
\label{e:beta+beta-}
\]
We can then take the logarithm of Eq.~\eref{e:beta+beta-} and apply the Cauchy projectors (see Section~\ref{A:RHPsoln}), obtaining 
\[
\nonumber
\log\beta^\pm(\zeta,z) = \mp\frac1{2\pi i}\int_\Sigma \frac{\log[1+|b(\zeta',z)|^2]}{\zeta-\zeta'}\d\zeta'\,,
\]
where we used that $\det S=1$, implying
$1/(s_{1,1}s_{2,2}) = 1 - b\~b = 1 + |b|^2$ $\forall \zeta\in\Sigma$ by the definition~\eref{e:reflcoeffdef} and symmetries~\eref{e:bsymm} of the reflection coefficients.
Now, using definition~\eref{e:betadef} to replace $\beta^\pm$ with $s_{1,1}$ and $s_{2,2}$, 
we finally obtain the so-called ``trace'' formulae~\eref{e:trace1} and~\eref{e:trace2},
which express the analytic scattering coefficients
in terms of the discrete eigenvalues and the reflection coefficient.
For reflectionless solutions $s_{1,2}(\zeta,z)=0$, $\forall \zeta\in\Sigma$,
so the integrals in Eqs.~\eref{e:trace1} and~\eref{e:trace2} are identically zero.

Now note that 
\[
\int_{q_o}^\infty \log[1+|b(\zeta',z)|^2]\, \d\zeta'/\zeta' = - \int_{-q_o}^0 \log[1+|b(\zeta',z)|^2]\, \d\zeta'/\zeta'\,,
\nonumber
\]
because $|b(\zeta',z)|= |b(-q_o^2/\zeta',z)|$ thanks to the symmetry~\eref{e:bsymm}.
A similar relation holds between the integral from $-\infty$ to $-q_o$ and that from 0 to $q_o$
and for those in the upper/lower semicircle of radius $C_o$.
However, due to the orientation of the continuous spectrum,
these individual integrals do not cancel with each other, but they add together instead.
Recalling the asymptotics~\eref{e:Sk20asymp}, 
we then obtain the so-called ``theta'' condition~\eref{e:theta}.

\subsection{Explicit evaluation of $R_{\pm,d}$}
\label{A:Rpm_explicit}

In this appendix, we evaluate the integral~\eref{e:R_d_axis} that defines $R_{\pm,d}$.
Recall that the auxiliary matrix $R_{\pm}$ has the same value when the Hilbert transform is computed on either $\lambda'$-sheet,
so we will compute this matrix on sheet I, i.e., we choose $\nu = 1$ and $\rho_{-}(\lambda,z) = h_-\,\sigma_3$ from Eqs.~\eref{e:rhopm=cdotsigma1},~\eref{e:rhopm=cdotsigma2},~\eref{e:rhopm=cdotsigma3} and~\eref{e:rhopm=cdotsigma4}.
We start by rewriting the integral~\eref{e:R_d_axis} for convenience.
Also recall from the main text, we know in inhomogeneous broadening the auxiliary quantity $w_- = 0$ in Eq.~\eref{e:R_d_axis}.

\paragraph{Firstly, we consider the case where $\lambda\in\Complex\backslash\Real$.}

Recall that the Hilbert transform is given by Eq.~\eref{e:Hilbert}.
There is no singularity on the integration contour,
and the principal value is not needed. 
To compute the integral, it is convenient to write things in terms of the uniformization variable $\zeta$.
Recall that $\lambda= (\zeta-q_o^2/\zeta)/2$ and $\gamma = (\zeta+q_o^2/\zeta)/2$ from Eq.~\eref{e:unifinverse}.  
Letting $s=\zeta'$ it is $d\lambda' = \half(1+q_o^2/s^2)\,\d s$
and, taking the principal branch of the square root, the integration contour is $L_o = (-\infty,-q_o)\cup(q_o,\infty)$.
Moreover,
\[
\frac1{\lambda'-\lambda} = \frac{2s}{\Delta_2(s)}\,,\qquad
\frac1{\lambda'^2+\epsilon^2}= \frac{4s^2}{\Delta_4(s)}\,,
\nonumber
\]
where
\[
\nonumber
\Delta_2(s) = s^2- (\zeta- q_o^2/\zeta)\,s - q_o^2\,,\qquad
\Delta_4(s) = (s^2-2i\epsilon s - q_o^2)(s^2+2i\epsilon s - q_o^2)\,. 
\]
Collecting all the pieces,
\[
R_{-,d}(\zeta,z) = \sigma_3\,\frac{2\epsilon h_-}{\pi\gamma\zeta} \int_{L_o}\frac{s[(\zeta s + q_o^2)^2 - (\zeta - s)^2q_o^2]}{\Delta_2(s)\Delta_4(s)}\,\d s\,.
\label{e:Rpm2}
\]
The two roots of $\Delta_2(s)$ are
\[
\nonumber
s= \lambda \pm \gamma = \half[\zeta - q_o^2/\zeta \pm(\zeta + q_o^2/\zeta)] = \{\zeta, - q_o^2/\zeta\}\,,
\]
while the four roots of $\Delta_4(s)$ are
\[
\nonumber
s = i\epsilon \pm \sqrt{q_o^2-\epsilon^2}\,,\qquad s = -i\epsilon \pm \sqrt{q_o^2 - \epsilon^2}\,.
\]
Let $s_1,\dots,s_6$ denote the zeros of $\Delta_2(s)$ and $\Delta_4(s)$, 
and let $f(s)$ denote the numerator of the integrand in Eq.~\eref{e:Rpm2}.
Expanding the integrand in partial fractions we have:
\[
\nonumber
\frac{f(s)}{\Delta_2(s)\Delta_4(s)} = \sum_{j=1}^6 \frac{f_j}{s-s_j}\,,\qquad
f_j = \Res\limits_{s=s_j}\bigg[\frac{f(s)}{\Delta_2(s)\Delta_4(s)}\bigg] 
= f(s_j)\bigg/\mathop{\prod\nolimits'\kern-0.2em}\limits_{m=1}^6\,(s-s_m)\,,
\]
where the prime indicates that the term $m=j$ is omitted from the product.
Hence
\[
R_{-,d}(\zeta,z) = \sigma_3\,\frac{2\epsilon h_-}{\pi\gamma\zeta} \sum_{j=1}^6 f_j \, [\log(s-s_j)]_{L_o}\,.
\nonumber
\]
After some tedious but straightforward algebra, we then get Eq.~\eref{e:Rd-explicit_NZBG}.

\paragraph{Secondly, we consider the case where $\lambda\in\Real$.}

In this case, there is a singularity in the integral and the principal value becomes necessary.
Then we use the following relationship to compute the integral in Eq.~\eref{e:R_d_axis}.
\[
\label{e:Rrelation}
\pvint_\Real \frac{\gamma}{\gamma'}\rho_{\pm,d}' g(\lambda')\frac{\d \lambda'}{\lambda'-\lambda}
= \int_L \frac{\gamma}{\gamma'}\rho_{\pm,d}' g(\lambda')\frac{\d \lambda'}{\lambda'-\lambda}
+ \pi i\, \Res\limits_{\lambda'=\lambda}\bigg[\frac{\gamma}{\gamma'}\rho_{\pm,d}' g(\lambda')\frac{1}{\lambda'-\lambda}\bigg]\,,
\]
where the contour $L$ is: $L = (-\infty,\lambda-r)\cup\{z=r\, \e^{i\theta}| \, \theta\in[0,\pi]\}\cup(\lambda+r,\infty)$,
provided $a$ is sufficiently small and there are no singularities on the contour.
We will also use $\lambda'$ on sheet I.
The integral on the RHS of Eq.~\eref{e:Rrelation} can be computed normally, because $\lambda$ is not on the contour. 
So this integral is exactly the same as the one we computed in the first case where $\lambda\in\Complex\backslash\Real$.
Thus,
\[
\nonumber
\int_L \frac{\gamma}{\gamma'}\rho_{\pm,d}' g(\lambda')\frac{\d \lambda'}{\lambda'-\lambda}
=  \sigma_3\,\frac{\epsilon h_-}{\pi(\epsilon^2+\lambda^2)}
\big[\, \Theta(\lambda) - \gamma\Theta(i\epsilon)\big/(q_o^2-\epsilon^2)^{1/2} \,\big]\,,
\]
The second term on the RHS of Eq.~\eref{e:Rrelation} can be computed easily,
\[
\nonumber
\pi i\, \Res_{\lambda'=\lambda}\bigg[\frac{\gamma}{\gamma'}\rho_{\pm,d}' g(\lambda')\bigg] = \pi h_- i g(\lambda) \sigma_3\,.
\]
Therefore, by using the relationship~\eref{e:Rrelation}, 
we obtain Eq.~\eref{e:R-dreal} for $\lambda\in\Real$.

\subsection{IST with branch cut outside}
\label{s:2ndIST}

In this appendix, we will formulate an alternative version of IST to solve the same problem~\eref{e:drhodt} and~\eref{e:dqdz}.
The idea of the new version is to use a different branch cut in the complex $\lambda$-plane.
As shown later, this version of IST is equivalent to the one presented in the main text.

Recall that the eigenvalue $\gamma(\lambda)$ was defined in the main text with
$\gamma(\lambda)=\sign(\lambda)\sqrt{q_o^2+\lambda^2}$ for $\lambda\in\Real$.
Alternatively,
one could define it as $\gamma(\lambda)=\sqrt{q_o^2+\lambda^2}$ for $\lambda\in\Real$ (i.e., without the signum function), 
and then formulate the IST for MBEs in a similar way.

\paragraph{Background states.}

As before, we first investigate whether there exist exact 
``constant'' solutions of the MBEs~\eref{e:drhodt} and~\eref{e:dqdz} with NZBC. 
So suppose for now that $q(t,z) = q_o(z)$ $\forall t\in\Real$, with $q_o\ne0$.
Equation~\eref{e:rhobackground} is still the solution of Eq.~\eref{e:drhodt}, i.e., 
$\rho(t,z,\lambda) = \e^{X_ot}\,C\,\e^{-X_ot}$.
The eigenvalues of $X_o$ are still $\pm i\gamma$, where $\gamma^2 = \lambda^2 + q_o^2$.
Consistently with the above discussion, we now take 
\vspace*{-0.4ex}
\[
\nonumber
\gamma(\lambda) = (\lambda^2 + q_o^2)^{1/2}\,,\qquad
\lambda \in \Real\,. 
\]
With this choice, $\gamma$ is always continuous along the continuous spectrum.
The branching structure of $\gamma(\lambda)$ will be discussed later.
We can write an eigenvector matrix of $X_o$ as Eq.~\eref{e:Ybackground}, i.e.,
$Y_o = I + (i/\zeta) \sigma_3Q_0$, 
where we use the short notation~\eref{e:uniformization} again.
Thus, we again have Eq.~\eref{e:rhobackgroundexplicit}, i.e., 
$\rho(t,z,\lambda) = Y_o\,\e^{i\gamma t\sigma_3}\rho_o\,\e^{-i\gamma t\sigma_3}Y_o^{-1}$, 
where $\rho_o = Y_o^{-1}C Y_o$.
We again express all $\lambda$ dependence in terms of $\zeta$,
noting that the inverse transformation to Eq.~\eref{e:uniformization} is the same as Eq.~\eref{e:unifinverse}.

Using similar arguments as in Section~\ref{s:background}, 
we can again write  
$\rho(t,z,\zeta) = \@h\cdot\brho$, 
where $\brho = (\rho_1,\rho_2,\rho_3)^T$, 
$\@h\cdot\@h = 1$ and $\rho_j(t,z,\zeta)$ is still given by Eq.~\eref{e:rhojbackground}, i.e., 
$\rho_j(t,z,\zeta) = Y_o\,\e^{i\gamma t \sigma_3}\sigma_j\,\e^{-i\gamma t\sigma_3} Y_o^{-1}$,
with $\rho_j^\dag= \rho_j$, $\tr\rho_j=0$, and $\det\rho_j = -1$.
Explicitly, we still have formulas~\eref{e:rhojbackgroundexplicit1},~\eref{e:rhojbackgroundexplicit2} and~\eref{e:rhojbackgroundexplicit3}.

Now we insert this behavior into Eq.~\eref{e:dqdz} as before.
Direct calculations yield Eqs.~\eref{e:[J,rho]1},~\eref{e:[J,rho]2} and~\eref{e:[J,rho]3} again.
By a similar discussion as in the main text, 
we know that only the commutator in Eq.~\eref{e:[J,rho]3} is $t$-independent.
Moreover, we have $\int[\sigma_3,\rho_3]g(\lambda)\d\lambda = -iw_o[\sigma_3,Q_0]$, 
where $w_o = \int g(\lambda)/\gamma\,\d\lambda$.
Then, it is easy to compute the corresponding integral 
\[
\nonumber
w_0 = 2/\big[\pi(q_o^2-\epsilon^2)^{1/2}\big]\arccos(\epsilon/q_o)\,,
\]
with the choice of the detuning function $g(\lambda)$ as in Eq.~\eref{e:Lorentzian}.
Notice that due to different definitions of $\gamma$, this integral differs from the one in the main text.
It follows that the only self-consistent solutions are
\[
\nonumber
q(t,z) = q_o \exp\big\{2i \,h_3\, z/[\pi(q_o^2-\epsilon^2)^{1/2}]\arccos(\epsilon/q_o)\big\}\,,\qquad
\rho(t,z,\zeta) = h_3(\lambda \sigma_3-i Q_0)/\gamma\,.
\]
These expressions differ from those in Eq.~\eref{e:backgroundstates} for all $z>0$.

\paragraph{Formulation of the IST.}

As before, we define the quantity
\[
\label{e:defgamma2}
\gamma(\lambda)=(q_o^2+\lambda^2)^{1/2}\,.
\]
Now however we take the branch cut on $i(-\infty,-q_o]\cup i[q_o,\infty)$.
This is done by taking $\lambda+i q_o=r_1\e^{i\theta_1}$ and $\lambda-i q_o = r_2\e^{i\theta_2}$, 
so that
\vspace*{-1ex}
\[
\nonumber
\gamma(\lambda) = \sqrt{r_1 r_2}\e^{i\Theta}\,,\qquad
\Theta = \frac{\theta_1+\theta_2}{2} +m\pi\,,
\]
where $m=0,1$ indicates the first or the second sheet, respectively, 
and $-\pi/2\le\theta_1< 3\pi/2$, $-3\pi/2<\theta_2\le \pi/2$.
It is easily verified that with this choice the discontinuity of $\gamma$ occurs on the segment $i(-\infty,-q_o]\cup [q_o,\infty)$.
As before, the Riemann surface is obtained by gluing the two sheets of the complex $\lambda$-plane along the cut by defining the uniformization variable $\zeta=\lambda+\gamma$,
which maps the first sheet into the right half plane and the second one into the left half plane.
The continuous spectrum is the same as before, i.e., $\lambda\in\Real\cup i[-q_o,q_o]$.
On the $\zeta$-plane, the continuous spectrum is also the same, $\zeta\in\Real\cup C_o$.
The left half of the first sheet and the right half of the second sheet are mapped into the interior of $C_o$.
The branch cut becomes the imaginary axis on the $\zeta$-plane.
Moreover, the inverse transformations are the same, i.e., Eq. \eref{e:unifinverse}.

Differently from before, however, 
for $\lambda\in\Real$ we have $\gamma=\sqrt{q_o^2+\lambda^2}$, which has no discontinuity at $\lambda=0$.
If one now takes the limit $q_o\to0$, it is obvious that $\gamma=\pm|\lambda|$ for $\lambda\in\Real$ (where the $\pm$ values apply on sheet I or II respectively).
Thus, 
\textit{this formulation of the IST does not reduce to the one with ZBCs in the limit $q_o\to0$ directly}.
In fact, we will show below that this formulation of the IST yields different solutions of the MBEs.

Even though the new $\gamma$ is different from the one in the main text,
for $\lambda\in\Complex$, $\gamma(\lambda)$ is still given by definition~\eref{e:defgamma}. 
This means that the calculations in the two formulations of the IST are very similar.
In fact, the direct and inverse problem are identical in both versions.
This is because 
all of the IST was formulated using the uniformization variable $\zeta = \lambda + \gamma(\lambda)$, and 
the values of $\gamma(\lambda)$ for $\zeta\in\Complex$ are exactly the same in the two formulations.
Essentially, the definition of $\gamma(\lambda)$ amounts to just switching the left half planes between sheet I and sheet II.

In other words,
we can use the same definition of the regions $\Gamma^\pm$ of the complex $\zeta$-plane from definition~\eref{e:Cpmdef}. 
It is then easy to show that the analyticity properties of the columns of the eigenfunctions $\mu_{\pm}$ in the complex $\zeta$-plane are still the same, 
namely Eq.~\eref{e:analyticregion}.
The analyticity properties of the scattering data 
and the symmetries for eigenfunctions and scattering data
are also the same as before.

The only part of the IST that is affected by the change is the propagation,
because it involves integrals over $\lambda\in\Real$.
It is easy to repeat the calculation in Section~\ref{s:boundaryvalues} and \ref{s:time} 
and obtain the integrals for the boundary data $w_\pm$ and $R_{\pm,d}$.
The resulting integrals are formally the same as Eqs.~\eref{e:wpmdef} and~\eref{e:R_d_axis}.
In other words, as long as the explicit expression of $\gamma$ is not used,
the formulas are the same.

\paragraph{Equivalence between two versions of IST.}

In this part, we will prove that the two versions of IST are equivalent,
in the sense that one could obtain the same solutions simultaneously by using different boundary conditions in two versions of IST.

To begin with, it is convenient to introduce a new notation in this proof.
The superscript $\I$ or $\II$ denotes the variable or function in the first or second version of IST, respectively.
For example, $\gamma^\I$ denotes $\gamma$ in the first IST (i.e., the IST presented in the main text) 
and $\rho_-^{\II}$ denotes the boundary condition $\rho_-$ in the second version 
(i.e., the IST discussed in this appendix).

Firstly, one should notice that the solutions of MBEs are unique.
More precisely, the solutions are uniquely determined by the input pulse $Q(t,0)$,
together with the initial state $\rho_-(\zeta,z)$ which is defined by Eq.~\eref{e:rhopmasymp}.
Thus, to prove the equivalence between the two versions of IST, 
it is sufficient to find both sets of data in the two versions of IST, respectively,
that produce the same solution.

Secondly, one should also notice that, 
both quantities $Q(t,0)$ and $\rho_-(\zeta,z)$ are defined with $\lambda\in\Real$. 
So we can focus on real values of $\lambda$ (or $\gamma$, $\zeta$) instead of complex values.
Moreover, because the fundamental difference between the two versions of IST comes from the definition of $\gamma$ with negative values of $\lambda$
(or equivalently, how to take the branch cut), 
it is sufficient to only consider the case where $\lambda<0$.

Now, we are ready to begin our proof.
For any given solution $Q(t,z)$ and $\rho(t,z,\lambda)$,
it is obvious that the corresponding inputs in the two versions of IST must be the same.
[Recall that the input pulse is $Q(t,0)$.]
It is then sufficient to consider the initial data $\rho_-^\I(\zeta^\I,z)$ and $\rho_-^{\II}(\zeta^{\II},z)$.
Let $q_- = \lim_{t\to-\infty}q(t,z)$ as before, and let the density matrix be
\vspace{-1ex}
\[
\nonumber
\rho(t,z,\lambda) = h_1(t,z,\lambda)\sigma_1 + h_2(t,z,\lambda)\sigma_2 + h_3(t,z,\lambda)\sigma_3\,,
\]
where $\sigma_j$ are Pauli matrices from Eq.~\eref{e:Paulidef} and $h_j\in\Real$ for $j= 1,2,3$.
By using relation~\eref{e:rhopmasymp}, 
one can compute the corresponding initial states $\rho_-^\I$ and $\rho_-^\II$.
Moreover, since both $\rho_-^\I$ and $\rho_-^\II$ are Hermitian matrices, 
it is sufficient to compute the $(1,1)$ and $(1,2)$ components.
Consequently, the elements are (where $\gamma_o = \sqrt{q_o^2+\lambda^2}$ for $\lambda<0$)
\begin{gather}
\label{e:2setbc1}
D_-^\I(\zeta^\I,z) = -\big[2 h_3 \lambda -i h_1 (q_- -q_-^*) + h_2 (q_- +q_-^*)\big]/(2 \gamma_o)\,,\\
\label{e:2setbc2}
D_-^{\II}(\zeta^{\II},z) = \big[2 h_3 \lambda -i h_1 (q_- - q_-^*) + h_2 (q_- +q_-^*)\big]/(2 \gamma_o)\,,\\
\label{e:2setbc3}
P_-^\I(\zeta^\I,z) =
-\frac{i q_-}{\gamma_o } \e^{2 i \gamma_o  t} h_3
+\frac{\e^{2 i \gamma_o  t}}{2 \gamma_o (\gamma_o +\lambda )}
\bigg[ 
\bigg(\frac{q_-^2}{q_o^2} 2\lambda(\gamma_o +\lambda )+q_-^2\bigg)( h_1 + i h_2)+q_o^2( h_1 - i h_2)
\bigg]\,,\\
\label{e:2setbc4}
P_-^{\II}(\zeta^{\II},z) = \frac{i q_-}{\gamma_o } \e^{-2 i \gamma_o  t} h_3
+\frac{\e^{-2 i \gamma_o  t}}{2 \gamma_o  (\gamma_o +\lambda )} 
[
(2 \lambda  (\gamma_o +\lambda ) + q_o^2)( h_1 - i h_2)+q_-^2( h_1 + i h_2)
]\,.
\end{gather}
Thus, the two sets of initial states for the medium are found.

By comparing the above expressions~\eref{e:2setbc1},~\eref{e:2setbc2},~\eref{e:2setbc3} and~\eref{e:2setbc4},
we obtain the following relationship between the two sets of data that produce the same solution with $\lambda<0$
\[
\label{e:bcrelation1}
\rho_{-,d}^\I = -\rho_{-,d}^{\II}\,,\qquad
\rho_{-o}^\I = \frac{q_-}{q_-^*}(\rho_{-o}^{\II})^*\,,
\]
or the reversion
\[
\label{e:bcrelation2}
\rho_{-,d}^{\II} = -\rho_{-,d}^{\I}\,,\qquad
\rho_{-o}^{\II} = \frac{q_-}{q_-^*}(\rho_{-o}^{\I})^*\,,
\]
where subscript $d$ or $o$ denotes the diagonal or off-diagonal part of the matrix, respectively, as before.
Recall that for positive values of $\lambda$, the two sets of data are exactly the same.
Therefore, if one solution is produced by either one of the versions of IST, 
the relationships~\eref{e:bcrelation1} and~\eref{e:bcrelation2} ensure that the same solution can also be produced by using the other version.
Thus the two versions of IST have the same solutions set and the equivalence is proved.

\makeatletter
\def\@biblabel#1{#1.}
\makeatother


\end{document}